\documentclass[a4paper,11pt]{article}
\pdfoutput=1 

\usepackage{jheppub} 
\usepackage[T1]{fontenc} 
\usepackage[english]{babel}
\usepackage{graphicx} 
\usepackage{slashed}
\usepackage{caption}
\usepackage{amsmath}
\makeatletter
  \def\my@tag@font{\normalsize}
  \def\maketag@@@#1{\hbox{\m@th\normalfont\my@tag@font#1}}
  \let\amsmath@eqref\eqref
  \renewcommand\eqref[1]{{\let\my@tag@font\relax\amsmath@eqref{#1}}}
\makeatother
\usepackage{amsthm}
\usepackage{amsfonts}
\usepackage{mathtools}
\usepackage{subcaption}
\usepackage{tikz} 
\usepackage{listings}
\usepackage{tikz-3dplot}
\usepackage{placeins}
\usepackage{scalerel}
\usepackage{verbatim}
\usepackage{listingsutf8}
\usepackage{framed}
\usepackage{minibox}
\usepackage{float}
\usepackage{wrapfig}
\usepackage{longtable}
\usepackage{titlesec}
\usepackage[shortlabels]{enumitem}
\usepackage{multirow}

\usepackage{braket}
\usepackage{cases}
\usepackage{physics}
\usepackage{dsfont}
\usepackage{caption}
\usepackage{array}
\usepackage{hhline}
\usepackage{verbatim}
\usepackage{empheq}
\usepackage{mathrsfs}
\usepackage{xcolor}
\usepackage{layout}
\usepackage[mathscr]{euscript}

\usepackage{tcolorbox}
\usepackage{amstext} 
\usepackage{array}
\newcolumntype{L}{>{$}l<{$}}
\newcolumntype{C}{>{$}c<{$}}

\DeclarePairedDelimiter\floor{\lfloor}{\rfloor}

\newtheorem{theorem}{Theorem}[section]

\newcommand{\sdet}{\text{sdet}}

\theoremstyle{plain}
\theoremstyle{plain}

\theoremstyle{definition}

\makeatletter
\def\@fpheader{\relax}
\makeatother

\title{Superfield twist-$2$ operators in $\mathcal{N}=1$ SCFTs and their renormalization-group improved generating functional in $\mathcal{N}=1$ SYM theory}

\author[a,b]{Giacomo Santoni}
\author[a,b]{Francesco Scardino}

\affiliation[a]{Physics Department, INFN Roma1, \\
	Piazzale A. Moro 2, Roma, I-00185, Italy}
\affiliation[b]{Physics Department, Sapienza University,\\Piazzale A. Moro 2, Roma, I-00185, Italy}

\emailAdd{giacomo.santoni@uniroma1.it}
\emailAdd{francesco.scardino@uniroma1.it}

\abstract{ \sloppy We provide a new construction of superfield collinear twist-$2$ operators as infinite-dimensional, irreducible representations of the collinear superconformal algebra in $\mathcal{N}=1$ superconformal field theories. As an application, we realize the above representations in terms of free superfields, in a manifestly gauge-invariant and supersymmetric-covariant fashion, in the zero coupling limit of $\mathcal{N}=1$ supersymmetric Yang-Mills (SYM) theory. This realization makes manifest their mixing and renormalization properties at one loop. We also extend to the superfield formalism the perturbative and nonperturbative techniques in \cite{Bochicchio:2021nup, Bochicchio:2022uat, BPS41, BPS42, Bochicchio:2016toi, Bochicchio:2024gtn, BPSL} to a large class of supersymmetric theories that are superconformal in the zero-coupling limit. Specifically, we compute the generating functional of superfield twist-$2$ operators in $\mathcal{N}=1$ SU($N$) SYM theory in the zero coupling limit. We also work out in a closed form the corresponding asymptotic renormalization-group improved generating functional in Euclidean superspace and its planar and leading nonplanar large-$N$ expansion. We verify -- as originally predicted in \cite{Bochicchio:2016toi} and verified in the component formalism \cite{BPS41, BPS42, Bochicchio:2024gtn, BPSL} -- that the leading nonplanar asymptotic RG-improved generating functional matches the structure of logarithm of a functional superdeterminant of the corresponding nonperturbative object, which it should be asymptotic to at short distances because of the asymptotic freedom. Hence, our large-$N$ computation sets strong ultraviolet asymptotic constraints on the nonperturbative solution of large-$N$ $\mathcal{N} = 1$ SYM theory that may be a pivotal guide for the search of such a solution.}

\begin{document} 
	
	\definecolor{c969696}{RGB}{150,150,150}
	\maketitle 
	\flushbottom

\section{Introduction}

Twist-$2$ operators are fundamental for the study of deep inelastic scattering in QCD (\cite{Braun:2003rp} and references therein) because they dominate the operator product expansions on the light-cone \cite{Braun:2003rp}. Besides, in the zero-coupling limit they transform under irreducible representations of the conformal group\footnote{These representations actually extend to the order $g^2$ of perturbation theory in the so-called conformal renormalization scheme \cite{Braun:2003rp}.} and, being conserved, are the Noether currents of higher-spin symmetries \cite{Todorov:2012xx, Mikhailov:2002bp}. The above conformal properties also make the computation of their one-loop anomalous dimensions especially simple \cite{Belitsky:2007jp, Balitsky:1987bk, Belitsky:1998gc}. \par 
More recently, the short-distance asymptotics of the generating functional of Euclidean correlators of single-trace twist-$2$ operators has played a central role in constraining the yet-to-come nonperturbative solution of large-$N$ SU($N$) YM theory \cite{Bochicchio:2022uat, Bochicchio:2016toi, Bochicchio:2024gtn, BPSL} and $\mathcal{N}=1$ SYM theory \cite{BPS41,BPS42}. Remarkably, the above generating functionals have the structure of the logarithm of a functional (super-)determinant. Moreover, the above structure has a nonperturbative interpretation in terms of the glueball/gluinoball one-loop effective action at large $N$ as originally predicted in \cite{Bochicchio:2016toi} and verified in the component formalism \cite{BPS41,BPS42, Bochicchio:2024gtn, BPSL}. In addition, the aforementioned structure nicely intertwines with the topology of leading nonplanar diagrams in the large-$N$ expansion of the SU($N$) theory as opposed to the U($N$) one \cite{Bochicchio:2024gtn, BPSL}.  \par
With the goal of extending the results of \cite{Bochicchio:2021nup, Bochicchio:2022uat, BPS41, BPS42, Bochicchio:2016toi, Bochicchio:2024gtn, BPSL} to supersymmetric theories, possibly including matter fields, it is of great interest to study and generalize the construction of twist-$2$ operators in $\mathcal{N}=1$ SYM theory \cite{BPS41,BPS42} in a formalism that makes supersymmetry manifest. This endeavour leads, in the present paper, to several technical developments. \par
First, following \cite{Bochicchio:2022uat, BPS41, BPS42, Bochicchio:2016toi, Bochicchio:2024gtn, BPSL} we work out the general structure of the generating functional of the large-$N$ connected correlators of glueball/gluinoball superfields in $\mathcal{N}=1$ SYM theory. We show in the superfield formalism that its leading nonplanar part can be expressed as the logarithm of a functional superdeterminant, according to the previous result in the component formalism \cite{BPS41, BPS42}. \par
Second, we construct the $\mathcal{N}= 1$ supersymmetric generalization of twist-$2$ operators. Though this problem is not new in the literature \cite{Belitsky:1998gu, Derkachov:2000ne, Belitsky:2003sh, Belitsky:2004sc, Belitsky:2005gr, Belitsky:2006cp}, one of the original contributions of the present work is the construction of the above operators in terms of superfields, in a manifestly gauge-invariant and supersymmetric-covariant fashion. Indeed, contrary to the previous approaches, our construction employs a covariant superfield instead of a (possibly non-local) light-cone superfield \cite{Belitsky:2004sc}, whose construction relies on the light-cone gauge $A_+=0$. Our approach has the advantage to yield an extremely compact expression of the superconformal multiplets, whose elements are embedded inside a unique superfield, and can be extracted by differentiating with respect to the superspace coordinates. Furthermore, our approach makes the renormalization properties \cite{Belitsky:1998gu} of the twist-$2$ operators manifest. \par 
In fact, the construction of composite superfield composite collinear twist-$2$ operators is deeply tied to the direct-sum decomposition of the tensor product of two irreducible representations of the collinear superconformal algebra, isomorphic to the superalgebra $\mathfrak{sl}(2|1)$. We perform this task in full generality, with minimal assumptions. Indeed, one of our new results is the computation of the Clebsch-Gordan coefficients for the tensor product of two possibly non-chiral representations. To the best of our knowledge, Clebsch-Gordan coefficients are presently known only for either finite-dimensional representations \cite{GOTZ2007829, Marcu:1979sg, Frappat:1996pb, 10.1063/1.523149} of $\mathfrak{sl}(2|1)$ that have no use in this context, or for chiral representations \cite{Belitsky:1998gu, Derkachov:2000ne, Belitsky:2003sh, Belitsky:2004sc, Belitsky:2005gr, Belitsky:2006cp}. The direct-sum decomposition of general non-chiral representations is not a mere mathematical curiosity, since it also makes possible to construct higher-twist operators from an arbitrary number of collinear superconformal primaries by iterating the procedure to construct the superfield twist-$2$ operators.\par
Third, we construct free field realizations of the above representations of superfield twist-$2$ operators that are bilinear in the fundamental superfields.\par 
Fourth, we work out the generating functionals of the corresponding connected superconformal correlators.\par
Fifth, we explicitly compute the above generating functional in the zero coupling limit of $\mathcal{N}=1$ SYM theory in a manifestly supersymmetric form.\par
Sixth, we re-derive the renormalization properties of the supersymmetric twist-$2$ operators first found in \cite{Belitsky:1998gu} by employing our new superfield formalism that makes them immediately apparent. \par
Seventh, we work out the short-distance asymptotics of the RG-improved generating functionals in Euclidean superspace in $\mathcal{N}= 1$ SU($N$) SYM theory, in a renormalization scheme first introduced in Ref. \cite{Bochicchio:2021xyy}, and referred to as \emph{non-resonant diagonal} \cite{Bochicchio:2022uat,Bochicchio:2021nup,BPS41,BPS42, Bochicchio:2021xyy} where the mixing of the renormalized superfield twist-$2$ operators is reduced to the multiplicatively renormalizable case.
\par Finally, we verify that the leading nonplanar asymptotic RG-improved generating functional matches the structure of the logarithm of a functional superdeterminant of the corresponding nonperturbative object -- as originally predicted in \cite{Bochicchio:2016toi} and verified in the component formalism \cite{BPS41,BPS42, Bochicchio:2024gtn, BPSL} -- arising from the glueball/gluinoball one-loop effective action, which it should be asymptotic to at short distances because of the asymptotic freedom. Hence, by creating a bridge between perturbative and nonpertubative physics, our results strongly constrain the yet-to-come nonperturbative solution of large-$N$ $\mathcal{N}= 1$ SYM theory and may be an essential guide for the search of such solution.

\section{Plan of the paper}
In section \ref{sec:nonperturbative}, we recall the structure of the nonperturbative generating functional of the correlators in large-$N$ $\mathcal{N} = 1$ SYM arising from the glueball/gluinoball one-loop effective action, and adapt it to the superfield formalism. \par
In section \ref{sec:sl21}, we construct the representations of the collinear superconformal algebra with the highest-weight technique, and find the direct-sum decomposition of the product of two such representations, including the computation of their Clebsch-Gordan coefficients. \par
In section \ref{sec:genfreescft}, we calculate the abstract generating functionals of bilinear operators made of free (super)fields of both bosonic and fermionic statistics.\par
In section \ref{sec:appliedcorrelators}, we concretely compute the above generating functionals in terms of superfields in $\mathcal{N}=1$ superconformal field theories arising as the zero-coupling limit of supersymmetric gauge theories.\par
In section \ref{sec:sym}, we apply our results to $\mathcal{N}=1$ SYM theory by deriving a manifestly supersymmetric form of the conformal generating functional of superfield twist-$2$ operators. We remark that our superconformal generating functional is the logarithm of a functional (super-)determinant, according to the previous computations in pure YM theory \cite{Bochicchio:2021nup} and in $\mathcal{N}=1$ SYM theory in the component formalism \cite{BPS41, BPS42}. Moreover, in subsection \ref{sec:check}, we verify for a certain spin tower of twist-$2$ operators that our result for the supersymmetric generating functional coincides with its component version in Refs. \cite{BPS41, BPS42}. \par
In section \ref{sec:rgimprove}, we derive the renormalization properties of twist-$2$ operators in $\mathcal{N}=1$ SYM theory. Besides, we explicitly compute the short-distance asymptotics of the RG-improved generating functional of superfield twist-$2$ operators in $\mathcal{N}=1$ SYM theory that inherits the structure of the logarithm of a functional superdeterminant as well.\par
In section \ref{sec:interpretation}, we verify that indeed the above asymptotic RG-improved generating functional matches the structure of the logarithm of a functional superdeterminant of the corresponding nonperturbative object arising from the glueball/gluinoball one-loop effective action recalled in section \ref{sec:nonperturbative}. \par
In section \ref{sec:conclusions}, we draw the conclusions of our work. \par
In appendix \ref{app:conventions}, we fix the notations and conventions that we follow throughout the paper.\par
In appendix \ref{app:euclidean}, we fix the notations and conventions regarding the analytic continuation to Euclidean superspace.\par 
In appendix \ref{app:supconfcorr}, we compute the $2$-point correlators implied by the superconformal symmetry in the coordinate and momentum representation.\par
In appendix \ref{app:supermatrix}, we work out some useful identities about superdeterminants.\par
In appendix \ref{app:sl2}, we employ our techniques of section \ref{sec:sl21} to re-derive in our language the results in the non-supersymmetric theory \cite{Braun:2003rp}.\par
In appendix \ref{app:cjj}, we provide a proof of the identities involving the superconformal polynomials in section \ref{sec:sl21}.
\par In appendix \ref{app:sqcd}, we use the techniques of this work to construct the superfield twist-$2$ operators built by chiral matter superfields in $\mathcal{N}=1$ SQCD.

\section{Nonperturbative effective action of $\mathcal{N}=1$ SYM theory}
\label{sec:nonperturbative}
In this section we follow the treatment in Refs. \cite{BPS41, BPS42}, first recalling the general structure of the large-$N$ effective action in components \cite{BPS41, BPS42} in subsection \ref{sec:npcomponent} and then extending it to the superfield formalism in subsection \ref{sec:npsuperfield}.

\subsection{Effective action in components}
\label{sec:npcomponent}
It has been known for almost fifty years that $\mathcal{N}=1$ SU($N$) SYM theory admits 't Hooft large-$N$ topological expansion \cite{tHooft:1973alw} for the $n$-point connected correlators of gauge-invariant single-trace operators. The corresponding Feynman diagrams in 't Hooft double-line representation -- after a suitable gluing of reversely oriented lines -- are topologically classified \cite{tHooft:1973alw,Veneziano:1976wm} by the sum on the genus $g$ of $n$-punctured closed Riemann surfaces, where each topology is weighted by a factor $N^{\chi}$, with $\chi=2-2g-n$ the Euler characteristic of the Riemann surface. \par
Consequently, the corresponding nonperturbative large-$N$ effective theory involves an infinite number of weakly interacting glueballs and gluinoballs with coupling of order $\frac{1}{N}$ \cite{tHooft:1973alw,Migdal:1977nu,Witten:1979kh} and masses proportional to the RG-invariant scale $\Lambda_{SYM}$. \par
Let us denote as $O_s(x^E)$ and $M_s(x^E)$ two sets of gauge-invariant single-trace composite operators in $\mathcal{N}=1$ SU($N$) SYM theory. The bosonic operators $O_s$ create glueball states, while the fermionic operators $M_s$ create gluinoball states. The generating functional of the correlators of $O_s$ and $M_s$ reads
\begin{align}
	\mathcal{Z}^E[J_{O},J_{M}]=\frac{1}{\mathcal{Z}^E} \int [dA][d\chi]\, e^{-  S_{SYM}+ \sum_s\int J_{O_s}O_s+  J_{M_s}M_s}
\end{align}
where in the path-integral we denoted the integration measure over the gauge field as $[dA]$ and the integration measure over the gluino fields as $[d\chi]$. Here and in the rest of this subsection, the symbol $\int$ denotes the integration measure over Euclidean spacetime $\int d^4x^E$. The connected generating functional $\mathcal{W}^E[J_{O},J_{M}]=\text{log}\ \mathcal{Z}^E[J_{O},J_{M}]$ admits the large-$N$ expansion
\begin{align}
	\label{eq:thooft0}
	\mathcal{W}^E[J_{O},J_{M}]=\mathcal{W}^E_{\text{sphere}}[J_{O},J_{M}]+\mathcal{W}^E_{\text{torus}}[J_{O},J_{M}]+ \cdots
\end{align}
Nonperturbatively, $\mathcal{W}^E_{\text{sphere}}[J_{O},J_{M}]$, which perturbatively is the ('t Hooft-)planar contribution \cite{tHooft:1973alw}, is a sum of tree diagrams involving glueball/gluinoball propagators and vertices, while $\mathcal{W}^E_{\text{torus}}[J_{O},J_{M}]$, which perturbatively is the leading-non('t Hooft-)planar contribution, is a sum of glueball/gluinoball one-loop diagrams.
\par Nonperturbatively, $\mathcal{W}^E_{\text{torus}}[J_O,J_M]$ should have the structure of the logarithm of a functional (super)determinant, as it has been originally predicted in the pure YM case \cite{Bochicchio:2016toi} on the basis of fundamental principles, and subsequently asymptotically verified \cite{Bochicchio:2022uat,BPSL,Bochicchio:2024gtn}.
\par  Indeed, in analogy with the pure YM case \cite{Bochicchio:2016toi,BPSL,Bochicchio:2024gtn}, in the yet-to-come nonperturbative solution of large-$N$ $\mathcal{N}=1$ SYM theory, the very same correlators should be computed by the correlators of bosonic and fermionic glueball/gluinoball fields $\Phi$ and $\Psi$ with an infinite number of components, the corresponding generating functional being schematically \cite{Bochicchio:2016toi,BPSL,Bochicchio:2024gtn}
\begin{align}
	\mathcal{Z}^E_{\text{glueball/gluinoball}}[J_{\Phi},J_{\Psi}]= \frac{1}{\mathcal{Z}^{E}_{\text{glueball/gluinoball}}} \int[d\Phi] [d\Psi]\ e^{-S_{\text{glueball/gluinoball}}(\Phi,\Psi)+\int \Phi \ast_1 J_{\Phi}+\Psi \ast'_1 J_{\Psi}}
\end{align}
with \cite{Bochicchio:2016toi,BPSL,Bochicchio:2024gtn}
\begin{align}
\label{eq:effspace}
	S_{\text{glueball/gluinoball}}(\Phi,\Psi) =\frac{1}{2}\int \Bigg[& \Phi\ast_2(-\Delta+M^2)\Phi + \Psi\ast'_2(-\Delta+M^2)\Psi \nonumber \\
	&+\frac{1}{3N}\Phi \ast_3\Phi\ast_3\Phi+\frac{1}{N}\Psi \ast'_3\Phi\ast'_3\Psi+\cdots \Bigg]
\end{align}
where $\ast_2$, $\ast_1$ and $\ast'_2$, $\ast'_1$ are fixed below, the ellipses and $\ast_3$, $\ast'_3$ respectively stand for $n$-glueball/gluinoball vertices with $n>3$ and some presently unknown operation on the glueball/gluinoball fields that, by assuming locality, Euclidean invariance may involve derivatives. Hence, nonperturbatively the connected generating functional $\mathcal{W}^E_{\text{glueball/gluinoball}}[J_{\Phi},J_{\Psi}] = \log\mathcal{Z}^E_{\text{glueball/gluinoball}}[J_{\Phi},J_{\Psi}]$ reads to one loop of glueballs/gluinoballs \cite{Bochicchio:2016toi,BPSL,Bochicchio:2024gtn}
	\begin{equation}\label{eq:glueballW0}
						\resizebox{0.98\textwidth}{!}{%
			$\begin{aligned}
		\mathcal{W}^E_{\text{glueball/gluinoball}}[J_{\Phi},J_{\Psi}] =& -S_{\text{glueball/gluinoball}}(\Phi_J,\Psi_J)+\int \Phi_J \ast_1 J_{\Phi}+\int \Psi_J \ast'_1 J_{\Psi} + \cdots \\
		&+\frac{1}{2}\log\text{sdet}
		\begin{pmatrix}\ast'_2(-\Delta+M^2)+\frac{1}{N}\ast'_3\Phi_J\ast'_3& \frac{1}{N}\ast'_3\ast'_3\Psi_J\\ 
			\frac{1}{N}\ast'_3\ast'_3\Psi_J&\ast_2(-\Delta+M^2)+\frac{1}{N}\ast_3\Phi_J\ast_3 \end{pmatrix}
						\end{aligned}$
	} 
	\end{equation}
where $\Phi_J$, $\Psi_J$ are determined by
\begin{align}
	&\frac{\delta S_{\text{glueball/gluinoball}}}{\delta\Phi}\Big\rvert_{\Phi_J}=\ast_1 J_{\Phi}, \ && \frac{\delta S_{\text{glueball/gluinoball}}}{\delta\Psi}\Big\rvert_{\Psi_J}=\ast'_1 J_{\Psi}
\end{align}
The superdeterminant in Eq. \eqref{eq:glueballW0} can be computed in terms of a supertrace by means of the identity $\text{log sdet}(X)=\text{str log}(X)$. The supertrace of a supermatrix is defined as \cite{ZinnJustin2021QuantumFT}
\begin{align}
\label{eq:strabcd}
\text{str}\begin{pmatrix} 
        A & B \\
        C & D
    \end{pmatrix}=\mathrm{tr}(A)-\mathrm{tr}(D)
\end{align}
In our case, the submatrices 
\begin{align}
    &A_{s_1,s_2}(x^E_1; x^E_2),  && D_{s_1,s_2}(x^E_1; x^E_2)
\end{align}
appearing in the supertrace are computed by expanding the logarithm in Eq. \eqref{eq:glueballW0}. Since they depend both on the discrete indices and the Euclidean spacetime coordinates, their trace is defined as
\begin{align}
    &\mathrm{tr}(A)=\sum_s \int d^4x^E\ A_{s,s}(x^E; x^E)\ , && \mathrm{tr}(D)=\sum_s \int d^4x^E\ D_{s,s}(x^E; x^E)
\end{align}
The dictionary between $\mathcal{W}^E[J_{O},J_{M}]$ and $\mathcal{W}^E_{\text{glueball/gluinoball}}[J_{\Phi},J_{\Psi}]$ is obtained by matching the corresponding
spectral representations -- as a sum of free propagators with residues $R_{sm}$, $R'_{sm}$ -- for the $2$-point correlators at $N =\infty$ \cite{Bochicchio:2024gtn, BPSL} of $O_s$, $M_s$, respectively, that, by fixing $\ast_2$, $\ast'_2$ according to the canonical normalization of the glueball/gluinoball kinetic term, uniquely determines the coupling of $J_{\Phi}$, $J_{\Psi}$ to the tower of glueball/gluinoball fields $\Phi \ast_1 J_{\Phi} = \sum_{sm} \Phi_{sm} \sqrt{R_{sm}} J_{\Phi_s}$, $\Psi \ast'_1 J_{\Psi} = \sum_{sm} \Psi_{sm} \sqrt{R'_{sm}} J_{\Psi_s}$, respectively.

\subsection{Effective action in superfields}
\label{sec:npsuperfield}
Since in the present paper we write down $\mathcal{N}=1$ SU($N$) SYM theory in the superspace formalism, it is convenient to rewrite schematically the large-$N$ effective action of subsection \ref{sec:npcomponent} by means of superfields as well.
\par Let us denote as $\tilde{O}_s(x^E,\theta^E,\bar\theta^E)$ and $\tilde{M}_s(x^E,\theta^E,\bar\theta^E)$ two sets of bosonic and fermionic gauge-invariant single-trace composite \emph{superfields} in Euclidean $\mathcal{N}=1$ SU($N$) SYM theory (see Appendix \ref{app:euclidean} for definitions and conventions on Euclidean superspace). We assume that $\tilde{O}_s$ and $\tilde{M}_s$ are non-chiral superfields\footnote{See Appendix \ref{app:repchir} for the definition of chirality in supersymmetric theories. We choose $\tilde{O}_s$ and $\tilde{M}_s$ to be non-chiral because the twist-$2$ operators that we construct in section \ref{sec:sym} are non-chiral.} whose components are operators connected by supersymmetry transformations that interpolate glueballs and gluinoballs, depending on their statistics. The generating functional of the correlators of $\tilde{O}_s$ and $\tilde{M}_s$ reads
\begin{align}
	\mathcal{Z}^E[J_{\tilde{O}},J_{\tilde{M}}]=\frac{1}{\mathcal{Z}^E} \int [dV]\, e^{-  S_{SYM}+ \sum_s\int J_{\tilde{O}_s}\tilde{O}_s+  J_{\tilde{M}_s}\tilde{M}_s}
\end{align}
where we denoted the integration measure over the vector superfield (see section \ref{sec:sym}) as $[dV]$. Here and in the rest of this subsection, the symbol $\int$ denotes the combined spacetime and Berezin integration $\int d^4x^E d^2\theta^E d^2\bar\theta^E$ in the whole Euclidean superspace. The connected generating functional $\mathcal{W}^E[J_{\tilde{O}},J_{\tilde{M}}]=\text{log}\ \mathcal{Z}^E[J_{\tilde{O}},J_{\tilde{M}}]$ in the superfield formalism admits a large-$N$ expansion as well
\begin{align}
	\label{eq:thooft}
	\mathcal{W}^E[J_{\tilde{O}},J_{\tilde{M}}]=\mathcal{W}^E_{\text{sphere}}[J_{\tilde{O}},J_{\tilde{M}}]+\mathcal{W}^E_{\text{torus}}[J_{\tilde{O}},J_{\tilde{M}}]+ \cdots
\end{align}
In analogy with subsection \ref{sec:npcomponent}, we get for the corresponding nonperturbative effective action in the superfield formalism in terms of bosonic and fermionic glueball/gluinoball non-chiral superfields $\widetilde{\Phi}$ and $\widetilde{\Psi}$ with an infinite number of components. Schematically,
\begin{align}
	\mathcal{Z}^E_{\text{glueball/gluinoball}}[J_{\widetilde{\Phi}},J_{\widetilde{\Psi}}]= \frac{1}{\mathcal{Z}^{E}_{\text{glueball/gluinoball}}} \int[d\widetilde{\Phi}] [d\widetilde{\Psi}]\ e^{-S_{\text{glueball/gluinoball}}(\widetilde{\Phi},\widetilde{\Psi})+\int \widetilde{\Phi} \ast_1 J_{\widetilde{\Phi}}+\widetilde{\Psi} \ast'_1 J_{\widetilde{\Psi}}}
\end{align}
with \cite{BPS41, BPS42, Bochicchio:2016toi,BPSL,Bochicchio:2024gtn}
\begin{align}
	S_{\text{glueball/gluinoball}}(\widetilde{\Phi},\widetilde{\Psi}) =\frac{1}{2}\int \Bigg[& \widetilde{\Phi}\ast_2(-\Delta+M^2)\widetilde{\Phi} + \widetilde{\Psi}\ast'_2(-\Delta+M^2)\widetilde{\Psi} \nonumber \\
	&+\frac{1}{3N}\widetilde{\Phi} \ast_3\widetilde{\Phi}\ast_3\widetilde{\Phi}+\frac{1}{N}\widetilde{\Psi} \ast'_3\widetilde{\Phi}\ast'_3\widetilde{\Psi}+\cdots \Bigg]
\end{align}
where the corresponding symbols are defined analogously to subsection \ref{sec:npcomponent}. Hence, nonperturbatively the connected generating functional $\mathcal{W}^E_{\text{glueball/gluinoball}}[J_{\widetilde{\Phi}},J_{\widetilde{\Psi}}] = \log\mathcal{Z}^E_{\text{glueball/gluinoball}}[J_{\widetilde{\Phi}},J_{\widetilde{\Psi}}]$ reads to one loop of glueballs/gluinoballs \cite{BPS41, BPS42, Bochicchio:2016toi,BPSL,Bochicchio:2024gtn}
	\begin{equation}\label{eq:glueballW}
						\resizebox{0.98\textwidth}{!}{%
			$\begin{aligned}
		\mathcal{W}^E_{\text{glueball/gluinoball}}[J_{\widetilde{\Phi}},J_{\widetilde{\Psi}}] =& -S_{\text{glueball/gluinoball}}(\widetilde{\Phi}_J,\widetilde{\Psi}_J)+\int \widetilde{\Phi}_J \ast_1 J_{\widetilde{\Phi}}+\int \widetilde{\Psi}_J \ast'_1 J_{\widetilde{\Psi}} + \cdots \\
		&+\frac{1}{2}\log\text{sdet}
		\begin{pmatrix}\ast'_2(-\Delta+M^2)+\frac{1}{N}\ast'_3\widetilde{\Phi}_J\ast'_3& \frac{1}{N}\ast'_3\ast'_3\widetilde{\Psi}_J\\ 
			\frac{1}{N}\ast'_3\ast'_3\widetilde{\Psi}_J&\ast_2(-\Delta+M^2)+\frac{1}{N}\ast_3\widetilde{\Phi}_J\ast_3 \end{pmatrix}
						\end{aligned}$
	} 
	\end{equation}
where $\widetilde{\Phi}_J$, $\widetilde{\Psi}_J$ are determined by
\begin{align}
	&\frac{\delta S_{\text{glueball/gluinoball}}}{\delta\widetilde{\Phi}}\Big\rvert_{\widetilde{\Phi}_J}=\ast_1 J_{\widetilde{\Phi}} \ , && \frac{\delta S_{\text{glueball/gluinoball}}}{\delta\widetilde{\Psi}}\Big\rvert_{\widetilde{\Psi}_J}=\ast'_1 J_{\widetilde{\Psi}}
\end{align}
A subtlety arises in the evaluation of the superdeterminant in Eq. \eqref{eq:glueballW}, since the submatrices in Eq. \eqref{eq:glueballW} are now actually valued in superfields. The corresponding entries of Eq. \eqref{eq:strabcd} read
\begin{align}
    &A_{s_1,s_2}(x^E_1,\theta^E_1,\bar{\theta}^E_1; x^E_2,\theta^E_2,\bar{\theta}^E_2),  && D_{s_1,s_2}(x^E_1,\theta^E_1,\bar{\theta}^E_1; x^E_2,\theta^E_2,\bar{\theta}^E_2)
\end{align}
with traces defined by
\begin{align}
    &\mathrm{tr}(A)=\sum_s \int d^4x^Ed^2\theta^Ed^2\bar{\theta}^E\ A_{s,s}(x^E,\theta^E,\bar{\theta}^E; x^E,\theta^E,\bar{\theta}^E) \nonumber \\
    &\mathrm{tr}(D)=\sum_s \int d^4x^Ed^2\theta^Ed^2\bar{\theta}^E\ D_{s,s}(x^E,\theta^E,\bar{\theta}^E; x^E,\theta^E,\bar{\theta}^E)
\end{align}
The dictionary between $\mathcal{W}^E[J_{\tilde{O}},J_{\tilde{M}}]$ and $\mathcal{W}^E_{\text{glueball/gluinoball}}[J_{\widetilde{\Phi}},J_{\widetilde{\Psi}}]$ is analogous to the one in the component formalism.

\section{The $\mathfrak{sl}(2|1)$ superalgebra}
\label{sec:sl21}

\subsection{Introduction}
\label{sec:sl21intro}

Suppose we have a $\mathcal{N}=1$ superconformal field theory in a superspace with coordinates $x^{\mu}, \theta^{\alpha},\bar{\theta}^{\dot{\alpha}}$ (see appendix \ref{app:spinors} for notations and conventions on spinors). We define the \emph{light-cone} to be the surface (see appendix \ref{app:notation} for notations and conventions on the light-cone)
\begin{equation}
	\label{eq:lc}
	(x^{+}, x^{-}, \theta^1, \bar{\theta}^{\dot{1}}) \quad \text{with all other coordinates being $0$}
\end{equation}
This surface is closed under the action of the \emph{collinear superconformal algebra}, which is defined as the superconformal algebra projected onto the light-cone directions. Its generators are related to those of the full superconformal algebra (see appendix \ref{app:repchir} for notation and conventions) by \cite{Belitsky:2004sc}
\begin{equation}
	\label{eq:collsupalg}
	\begin{gathered}
		\mathbf{V}_a=\begin{pmatrix}
			\mathbf{V}_+ \\
			\mathbf{V}_-
		\end{pmatrix}=
		\begin{pmatrix}
			\frac{i\varrho}{2}\mathbf{Q}_1 \\
			\frac{1}{2\varrho}\bar{\mathbf{S}}_{\dot{2}}
		\end{pmatrix}\ , \qquad 
		\mathbf{W}_a=\begin{pmatrix}
			\mathbf{W}_+ \\
			\mathbf{W}_-
		\end{pmatrix}=
		\begin{pmatrix}
			-\frac{\varrho}{2}\bar{\mathbf{Q}}_{\dot{1}} \\
			-\frac{1}{2\varrho}\mathbf{S}_2
		\end{pmatrix} \\
		\mathbf{L}_+=\mathbf{L}_1+i\mathbf{L}_2=-i\mathbf{P}_+\ , \qquad  \mathbf{L}_-=\mathbf{L}_1-i\mathbf{L}_2=\frac{i}{2}\mathbf{K}_-\ , \qquad \mathbf{L}=\mathbf{L}_3=\frac{i}{2}(\mathbf{D}+\mathbf{M}_{-+}) \\
		\mathbf{B}=-\frac{3}{4}\mathbf{R}+\frac{1}{2}\mathbf{M}_{12}\ , \qquad \mathbf{E}=\frac{i}{2}(\mathbf{D}-\mathbf{M}_{-+})
	\end{gathered}
\end{equation}
with $\varrho=2^{1/4} $. The commutation rules of this subalgebra are in Eq. \eqref{eq:supercomm}, with the \emph{collinear twist} $\mathbf{E}$ and $\textbf{P}_-$ commuting with all the other generators. This algebra, is isomorphic to the $\mathbb{Z}_2$-graded algebra $\mathfrak{sl}(2|1)$ \cite{Frappat:1996pb}. Any superfield $\Phi(x,\theta,\bar{\theta})$ transforming irreducibly under the algebra \eqref{eq:collsupalg} is characterized by the numbers
\begin{equation}
	\label{eq:twist}
	\begin{gathered}
		\left[\mathbf{L},\Phi(0)\right]=j\Phi(0)\ , \qquad j=\frac{D+s}{2} \\
		\left[\mathbf{E},\Phi(0)\right]=\frac{\tau}{2}\Phi(0)\ , \qquad \tau=D-s \\
		\left[\mathbf{B},\Phi(0)\right]=b\Phi(0)\ , \qquad  b= \frac{3}{4}r+\frac{h}{2}
	\end{gathered}    
\end{equation}
where $s$ is the spin projection along the light-cone directions, $h$ is the \emph{helicity} and $D$ is the canonical dimension. $j$ is called collinear conformal spin and $\tau$ is called \emph{collinear twist} and $b$ is called \emph{$b$-charge}. If a superfield satisfies the further conditions
\begin{equation}
    \left[\textbf{L}_-,\Phi(0)\right]=\left[\textbf{V}_-,\Phi(0)\right\}=\left[\textbf{W}_-,\Phi(0)\right\}=0
\end{equation}
it is called \emph{collinear superconformal primary}.
\par As it will be shown in the next subsections, given a collinear superconformal primary $\Phi(0)$, the operators
\begin{equation}
\label{eq:vectors}
    \begin{gathered}
        \left[\mathbf{P}_+,...\left[\mathbf{P}_+,\Phi(0)\right]...\right] \\
         \left[\mathbf{P}_+,...\left[\mathbf{P}_+,\left[\mathbf{Q}_1,\Phi(0)\right\}\right]...\right] \\
         \left[\mathbf{P}_+,...\left[\mathbf{P}_+,\left[\bar{\mathbf{Q}}_{\dot{1}},\Phi(0)\right\}\right]...\right]  \\
         \left[\mathbf{P}_+,...\left[\mathbf{P}_+,\left[\mathbf{Q}_1,\left[\bar{\mathbf{Q}}_{\dot{1}},\Phi(0)\right\}\right\}\right]...\right]
    \end{gathered}
\end{equation}
form a $\mathbb{Z}_2$-graded vector space that is closed under the adjoint action of the algebra \eqref{eq:collsupalg}. In other words, these objects furnish a representation of the collinear superconformal algebra. The superfield translated along the light-cone directions
\begin{equation}
\label{eq:suptransl}
    \Phi(x^+,x^-,\theta^1,\bar{\theta}^{\dot{1}})=e^{+i\left(x^+\mathbf{P}_++x^-\mathbf{P}_-+\theta^{1}\mathbf{Q}_{1}+\bar{\mathbf{Q}}_{\dot{1}}\bar{\theta}^{\dot{1}}\right)}\Phi(0)e^{-i\left(x^+\mathbf{P}_++x^-\mathbf{P}_-+\theta^{1}\mathbf{Q}_{1}+\bar{\mathbf{Q}}_{\dot{1}}\bar{\theta}^{\dot{1}}\right)}
\end{equation}
can be seen as a generating function for the operators \eqref{eq:vectors}. On this generating function, the generators \eqref{eq:collsupalg} act by differentiation on $x^+,\theta^1,\bar{\theta}^{\dot{1}}$ (see appendix \ref{app:repchir} for more details).
\par The goal of this section is to find the direct sum decomposition and the corresponding Clebsch-Gordan coefficients of a tensor product of two representations of \eqref{eq:collsupalg} i.e. a rule to construct, from two collinear superconformal primaries $\Phi_1(0)$, $\Phi_2(0)$, a new collinear superconformal primary that is bilinear in its constituent superfields.
\par (Super)conformal field theories enjoy the operator-state correspondence \cite{DiFrancesco:1997nk}, according to which any vector $\Psi$ in the Hilbert space of states can be obtained by acting on a (super)conformally-invariant vacuum $\Psi_{\text{vac}}$ with some local operator evaluated at the origin. In this section we will study the realization of the collinear superconformal algebra on the Hilbert space of states of the theory, keeping in mind that the operator-state correspondence ensures that any representation-theoretical result we will obtain applies also to the local operators.
The superconformal invariance of the vacuum allows us to write
\begin{equation}
\label{eq:sl21dict}
    \Psi=\Phi(0)\Psi_{\text{vac}}\ , \qquad \textbf{P}_+\Psi=\left[\textbf{P}_+, \Phi(0)\right]\Psi_{\text{vac}}
\end{equation}
and so on. The Clebsch-Gordan coefficients for the $\mathfrak{sl}(2|1)$ representations will be found through the highest weight technique \cite{Procesi2006LieGA}.
The main new results of this section are the Clebsch-Gordan coefficients for \emph{general} representations of the $\mathfrak{sl}(2|1)$ algebra and a new, concise form of the Clebsch-Gordan coefficients for the composition of a class of representations called \emph{chiral} representations, that will be defined below. These results are not only general, but also easy to generalize even further, since they also allow us to construct collinear superconformal primary operators also from the product of three or more primaries.

\subsection{Generators and (anti)commutators}
\label{sec:gencomm}
The  Lie superalgebra $\mathfrak{sl}(2|1)$ consists of four even generators $\mathbf{L}_{i=1,2,3}, \mathbf{B}$ and four odd generators $\mathbf{V}_{a=1,2}, \mathbf{W}_{a=1,2}$. The commutation rules that define the algebra are \cite{Derkachov:2000ne, 10.1063/1.523980}
\begin{equation}
\label{eq:supercomm}
    \begin{gathered}
        [\mathbf{L}_i,\mathbf{L}_j]=i\varepsilon_{ijk}\mathbf{L}_k\ , \qquad [\mathbf{B},\mathbf{L}_i]=0 \\
        [\mathbf{L}_i,\mathbf{V}_a]=\frac{1}{2}(\sigma_i)_{ba}\mathbf{V}_b\ , \qquad [\mathbf{L}_i,\mathbf{W}_a]=\frac{1}{2}(\sigma_i)_{ba}\mathbf{W}_b \\
        [\mathbf{B},\mathbf{V}_a]=+\frac{1}{2}\mathbf{V}_a\ , \qquad [\mathbf{B},\mathbf{W}_a]=-\frac{1}{2}\mathbf{W}_a \\
        \{\mathbf{V}_a,\mathbf{V}_b\}=\{\mathbf{W}_a,\mathbf{W}_b\}=0\ , \qquad \{\mathbf{V}_a,\mathbf{W}_b\}=(i\sigma_2\sigma_i)_{ab}\mathbf{L}_i+(i\sigma_2)_{ab}\mathbf{B}
    \end{gathered}
\end{equation}
where the $\sigma_{i=1,2,3}$ are the Pauli matrices. For our purposes, it is useful to change the basis by introducing $\mathbf{L}_{\pm}=\mathbf{L}_1\pm i\mathbf{L}_2$, $\mathbf{L}=\mathbf{L}_3$, $\mathbf{V}_\pm=\mathbf{V}_{1,2}$, $\mathbf{W}_\pm=\mathbf{W}_{1,2}$. In this basis, the last anticommutator in \eqref{eq:supercomm} takes the form
\begin{equation}
    \{\mathbf{V}_a,\mathbf{W}_b\}=\begin{pmatrix}
        +\mathbf{L}_+ & -\mathbf{L}+\mathbf{B} \\
        -\mathbf{L}-\mathbf{B} & -\mathbf{L}_-
    \end{pmatrix}
\end{equation}
The generators $\mathbf{L}_i, \mathbf{B}$ form a $\mathfrak{sl}(2)\oplus \mathfrak{u}(1)$ subalgebra, and the generators $\mathbf{L}_+,\mathbf{V}_+, \mathbf{W}_+$ form a one-dimensional super-Poincaré algebra. The quadratic Casimir element is
\begin{equation}
	\label{casimir}
    \mathbb{C}^2=\mathbb{L}^2-\mathbf{B}^2+\mathbf{V}_+\mathbf{W}_-+\mathbf{W}_+\mathbf{V}_-
\end{equation}
where $\mathbb{L}^2=\mathbf{L}_+\mathbf{L}_-+\mathbf{L}^2$ is the quadratic Casimir of the $\mathfrak{sl}(2)\oplus \mathfrak{u}(1)$ subalgebra.

\subsection{Representations}
\label{sec:sl21rep}

\paragraph{Abstract construction}
Each representation of $\mathfrak{sl}(2|1)$ is uniquely identified by two real numbers $j$ and $b$. The basis vectors of the representation $[j,b]$ are denoted as
\begin{equation}
    \Psi_{j,b; \mathscr{J}, \mathscr{L}, \mathscr{B}}
\end{equation}
The representation $[j,b]$ has a highest weight vector $\Psi_{j,b; j,j,b}$ satisfying
\begin{align}
\label{eq:hw}
    \mathbf{L}_-\Psi_{j,b; j,j,b}=\mathbf{V}_-\Psi_{j,b; j,j,b}=\mathbf{W}_-\Psi_{j,b; j,j,b}=0 \nonumber \\
    \qquad \mathbf{L}\Psi_{j,b; j,j,b}=j\Psi_{j,b; j,j,b}\ , \qquad \mathbf{B}\Psi_{j,b; j,j,b}=b\Psi_{j,b; j,j,b}
\end{align}
In a generic vector $\Psi_{j,b; \mathscr{J}, \mathscr{L}, \mathscr{B}}\in [j,b]$, the numbers $\mathscr{J}$, $\mathscr{L}$ and $\mathscr{B}$ denote, respectively, the eigenvalues of
\begin{align}
    & \mathbb{L}^2\Psi_{j,b;\mathscr{J}, \mathscr{L}, \mathscr{B}}=\mathscr{J}(\mathscr{J}-1)\Psi_{j,b;\mathscr{J}, \mathscr{L}, \mathscr{B}} \nonumber \\
    &\mathbf{L}\Psi_{j,b;\mathscr{J}, \mathscr{L}, \mathscr{B}}=\mathscr{L}\Psi_{j,b;\mathscr{J}, \mathscr{L}, \mathscr{B}} \nonumber \\
    &\mathbf{B}\Psi_{j,b;\mathscr{J}, \mathscr{L}, \mathscr{B}}=\mathscr{B}\Psi_{j,b;\mathscr{J}, \mathscr{L}, \mathscr{B}}
\end{align}
The rest of the representation $[j,b]$ can be constructed from the highest weight by using the generators $\mathbf{L}_{\pm}, \mathbf{V}_{\pm}, \mathbf{W}_{\pm}$ as ladder operators for $\mathbf{L}$ and $\mathbf{B}$. Since $\mathfrak{sl}(2|1)$ is a superalgebra the vector space $[j,b]$ is $\mathbb{Z}_2$-graded. The action of the supersymmetri generators $\mathbf{V}_+$ and $\mathbf{W}_+$ on the highest weight creates four vectors that are annihilated by $\textbf{L}_-$
\begin{align}
\label{eq:sl2prim}
    & \Psi_{j,b;j,j,b} \nonumber\\    
    &\Psi_{j,b;j+\frac{1}{2},j+\frac{1}{2},b+\frac{1}{2}}=\mathbf{V}_+\Psi_{j,b;j,j,b} \nonumber\\
    &\Psi_{j,b;j+\frac{1}{2},j+\frac{1}{2},b-\frac{1}{2}}= \mathbf{W}_+\Psi_{j,b;j,j,b}\nonumber\\
    &\Psi_{j,b;j+1,j+1,b}=\left(\frac{b+j}{2j}\mathbf{W}_+\mathbf{V}_++\frac{b-j}{2j}\mathbf{V}_+\mathbf{W}_+\right)\Psi_{j,b;j,j,b}    
\end{align}
We call these vectors \emph{supersymmetric descendants} of $\Psi_{j,b;j,j,b}$ or, equivalently, $\mathfrak{sl}(2)\oplus \mathfrak{u}(1)$\emph{-highest weight vectors}, since they are the highest weights of the $\mathfrak{sl}(2)\oplus \mathfrak{u}(1)$-modules inside $[j,b]$. All the other vectors of $[j,b]$ are constructed by repeatedly applying $\mathbf{L}_+$
\begin{align}
\label{eq:sl21desc}
    & \Psi_{j,b;j,j+n,b}=\mathbf{L}_+^n\Psi_{j,b;j,j,b} \nonumber\\    
    &\Psi_{j,b;j+\frac{1}{2},j+\frac{1}{2}+n,b+\frac{1}{2}}=\mathbf{L}_+^n\Psi_{j,b;j+\frac{1}{2},j+\frac{1}{2},b+\frac{1}{2}} \nonumber\\
    &\Psi_{j,b;j+\frac{1}{2},j+\frac{1}{2}+n,b-\frac{1}{2}}= \mathbf{L}_+^n\Psi_{j,b;j+\frac{1}{2},j+\frac{1}{2},b-\frac{1}{2}} \nonumber\\
    &\Psi_{j,b;j+1,j+1+n,b}=\mathbf{L}_+^n\Psi_{j,b;j+1,j+1,b}  
\end{align}
We call these vectors \emph{conformal descendants} of $\mathfrak{sl}(2)\oplus \mathfrak{u}(1)$\emph{-descendants} of the vectors in Eq. \eqref{eq:sl2prim}. The quadratic Casimir element defined in Eq. \eqref{casimir} takes the value
\begin{equation}
    \mathbb{C}^2\Psi_{j,b;j,j,b}=(j^2-b^2)\Psi_{j,b;j,j,b}
\end{equation}
in the representation $[j,b]$.\par
Chiral representations (see appendix \ref{app:repchir} for more details on this notion) are defined by one of the following conditions on the highest weight
\begin{equation}
\label{eq:chirality}
        \mathbf{W}_+\Psi_{j,b;j,j,b}^{(L)}=0  \ , \qquad \mathbf{V}_+\Psi_{j,b;j,j,b}^{(R)}=0
\end{equation}
These two conditions define \emph{left-handed} and \emph{right-handed} chiral representations respectively. The anticommutators $\{\mathbf{V}_\pm,\mathbf{W}_\mp\}=-\mathbf{L}\pm \mathbf{B}$ imply that the conditions \eqref{eq:chirality} can be satisfied consistently only if $j\pm b=0$. Therefore, chiral representations are labelled by $[j,\mp j]$ and the space consists only of the vectors
\begin{align}
    & \Psi_{j,-j;j,j+n,-j}^{(L)}=\mathbf{L}_+^n\Psi_{j,-j;j,j,-j}^{(L)} && \Psi_{j,j;j,j+n,j}^{(R)}=\mathbf{L}_+^n\Psi_{j,b;j,j,+j}^{(R)} \nonumber \\
    & \Psi_{j,-j;j+\frac{1}{2},j+\frac{1}{2}+n,-j+\frac{1}{2}}^{(L)}=\mathbf{L}_+^n\mathbf{V}_+ \Psi_{j,-j;j,j,-j}^{(L)} && \Psi_{j,j;j+\frac{1}{2},j+\frac{1}{2}+n,j-\frac{1}{2}}^{(R)}=\mathbf{L}_+^n\mathbf{W}_+\Psi_{j,j;j,j,+j}^{(R)}
\end{align}
The quadratic Casimir element vanishes on chiral representations.

\paragraph{Representation by differential operators}
We now construct a representation on the space of functions in superspace. The generating function of the descendants for the representation $[j,b]$ is now defined as
\begin{align}
\label{eq:superfunctional}
    &\mathcal{F}_{j,b}(s,\eta,\bar\eta)= e^{-s\mathbf{L}_++\eta\mathbf{V}_++\bar\eta\mathbf{W}_+}\Psi_{j,b;j,j,b} \nonumber\\
    &=  e^{-s\mathbf{L}_+}\left[1+\eta\mathbf{V}_++\bar{\eta}\mathbf{W}_++\eta\bar{\eta}\left(\frac{b+j}{2j}\mathbf{W}_+\mathbf{V}_++\frac{b-j}{2j}\mathbf{V}_+\mathbf{W}_+\right)-\frac{b}{2j}\eta\bar{\eta}\mathbf{L}_+\right]\Psi_{j,b;j,j,b}
\end{align}
where in the second lines we have expanded with respect to the Grassmann variables $\eta$ and $\bar{\eta}$.
The resulting action of the generators as differential operators on the super-coordinates $(s,\eta,\bar\eta)$ is
\begin{align}
\label{eq:differential}
    & \mathbf{L}_+\mathcal{F}_{j,b}=L_-\mathcal{F}_{j,b} \qquad  && L_-=-\partial_s \nonumber\\
    & \mathbf{V}_+\mathcal{F}_{j,b}=W_-\mathcal{F}_{j,b} \qquad  && W_-=\partial_{\eta}+ \frac{1}{2}\bar\eta\partial_s \nonumber\\
    & \mathbf{W}_+\mathcal{F}_{j,b} =V_-\mathcal{F}_{j,b} \qquad && V_-=\partial_{\bar\eta}+ \frac{1}{2}\eta\partial_s \nonumber\\
    & \mathbf{L}_-\mathcal{F}_{j,b} =L_+\mathcal{F}_{j,b} \qquad && L_+= s^2+s(\eta\partial_{\eta}+\bar\eta\partial_{\bar\eta})+2js+b\eta\bar\eta \nonumber\\
    & \mathbf{V}_-\mathcal{F}_{j,b} =W_+\mathcal{F}_{j,b} \qquad && W_+=sW_-+\frac{1}{2}\bar\eta\eta\partial_{\eta}+(j+b)\bar\eta \nonumber\\
    & \mathbf{W}_-\mathcal{F}_{j,b}=V_+\mathcal{F}_{j,b} \qquad && V_+=sV_-+\frac{1}{2}\eta\bar\eta\partial_{\bar\eta}+(j-b)\eta \nonumber\\
    & \mathbf{L}\mathcal{F}_{j,b} =L \mathcal{F}_{j,b} \qquad && L=s\partial_s+\frac{1}{2}(\eta\partial_{\eta}+\bar\eta\partial_{\bar\eta})+j \nonumber\\
    &\mathbf{B} \mathcal{F}_{j,b}=B \mathcal{F}_{j,b} \qquad && B=\frac{1}{2}\eta\partial_{\eta}-\frac{1}{2}\bar\eta\partial_{\bar\eta}+b \nonumber\\
\end{align}
The correspondence $\mathbf{L}_{\pm}\leftrightarrow L_\mp$, $\mathbf{V}_{\pm}\leftrightarrow W_\mp$, $\mathbf{W}_{\pm}\leftrightarrow V_\mp$ is needed to leave unchanged the commutation rules between the generators of the differential representation \footnote{This redefinition is also employed for the $\mathfrak{sl}(2)$ algebra in Ref. \cite{Braun:2003rp}.}. Integrating these infinitesimal transformations, one finds the finite transformation laws
\begin{align}
\label{eq:supfinite}
    &e^{\lambda L_-}\mathcal{F}_{j,b}(s,\eta,\bar\eta)=\mathcal{F}_{j,b}(s-\lambda,\eta,\bar\eta) \nonumber\\
    &e^{\epsilon W_-}\mathcal{F}_{j,b}(s,\eta,\bar\eta)=  \mathcal{F}_{j,b}\left(s+\frac{\epsilon\bar\eta}{2},\eta+\epsilon,\bar{\eta}\right) \nonumber\\
    &e^{\epsilon V_-}\mathcal{F}_{j,b}(s,\eta,\bar\eta)=  \mathcal{F}_{j,b}\left(s+\frac{\epsilon\eta}{2},\eta,\bar{\eta}+\epsilon\right) \nonumber\\
    &e^{\lambda L_+}\mathcal{F}_{j,b}(s,\eta,\bar\eta)=\frac{1}{\left[1-\lambda\left(s+\frac{b}{2j}\eta\bar\eta\right)\right]^{2j}}\mathcal{F}_{j,b}\left(\frac{s}{1-\lambda s},\frac{\eta}{1-\lambda s},\frac{\bar{\eta}}{1-\lambda s}\right) \nonumber\\
    &e^{\epsilon W_+}\mathcal{F}_{j,b}(s,\eta,\bar\eta)=(1+\epsilon\bar\eta)^{j+b}\mathcal{F}\left(\frac{s}{1-\frac{\epsilon\bar\eta}{2}},\frac{\eta+\epsilon s}{1-\frac{\epsilon\bar\eta}{2}},\bar\eta\right)  \nonumber\\
    &e^{\epsilon V_+}\mathcal{F}_{j,b}(s,\eta,\bar\eta)=  (1+\epsilon\eta)^{j-b}\mathcal{F}\left(\frac{s}{1-\frac{\epsilon\eta}{2}},\eta, \frac{\bar\eta+\epsilon s}{1-\frac{\epsilon\eta}{2}}\right) \nonumber\\
    &e^{\lambda L}\mathcal{F}_{j,b}(s,\eta,\bar\eta) = \lambda^j\mathcal{F}_{j,b}\left(\lambda s, \lambda^{1/2}\eta,\lambda^{1/2}\bar\eta\right) \nonumber\\
    &e^{\lambda B}\mathcal{F}_{j,b}(s,\eta,\bar\eta) = \lambda^b\mathcal{F}_{j,b}\left(s, \lambda^{1/2}\eta,\lambda^{-1/2}\bar\eta\right)
\end{align}
which are easily obtained by combining eqs. \eqref{eq:superfunctional} and \eqref{eq:differential},
with $\lambda$ and $\epsilon$ being  bosonic and fermionic parameters respectively of the finite transformations.
This representation encodes the \emph{right}-action of the algebra on the group elements. The \emph{left}-action of the generators $\mathbf{V}_+$, $\mathbf{W}_+$ is encoded in the chiral covariant derivatives $\mathcal{D}$, $\bar{\mathcal{D}}$ defined as
\begin{equation}
    e^{-s\mathbf{L}_++\eta\mathbf{V}_++\bar\eta\mathbf{W}_+} e^{\zeta\mathbf{V}_++\bar\zeta\mathbf{W}_+}= e^{\zeta \mathcal{D}+\bar\zeta\bar{\mathcal{D}}} e^{-s\mathbf{L}_++\eta\mathbf{V}_++\bar\eta\mathbf{W}_+}
\end{equation}
where $ \zeta$ and $\bar{\zeta}$ are odd variables. The Baker-Campbell-Hausdorff formula yields, as a result
\begin{align}
\label{eq:susyder}
    \mathcal{D}=W_-+\bar\eta L_- = \partial_{\eta}-\frac{1}{2}\bar\eta\partial_s \nonumber\nonumber\\
    \bar{\mathcal{D}}=V_-+\eta L_-=\partial_{\bar\eta}-\frac{1}{2}\eta\partial_s
\end{align}
with the anticommutator
\begin{equation}
    \left\{\mathcal{D}, \bar{\mathcal{D}}\right\}=-\partial_s
\end{equation}
We introduce the quantities $s_{L,R}=s\mp\frac{1}{2}\eta\bar\eta$ with the property
\begin{equation}
    \bar{\mathcal{D}} s_L=\bar{\mathcal{D}}\eta=\mathcal{D}s_R=\mathcal{D}\bar\eta=0
\end{equation}
Imposing the conditions \eqref{eq:chirality} on \eqref{eq:superfunctional}, one obtains generating functions of the form
\begin{align}
\label{eq:components}
        \mathcal{F}_{j,-j}(s_L,\eta)=\mathcal{F}^{(0)}_{j,-j}(s_L)+\eta\mathcal{F}^{(1)}_{j,-j}(s_L)\nonumber\\
        \bar{\mathcal{F}}_{j,+j}(s_R,\bar\eta)=\bar{\mathcal{F}}^{(0)}_{j,+j}(s_R)+\bar\eta\bar{\mathcal{F}}^{(1)}_{j,+j}(s_R)
\end{align}
where\footnote{In our notation, the subscript indices of the generating functions for the descendants denote the collinear conformal spin and the $b$-charge \emph{of the superconformal primary}, and not of the descendant.}
\begin{align}
\label{eq:components2}
        \mathcal{F}^{(0)}_{j,-j}(s)\equiv e^{-s\mathbf{L}_+}\Psi_{j,-j;j,j,-j}\ , \qquad \mathcal{F}^{(1)}_{j,-j}(s)=e^{-s\mathbf{L}_+}\mathbf{V}_+\Psi_{j,-j;j,j,-j} \nonumber\\
        \bar{\mathcal{F}}^{(0)}_{j,+j}(s)\equiv e^{-s\mathbf{L}_+}\Psi_{j,j;j,j,j}\ , \qquad\bar{\mathcal{F}}^{(1)}_{j,+j}(s)=e^{-s\mathbf{L}_+}\mathbf{W}_+\Psi_{j,j;j,j,j}
\end{align}
Before concluding this subsection, we show that the chiral covariant derivatives $\mathcal{D}$, $\bar{\mathcal{D}}$ allow us to extract the $\mathfrak{sl}(2)$-highest weights \eqref{eq:sl2prim} from a general generating function \eqref{eq:superfunctional} as follows
\begin{align}
\label{eq:extr}
    &\mathcal{F}\rvert_{s=\eta=\bar\eta=0}=\Psi_{j,b;j,j,b} \nonumber\\
    &\mathcal{D}\mathcal{F}\rvert_{s=\eta=\bar\eta=0}=\Psi_{j,b;j+\frac{1}{2},j+\frac{1}{2},b+\frac{1}{2}} \nonumber\\
    &\bar{\mathcal{D}}\mathcal{F}\rvert_{s=\eta=\bar\eta=0}=\Psi_{j,b;j+\frac{1}{2},j+\frac{1}{2},b+\frac{1}{2}} \nonumber\\
    &\left(\frac{b-j}{2j}\mathcal{D}\bar{\mathcal{D}}+\frac{b+j}{2j}\bar{\mathcal{D}}\mathcal{D}\right)\mathcal{F}\rvert_{s=\eta=\bar\eta=0}=\Psi_{j,b;j+1,j+1,b}
\end{align}

\subsection{Direct sum decomposition}
\label{app:sl21compapp}
We are looking for the direct sum decomposition of a tensor product of two representations of $\mathfrak{sl}(2|1)$. To achieve this goal we introduce another realization of the representations on a space of polynomials. This realization was defined for the first time in Ref. \cite{Derkachov:2000ne}. Our method is far from new in representation theory, see e.g. \cite{Bargmann:1977gy} and reference therein. Roughly speaking, it is the same as finding the Clebsch-Gordan coefficients of the algebra $\mathfrak{so}(3)$ using traceless symmetric tensors.
\par In the special case where the tensor product of two copies of the same representation $[j,b]\otimes [j,b]$ is involved, we will consider the graded-symmetrized vector space
\begin{equation}
\label{eq:assumption}
    \mathcal{S}\left([j,b]\otimes [j,b]\right)=\left\{\frac{1}{2}\left(\Psi_1\otimes \Psi_2-(-1)^{|\Psi_1||\Psi_2|}\Psi_2\otimes \Psi_1\right)\Big\rvert \Psi_1,\Psi_2\in [j,b] \right\}
\end{equation}
where $|\Psi_1|, |\Psi_2|\in \{0,1\}$ are the $\mathbb{Z}_2$-gradings on the vectors $\Psi_1$ and $\Psi_2$. This choice is necessary to have a sensible field theory interpretation of our results. It reflects the possibility to exchange fields inside a product e.g. $\phi_1(x_1)\phi_2(x_2)=\phi_2(x_2)\phi_1(x_1)$ for a pair of bosonic fields. We will discuss the consequences of this assumption case by case later in this subsection.
\par The reader who is not familiar with these techniques is encouraged to read appendix \ref{sec:compsl2} in which this same procedure is implemented in the easier case of the algebra $\mathfrak{sl}(2)$.

\subsubsection{The polynomial realization}
It is convenient to introduce the new variable
\begin{equation}
    t=s+\frac{b}{2j}\eta\bar\eta
\end{equation}
and express the infinitesimal transformations \eqref{eq:differential} as
\begin{align}
    & L_+=t^2\partial_t+t(\eta\partial_{\eta}+\bar\eta\partial_{\bar\eta}) +2jt \  \qquad &&  L_-=-\partial_t\nonumber \\
    & L=t\partial_t+\frac{1}{2}(\eta\partial_{\eta}+\bar\eta\partial_{\bar\eta})+j\  \qquad && B=\frac{1}{2}\eta\partial_{\eta}-\frac{1}{2}\bar\eta\partial_{\bar\eta}+b \nonumber\\
    & V_+=tV_-+\frac{j-b}{2j}\eta\bar\eta\partial_{\bar\eta}+(j-b)\eta\  \qquad && V_-=\partial_{\bar\eta}+ \frac{j-b}{2j}\eta\partial_t\nonumber \\
    & W_+=tW_-+\frac{j+b}{2j}\bar\eta\eta\partial_{\eta}+(j+b)\bar\eta\  \qquad &&
    W_-=\partial_{\eta}+ \frac{j+b}{2}\bar\theta\partial_s
\end{align}
Consequently, the chiral covariant derivatives take the form
\begin{equation}
    \mathcal{D}=\partial_{\eta}-\frac{j-b}{2j}\bar\eta \partial_t\ , \qquad \bar{\mathcal{D}}=\partial_{\bar\eta}-\frac{j+b}{2j}\eta \partial_t
\end{equation}
\sloppy In this variables, we can construct a representation $[j,b]$ on the vector space of polynomials in $t,\eta,\bar{\eta}$.\par
The rules and notation are the same of appendix \ref{sec:compsl2}: \par 
The polynomial corresponding to the vector $\Psi_{j,b;\mathscr{J},\mathscr{L},\mathscr{B}}\in [j,b]$ is denoted as $\mathcal{P}_{j,b;\mathscr{J},\mathscr{L},\mathscr{B}}(s,\eta,\bar{\eta})$. The vector $\Psi_{j_1,b_1;\mathscr{J}_1,\mathscr{L}_1,\mathscr{B}_1}\otimes\Psi_{j_2,b_2;\mathscr{J}_2,\mathscr{L}_2,\mathscr{B}_2} \in [j_1,b_1]\otimes [j_2,b_2]$ is represented by a product of polynomials $\mathcal{P}_{j_1,b_1;\mathscr{J}_1,\mathscr{L}_1,\mathscr{B}_1}(t_1,\eta_1,\bar{\eta}_1) \mathcal{P}_{j_2,b_2;\mathscr{J}_2,\mathscr{L}_2,\mathscr{B}_2}(t_2,\eta_2,\bar{\eta}_2)$. Recall that we are dealing with graded objects, and their order is not arbitrary.
\par The action of a generator $G\in \mathfrak{sl}(2|1)$ on a product of two polynomials depends on the $\mathbb{Z}_2$-grading of the representations they belong to. Denoting as $(-1)^{|\Psi_1|} $ the $\mathbb{Z}_2$-grading of the highest weight $\Psi_{j_1,b_1;j_1,j_1,b_1}\in[j_1,b_1]$ and as $(-1)^{|G|}$ the $\mathbb{Z}_2$-grading of the generator $G$, we \emph{define}
\begin{equation}
    G\mathcal{P}_{j_1,b_1;\mathscr{J}_1,\mathscr{L}_1,\mathscr{B}_1}\mathcal{P}_{j_2,b_2;\mathscr{J}_2,\mathscr{L}_2,\mathscr{B}_2}=\left(G^{(1)}+(-1)^{|\Psi_1||G|}G^{(2)}\right)\mathcal{P}_{j_1,b_1;\mathscr{J}_1,\mathscr{L}_1,\mathscr{B}_1}\mathcal{P}_{j_2,b_2;\mathscr{J}_2,\mathscr{L}_2,\mathscr{B}_2}
\end{equation}
where $G^{(1)}$ is the generator acting on the first polynomial and $G^{(2)}$ on the second, both from the left. The factor $(-1)^{|\Psi_1||G|}\in \mathbb{Z}_2$ has been introduced by hand to mimic a property of $\mathbb{Z}_2$-graded vector spaces: if $v_1,v_2$ are graded vectors and $A, B$ are graded matrices, then $(A\otimes B)(v_1\otimes v_2)=(-1)^{|B||v_1|}(A v_1\otimes Bv_2)$ \footnote{See Refs. \cite{Scheunert:1976wi, Kac:1977qb} for more properties of graded vector spaces}. There is no trouble in doing this, as long as the generators $\left(G^{(1)}+(-1)^{|\Psi_1||G|}G^{(2)}\right)$ continue satisfying the (anti)commutation rules of $\mathfrak{sl}(2|1)$.
\par We now turn to the explicit construction of the polynomials in a representation $[j,b]$. As in section \ref{app:sl21compapp}, the highest weight of this representation, which must be annihilated by $L_-,\ V_-,\ W_-$, can be only the constant polynomial, which we  normalize to unity. The descendants are obtained by repeatedly applying the creation operators $L_+,\ V_+,\ W_+$ as in \eqref{eq:sl2prim}. In the end, one obtains the monomials
\begin{align}
\label{eq:monomials}
    & \mathcal{P}_{j,b;j,j+n,b}=(2j)_nt^n \nonumber \\
    & \mathcal{P}_{j,b;j+\frac{1}{2},j+\frac{1}{2}+n, b+\frac{1}{2}}=(j-b)(2j+1)_nt^n\eta \nonumber \\
    & \mathcal{P}_{j,b;j+\frac{1}{2},j+\frac{1}{2}+n, b-\frac{1}{2}}=(j+b)(2j+1)_nt^n\bar\eta \nonumber \\
    & \mathcal{P}_{j,b;j+1,j+1+n,b}=(b^2-j^2)\frac{2j+1}{2j}(2j+2)_nt^n\eta\bar\eta
\end{align}
where we used the notation $(a)_n\equiv \Gamma(a+n)/\Gamma(a)$. This representation makes the search for the direct sum decomposition of the tensor product two generic representations particularly easy.

\subsubsection{General case}
Suppose we have two representations $[j_1,b_1]$ and $[j_2,b_2]$, and that the $\mathbb{Z}_2$-grading of highest-weight vector $\Psi_{j_1, b_1; j_1,j_1,b_1}$ of the first representation is $(-1)^{|\Psi_1|}$. The polynomials $\mathcal{P}$ in $t_{1,2}$, $\eta_{1,2}$, $\bar{\eta}_{1,2}$ corresponding to the highest weights in $[j_1,b_1]\otimes [j_1,b_2]$ must satisfy the three conditions
\begin{equation}
\label{eq:conditions}
    \begin{gathered}
        \left(L_-^{(1)}+L_-^{(2)}\right)\mathcal{P}=\left(V_-^{(1)}+(-1)^{|\Psi_1|}V_-^{(2)}\right)\mathcal{P}=\left(W_-^{(1)}+(-1)^{|\Psi_1|}W_-^{(2)}\right)\mathcal{P}=0
    \end{gathered}
\end{equation}
Relabeling the variables $\eta_{1,2}$ as $\eta_{1,2}^+$ and $\bar\eta_{1,2}$ as $\eta_{1,2}^-$ and the difference $t_{12}=t_1-t_2$ for convenience, we see that there are six independent towers of polynomials satisfying these requirements
\begin{align}
\label{eq:lowestw}
        & \mathcal{P}^{\pm}_{I}=\left(t_{12}\pm\frac{j_1\mp b_2}{2j_1}\eta_1^+\eta_1^-\pm\frac{j_2\pm b_2}{2j_2}\eta_2^+\eta_2^-+(-1)^{|\Psi_1|}\eta_1^{\mp}\eta_2^{\pm}\right)^n \nonumber\\
        & \mathcal{P}^{\pm}_{II}=\left(t_{12}\pm\frac{j_1\mp b_2}{2j_1}\eta_1^+\eta_1^-\pm\frac{j_2\pm b_2}{2j_2}\eta_2^+\eta_2^-+(-1)^{|\Psi_1|}\eta_1^{\mp}\eta_2^{\pm}\right)^n\left(\eta_1^{\pm}-(-1)^{|\Psi_1|}\eta_2^{\pm}\right) \nonumber\\
        & \mathcal{P}^{\pm}_{III}=\left(t_{12}\pm\frac{j_1\mp b_2}{2j_1}\eta_1^+\eta_1^-\pm\frac{j_2\pm b_2}{2j_2}\eta_2^+\eta_2^-+(-1)^{|\Psi_1|}\eta_1^{\mp}\eta_2^{\pm}\right)^n\left(\eta_1^{\mp}-(-1)^{|\Psi_1|}\eta_2^{\mp}\right)
\end{align}
We again proceed as in appendix \ref{sec:compsl2} and apply the generators $L$, $B$ to identify the representation to which these polynomials belong. We find that
\begin{align}
\label{eq:ppsipm}
    &\mathcal{P}^{\pm}_{I} \longleftrightarrow {^{\pm}\Psi}_{j+n,b;j+n,j+n,b}^{j_1,b_1;j_2,b_2} \nonumber \\
    &\mathcal{P}^{\pm}_{II} \longleftrightarrow {^{\pm}\Psi}_{j+\frac{1}{2}+n,b\pm\frac{1}{2};j+\frac{1}{2}+n,j+\frac{1}{2}+n,b\pm\frac{1}{2}}^{j_1,b_1;j_2,b_2} \nonumber \\
    &\mathcal{P}^{\pm}_{III} \longleftrightarrow {^{\pm}\Psi}_{j+\frac{1}{2}+n,b\mp\frac{1}{2};j+\frac{1}{2}+n,j+\frac{1}{2}+n,b\mp \frac{1}{2}}^{j_1,b_1;j_2,b_2}
\end{align}
where $j=j_1+j_2$ and $b=b_1+b_2$. The $^{\pm}$ have been used to label distinct vectors in $[j_1,b_1]\otimes [j_2,b_2]$ transforming under the same $\mathfrak{sl}(2|1)$ representation, and have nothing to do with their $\mathbb{Z}_2$-grading, chirality, or any other intrinsic property of the representation. The upper indices of the vectors in Eq. \eqref{eq:ppsipm} indicate that these vectors belong to the tensor product of representations $[j_1,b_1]\otimes [j_2,b_2]$.
We thus infer that
\begin{align}
    [j_1,b_1]\otimes[j_2,b_2]&=\bigoplus_{n=0}^{\infty}[j+n,b]^+\oplus [j+\textstyle\frac{1}{2}+n,b+\frac{1}{2}]^+\oplus[j+\frac{1}{2}+n,b-\frac{1}{2}]^+\nonumber\\
    &\oplus\bigoplus_{n=0}^{\infty}[j+n,b]^-\oplus [j+\textstyle\frac{1}{2}+n,b-\frac{1}{2}]^-\oplus[j+\frac{1}{2}+n,b+\frac{1}{2}]^-
\end{align}
To find the Clebsch-Gordan coefficients for this tensor product we have to expand Eq. \eqref{eq:lowestw} in a sum of monomials, and identify in each of them the monomials in Eq.\eqref{eq:monomials}, in analogy with Eq. \eqref{eq:ppp} of appendix \ref{sec:compsl2}.
\par Before writing the result of the calculation, we note that the first of the three conditions in \eqref{eq:conditions} implies that the vectors in Eq. \eqref{eq:ppsipm} can be written as a linear combinations of $\mathfrak{sl}(2)\oplus \mathfrak{u}(1)$-highest weight vectors in $[j_1,b_1]\otimes[j_2,b_2]$. For this reason, we introduce the following notation. Let $\Psi_{j_1,b_1;j_{\alpha},j_{\alpha},b_{\alpha}}\in [j_1,b_1]$ and $\Psi_{j_2,b_2;j_{\beta},j_{\beta},b_{\beta}}\in [j_2,b_2]$ be the $\mathfrak{sl}(2)\oplus \mathfrak{u}(1)$-highest weight vectors of two $\mathfrak{sl}(2)\oplus \mathfrak{u}(1)$-modules of $[j_1,b_1]$ and $[j_2,b_2]$ respectively, which means that
\begin{align}
    \mathbf{L}_-\Psi_{j_1,b_1;j_{\alpha},j_{\alpha},b_{\alpha}}=\mathbf{L}_-\Psi_{j_2,b_2;j_{\beta},j_{\beta},b_{\beta}}=0
\end{align}
and that
\begin{align}
    & (j_\alpha,b_{\alpha})\in \left\{\left(j_1,b_1\right), \left(j_1+\frac{1}{2},b_1+\frac{1}{2}\right),\left(j_1+\frac{1}{2},b_1-\frac{1}{2}\right), \left(j_1+1,b_1\right)\right\} \nonumber \\
    & (j_\beta,b_{\beta})\in \left\{\left(j_2,b_2\right), \left(j_2+\frac{1}{2},b_2+\frac{1}{2}\right),\left(j_2+\frac{1}{2},b_2-\frac{1}{2}\right), \left(j_2+1,b_2\right)\right\}
\end{align}
As discussed in appendix \ref{app:sl2}, from these two vectors it is possible to construct an infinite tower of $\mathfrak{sl}(2)\oplus \mathfrak{u}(1)$-highest weight vectors inside $[j_1,b_1]\otimes [j_2, b_2]$. We denote the $n$-th of these $\mathfrak{sl}(2)\oplus \mathfrak{u}(1)$-highest weight vectors
\begin{equation}
\label{eq:bracket}
	\left(\substack{j_{\alpha} \\ b_{\alpha}}\Big\rvert\substack{j_{\beta} \\ b_{\beta}}\right)_n=\Psi_{j_1,b_1; j_{\alpha},j_{\alpha},b_{\alpha}}\mathbb{P}^{j_{\alpha},j_{\beta}}_n(\overleftarrow{\mathbf{L}}_+, \overrightarrow{\mathbf{L}}_+)\Psi_{j_1,b_2;j_{\beta},j_{\beta},b_{\beta}}\ , \qquad n\in \mathbb{N}
\end{equation}
where $\mathbb{P}^{a,b}_n$ is the polynomial
\begin{equation}
\label{eq:symbol1}
	\mathbb{P}^{a,b}_n(x_1,x_2)=\sum_{n_1+n_2=n}\binom{n}{n_1}\frac{(-1)^{n_1}}{\Gamma(2a+n_1)\Gamma(2b+n_2)}x_1^{n_1}x_2^{n_2}
\end{equation}
Note that in \eqref{eq:bracket} the pairs $(j_{\alpha},b_{\alpha})$, $(j_{\beta},b_{\beta})$ uniquely identify the $\mathfrak{sl}(2)\oplus \mathfrak{u}(1)$-modules involved.
\par We pause for a moment to see how the polynomials \eqref{eq:symbol1} are related to the Jacobi polynomials $P^{(\alpha,\beta)}_n$ and the Gegenbauer polynomials $C_n^{\alpha}$, defined as
\begin{subequations}
\begin{equation}
    P^{(\alpha,\beta)}_n(z)=\sum_{k=0}^n\binom{n+\alpha}{k}\binom{n+\beta}{k+\beta}\left(\frac{z-1}{2}\right)^{k}\left(\frac{z+1}{2}\right)^{n-k}
\end{equation}
\begin{equation}
    C_n^{\alpha}(z)=\frac{\Gamma(n+2\alpha)\Gamma\left(\alpha+\frac{1}{2}\right)}{\Gamma(2\alpha)\Gamma\left(n+\alpha+\frac{1}{2}\right)}P^{\left(\alpha-\frac{1}{2},\alpha-\frac{1}{2}\right)}_n(z)
\end{equation}
\end{subequations}
The relation is
\begin{subequations}
\label{eq:jacgeg0}
\begin{equation}
    \mathbb{P}^{a,b}_n(x_1,x_2)=\frac{n!}{\Gamma(2a+n)\Gamma(2b+n)}(x_1+x_2)^nP_n^{(2a-1,2b-1)}\left(\frac{x_2-x_1}{x_2+x_1}\right)
\end{equation}
\begin{equation}
    \mathbb{P}^{\frac{1}{2}\left(\alpha+\frac{1}{2}\right),\frac{1}{2}\left(\alpha+\frac{1}{2}\right)}_n(x_1,x_2)=\frac{n!\Gamma(2\alpha)}{\Gamma\left(n+\alpha+\frac{1}{2}\right)\Gamma(2\alpha+n)\Gamma\left(\alpha+\frac{1}{2}\right)}(x_1+x_2)^nC_n^{\alpha}\left(\frac{x_2-x_1}{x_2+x_1}\right)
\end{equation}
\end{subequations}
For the details on the properties and use of this polynomial see Ref. \cite{Braun:2003rp}, the appendices of Refs. \cite{Bochicchio:2021nup, Bochicchio:2022uat} and also appendix \ref{app:sl2}.
\par We are now ready to show the direct sum decomposition for a tensor product two general $\mathfrak{sl}(2\rvert1)$ representations
\begin{subequations}
	\label{eq:generalcomplaw}
	\begin{align}
			^{\pm}\Psi_{j+n,b;j+n,j+n,b}^{j_1,b_1;j_2,b_2}=&  \left(\substack{j_1 \\ b_1}\Big\rvert\substack{j_2 \\ b_2}\right)_n\pm n\frac{2j_1}{j_1\pm b_1} \left(\substack{j_1+1 \\ b_1}\Big\rvert\substack{j_2 \\ b_2}\right)_{n-1}\pm n\frac{2j_2}{j_2\mp b_2} \left(\substack{j_1 \\ b_1}\Big\rvert\substack{j_2+1 \\ b_2}\right)_{n-1}\nonumber \\
			+&n(-1)^{\rvert\Psi_1\rvert+1}\frac{2j_1}{j_1\pm b_1}\frac{2j_2}{j_2\mp b_2}\left(\substack{j_1+\frac{1}{2} \\ b_1\mp \frac{1}{2}}\Big\rvert\substack{j_2+\frac{1}{2} \\ b_2\pm\frac{1}{2}}\right)_{n-1} \nonumber\\
			+&n(n-1)\frac{2j_1}{j_1\pm b_1}\frac{2j_2}{j_2\mp b_2}\left(\substack{j_1+1 \\ b_1}\Big\rvert\substack{j_2+1 \\ b_2}\right)_{n-2}
	\end{align}
	\begin{align}
			^{\pm}\Psi_{j+\frac{1}{2}+n,b\pm\frac{1}{2};j+\frac{1}{2}+n,j+\frac{1}{2}+n,b\pm\frac{1}{2}}^{j_1,b_1;j_2,b_2}=& \frac{2j_1}{j_1\mp b_1}\left(\substack{j_1+\frac{1}{2} \\ b_1\pm\frac{1}{2}}\Big\rvert\substack{j_2 \\ b_2}\right)_{n}+ (-1)^{\rvert\Psi_1\rvert+1}\frac{2j_2}{j_2\mp b_2}\left(\substack{j_1 \\ b_1}\Big\rvert\substack{j_2+\frac{1}{2} \\ b_2\pm\frac{1}{2}}\right)_{n}\nonumber \\
			\pm & n(-1)^{\rvert\Psi_1\rvert}\frac{2j_1}{j_1\mp b_1}\frac{2j_2}{j_2\mp b_2} \left(\substack{j_1+1 \\ b_1}\Big\rvert\substack{j_2+\frac{1}{2} \\ b_2\pm \frac{1}{2}}\right)_{n-1}\nonumber \\
			\pm & n \frac{2j_1}{j_1\mp b_1}\frac{2j_2}{j_2\mp b_2} \left(\substack{j_1+\frac{1}{2} \\ b_1\pm \frac{1}{2}}\Big\rvert\substack{j_2+1 \\ b_2}\right)_{n-1}
	\end{align}
	\begin{align}
			^{\pm}\Psi_{j+\frac{1}{2}+n,b\mp\frac{1}{2};j+\frac{1}{2}+n,j+\frac{1}{2}+n,b\mp \frac{1}{2}}^{j_1,b_1;j_2,b_2}=& \frac{2j_1}{j_1\pm b_1}\left(\substack{j_1+\frac{1}{2} \\ b_1\mp \frac{1}{2}}\Big\rvert\substack{j_2 \\ b_2}\right)_{n}+(-1)^{\rvert\Psi_1\rvert+1}\frac{2j_2}{j_2\pm b_2}\left(\substack{j_1 \\ b_1}\Big\rvert\substack{j_2+\frac{1}{2} \\ b_2\mp\frac{1}{2}}\right)_{n} \nonumber\\
			\mp & n\frac{2j_1}{j_1\pm b_1}\frac{2j_2}{j_2\pm b_2}\left(\substack{j_1+\frac{1}{2} \\ b_1\mp \frac{1}{2}}\Big\rvert\substack{j_2+1 \\ b_2}\right)_{n-1} \nonumber\\
			\mp & n (-1)^{\rvert\Psi_1\rvert} \frac{2j_1}{j_1\pm b_1}\frac{2j_2}{j_2\pm b_2} \left(\substack{j_1+1 \\ b_1}\Big\rvert\substack{j_2+\frac{1}{2} \\ b_2\mp \frac{1}{2}}\right)_{n-1}
	\end{align}
\end{subequations}
where the upper indices indicate that these vectors belong to the tensor product $[j_1,b_1]\otimes [j_2,b_2]$. Note that these expressions are singular when at least one of the representations satisfy the condition $j\pm b=0$. Actually, for our purposes this is the most interesting scenario, since $j\pm b=0$ is a necessary condition for a representation to be chiral. We shall elaborate about this in the next paragraphs. Due to the assumption in Eq. \eqref{eq:assumption} and to the symmetry properties of the polynomial \eqref{eq:symbol}, some of the terms appearing may vanish when some $j_{\alpha},j_{\beta}$ are equal.

\subsubsection{Chiral representations}
To work out the direct sum decomposition for a tensor product of two chiral representations, we have to impose the conditions $b_{1,2}\pm j_{1,2}=0$ from the beginning and the polynomials must satisfy additional chirality conditions. Again, we relabel the chiral covariant derivatives as $\mathcal{D}=\mathcal{D}^+$ and $\bar{\mathcal{D}}=\mathcal{D}^-$ for later convenience, and also write
\begin{equation}
\label{eq:jjbar}
    j=j_1+j_2\ , \qquad \bar{j}=j_1-j_2
\end{equation}
We have the following cases: \\

\textbf{Same chirality} ($j_1\pm b_1=j_2\pm b_2=0$). The polynomials must satisfy the additional conditions
\begin{equation}
    (\mathcal{D}^{\mp})^{(1)}\mathcal{P}= (\mathcal{D}^{\mp})^{(2)}\mathcal{P}=0
\end{equation}
The only available polynomials are
\begin{equation}
    \begin{gathered}
      \mathcal{P}_n^{(\text{same})}=t_{12}^n\left(\eta_1^{\pm}-(-1)^{|\Psi_1|}\eta_2^{\pm}\right)
    \end{gathered}
\end{equation}
This implies the direct sum decomposition
\begin{equation}
    [j_1,\mp j_1]\otimes [j_1, \mp j_2]=\bigoplus_{n=0}^{\infty}[j+n+\frac{1}{2},\mp j \pm \frac{1}{2}]
\end{equation}
and the decomposition
\begin{subequations}
    \begin{align}
    \label{eq:samechir++}
         &\Psi^{j_1,-j_1;j_2,-j_2}_{j+\frac{1}{2}+n,-j+\frac{1}{2};j+\frac{1}{2}+n,j+\frac{1}{2}+n, -j+\frac{1}{2}}= \nonumber\\
         =&\Psi_{j_1,-j_1;j_1,j_1,-j_1}\left[\overleftarrow{\mathbf{V}}_+\mathbb{P}^{j_1+\frac{1}{2},j_2}_n(\overleftarrow{\mathbf{L}}_+,\overrightarrow{\mathbf{L}}_+)-(-1)^{|\Psi_1|} \mathbb{P}^{j_1,j_2+\frac{1}{2}}_n(\overleftarrow{\mathbf{L}}_+,\overrightarrow{\mathbf{L}}_+)\ \overrightarrow{\mathbf{V}}_+\right]\Psi_{j_2,-j_2;j_2,j_2,-j_2}
\end{align} 
\begin{align}
    \label{eq:samechir--}
         &\Psi^{j_1,j_1;j_2,j_2}_{j+\frac{1}{2}+n,+j-\frac{1}{2};j+\frac{1}{2}+n,j+\frac{1}{2}+n, +j-\frac{1}{2}}= \nonumber\\
         =&\Psi_{j_1,+j_1;j_1,j_1,+j_1}\left[\overleftarrow{\mathbf{W}}_+\mathbb{P}^{j_1+\frac{1}{2},j_2}_n(\overleftarrow{\mathbf{L}}_+,\overrightarrow{\mathbf{L}}_+)-(-1)^{|\Psi_1|} \mathbb{P}^{j_1,j_2+\frac{1}{2}}_n(\overleftarrow{\mathbf{L}}_+,\overrightarrow{\mathbf{L}}_+)\ \overrightarrow{\mathbf{W}}_+\right]\Psi_{j_2,+j_2;j_2,j_2,+j_2}
\end{align} 
\end{subequations} \\

\textbf{Opposite chirality} ($j_1\pm b_1=j_2\mp b_2=0$). The polynomials must satisfy the additional conditions
\begin{equation}
    (\mathcal{D}^{\mp})^{(1)}\mathcal{P}= (\mathcal{D}^{\pm})^{(2)}\mathcal{P}=0
\end{equation}
The only available polynomials are
\begin{equation}
    \begin{gathered}
        \mathcal{P}_n^{(\text{opp})}=\left(t_{12}+(-1)^{|\Psi_1|}\eta_1^{+}\eta_2^{-}\right)^n
    \end{gathered}
\end{equation}
This implies the direct sum decomposition
\begin{equation}
    [j_1,\mp j_1]\otimes [j_1, \pm j_2]=\bigoplus_{n=0}^{\infty}[j+n,\mp \bar{j} ]
\end{equation}
and the decomposition
\begin{subequations}
    \begin{align}
    \label{eq:oppchir+-}
         &\Psi^{j_1,-j_1;j_2,j_2}_{j+n,-\bar{j};j+n,j+n,-\bar{j}}= \nonumber\\
         =&\Psi_{j_1,-j_1;j_1,j_1,-j_1}\left[\mathbb{P}^{j_1,j_2}_n(\overleftarrow{\mathbf{L}}_+,\overrightarrow{\mathbf{L}}_+)-(-1)^{|\Psi_1|} n \overleftarrow{\mathbf{V}}_+\mathbb{P}^{j_1+\frac{1}{2},j_2+\frac{1}{2}}_{n-1}(\overleftarrow{\mathbf{L}}_+,\overrightarrow{\mathbf{L}}_+)\ \overrightarrow{\mathbf{W}}_+\right]\Psi_{j_2,+j_2;j_2,j_2,+j_2}
\end{align} 
    \begin{align}
    \label{eq:oppchir-+}
         &\Psi^{j_1,j_1;j_2,-j_2}_{j+n,\bar{j};j+n,j+n,\bar{j}}=\nonumber \\
         =&\Psi_{j_1,+j_1;j_1,j_1,+j_1}\left[\mathbb{P}^{j_1,j_2}_n(\overleftarrow{\mathbf{L}}_+,\overrightarrow{\mathbf{L}}_+)-(-1)^{|\Psi_1|} n \overleftarrow{\mathbf{W}}_+\mathbb{P}^{j_1+\frac{1}{2},j_2+\frac{1}{2}}_{n-1}(\overleftarrow{\mathbf{L}}_+,\overrightarrow{\mathbf{L}}_+)\ \overrightarrow{\mathbf{V}}_+\right]\Psi_{j_1,-j_2;j_2,j_2,-j_2}
\end{align} 
\end{subequations} \\

We also performed this calculation as in the first part of appendix \ref{sec:compsl2}, without working in any specific realization of the representations, and obtaining the same results.
\par Remarkably, the vectors in Eqs. \eqref{eq:samechir++}, \eqref{eq:samechir--}, \eqref{eq:oppchir+-} and \eqref{eq:oppchir-+}  can be put in a more compact form by means of the identities
\begin{align}
\label{eq:uvw}
    & \mathbf{V}_+\Psi=0\quad \implies \quad  & \mathbf{L}_+^n\Psi=(\mathbf{V}_++\mathbf{W}_+)^{2n}\Psi\ , \quad \mathbf{W}_+\mathbf{L}_+^n\Psi=(\mathbf{V}_++\mathbf{W}_+)^{2n+1}\Psi\nonumber \\
    & \mathbf{W}_+\Psi=0\quad \implies \quad  & \mathbf{L}_+^n\Psi=(\mathbf{V}_++\mathbf{W}_+)^{2n}\Psi\ , \quad \mathbf{V}_+\mathbf{L}_+^n\Psi=(\mathbf{V}_++\mathbf{W}_+)^{2n+1}\Psi
\end{align}
We introduce the generator $\mathbf{U}_+=\mathbf{V}_++\mathbf{W}_+$ and write concisely
\begin{subequations}
\label{eq:cjj}
    \begin{align}
    \label{eq:cjja}
        &\Psi^{j_1,\mp j_1;j_2\mp j_2}_{j+\frac{1}{2}+n,\mp j\pm\frac{1}{2};j+\frac{1}{2}+n,j+\frac{1}{2}+n,\mp j\pm\frac{1}{2}}= \nonumber  \\
        &\qquad =n!\ (-1)^{|\Psi_1|+1}\Psi_{j_1,\mp j_1;j_1,j_1,\mp j_1}\mathbb{C}^{j_1,j_2}_{2n+1}(\overleftarrow{\mathbf{U}}_+,\overrightarrow{\mathbf{U}}_+)\Psi_{j_2,\mp j_2;j_2,j_2,\mp j_2}
    \end{align}
and
    \begin{align}
    \label{eq:cjjb}
        &\Psi^{j_1,\mp j_1;j_2\pm j_2}_{j+n,\mp \bar{j};j+n,j+n,\mp \bar{j}} =
        \nonumber\\
        &\qquad =n!\ \Psi_{j_1,\mp j_1;j_1,j_1,\mp j_1}\mathbb{C}^{j_1,j_2}_{2n}(\overleftarrow{\mathbf{U}}_+,\overrightarrow{\mathbf{U}}_+)\Psi_{j_2,\pm j_2;j_2,j_2,\pm j_2}
    \end{align}
\end{subequations}
The polynomial $\mathbb{C}^{j_1,j_2}_n(\alpha,\beta)$ is defined as
\begin{equation}
	\label{eq:polynomialC}
    \mathbb{C}^{j_1,j_2}_n(\alpha,\beta)=\sum_{k_1+k_2=n}\frac{(-1)^{\lfloor\frac{k_1+1-|\Psi_1|}{2}\rfloor}\alpha^{k_1}\beta^{k_2}}{\Gamma\left(1+\lfloor\frac{k_1}{2}\rfloor\right)\Gamma\left(1+\lfloor\frac{k_2}{2}\rfloor\right)\Gamma\left(2j_1+\lfloor\frac{k_1+1}{2}\rfloor\right)\Gamma\left(2j_2+\lfloor\frac{k_2+1}{2}\rfloor\right)}
\end{equation}
with $\alpha,\beta$ being odd variables squaring to some even variable. The proof of this statement in one of the four possible cases can be found in appendix \ref{app:cjj}. The importance of this result in the context of the present work should not be underestimated. It is thanks to the existence of this polynomial that we can write the generating functionals of sections \ref{sec:genfreescft}, \ref{sec:appliedcorrelators}, \ref{sec:sym} in an elegant closed form. When $j_1=j_2$, instead of the ordinary tensor product of two representations, we consider the graded-symmetrized vector space in Eq. \eqref{eq:assumption}. Consequently, if the two representations are $\mathbb{Z}_2$-even, Eq. \eqref{eq:cjja} is nonzero only for $n$ odd while Eq. \eqref{eq:cjja} is nonzero only for $n$ even. If the two representations are $\mathbb{Z}_2$-odd, Eq. \eqref{eq:cjja} is nonzero only for $n$ even while Eq. \eqref{eq:cjja} is nonzero only for $n$ odd.

\subsubsection{Chiral supersymmetric descendants and generating functions}
\label{sec:desc}
The repeated application of $\mathbf{V}_+$ and $\mathbf{W}_+$ according to \eqref{eq:sl2prim} allows us to extract the $\mathfrak{sl}(2)\oplus \mathfrak{u}(1)$-highest weight vectors inside the $\mathfrak{sl}(2|1)$ multiplet. In this subsection, we use the following condensed notation 
\begin{equation}
\label{eq:condensed0}
    \Psi_i\equiv \Psi_{j_i, - j_1;j_i, j_i, - j_i}\ , \qquad , \bar{\Psi}_i\equiv \Psi_{j_i, + j_1;j_i, j_i, + j_i} \qquad \Psi_1\mathbb{P}^{j_1,j_2}_n \bar{\Psi}_2\equiv \Psi_1\mathbb{P}^{j_1,j_2}_n(\overleftarrow{\mathbf{L}}_+,\overrightarrow{\mathbf{L}}_+)\bar{\Psi}_2
\end{equation}
and so on. We list the $\mathfrak{sl}(2)$-highest weight vectors as follows. If $\Psi$ is a $\mathfrak{sl}(2|1)$-highest weight vectors in Eqs. \eqref{eq:samechir++}, \eqref{eq:samechir--}, \eqref{eq:oppchir+-}, \eqref{eq:oppchir-+}, then following Eq. \eqref{eq:sl2prim} we write
\begin{align}
\label{eq:susydesc}
         &\Psi_0 = \Psi \nonumber\\
         &\Psi_V = \textbf{V}_+\Psi  \nonumber\\
         &\Psi_W = \textbf{W}_+\Psi \nonumber\\
         &\Psi_{VW} = \left(\frac{b+j}{2j}\textbf{W}_+\textbf{V}_++\frac{b-j}{2j}\textbf{V}_+\textbf{W}_+\right)\Psi 
\end{align} 
We also show the component expansion of the generating function of a representation. As explained in Eq. \eqref{eq:differential}, the generating function is obtained by applying $e^{-s\mathbf{L}_++\eta\mathbf{V}_++\bar{\eta}\mathbf{W}_+}$ to the highest weight vectors that we found, and their components are obtained by applying $e^{-s\textbf{L}_+}$ to the supersymmetric descendants \eqref{eq:susydesc}. 
Again, we  use a condensed notation, for example
\begin{equation}
	\label{eq:chiralgenf}
	\mathcal{F}_1\mathbb{C}_{2n+1}^{j_1,j_2}\mathcal{F}_2 \equiv \mathcal{F}_1(s_L,\eta)\mathbb{C}_{2n+1}^{j_1,j_2} (\overleftarrow{\mathcal{D}}+\overleftarrow{\bar{\mathcal{D}}},\overrightarrow{\mathcal{D}}+\overrightarrow{\bar{\mathcal{D}}})\mathcal{F}_2(s_L.\eta)
\end{equation}	
which decomposes in many terms, we show as an example the product
\begin{equation}
	\mathcal{F}_1^{(0)}\mathbb{P}_n^{j_1,j_2}\mathcal{F}_2^{(0)}\equiv \mathcal{F}_1^{(0)}(s)\mathbb{P}_n^{j_1,j_2}(\overleftarrow{\partial}_s,\overrightarrow{\partial}_s)\mathcal{F}_2^{(0)}(s)
\end{equation}
where the symbols $\mathcal{F}_i\equiv \mathcal{F}_{j_i,- j_i}$ denote the generating functions of the elementary representations. $\bar{\mathcal{F}}_i\equiv \bar{\mathcal{F}}_{j_i,+j_i} $ and the components $\mathcal{F}_i^{(0)}, \mathcal{F}_1^{(1)}$ are defined in Eq. \eqref{eq:components2}.\\

\noindent \textbullet\quad For the highest weight vector in Eq. \eqref{eq:samechir++} we have the descendants
\begin{align}
         &\Psi_0=  \Psi_1\left[\overleftarrow{\textbf{V}}_+\mathbb{P}_n^{j_1+\frac{1}{2},j_2}-(-1)^{|\Psi_1|}\mathbb{P}_n^{j_1,j_2+\frac{1}{2}} \overrightarrow{\mathbf{V}}_+\right]\Psi_2 \nonumber\\
         &\Psi_V=  -(-1)^{|\Psi_1|}(2j_1+2j_2+n)\ \Psi_1\overleftarrow{\textbf{V}}_+\mathbb{P}_n^{j_1+\frac{1}{2},j_2+\frac{1}{2}}\overrightarrow{\mathbf{V}}_+\Psi_2  \nonumber\\
         &\Psi_W=  -\Psi_1\mathbb{P}_{n+1}^{j_1,j_2}\Psi_2 \nonumber\\
         &\Psi_{VW}= +\frac{2j_1+2j_2+n}{2(j_1+j_2+n+\frac{1}{2})}  \nonumber\\
         &\qquad\quad\Psi_1\left[(2j_1+n+1)\ \overleftarrow{\textbf{V}}_+\mathbb{P}_{n+1}^{j_1+\frac{1}{2},j_2}+(2j_2+n+1)(-1)^{|\Psi_1|}\mathbb{P}_{n+1}^{j_1,j_2+\frac{1}{2}} \overrightarrow{\mathbf{V}}_+\right]\Psi_2
\end{align}
and its generating function is
\begin{equation}
\label{eq:genfun++}
	\resizebox{1.00\textwidth}{!}{
		$\begin{aligned}
    &n!(-1)^{n+|\Psi_1|+1}\ \mathcal{F}_1 \mathbb{C}^{j_1,j_2}_{2n+1}\mathcal{F}_2= \\
    &\quad +\left[\mathcal{F}_1^{(1)}\mathbb{P}_n^{j_1+\frac{1}{2},j_2}\mathcal{F}_2^{(0)}-(-1)^{|\Psi_1|} \mathcal{F}_1^{(0)}\mathbb{P}_n^{j_1,j_2+\frac{1}{2}}\mathcal{F}_1^{(1)}\right] \\
    &\quad+ \eta\ (-1)^{|\Psi_1|+1}(2j_1+2j_2+n)\ \mathcal{F}_1^{(1)} \mathbb{P}_n^{j_1+\frac{1}{2},j_2+\frac{1}{2}} \mathcal{F}_2^{(1)} \\
    &\quad+\bar{\eta}\ \mathcal{F}_1^{(0)} \mathbb{P}_{n+1}^{j_1,j_2} \mathcal{F}_2^{(0)} \\
    &\quad- \eta\bar{\eta}\ \frac{2j_1+2j_2+n}{2(j_1+j_2+n+\frac{1}{2})} \left[(2j_1+n+1)\ \mathcal{F}_1^{(1)}\mathbb{P}_{n+1}^{j_1+\frac{1}{2},j_2}\mathcal{F}_2^{(0)} +(2j_2+n+1)(-1)^{|\Psi_1|} \mathcal{F}_1^{(0)}\mathbb{P}_{n+1}^{j_1,j_2+\frac{1}{2}}\mathcal{F}_2^{(1)} \right]   \\
    &\quad+\eta\bar{\eta}\frac{-j_1-j_2+\frac{1}{2}}{2\left(j_2+j_2+\frac{1}{2}\right)}\partial_s\left[\mathcal{F}_1^{(1)}\mathbb{P}_n^{j_1+\frac{1}{2},j_2}\mathcal{F}_2^{(0)}-(-1)^{|\Psi_1|} \mathcal{F}_1^{(0)}\mathbb{P}_n^{j_1,j_2+n+\frac{1}{2}}\mathcal{F}_1^{(1)}\right]
\end{aligned}$
}
\end{equation}
\textbullet\quad For the highest weight vector in Eq. \eqref{eq:samechir--} we have
\begin{align}
         &\Psi_0 =  \bar{\Psi}_1\left[\overleftarrow{\textbf{W}}_+\mathbb{P}_n^{j_1+\frac{1}{2},j_2} -(-1)^{|\Psi_1|}\mathbb{P}_n^{j_1,j_2+\frac{1}{2}}  \overrightarrow{\mathbf{W}}_+\right]\bar{\Psi}_2 \nonumber\\
         &\Psi_V = -\bar{\Psi}_1\mathbb{P}_{n+1}^{j_1,j_2}\bar{\Psi}_2  \nonumber\\
         &\Psi_W =  -(-1)^{|\Psi_1|}(2j_1+2j_2+n)\ \bar{\Psi}_1\overleftarrow{\textbf{W}}_+\mathbb{P}_n^{j_1+\frac{1}{2},j_2+\frac{1}{2}}\overrightarrow{\textbf{W}}_+\bar{\Psi}_2  \nonumber\\
         &\Psi_{VW} =-\frac{2j_1+2j_2+n}{2(j_1+j_2+n+\frac{1}{2})}  \nonumber\\
         &  \bar{\Psi}_1\left[(2j_1+n+1)\ \overleftarrow{\textbf{W}}_+\mathbb{P}_{n+1}^{j_1+\frac{1}{2},j_2}+(2j_2+n+1)(-1)^{|\Psi_1|}\mathbb{P}_{n+1}^{j_1,j_2+\frac{1}{2}} \overrightarrow{\mathbf{W}}_+\right]\bar{\Psi}_2  \nonumber\\
\end{align}
and the generating function is
\begin{equation}
	\label{eq:genfun--}
	\resizebox{1.00\textwidth}{!}{
		$\begin{aligned}
    & n!(-1)^{n+|\Psi_1|+1}\ \bar{\mathcal{F}}_1\mathbb{C}^{j_1,j_2}_{2n+1}\bar{\mathcal{F}}_2= \\
     &\quad \left[\bar{\mathcal{F}}_1^{(1)}\mathbb{P}_n^{j_1+\frac{1}{2},j_2}\bar{\mathcal{F}}_2^{(0)}-(-1)^{|\Psi_1|} \bar{\mathcal{F}}_1^{(0)}\mathbb{P}_n^{j_1,j_2+\frac{1}{2}}\bar{\mathcal{F}}_1^{(1)}\right] \\
    &\quad+ \eta\ \bar{\mathcal{F}}_1^{(0)} \mathbb{P}_{n+1}^{j_1,j_2} \bar{\mathcal{F}}_2^{(0)}  \\
    &\quad +\bar{\eta}\ (-1)^{|\Psi_1|+1}(2j_1+2j_2+n)\ \bar{\mathcal{F}}_1^{(1)} \mathbb{P}_n^{j_1+\frac{1}{2},j_2+\frac{1}{2}} \bar{\mathcal{F}}_2^{(1)} \\
    &\quad+ \eta\bar{\eta}\ \frac{2j_1+2j_2+n}{2(j_1+j_2+n+\frac{1}{2})} \left[(2j_1+n+1)\ \bar{\mathcal{F}}_1^{(1)}\mathbb{P}_{n+1}^{j_1+\frac{1}{2},j_2}\bar{\mathcal{F}}_2^{(0)} +(2j_2+n+1)(-1)^{|\Psi_1|} \bar{\mathcal{F}}_1^{(0)}\mathbb{P}_{n+1}^{j_1,j_2+\frac{1}{2}}\bar{\mathcal{F}}_2^{(1)} \right]  \\
    &\quad+ \eta\bar{\eta}\frac{j_1+j_2-\frac{1}{2}}{2\left(j_2+j_2+n+\frac{1}{2}\right)}\partial_s\left[\bar{\mathcal{F}}_1^{(1)}\mathbb{P}_n^{j_1+\frac{1}{2},j_2}\bar{\mathcal{F}}_2^{(0)}-(-1)^{|\Psi_1|} \bar{\mathcal{F}}_1^{(0)}\mathbb{P}_n^{j_1,j_2+\frac{1}{2}}\bar{\mathcal{F}}_1^{(1)}\right]
\end{aligned}$
}
\end{equation}
\textbullet\quad For the highest weight vector Eq. \eqref{eq:oppchir+-} we have
\begin{align}
         &\Psi_0 =  \Psi_1\left[\mathbb{P}^{j_1,j_2}_n -(-1)^{|\Psi_1|}n\ \overleftarrow{\textbf{V}}_+\mathbb{P}^{j_1+\frac{1}{2},j_2+\frac{1}{2}}_{n-1} \overrightarrow{\mathbf{W}}_+\right]\bar{\Psi}_2 \nonumber\\
         &\Psi_V =(2j_1+n)\ \Psi_1\overleftarrow{\textbf{V}}_+ \mathbb{P}^{j_1+\frac{1}{2},j_2}_{n} \bar{\Psi}_2 \nonumber\\
         &\Psi_W= (-1)^{|\Psi_1|}(2j_2+n)\ \Psi_1\mathbb{P}^{j_1,j_2+\frac{1}{2}}_{n} \overrightarrow{\mathbf{W}}_+\bar{\Psi}_2 \nonumber\\
         &\Psi_{VW}=-\frac{(2j_1+n)(2j_2+n)}{2(j_1+j_2+n)}\ \Psi_1\left[ \mathbb{P}^{j_1,j_2}_{n+1} +(-1)^{|\Psi_1|}(2j_1+2j_2+n)\ \overleftarrow{\textbf{V}}_+\mathbb{P}^{j_1+\frac{1}{2},j_2+\frac{1}{2}}_{n}  \overrightarrow{\mathbf{W}}_+\right]\bar{\Psi}_2
\end{align}
The generating function is
\begin{align}
\label{eq:genfun+-}
        &n!(-1)^{n}\mathcal{F}_1\mathbb{C}_{2n}^{j_1,j_2}\bar{\mathcal{F}}_2 = \nonumber\\
        &\quad \left[\mathcal{F}_1^{(0)}\mathbb{P}_n^{j_1,j_2}\bar{\mathcal{F}}_2^{(0)}+(-1)^{|\Psi_1|}n\mathcal{F}_1^{(1)} \mathbb{P}_{n+1}^{j_1+\frac{1}{2},j_2+\frac{1}{2}} \bar{\mathcal{F}}_2^{(1)}\right] \nonumber\\
        &\quad+\eta\ (2j_1+n)\mathcal{F}_1^{(1)}\mathbb{P}^{j_1+\frac{1}{2},j_2}_{n}\bar{\mathcal{F}}_2^{(0)} \nonumber\\
        &\quad+\bar{\eta}\ (-1)^{|\Psi_1|}(2j_2+n)\mathcal{F}_1^{(0)}\mathbb{P}^{j_1,j_2+\frac{1}{2}}_{n}\bar{\mathcal{F}}_2^{(1)} \nonumber\\
        &\quad+\eta\bar{\eta}\ \frac{(2j_1+n)(2j_2+n)}{2(j_1+j_2+n)}\left[\mathcal{F}_1^{(0)} \mathbb{P}^{j_1,j_2}_{n+1}\bar{\mathcal{F}}_2^{(0)} -(-1)^{|\Psi_1|}(2j_1+2j_2+n)\ \mathcal{F}_1^{(1)}\mathbb{P}^{j_1+\frac{1}{2},j_2+\frac{1}{2}}_{n}  \bar{\mathcal{F}}_2^{(1)}\right]  \nonumber\\
        &\quad+\eta\bar{\eta}\frac{-j_1+j_2}{2\left(j_1+j_2+n\right)}\partial_s\left[\mathcal{F}_1^{(0)}\mathbb{P}_n^{j_1,j_2}\bar{\mathcal{F}}_2^{(0)}+(-1)^{|\Psi_1|}n\mathcal{F}_1^{(1)} \mathbb{P}_{n+1}^{j_1+\frac{1}{2},j_2+\frac{1}{2}} \bar{\mathcal{F}}_2^{(1)}\right]
\end{align}
\textbullet\quad For the highest weight vector in Eq. \eqref{eq:oppchir-+} we have
\begin{align}
          &\Psi_0 = \bar{\Psi}_1\left[\mathbb{P}^{j_1,j_2}_n -(-1)^{|\Psi_1|}n\ \overleftarrow{\textbf{W}}_+\mathbb{P}^{j_1+\frac{1}{2},j_2+\frac{1}{2}}_{n-1}  \overrightarrow{\mathbf{V}}_+\right]\Psi_2 \nonumber\\
         &\Psi_V =(-1)^{|\Psi_1|}(2j_2+n)\ \bar{\Psi}_1\mathbb{P}^{j_1,j_2+\frac{1}{2}}_{n} \overrightarrow{\mathbf{V}}_+\Psi_2 \nonumber\\
         &\Psi_W = (2j_1+n)\ \bar{\Psi}_1\overleftarrow{\textbf{W}}_+\mathbb{P}^{j_1+\frac{1}{2},j_2}_{n}\Psi_2 \nonumber\\
         &\Psi_{VW} = +\frac{(2j_1+n)(2j_2+n)}{2(j_1+j_2+n)}\ \bar{\Psi}_1\left[ \mathbb{P}^{j_1,j_2}_{n+1} +(-1)^{|\Psi_1|}(2j_1+2j_2+n)\ \overleftarrow{\textbf{W}}_+\mathbb{P}^{j_1+\frac{1}{2},j_2+\frac{1}{2}}_{n} \overrightarrow{\mathbf{V}}_+\right]\Psi_2
\end{align}
and its generating function is
\begin{align}
\label{eq:genfun-+}
        &n!(-1)^{n}\bar{\mathcal{F}}_1\mathbb{C}_{2n}^{j_1,j_2}\mathcal{F}_2 = \nonumber\\
        &\quad \left[\bar{\mathcal{F}}_1^{(0)}\mathbb{P}_n^{j_1,j_2}\mathcal{F}_2^{(0)}+(-1)^{|\Psi_1|}n\bar{\mathcal{F}}_1^{(1)} \mathbb{P}_{n+1}^{j_1+\frac{1}{2},j_2+\frac{1}{2}} \mathcal{F}_2^{(1)}\right] \nonumber\\
        &\quad+\eta\ (2j_2+n)(-1)^{|\Psi_1|} \bar{\mathcal{F}}_1^{(0)}\mathbb{P}^{j_1,j_2+\frac{1}{2}}_{n} \mathcal{F}_2^{(1)} \nonumber\\
        &\quad+\bar{\eta}\ (2j_1+n)\bar{\mathcal{F}}_1^{(1)}\mathbb{P}^{j_1+\frac{1}{2},j_2}_{n}\mathcal{F}_2^{(0)} \nonumber\\
        &\quad-\eta\bar{\eta}\ \frac{(2j_1+n)(2j_2+n)}{2(j_1+j_2+n)}\left[\bar{\mathcal{F}}_1^{(0)} \mathbb{P}^{j_1,j_2}_{n+1}\mathcal{F}_2^{(0)} -(-1)^{|\Psi_1|}(2j_1+2j_2+n)\ \bar{\mathcal{F}}_1^{(1)}\mathbb{P}^{j_1+\frac{1}{2},j_2+\frac{1}{2}}_{n}   \mathcal{F}_2^{(1)}\right]  \nonumber\\
        &\quad+\eta\bar{\eta}\frac{+j_1-j_2}{2\left(j_1+j_2+n\right)}\partial_s\left[\bar{\mathcal{F}}_1^{(0)}\mathbb{P}_n^{j_1,j_2}\mathcal{F}_2^{(0)}+(-1)^{|\Psi_1|}n\bar{\mathcal{F}}_1^{(1)} \mathbb{P}_{n+1}^{j_1+\frac{1}{2},j_2+\frac{1}{2}} \mathcal{F}_2^{(1)}\right]
\end{align}

\subsection{Field realization}
\label{sec:interp}
To go back to field theory in superspace we recall from section \ref{sec:sl21intro} that in the notation introduced in subsection \ref{sec:gencomm} a highest weight vector $\Psi_{j,b; j,j,b}$ formally corresponds to a collinear superconformal primary field $\Phi_{j,b}(0)$ evaluated at the origin, with collinear conformal spin $j$ and $b$-charge $b$
\begin{equation}
    \Psi_{j,b;j,j,b}= \Phi_{j,b}(0)\Psi_{\text{vac}}
\end{equation}
For the descendants, Eqs. \eqref{eq:collsupalg} and \eqref{eq:sl21dict} entail the correspondence
\begin{align}
\label{eq:desc2}
        & \mathbf{L}_+^n\Psi_{j,b;j,j,b} =  (-i)^n \underbrace{\left[\mathbf{P}_+,...\left[\mathbf{P}_+,\Phi_{j,b}(0)\right]...\right]}_{n \text{ commutators with } \textbf{P}_+}\Psi_{\text{vac}} \nonumber\\
        & \mathbf{L}_+^n\mathbf{V}_+\Psi_{j,b;j,j,b} = (-i)^n\left(\frac{i\varrho}{2}\right)\underbrace{\left[\mathbf{P}_+,...\left[\mathbf{P}_+,\left[\mathbf{Q}_1,\Phi_{j,b}(0)\right\}\right]...\right]}_{n \text{ commutators with } \textbf{P}_+}\Psi_{\text{vac}} \nonumber\\
        & \mathbf{L}_+^n\mathbf{W}_+\Psi_{j,b;j,j,b} = (-i)^n\left(-\frac{\varrho}{2}\right)\underbrace{\left[\mathbf{P}_+,...\left[\mathbf{P}_+,\left[\bar{\mathbf{Q}}_{\dot{1}},\Phi_{j,b}(0)\right\}\right]...\right]}_{n \text{ commutators with } \textbf{P}_+}\Psi_{\text{vac}}  \nonumber\\
        & \mathbf{L}_+^n\mathbf{V}_+\mathbf{W}_+\Psi_{j,b;j,j,b} = (-i)^n\left(\frac{i\varrho}{2}\right)\left(-\frac{\varrho}{2}\right)\underbrace{\left[\mathbf{P}_+,...\left[\mathbf{P}_+,\left[\mathbf{Q}_1,\left[\bar{\mathbf{Q}}_{\dot{1}},\Phi_{j,b}(0)\right\}\right\}\right]...\right]}_{n \text{ commutators with } \textbf{P}_+}\Psi_{\text{vac}}
\end{align}
From \eqref{eq:suptransl} it follows that the generating function \eqref{eq:differential} corresponds to\footnote{For the sake of brevity we write, as arguments of a (super)field, only the coordinates that are nonzero e.g. $\Phi_{j,b}(x^+,\theta^1,\bar{\theta}^{\dot{1}}) \equiv \Phi_{j,b}(x^+, x^-=0, x_{\perp}=0,\theta^1,\bar{\theta}^{\dot{1}}, \theta^2=0,\bar{\theta}^{\dot{2}}=0)$.}
\begin{equation}
\label{eq:fieldcorrespondence}
    \begin{gathered}
        \mathcal{F}_{j,b}(s,\eta,\bar{\eta})= \Phi_{j,b}(x^+,\theta^1,\bar{\theta}^{\dot{1}})\Psi_{\text{vac}}
    \end{gathered}
\end{equation}
provided that we identify
\begin{equation}
\label{eq:units}
    \begin{gathered}
        s= x^+\ , \qquad \eta=\frac{2}{\varrho}\theta^1\ , \qquad \bar{\eta}=\frac{2i}{\varrho}\bar{\theta}^{\dot{1}}    \\
        \mathcal{D}=\frac{\varrho}{2}D_1\ , \qquad \bar{\mathcal{D}}=\frac{i\varrho}{2}\bar{D}_{\dot{1}}
    \end{gathered}
\end{equation}
where again $\varrho=2^{1/4}$. Thanks to this correspondence we know that, after passing to the units \eqref{eq:units} the collinear superconformal algebra \eqref{eq:collsupalg} acts on a superfield $\Phi_{j,b}(x^+,x^-,\theta^1,\bar{\theta}^{\dot{1}})$ living on the light-cone as in Eq. \eqref{eq:differential}, with the $x^-$ coordinate being inert under these transformations. \par
Given two chiral collinear superconformal primary operators $\Phi_{j_1,\mp j_1}(0)$ and $\Phi_{j_2,\mp j_2}(0)$, we can translate to the units $x^+,\theta^1,\bar{\theta}^{\dot{1}}$ the expressions of \ref{sec:desc} and immediately infer that the operators
\begin{equation}
\label{eq:primaryreal}
\begin{gathered}
    \Phi_{j_1,\mp j_1}(0) \mathbb{C}^{j_1,j_2}_{2n+1}(\overleftarrow{D}_1+i\overleftarrow{\bar{D}}_{\dot{1}},\overrightarrow{D}_1+i\overrightarrow{\bar{D}}_{\dot{1}})\ \Phi_{j_2,\mp j_2}(0) \\
    \Phi_{j_1,\mp j_1}(0)\mathbb{C}^{j_1,j_2}_{2n+1}(\overleftarrow{D}_1+i\overleftarrow{\bar{D}}_{\dot{1}},\overrightarrow{D}_1+i\overrightarrow{\bar{D}}_{\dot{1}})\ \Phi_{j_2,\pm j_2}(0)
\end{gathered}
\end{equation}
are two-particle collinear superconformal primaries. Translating these operators along the light-cone, one obtains four objects that transforms irreducibly under the representation \eqref{eq:differential} (after passing to the units $s,\eta,\bar{\eta}$ in Eq. \eqref{eq:units}) and that can be seen as generating functions for the descendants \eqref{eq:desc2}.
\par One can also lift these operators to the whole superspace. In this case the corresponding superfields will be lifted to representations of the whole superconformal algebra. This is what we will do in sections \ref{sec:appliedcorrelators}, \ref{sec:sym}. This lifting is necessary to study the renormalization properties of these operators to order $g^2$, in which the theory is still superconformal \cite{Braun:2003rp}. The results of subsection \ref{sec:desc} allow us to easily extract the component fields of the operators in \eqref{eq:primaryreal} when the only nonzero odd coordinates are $\theta^1$ and $\bar{\theta}^{\dot{1}}$. We shall perform this calculation in subsection \ref{sec:components}.

\section{Generating functionals}
\label{sec:genfreescft}

\subsection{Introduction}
In the previous section, we formulated a recipe to construct towers of local collinear superconformal primary superfields in $\mathcal{N}=1$ supersymmetric field theories out of two local primary superfields. \par
In Refs. \cite{Bochicchio:2021nup,Bochicchio:2022uat, BPS41, BPS42}  it was found that the generating functional of connected correlators of bilinear operators made of free fields has the form of the logarithm of a functional superdeterminant of a Fredholm-type operator. So far, this object has been explicitly found only in some particular cases, namely YM theory, QCD, and $\mathcal{N}=1$ SYM theory in ordinary spacetime. In this section, we are going to generalize this construction to the operators \eqref{eq:primaryreal} in a superconformal field theory in superspace constructed out of free superfields. \par
The generating functional of connected correlators in superspace for the operators constructed in \ref{sec:interp} and realized in a free superconformal theory is shown in the subsections \ref{sec:dictionary}, \ref{sec:ordinary}. Since working directly with the bilinear operators \eqref{eq:primaryreal} is notationally demanding, in the two subsections \ref{sec:functional}, \ref{sec:correlators} we preliminarly derive some general formulae involving Gaussian functional integrals, involving both bosonic and fermionic variables, using an abstract notation. These formulae extend and generalize to superspace several identities that were employed in the previous works on the subject \cite{Bochicchio:2021nup,Bochicchio:2022uat, BPS41,BPS42}.

\subsection{Generating functionals for bilinear operators}
\label{sec:functional}
In this subsection we compute the generating functional of connected correlators of bilinear operators made of free fields. We use an abstract notation, in which the bosonic fields are $\phi_{ia},\bar{\phi}_{ia}$ and the fermionic fields are $\psi_{ia},\bar{\psi}_{ia}$. The subscripts $i$ and $a$ denote two kinds of indices the field may carry: the indices $i,j,\ell,...$ mimic the superspace coordinates, while the indices $a,b,..$ mimic the discrete indices. In order to construct the analogous of the polynomials $\mathbb{C}_n^{j_1j_2}$, we introduce the matrices $ {({C_n})_i}^{j\ell}$ acting on the two-boson or two-fermion monomials as
\begin{equation}
\label{eq:decomposition}
\begin{gathered}
    \sum_{j,\ell}{({C_n})_i}^{j\ell}\ \phi_{ja}^{(1)}\phi_{\ell b}^{(2)}=\sum_{j,\ell,k} {({u_{n,k})}_i}^j\phi_{j a}^{(1)}\ {(v_{n,k})_i}^{\ell}\phi_{l b}^{(2)} \\
    \sum_{j,\ell}{({C_n})_i}^{j\ell}\ \psi_{j a}^{(1)}\psi_{\ell b}^{(2)}=\sum_{j,\ell, k} {({u_{n,k})_i}^j}\psi_{j a}^{(1)}\ {(v_{n,k})_i}^{\ell}\psi_{\ell b}^{(2)}
\end{gathered}
\end{equation}
where ${({u_{n,k})}_i}^j, {({v_{n,k})}_i}^j$ are matrices that mimic the $(D_1+i\bar{D}_{\dot{1}})^k$ and $(D_1+i\bar{D}_{\dot{1}})^{n-k}$ in the $\mathbb{C}_n^{j_1j_2}$. In particular the indices $n,k$ keep track of the number of derivatives in each matrix, while the indices $i,j$ represent the action of the derivative on the superspace coordinates a certain field. \par 
Note that ${({u_{n,k})}_i}^j, {({v_{n,k})}_i}^j$ act on the fields from the left. The right action is \emph{defined} as
\begin{align}
    & \sum_{j}{({u_{n,k})}_i}^j\phi_{j a}^{(1)}=(-1)^{\lfloor\frac{k}{2}\rfloor}\sum_j\phi_{j a}^{(1)} \ ^{j}{_{i}({u_{n,k})}} \nonumber \\
    & \sum_j{({v_{n,k})}_i}^j\phi_{j a}^{(1)}=(-1)^{\lfloor\frac{n-k}{2}\rfloor}\sum_j\phi_{j a}^{(1)} \ {^j}{_{i}({v_{n,k}})} \nonumber \\
    & \sum_j{({u_{n,k})}_i}^j\psi_{j a}^{(1)}=(-1)^{|u_{n,k}|+\lfloor\frac{k}{2}\rfloor}\sum_j\psi_{j a}^{(1)}\ ^{j}{_{i}({u_{n,k})}} \nonumber \\
    & \sum_j{({v_{n,k})}_i}^j\psi_{j a}^{(1)}=(-1)^{|v_{n,k}|+\lfloor\frac{n-k}{2}\rfloor}\sum_j\psi_{j a}^{(1)}\ ^{j}{_{i}({v_{n,k})}}
\end{align}
where  the notation $(-1)^{\rvert \cdot\rvert}$ represents the statistics of a certain object, here of $u$ and $v$, and the floor function $\floor{x}$ is the largest integer $k\leq x$.\par
These definitions do not coincide with the supertransposition defined in super-linear algebra literature (see e.g. \cite{DeWitt:2012mdz}), and in the present work they have been postulated only to mimic the properties of objects like $(D_1+i\bar{D}_{\dot{1}})^k$, which, when acting on a function $f(x,\theta,\bar{\theta})$ satisfy $(D_1+i\bar{D}_{\dot{1}})^kf(x,\theta,\bar{\theta})=f(x,\theta,\bar{\theta})(\overleftarrow{D}_1+i\overleftarrow{\bar{D}}_{\dot{1}})^k(-1)^{k|f|+\lfloor\frac{k}{2}\rfloor}$.  We also assume the symmetry properties of the polynomials
\begin{align}
	\label{eq:symmetry}
    &\sum_{j,\ell}{({C_n})_i}^{j\ell}\ \phi_{j a}^{(1)}\phi_{\ell b}^{(2)}=(-1)^{\lfloor\frac{n+1}{2}\rfloor}\sum_{j,\ell}{({C_n})_i}^{\ell j}\ \phi_{\ell b}^{(2)}\phi_{j a}^{(1)} \nonumber\\
    &\sum_{j,\ell}{({C_n})_i}^{j\ell}\ \psi_{j a}^{(1)}\psi_{\ell b}^{(2)}=(-1)^{\lfloor\frac{n+1}{2}\rfloor+1}\sum_{j,\ell}{({C_n})_i}^{\ell j}\ \psi_{\ell b}^{(2)}\psi_{j a}^{(1)}
\end{align}
which are analogous to those found in appendix \ref{sec:csymm}. We assume the statistics of the ${({C_n})_i}^{j\ell}$ to be the same as of the polynomials in Eq. \eqref{eq:polynomialC}
\begin{align}
\label{eq:statistics}
    &(-1)^{|u_{n,k}|}=(-1)^k ,\qquad\,(-1)^{|v_{n,k}|} = (-1)^{n-k}\nonumber\\
     &(-1)^{|C_n|}=(-1)^{|u_{n,k}|}\,(-1)^{|v_{n,k}|}=(-1)^k (-1)^{n-k}=(-1)^n
\end{align}
notice that the statistics of $C_n$ is by definition the product of the statistics of its components $u$ and $v$. We also introduce a set of matrices $(t^{\alpha})^{ab}$ acting on the $a,b,...$ indices of the superfields. We choose a basis in which each of these matrices is either symmetric or antisymmetric
\begin{equation}
\label{eq:transpose}
    (t^{\alpha})^{ba}=(-1)^{|t^{\alpha}|}(t^{\alpha})^{ab}
\end{equation}
The matrices $t^{\alpha}$ have even ($+1$) statistics.

\paragraph{Bosonic case}
Let us consider a theory of bosonic free fields $\phi$ and $\bar{\phi}$, with propagator $\expval{\phi_{ai}\bar\phi_{bj}}=\Delta^{-1}_{ai, bj}$ and the bilinear operators that we denote, omitting the $i,j$ and $a,b$ indices, as
\begin{equation}
    \mathcal{O}_n^{\alpha}=(C_{2n}\otimes t^{\alpha})\cdot \bar\phi \phi\ , \qquad \mathcal{S}_n^{\alpha}=(C_{2n+1}\otimes t^{\alpha})\cdot \phi \phi\ , \qquad \bar{\mathcal{S}}_n^{\alpha}=(C_{2n+1}\otimes t^{\alpha})\cdot\bar\phi\bar\phi
\end{equation}
The symmetry properties of Eqs. \eqref{eq:symmetry} and \eqref{eq:transpose} require that in the operators $\mathcal{S}_n^{\alpha}$ and $\bar{\mathcal{S}}_n^{\alpha}$ the $t^{\alpha}$ are chosen so that
\begin{equation}
\label{eq:tssbarbos}
    (-1)^{n+1+|t^{\alpha}|}=+1 \qquad \text{for}\  \mathcal{S}_n^{\alpha}\ ,  \bar{\mathcal{S}}_n^{\alpha}
\end{equation}
No similar condition is required for the $\mathcal{O}_n^{\alpha}$. The generating functional for the correlators of arbitrary strings of these operators is
\begin{equation}
\label{eq:zeta}
    \mathcal{Z}[J]=\int[d\bar\phi][d\phi]\ \mathrm{exp}\ S(\phi,\bar\phi, J) 
\end{equation}
with:
\begin{equation}
\label{eq:zetaphi}
  S(\phi,\bar\phi, J)= -\sum_{i,j,a,b}\bar\phi_{ia}\ \Delta^{ia,jb}\phi_{jb}+\sum_{i,n,\alpha}\left[(\mathcal{O}_n^{\alpha})_i(J_{\mathcal{O}_n^{\alpha}})^i+(\mathcal{S}_n^{\alpha})_i(J_{\mathcal{S}_n^{\alpha}})^i+(\bar{\mathcal{S}}_n^{\alpha})_i(\bar{J}_{\mathcal{S}_n^{\alpha}})^i\right]
\end{equation}
Here and in the rest of this section, the symbol $\Delta^{ai,bj}$ represents the quadratic kernel of some kinetic term, and is not necessarily a Laplacian operator. If the theory has a gauge symmetry, some gauge-fixing is intended. Note that the external currents are \emph{on the right} and that they also possess an index $i$. Since the action must be even, Eq. \eqref{eq:statistics} require that we choose
\begin{equation}
\label{eq:jstatistics}
  (-1)^{|J_{\mathcal{O}_n^{\alpha}}|}=1\ , \qquad (-1)^{|J_{\mathcal{S}_n^{\alpha}}|} = (-1)^{|\bar{J}_{\mathcal{S}_n^{\alpha}}|}=-1
\end{equation}
Using the decomposition \eqref{eq:decomposition}, displacing the currents between the $u$'s and the $v$'s, and symmetrizing, the exponent takes the form
\begin{align}
    &S(\phi,\bar\phi, J)= -\bar\phi\ \Delta\phi \nonumber \\
    &+\sum_{n,k}\Bigg[\phi\ u_{2n+1,k}^{\mathrm{T}} (-1)^{k+1+\lfloor\frac{k}{2}\rfloor} t^{\alpha}J_{\mathcal{S}_n^{\alpha}} v_{2n+1,k}\phi+\bar{\phi}\ u_{2n+1,k}^{\mathrm{T}} (-1)^{k+1+\lfloor\frac{k}{2}\rfloor} t^{\alpha}\bar{J}_{\mathcal{S}_n^{\alpha}} v_{2n+1,k}\bar{\phi} \nonumber  \\
    &+\bar\phi\ u_{2n,k}^{\mathrm{T}}(-1)^{\lfloor\frac{k}{2}\rfloor}t^{\alpha}\frac{J_{\mathcal{O}_n^{\alpha}}}{2}\ v_{2n,k}\ \phi+\phi\ u_{2n,k}^{\mathrm{T}}(-1)^{n+|t^{\alpha}|+\lfloor\frac{k}{2}\rfloor}t^{\alpha}\frac{J_{\mathcal{O}_n^{\alpha}}}{2}\ v_{2n,k}\ \bar\phi\Bigg]
    \end{align}
where we used Eqs. \eqref{eq:statistics} and \eqref{eq:jstatistics} to displace the terms past each other, and the symmetry properties in Eqs. \eqref{eq:symmetry} and \eqref{eq:transpose} to symmetrize action. We have omitted the $i,j,...$ and $a,b,...$ indices to make the expression clearer, and $u_{n,k}^{\mathrm{T}}$ is the index-free notation for ${{^{i}}_j}(u_{n,k})$. 
\par In order to write this expression more compactly, we introduce a matrix notation for the fields and propagators
\begin{equation}
\label{eq:matrices}
    \Phi=\begin{pmatrix}
        \phi \\
        \bar\phi
    \end{pmatrix}\ , \qquad 
    \mathbf{\Delta}=\begin{pmatrix}
	& \Delta \\
	\Delta & 
\end{pmatrix} , \qquad 
     \mathbf{\Delta}^{-1}=\begin{pmatrix}
        & \Delta^{-1} \\
        \Delta^{-1} & 
    \end{pmatrix}\ 
\end{equation}
the components of the polynomials can also be organized as follows
\begin{equation}
\label{eq:uv}
    {U_{n,k}}=\begin{pmatrix}
	{u_{n,k}} & \\
	& {u_{n,k}}
\end{pmatrix}\ , \qquad 
{V_{n,k}}=\begin{pmatrix}
	{v_{n,k}} & \\
	&{ v_{n,k}}
\end{pmatrix} \ , \qquad 
\textbf{t}^{\alpha}=\begin{pmatrix}
     t^{\alpha} &  \\
      & t^{\alpha} \\
\end{pmatrix}
\end{equation}
the sources in matrix notation also read
\begin{equation}
\label{eq:jojs}
 J_{2n}^{\alpha}=\begin{pmatrix}
	& (-1)^{|t^{\alpha}|}\frac{J_{\mathcal{O}_n^{\alpha}}}{2} \\
	\frac{J_{\mathcal{O}_n^{\alpha}}}{2}  & \\
\end{pmatrix} \,\qquad
J_{2n+1}^{\alpha}=\begin{pmatrix}
	J_{\mathcal{S}_n^{\alpha}} & \\
	& \bar{J}_{\mathcal{S}_n^{\alpha}}
\end{pmatrix} 
\end{equation}
and we introduce a matrix $\mathcal{M}$ such that
\begin{equation}
\label{eq:m+}
   \mathcal{M}_{nk,n'k'}^+=\begin{cases}
	(-1)^{k+1} \delta_{nn'}\delta_{kk'}\mathds{1}_{2\times 2}\ , & n,n'\ \text{odd} \\
	\begin{pmatrix}
		(-1)^{\lfloor\frac{n}{2}\rfloor} & \\
		&  +1
	\end{pmatrix}\delta_{nn'}\delta_{kk'} \ , & n,n'\ \text{even} \\
	0\ , & \text{otherwise}
\end{cases}
\end{equation}
which allows us to write $\mathcal{Z}[J]$ as
\begin{equation}
    \mathcal{Z}_+[J]=\int [d\Phi]\ \mathrm{exp}\ S_+(\Phi, J) 
\end{equation}
with:
\begin{equation}
S(\Phi, J)=-\frac{1}{2}\Phi\Delta\Phi+\Phi\ U_{n,k}^{\mathrm{T}} (-1)^{\lfloor\frac{k}{2}\rfloor} \mathcal{M}^+_{nk,n'k'} \textbf{t}^{\alpha} J_{n'}^{\alpha}  V_{n',k'} \Phi
\end{equation}
we again used the index-free notation. The resulting generating functionals of correlators and of connected correlators are  
\begin{equation}
     \mathcal{Z}_+[J]=\mathrm{det}^{-\frac{1}{2}}\left({\delta_i}^{i'}{\delta_{a}}^b-2\sum_{\ell,j,a'}\sum_{n,k,n',k'}{(U_{n,k})_j}^{\ell} \mathbf{\Delta}^{-1}_{ia, \ell a'}\mathcal{M}^+_{nk,n'k'} (\textbf{t}^{\alpha})^{a'b}(J_{n'}^{\alpha})^j {(V_{n',k'})_{j}}^{i'}\right) 
\end{equation}
and
\begin{align}
	\label{eq:logdetinterm}
     \mathcal{W}_+[J]=&\mathrm{log}\ \mathcal{Z}_+[J] \nonumber \\
     =&-\frac{1}{2}\mathrm{tr}\ \mathrm{log}\left({\delta_i}^{i'}{\delta_{a}}^b-2\sum_{\ell,j,a'}\sum_{n,k,n',k'}{(U_{n,k})_j}^{\ell} \mathbf{\Delta}^{-1}_{ia, \ell a'}\mathcal{M}^+_{nk,n'k'} (\textbf{t}^{\alpha})^{a'b}(J_{n'}^{\alpha})^j {(V_{n',k'})_{j}}^{i'}\right)
\end{align}
notice that $\mathbf{\Delta}^{-1}$ is the $2\times 2$ matrix defined in Eq. \eqref{eq:matrices} and where we have interchanged $U$ and $ \mathbf{\Delta}^{-1}$ and restored the $i,j,\ell...$ indices. In Eq. \eqref{eq:logdetinterm} we used the property $\mathrm{log}\ \mathrm{det}= \mathrm{tr}\ \mathrm{log}$. All we have to do now is to mimic the procedure of Ref. \cite{Bochicchio:2022uat}, with the appropriate modifications due to the $\mathbb{Z}_2$-grading of the quantities: one has to expand the logarithm in a formal power series, and in each term displace the rightmost ${(V_{n,k})_{i}}^{j}$ on the left. In this way one can turn a trace in the indices $i$ in a trace in the indices $(i,n,k)$. However, since the $\mathbb{Z}_2$-grading of the ${(V_{n,k})_{i}}^{j}$ generally depends on the indices $n,k$, when ${(V_{n,k})_{i}}^{j}$ is odd the corresponding term changes takes an overall sign after the displacement. Hence, one finally ends up with a \emph{supertrace}, namely
{\small
\begin{align}
   &\mathcal{W}_+[J]= \nonumber \\
   =&-\frac{1}{2}\mathrm{str}\ \mathrm{log}\left({\delta_{i}}^{j}{\delta_a}^b\delta_{n_1k_1,n_2k_2}-2\sum_{i',j',a'}\sum_{n',k'}{(V_{n_1,k_1})_i}^{i'}{(U_{n',k'})_j}^{j'} \mathbf{\Delta}^{-1}_{i'a,j'a'}\mathcal{M}^+_{n'k',n_2,k_2}(\textbf{t}^{\alpha})^{a'b}(J_{n_2}^{\alpha})^j\right) \nonumber \\
   =&-\frac{1}{2}\mathrm{str}\ \mathrm{log}\left({\delta_{i}}^{j}{\delta_a}^b\delta_{n_1k_1,n_2k_2}-2\sum_{n',k',a'} (\mathbf{\Delta}^{-1}_{n_1k_1,n'k'})_{ia,ja'}\mathcal{M}^+_{n'k',n_2,k_2}(\textbf{t}^{\alpha})^{a'b}(J_{n_2}^{\alpha})^j\right)
\end{align}}
where we used the notation
\begin{equation}
    (\Delta^{-1}_{nk,n'k'})_{ia,jb}\equiv \sum_{i',j'}{(v_{n,k})_i}^{i'}{(u_{n',k'})_{j}}^{j'}\Delta^{-1}_{i'a,j'b}\ , \qquad \mathbf{\Delta}^{-1}_{nk,n'k'}=\begin{pmatrix}
        & \Delta^{-1}_{nk,n'k'} \\
        \Delta^{-1}_{nk,n'k'} & 
    \end{pmatrix}
\end{equation}
The supertrace is taken on the space of the indices $n,k,i$ and to the $2\times 2$ indices introduced with the matrices in Eq. \eqref{eq:matrices}. It is defined as 
\begin{equation}
\label{eq:str}
    \mathrm{str}\ X=\sum_{n,k,i,a}(-1)^{n-k}\mathrm{tr}_{2\times 2} X_{nkia,nkia}
\end{equation}
where $\mathrm{tr}$ is the partial trace over the $2\times 2$ indices. The factor $(-1)^{n-k}$ is the $\mathbb{Z}_2$-grading of the index $(n,k,i,a)$. Here, in the rest of this section and in the rest of this paper we will denote this quantity as $\mathrm{deg}(n,k,i,a)\in\{-1,+1\}$. Recall that the supertrace satisfies the familiar relation $\mathrm{str}\ \mathrm{log}(X)= \mathrm{log}\ \mathrm{sdet}(X)$ with the superdeterminant. More details on this procedure and on the appearance on the supertrace are shown in appendix \ref{app:supermatrix}. Using the definitions \eqref{eq:uv}, \eqref{eq:jojs}, \eqref{eq:m+} one finally finds
\begin{equation}
\label{eq:genfunbos}
\resizebox{1.00\textwidth}{!}{$
    \begin{split}
    & \mathcal{W}_+[J]=\mathrm{log}\ \mathcal{Z}_+[J]  \\
    =&-\frac{1}{2}\mathrm{str}\ \mathrm{log} \begin{pmatrix}
        \delta_{n_1k_1,n_2k_2}{\delta_{i}}^{j}{\delta_{a}}^{b}-\sum_n(\Delta^{-1}_{n_1\ k_1, 2n\ k_2})_{ia,ja'}(t^{\alpha})^{a'b}(J_{\mathcal{O}_n^{\alpha}})^j\delta_{2n,n_2} & 2\sum_n(\Delta^{-1}_{n_1\ k_1, 2n+1\ k_2})_{ia,ja'}(t^{\alpha})^{a'b}(\bar{J}_{\mathcal{S}_n^{\alpha}})^j(-1)^{k_2}\delta_{2n+1,n_2} \\
        2\sum_n(\Delta^{-1}_{n_1\ k_1, 2n+1\ k_2})_{ia,ja'}(t^{\alpha})^{a'b}(J_{\mathcal{S}_n^{\alpha}})^j(-1)^{k_2}\delta_{2n+1,n_2} & \delta_{n_1k_1,n_2k_2}{\delta_{i}}^j{\delta_{a}}^{b}-\sum_n(\Delta^{-1}_{n_1\ k_1, 2n\ k_2})_{ia,ja'}(t^{\alpha})^{a'b}(J_{\mathcal{O}_n^{\alpha}})^j(-1)^{n+|t^{\alpha}|}\delta_{2n,n_2}
    \end{pmatrix}
    \end{split}
    $}
\end{equation}
The residual trace over the $2\times 2$ indices can be computed with the rules in Eq. \eqref{eq:block}.

\paragraph{Fermionic case}
In the fermionic case, the procedure above can be carried out step by step, with minor modifications. Now, the bilinear operators are
\begin{equation}
    \mathcal{O}_n= (C_{2n}\otimes t^{\alpha})\cdot \bar\psi \psi\ , \qquad \mathcal{S}_n=(C_{2n+1}\otimes t^{\alpha})\cdot\psi \psi\ , \qquad \bar{\mathcal{S}}_n=(C_{2n+1}\otimes t^{\alpha})\cdot\bar\psi \bar\psi
\end{equation}
where, in analogy to the bosonic case, we choose the $t^{\alpha}$ that satisfy
\begin{equation}
\label{eq:tssbarferm}
    (-1)^{n+|t^{\alpha}|}=+1 \qquad \text{for}\  \mathcal{S}_n^{\alpha}\ ,  \bar{\mathcal{S}}_n^{\alpha}
\end{equation}
One has to use the integral \cite{Szabo:1996md}
\begin{equation}
    \int[d\Psi]\ e^{\frac{1}{2}\Psi A \Psi}=\mathrm{Pf}(A)
\end{equation}
where $\mathrm{Pf}(A)$ is the Pfaffian of the antisymmetric matrix $A$, which, as a polynomial in the matrix entries, enjoys the property $\mathrm{Pf}(A)^2=\mathrm{det}(A)$. Throughout the derivation, one obtains a generating functional analogous to that of Eq.\eqref{eq:logdetinterm}, except for the overall sign and for the appearance of the matrix
\begin{equation}
        \mathcal{M}_{nk,n'k'}^-=\begin{cases}
        \delta_{nn'}\delta_{kk'}\mathds{1}_{2\times 2}\ , & n,n'\ \text{odd} \\
        \begin{pmatrix}
            (-1)^{\lfloor\frac{n}{2}\rfloor+k+1} & \\
            &  (-1)^k
        \end{pmatrix}\delta_{nn'}\delta_{kk'} \ , & n,n'\ \text{even} \\
        0\ , & \text{otherwise}
    \end{cases}
\end{equation}
In the end, one ends up with
\begin{equation}
    \mathcal{W}_-[J]=+\frac{1}{2}\mathrm{str}\ \mathrm{log}\left({\delta_{i}}^{j}{\delta_a}^b\delta_{n_1k_1,n_2k_2}-2\sum_{n',k',a'} (\mathbf{\Delta}^{-1}_{n_1k_1,n'k'})_{ia,ja'}\mathcal{M}^-_{n'k',n_2,k_2}(\textbf{t}^{\alpha})^{a'b}(J_{n_2}^{\alpha})^j\right)
\end{equation}
or, more explicitly
\begin{equation}
\label{eq:genfunferm}
	\resizebox{1.00\textwidth}{!}{
	$
    \begin{aligned}
        &\mathcal{W}_-[J]=\mathrm{log}\ Z_-[J] \\
        =&+\frac{1}{2}\mathrm{str}\ \mathrm{log} \begin{pmatrix}
            \delta_{n_1k_1,n_2k_2}{\delta_{i}}^j{\delta_a}^b-\sum_n(\Delta^{-1}_{n_1\ k_1, 2n\ k_2})_{ia,ja'}(t^{\alpha})^{a'b}(J_{\mathcal{O}_n^{\alpha}})^j(-1)^{k_2}\delta_{2n,n_2} & -2\sum_n(\Delta^{-1}_{n_1\ k_1, 2n+1\ k_2})_{ia,ja'}(t^{\alpha})^{a'b}(\bar{J}_{\mathcal{S}_n^{\alpha}})^j\delta_{2n+1,n_2} \\
            -2\sum_n(\Delta^{-1}_{n_1 k_1, 2n+1\ k_2})_{ia,ja'}(t^{\alpha})^{a'b}(J_{\mathcal{S}_n^{\alpha}})^j\delta_{2n+1,n_2} & \delta_{n_1k_1,n_2k_2}{\delta_{i}}^j{\delta_a}^b-\sum_n(\Delta^{-1}_{n_1 k_1, 2n\ k_2})_{ia,ja'}(t^{\alpha})^{a'b}(J_{\mathcal{O}_n^{\alpha}})^j(-1)^{n+1+k_2+|t^{\alpha}|}\delta_{2n,n_2}
        \end{pmatrix}
    \end{aligned}
$}
\end{equation}
where again $\mathrm{deg}(n,k,i)=(-1)^{n-k}$. The residual trace over the $2\times 2$ indices can be computed with the rules in Eq. \eqref{eq:block}.

\subsection{Connected correlators}
\label{sec:correlators}
By differentiating the generating functionals, one can obtain the connected correlators between the operators $\mathcal{O}_n$, $\mathcal{S}_n$, $\bar{\mathcal{S}}_n$ of subsection \ref{sec:functional}. Before performing this computation, we present some simple identities that will be used. If $u_i$ are even variables, we have
\begin{equation}
    \frac{\partial}{\partial u_{i_1}}...\frac{\partial}{\partial u_{i_n}}\ u_{j_1}...u_{j_n}=\sum_{\sigma\in P_n}\delta_{i_1j_{\sigma(1)}}...\delta_{i_nj_{\sigma(n)}}
\end{equation}
where $P_n$ is the group of permutations of $n$ elements. If $\varepsilon_i$ and $\eta_i$ are odd variables, and $f$ is an analytic function, we have
\begin{equation}
    \left(- \frac{\partial}{\partial \eta_{i_1}}\right)...\left(- \frac{\partial}{\partial \eta_{i_n}}\right)f(\varepsilon\cdot \eta)=\varepsilon_{i_1}...\varepsilon_{i_n}\ f^{(n)}(\varepsilon\cdot\eta)
\end{equation}
Using this identity with $f(x)=x^n$ one can prove that
\begin{equation}
    \frac{\partial}{\partial \eta_{i_1}}... \frac{\partial}{\partial \eta_{i_n}}\eta_{j_1}...\eta_{j_n}=(-1)^{\frac{n(n-1)}{2}}\sum_{\sigma\in P_n}\mathrm{sgn}(\sigma)\delta_{i_1j_{\sigma(1)}}...\delta_{i_nj_{\sigma(n)}}
\end{equation}
\begin{align}
     \left(- \frac{\partial}{\partial \varepsilon_{i_1}}\right)\left(- \frac{\partial}{\partial \eta_{j_1}}\right)...&\left(- \frac{\partial}{\partial \varepsilon_{i_n}}\right)\left(- \frac{\partial}{\partial \eta_{j_n}}\right)\varepsilon_{k_1}\eta_{\ell_1}...\varepsilon_{k_n}\eta_{\ell_n} \nonumber \\
     =&(-1)^n\sum_{\sigma,\rho\in P_n}\mathrm{sgn}(\sigma)\ \mathrm{sgn}(\rho)\ \delta_{\ell_1j_{\sigma(1)}}...\delta_{\ell_nj_{\sigma(n)}}\ \delta_{k_1i_{\sigma(1)}}...\delta_{k_ni_{\sigma(n)}}
\end{align}
From these rules, it follows that given some \emph{odd} operators $F_n$ and currents $J_n$, the correct differentiation rule to obtain the correlators of the $F_n$ from its generating functional $\mathcal{Z}[J]=\displaystyle\int\mathrm{exp}\ (S+F_nJ_n)$ is
\begin{equation}
    \expval{F_{n_1}...F_{n_N}}=\frac{1}{\mathcal{Z}[0]}\left(-\frac{\partial}{\partial J_{n_1}}\right)...\left(-\frac{\partial}{\partial J_{n_N}}\right)\mathcal{Z}[J]\Bigg\rvert_{J=0}
\end{equation}
The same differentiation rule is valid for the generating functional of the connected correlators $\mathcal{W}[J]$. In the following paragraphs we will always omit the $i,j,\ell...$ and $a,b,...$ indices to have clearer expressions. The rule to recover them is the following
\begin{equation}
\begin{gathered}
    \Delta^{-1}_{n_1k_1, n_2k_2}t^{\alpha}\longrightarrow \left(\Delta^{-1}_{n_1k_1, n_2k_2}\right)_{i_1a_1,i_2b'}(t^{\alpha})^{b'a_2} \\
    t^{\alpha}\Delta^{-1}_{n_1k_1, n_2k_2}\longrightarrow (t^{\alpha})^{a_1b'}\left(\Delta^{-1}_{n_1k_1, n_2k_2}\right)_{i_1b',i_2a_2} \\
    J_{n_1}^{\alpha_1}\longrightarrow \left(J_{n_1}^{\alpha_1}\right)^{i_1}
\end{gathered}
\end{equation}
When, after this replacement, at least one lower and one upper $i_a$ index appear together, the sum over them over them is intended.

\paragraph{$\mathcal{O}_n^{\alpha}$ correlators in the bosonic theory}
Setting the currents $J_{\mathcal{S}}$, $\bar{J}_{\mathcal{S}}$ to zero, one obtains the generating functional
\begin{align}
\label{eq:ogenfun}
        \mathcal{W}_+[J_{\mathcal{O}}]=&-\frac{1}{2}\ \mathrm{log}\ \mathrm{sdet}\ \left[\delta_{n_1k_1, n_2k_2}-\Delta^{-1}_{2n_1\ k_1,2n_2\ k_2} t^{\alpha} J_{\mathcal{O}_{n_2^{\alpha}}}\right] \nonumber \\
        &-\frac{1}{2}\ \mathrm{log}\ \mathrm{sdet}\ \left[\delta_{n_1k_1, n_2k_2}-\Delta^{-1}_{2n_1\ k_1,2n_2\ k_2} (-1)^{n_2+|t^{\alpha}|} t^{\alpha} J_{\mathcal{O}_{n_2}^{\alpha}}\right] \nonumber \\
        =\sum_{M=1}^{\infty} \frac{1}{M} & \sum_{n_i,k_i,\alpha_i}\frac{1+(-1)^{\sum_i n_i+|t^{\alpha_i}|}}{2}(-1)^{k_1}\Delta^{-1}_{2n_1\ k_1, 2n_2\ k_2}t^{\alpha_2}...\Delta^{-1}_{2n_M\ k_M, 2n_1\ k_1}t^{\alpha_1}\ J_{\mathcal{O}_{n_1}^{\alpha_1}}...J_{\mathcal{O}_{n_M}^{\alpha_M}} 
\end{align}
Differentiating $M$ times one obtains the connected correlator
\begin{align}
    &\expval{\mathcal{O}_{n_1}^{\alpha_1}...\mathcal{O}_{n_M}^{\alpha_M}}_{\text{conn}}= \nonumber \\
    &\quad \frac{1}{M}\frac{1+(-1)^{\sum_i n_i+|t^{\alpha_i}|}}{2}\sum_{k_i}\sum_{\sigma\in P_M}(-1)^{k_{\sigma(1)}}\prod_{i=1}^M\Delta^{-1}_{2n_{\sigma(i)} k_{\sigma(i)}, 2n_{\sigma(i+1)} k_{\sigma(i+1)}}t^{\alpha_{\sigma(i+1)}}
\end{align}

\paragraph{$\mathcal{O}_n$ correlators in the fermionic theory}
The derivation is perfectly analogous to that of the previous paragraph. In the end, one obtains the result
\begin{align}
    &\expval{\mathcal{O}_{n_1}^{\alpha_1}...\mathcal{O}_{n_M}^{\alpha_M}}_{\text{conn}}=\nonumber\\
    &\quad-\frac{1}{M}\frac{1+(-1)^{M+\sum_i n_i+|t^{\alpha_i}|}}{2}\sum_{k_i}\sum_{\sigma\in P_M}(-1)^{k_{\sigma(1)}}\prod_{i=1}^M(-1)^{k_{\sigma(i)}}\Delta^{-1}_{2n_{\sigma(i)} k_{\sigma(i)}, 2n_{\sigma(i+1)} k_{\sigma(i+1)}}t^{\alpha_{\sigma(i+1)}}
\end{align}

\paragraph{$\mathcal{S}_n$, $\bar{\mathcal{S}}_n$ correlators in the bosonic theory}
Setting the currents $J_{\mathcal{O}}$ to zero, one obtains the generating functional
{\footnotesize
\begin{align}
    &\mathcal{W}_+[J_{\mathcal{S}},\bar{J}_{\mathcal{S}}]=\nonumber\\
    &-\frac{1}{2}\ \mathrm{log}\ \mathrm{sdet}\ \left[\delta_{n_1k_1,n_2k_2}-4\ \Delta^{-1}_{2n_1+1\ k_1, 2n+1\ k}t^{\alpha}\bar{J}_{\mathcal{S}_n^{\alpha}}(-1)^{k+1}\Delta^{-1}_{2n+1\ k,2n_2+1\ k_2}t^{\alpha_2}J_{\mathcal{S}_{n_2}^{\alpha_2}}(-1)^{k_2+1}\right]\nonumber \\
    &=\sum_{M=1}^{\infty}\frac{2^{2M-1}}{M}\sum_{\substack{n_i,k_i, \alpha_i \\ n_i',k_i',\alpha_i'}}(-1)^{k_1+1}\prod_{i=1}^M\Delta^{-1}_{2n_i+1\ k_i, 2n_i'+1\ k_i'}t^{\alpha_i'}\bar{J}_{\mathcal{S}_{n_i'}^{\alpha_i'}}(-1)^{k_i'}\Delta^{-1}_{2n_i'+1\ k_i',2n_{i+1}+1\ k_{i+1}}t^{\alpha_{i+1}}J_{\mathcal{S}_{n_{i+1}}^{\alpha_{i+1}}}(-1)^{k_{i+1}}
\end{align}}
Displacing all the currents to the left and differentiating, one obtains
\begin{align}
      &  \expval{\mathcal{S}_{n_1}^{\alpha_1}\bar{\mathcal{S}}_{n_1'}^{\alpha_1'}...\mathcal{S}_{n_M}^{\alpha_M}\bar{\mathcal{S}}_{n_M'}^{\alpha_M'}}_{\text{conn}}=+\frac{2^{2M-1}}{M}\sum_{k_i,k_i'}  \sum_{\rho,\sigma\in P_M}\mathrm{sgn}(\sigma)\mathrm{sgn}(\rho)\nonumber\\
        &\quad (-1)^{k_{\sigma(1)}}\prod_{i=1}^M\Delta^{-1}_{2n_{\sigma(i)}+1\ k_{\sigma(i)}, 2n_{\rho(i)}'+1\ k_{\rho(i)}'}t^{\alpha'_{\rho(i)}}\Delta^{-1}_{2n_{\rho(i)}'+1\ k_{\rho(i)}',2n_{\sigma(i+1)}+1\ k_{\sigma(i+1)}}t^{\alpha_{\sigma(i+1)}}
\end{align}

\paragraph{$\mathcal{S}_n$, $\bar{\mathcal{S}}_n$ correlators in the fermionic theory}
The derivation is perfectly analogous to that of the previous paragraph. In the end, one obtains the same selection rule for the correlators with a different number of $\mathcal{S}_n$ and $\bar{\mathcal{S}}_n$, and the result
\begin{align}
        &\expval{\mathcal{S}_{n_1}^{\alpha_1}\bar{\mathcal{S}}_{n_1'}^{\alpha_1'}...\mathcal{S}_{n_M}^{\alpha_M}\bar{\mathcal{S}}_{n_M'}^{\alpha_M'}}_{\text{conn}}=-\frac{2^{2M-1}}{M}\sum_{k_i,k_i'}  \sum_{\rho,\sigma\in P_M}\mathrm{sgn}(\sigma)\mathrm{sgn}(\rho)  \nonumber\\  &(-1)^{k_{\sigma(1)}}\prod_{i=1}^M(-1)^{k_{\sigma(i)}+k_{\rho(i)}'}\Delta^{-1}_{2n_{\sigma(i)}+1\ k_{\sigma(i)}, 2n_{\rho(i)}'+1\ k_{\rho(i)}'}t^{\alpha'_{\rho(i)}}\Delta^{-1}_{2n_{\rho(i)}'+1\ k_{\rho(i)}',2n_{\sigma(i+1)}+1\ k_{\sigma(i+1)}}t^{\alpha_{\sigma(i+1)}}
\end{align}

\subsection{From abstract notation to superspace}
\label{sec:dictionary}

We now see to what the quantities introduced above correspond when we have a field theory living in superspace. The variables $\phi,\bar\phi,\psi,\bar\psi$ are free chiral superfields transforming under some irreducible representation of the collinear superconformal algebra. We consider two free superfields with collinear conformal spin $j$, $\mathbb{Z}_2$-gradings $(-1)^{|\Phi|}=(-1)^{|\bar{\Phi}|}$, and write them as
\begin{equation}
\label{eq:phiphibar}
    \Phi_a(x_L,\theta)\ , \qquad \bar{\Phi}_a(x_R,\bar\theta)
\end{equation}
where $a$ denotes any other index (e.g. color or flavor) that is inert under the action of the collinear superconformal algebra. We indicate as
\begin{equation}
\label{eq:zetacoord}
\begin{gathered}
     Z=\left(x^{\mu},\theta^{\alpha},\bar{\theta}^{\dot{\alpha}}\right) \\
     \delta^{(8)}(Z_1,Z_2)=\delta^{(4)}(x_1-x_2)\delta^{(4)}(\theta_1-\theta_2)
\end{gathered}
\end{equation}
a general element of superspace and the delta function over superspace. The integration on superspace is defined as
\begin{equation}
    \int d^8Z=\int d^4x\ d^2\theta d^2\bar{\theta}
\end{equation}
where the $\int d^4x$ is an ordinary integration over spacetime, and the $\int d^2\theta d^2\bar{\theta}$ is the Berezin integration over the odd coordinates \cite{Buchbinder:1998qv}. Using this notation, we denote the two-point function as
\begin{equation}
    \expval{\Phi_a(Z_1)\bar\Phi_b(Z_2)}\equiv (\Delta^{-1})_{ab}(Z_1,Z_2)
\end{equation}
We choose the operators ${(u_{n,k})_i}^{i'}$ and ${(v_{n,k})_i}^{i'}$ to be
\begin{align}
    {(u_{n,k})_i}^{i'}\ \longleftrightarrow &\  \frac{1}{\Gamma(1+\lfloor\frac{k}{2}\rfloor)\Gamma(2j+\lfloor\frac{k+1}{2}\rfloor)}(-1)^{\lfloor\frac{k-|\Phi|}{2}\rfloor}(D_1+\bar{D}_{\dot{1}})^{k} \quad & (k=1,...,n) &&\nonumber \\
    {(v_{n,k})_i}^{i'}\ \longleftrightarrow &\  \frac{1}{\Gamma(1+\lfloor\frac{n-k}{2}\rfloor)\Gamma(2j+\lfloor\frac{n-k+1}{2}\rfloor)}(D_1+\bar{D}_{\dot{1}})^{n-k} \quad & (k=1,...,n) &&\nonumber \\
\end{align}
which means that
\begin{equation}
\label{eq:deltank}
\begin{alignedat}{2}
    (\Delta^{-1}_{nk,n'k'})_{ia,jb}\ \longleftrightarrow\ & \frac{1}{\Gamma(1+\lfloor\frac{n-k}{2}\rfloor)\Gamma(2j+\lfloor\frac{n-k+1}{2}\rfloor)}\frac{1}{\Gamma(1+\lfloor\frac{k'}{2}\rfloor)\Gamma(2j+\lfloor\frac{k'+1}{2}\rfloor)}(-1)^{\lfloor\frac{k'+1-|\Phi|}{2}\rfloor} && \\
     & (D_1^{(1)}+i\bar{D}_{\dot{1}}^{(1)})^{n-k}(D_1^{(2)}+i\bar{D}_{\dot{1}}^{(2)})^{k'} (\Delta^{-1})_{ab}(Z_1,Z_2) &&
\end{alignedat} 
\end{equation}
The relative order of the chiral covariant derivatives acting on $Z_1$ and $Z_2$  (with superscripts $^{(1,2)}$ respectively) is not arbitrary due to their odd statistics. The matrices $(t^{\alpha})^{ab}$ that act on the discrete indices of the superfields ar defined as in subsection \ref{sec:functional}.
\par In this dictionary, the abstract index $i$ corresponds to the superspace coordinate $Z$, while the index $a$ and the indices $(n,k)$  have the same meaning of subsection \ref{sec:functional}. The supertrace of a matrix $X$ possessing $(Z,a,n,k)$ and the $2\times 2$ indices introduced above is defined as
\begin{equation}
\label{eq:supertrace}
    \mathrm{str}\ X=\sum_{n,k,a}\int d^8Z\ \mathrm{deg}(n,k)\ \mathrm{tr}_{2\times 2}X_{nka,nka}(Z,Z)\ , \qquad \mathrm{deg}(n,k)\in\{-1,+1\}
\end{equation}
The residual trace over the $2\times 2$ indices can be computed with the rules in Eq. \eqref{eq:block}. A possible source of confusion is that although the spacetime coordinates $Z=(x^{\mu},\theta^{\alpha},\bar{\theta}^{\dot{\alpha}})$ are respectively even, odd and odd, they do not possess a $\mathbb{Z}_2$-grading as long as this definition of supertrace in Eq \eqref{eq:supertrace} is concerned. The $\mathbb{Z}_2$-grading associated to the statistics, and the $\mathbb{Z}_2$-grading as defined by the supertrace \eqref{eq:supertrace} are two independent notions. \\

To familiarize with this notation, it is useful to rewrite the two generating functionals \eqref{eq:genfunbos}, \eqref{eq:genfunferm} making the indices run in superspace
{\footnotesize
\begin{align}
\label{eq:supergenfun}
    &\mathcal{W}_{\pm}[J]=\nonumber\\
    &\mp\frac{1}{2}\mathrm{str}\ \mathrm{log}\left[\delta^{(8)}(Z_1,Z_2){\delta_{a}}^{b}\mathds{1}_{2\times 2}\delta_{n_1k_1,n_2k_2}-2 \sum_{n',k',a',\alpha} (\mathbf{\Delta}^{-1}_{n_1k_1,n'k'})_{aa'}(Z_1,Z_2)\mathcal{M}^{\pm}_{n'k',n_2k_2}(\textbf{t}^{\alpha})^{a'b}J_{n_2}^{\alpha}(Z_2)\right]
\end{align}}
where the upper sign is for bosonic theories and the lower sign is for fermionic theories. The supertrace is taken over all indices. The $2\times 2$  and superspace indices are not graded, while the grading of $(n,k)$ is $\mathrm{deg}(n,k)=(-1)^{n-k}$. Following the same passages in section 10 of Ref. \cite{Bochicchio:2021nup}, it is possible to express Eq. \eqref{eq:supergenfun} in momentum space. If we define
\begin{align}
\label{eq:fourier}
   & \int d^4x_1\ d^4x_2\   (\mathbf{\Delta}^{-1})_{ab}(x_1,\theta_1,\bar{\theta_1}; x_2,\theta_2,\bar{\theta_2})\ e^{-ip_1x_1-ip_2x_2} \nonumber\\
   &\equiv (2\pi)^4\delta^{(4)}(p_1+p_2)(\widetilde{ \mathbf{\Delta}}^{-1})_{ab}(p_1; \theta_1,\bar{\theta}_1, \theta_2,\bar{\theta}_2)
\end{align}
we obtain
{\small
\begin{align}
         \mathcal{W}_{\pm}[J]=  \mp &\frac{1}{2}\mathrm{str}\ \mathrm{log} \Bigg[ (2\pi)^4\delta^{(4)}(p_1-p_2)\delta^{(4)}(\theta_1-\theta_2){\delta_{a}}^{b}\mathds{1}_{2\times 2}\delta_{n_1k_1,n_2k_2}-\nonumber \\
         -&2\sum_{n',k',a',\alpha}( \widetilde{ \mathbf{\Delta}}^{-1}_{n_1k_1,n'k'})_{aa'}(p_1; \theta_1,\bar{\theta}_1, \theta_2,\bar{\theta}_2)\mathcal{M}^{\pm}_{n'k',n_2k_2}(\textbf{t}^{\alpha})^{a'b}\widetilde{J}_{n_2}^{\alpha}(p_1-p_2; \theta_1,\bar{\theta}_1, \theta_2,\bar{\theta}_2)
         \Bigg]
\end{align}}

\subsection{From superspace to ordinary space}
\label{sec:ordinary}
Let us consider the (unnormalized) generating functional of connected correlators of some set of composite operators $\mathcal{O}_n(Z)$
\begin{equation}
    \mathcal{W}[J]=\mathrm{log}\int [d\phi]\ \mathrm{exp}\left(S(\phi)+\int d^8Z\ \mathcal{O}_n^{\alpha}(Z)J_n^{\alpha}(Z)\right)
\end{equation}
The operators $\mathcal{O}_n^{\alpha}(Z)$ admit a component expansion (the order matters)
\begin{equation}
    \mathcal{O}_n^{\alpha}(Z)=\sum_A O_n^{\alpha, A}(x)e_A(\theta)
\end{equation}
where the $e_A(\theta)$ form a complete basis of monomials in the odd coordinates $\theta,\bar\theta$.
We choose the source, without any loss of generality, to be
\begin{equation}
\label{eq:currentchoice}
    J_n^{\alpha}(Z)=J_n^{\alpha}(x)\delta^{(4)}(\theta-\theta')
\end{equation}
and formally define
\begin{equation}
\label{eq:kdef}
    K_{n,A}^{\alpha}(x)\equiv e_A(\theta')J_n^{\alpha}(x)
\end{equation}
our generating functional can be rewritten as
\begin{equation}
     \mathcal{W}[K]=\mathrm{log}\int [d\phi]\ \mathrm{exp}\left(S(\phi)+\int d^4x\ O_n^A(x)K_{n, A}(x)\right)
\end{equation}
which generates the connected correlators of the $O_n^A(x)$ as
\begin{equation}
    \expval{O_{n_1}^{\alpha_1, A_1}(x_1)...O_{n_M}^{\alpha_M, A_M}(x_M)}_{\mathrm{conn}}=\left(\pm\frac{\partial}{\partial K_{n_1, A_1}^{\alpha_1}(x_1)}\right)...\left(\pm\frac{\partial}{\partial K_{n_M, A_M}^{\alpha_M}(x_M)}\right)\mathcal{W}[K]\bigg\rvert_{K=0}
\end{equation}
where the signs are positive for bosonic operators and negative for fermionic operators.
\par What is the form of the resulting generating functional? As a preliminary consideration, we note that the vectors $e_A(\theta)$ form an algebra
\begin{equation}
\label{eq:grassmannalgebra}
    e_A(\theta)e_B(\theta)=\sum_C{T_{AB}}^C e_C(\theta)
\end{equation}
where the ${T_{AB}}^C$ the structure constants of the algebra. To make it concrete, let us consider the case in which the operators in the generating functional are restricted to the light-cone \eqref{eq:lc}. We only have two odd coordinates $\theta^1$ and $\bar{\theta}^{\dot{1}}$. We choose the vectors $e_A(\theta)$ to be
\begin{equation}
\label{eq:esmall}
    e_1(\theta)=1\ , \qquad e_2(\theta)=\theta^1\ , \qquad e_3(\theta)=\bar\theta^{\dot{1}}\ , \qquad e_4(\theta)=\theta^1\bar\theta^{\dot{1}}
\end{equation}
Of course, this basis is not unique, and can be changed through any invertible linear transformation. The corresponding structure constants written in matrix form are
\begin{align}
	\label{eq:Talgebra}
    & T^1=\begin{pmatrix}
        +1 & 0 & 0 & 0 \\
        0 & 0 & 0 & 0 \\
        0 & 0 & 0 & 0 \\
        0 & 0 & 0 & 0 \\
    \end{pmatrix}\ , 
    && T^2=\begin{pmatrix}
        0 & +1 & 0 & 0 \\
        +1 & 0 & 0 & 0 \\
        0 & 0 & 0 & 0 \\
        0 & 0 & 0 & 0 \\
    \end{pmatrix} \nonumber \\
    &T^3=\begin{pmatrix}
        0 & 0 & +1 & 0 \\
        0 & 0 & 0 & 0 \\
        +1 & 0 & 0 & 0 \\
        0 & 0 & 0 & 0 \\
    \end{pmatrix}\ ,
    &&T^4=\begin{pmatrix}
        0 & 0 & 0 & +1 \\
        0 & 0 & +1 & 0 \\
        0 & -1 & 0 & 0 \\
        +1 & 0 & 0 & 0 \\
    \end{pmatrix}
\end{align}
Let us now consider the generating functional \eqref{eq:supergenfun}. To lighten the notation, we omit the color/flavor-like indices introduced in Eq. \eqref{eq:phiphibar}. We can choose the currents as in \eqref{eq:currentchoice} and decompose each kernel as (the order matters)
\begin{equation}
     (\mathbf{\Delta}^{-1}_{ns,n's'})_{aa'}(Z,Z')=\sum_{A,A'}e_A(\theta) (\mathbf{\Delta}^{-1}_{nk,n'k'})^{AA'}_{aa'}(x,x')e_{A'}(\theta')
\end{equation}
Plugging this expression into the generating functional, displacing the $e_A(\theta)$, by using the rules \eqref{eq:grassmannalgebra}, we end up with a new generating functional
\begin{equation}
\label{eq:genfunord}
\begin{split}
    &\mathcal{W}_{\pm}[K]=\mp\frac{1}{2}\mathrm{str}\ \mathrm{log}\Big[\delta^{(4)}(x_1-x_2){\delta^A}_B\mathds{1}_{2\times 2}\delta_{n_1k_1,n_2k_2}{\delta_a}^b \\
    -&2\sum_{A',C,n',k',a',\alpha}(-1)^{|e_{B}|(|K^{n_2}_C|+|e_{C}|)}  (\mathbf{\Delta}^{-1}_{n_1k_1,n'k'})^{AA'}_{aa'}(x_1,x_2)\mathcal{M}^{\pm}_{n'k',n_2k_2}(\textbf{t}^{\alpha})^{a'b}{T_{A'B}}^CK_{n_2,C}^{\alpha}(x_2)\Big]
\end{split}
\end{equation}
where the $\mathbb{Z}_2$-grading of the indices is $\mathrm{deg}(n,k,A)=(-1)^{n-k+|e_A|}$. A significant example of application of this method can be found in subsection \ref{sec:check}.

\section{Application to free SCFTs}
\label{sec:appliedcorrelators}

In this section, we apply the constructions of sections \ref{sec:sl21}, \ref{sec:genfreescft} to a superconformal field theory with bosonic and fermionic free fields in $\mathcal{N}=1$ superspace transforming irreducibly under arbitrary representations of the collinear superconformal algebra. The analysis of this section relies on the dictionaries described in subsections \ref{sec:interp}, \ref{sec:dictionary}.

\subsection{Superfields}
We consider a pair hermitian conjugate of chiral bosonic superfields $\Phi$, $\bar\Phi$ and a pair of hermitian conjugate chiral fermionic superfields $\Psi$, $\bar\Psi$ with additional discrete indices denoted by lowercase Latin letters $a,b,...$. We assume these superfields to be elementary, and we work in a gauge in which they are primaries under the collinear superconformal algebra. As in appendix \ref{app:supconfcorr}, we assume them to be the components with maximal spin along the light-cone of some superfields, and hence they transform irreducibly under the collinear superconformal group. They are free fields with nonzero two-point correlators\footnote{In Minkowski spacetime, the denominators of the propagators must be intended to include a negative positive imaginary infinitesimal $-i\varepsilon$ e.g. $x_{1\bar{2}}^2$ must be read as $x_{1\bar{2}}^2-i\varepsilon$ and $p^2$ as $p^2+i\varepsilon$. We will omit them in the rest of the paper.}
\begin{align}
\label{eq:prop}
     &\expval{\Phi_a(x_{L,1},\theta_1)\bar{\Phi}_b(x_{R,2},\bar{\theta}_2)}=C_{\Phi}\delta_{ab}\ ^{\Phi}\Delta^{-1}(Z_1,Z_2)=C_{\Phi}\delta^{ab}\frac{(x_{1\bar{2}}^-)^{\ell_{\Phi}}}{(x_{1\bar{2}}^2)^{2j_{\Phi}}} \\
     &\expval{\Psi_a(x_{L,1},\theta_1)\bar{\Psi}_b(x_{R,2},\bar{\theta}_2)}=C_{\Psi}\delta_{ab}\ ^{\Psi}\Delta^{-1}(Z_1,Z_2)=C_{\Psi}\delta_{ab}\frac{(x_{1\bar{2}}^-)^{\ell_{\Psi}}}{(x_{1\bar{2}}^2)^{2j_{\Psi}}}
\end{align}
The components of the superfields along the light-cone are
\begin{align}
\label{eq:lccomp}
    &\Phi\rvert_{\text{l.c.}}=\phi^{(1)}(x^+,x^{-})+\frac{2}{\varrho}\theta^1\ \phi^{(2)}(x^+,x^{-})\ , &&  \Psi\rvert_{\text{l.c.}}=\psi^{(1)}(x^+,x^{-}) +\frac{2}{\varrho}\theta^1\ \psi^{(2)}(x^+,x^{-}) \nonumber \\
    & \bar{\Phi}\rvert_{\text{l.c.}}=\bar{\phi}^{(1)}(x^+,x^{-})-\frac{2}{\varrho}\bar{\theta}^{\dot{1}}\ \bar{\phi}^{(2)}(x^+,x^{-}) \ ,  && \bar{\Psi}\rvert_{\text{l.c.}}=\bar{\psi}^{(1)}(x^+,x^{-}) -\frac{2}{\varrho}\bar{\theta}^{\dot{1}} \bar{\psi}^{(2)}(x^+,x^{-}) 
\end{align}
where to lighten the notation we omitted the remaining discrete indices. We remind the reader that in this paper $\varrho=2^{1/4}$ and that $x_{1\bar{2}}^{\mu}$ is the supertranslation invariant interval defined as
\begin{equation}
\begin{gathered}
    x_{1\bar{2}}^{\mu}=-x_{\bar{1}2}^{\mu}=x_1^{\mu}-x_2^{\mu}-i\theta_1^{\alpha}(\sigma^{\mu})_{\alpha\dot{\alpha}}\bar{\theta}_1^{\dot{\alpha}}-i\theta_2^{\alpha}(\sigma^{\mu})_{\alpha\dot{\alpha}}\bar{\theta}_2^{\dot{\alpha}}+2i\theta_1^{\alpha}(\sigma^{\mu})_{\alpha\dot{\alpha}}\bar{\theta}_2^{\dot{\alpha}}
\end{gathered}
\end{equation}
for future use we also define
\begin{equation}
    \theta_{12}^{\alpha}= \theta_{1}^{\alpha}- \theta_{2}^{\alpha} \ , \qquad \bar{\theta}_{12}^{\dot\alpha}=\bar{\theta}_{1}^{\dot\alpha}-\bar{\theta}_{2}^{\dot\alpha}
\end{equation}
See also Appendix \ref{app:supconfcorr}. Expanding the two-point correlators \eqref{eq:prop} in the odd coordinates, one can obtain the two-point correlators for the components
\begin{align}
\label{eq:propcomp}
         & \expval{\phi^{(1)}_a(x_1)\bar{\phi}^{(1)}_b(x_1)}=\delta_{ab}C_{\Phi}\frac{(x_{12}^-)^{\ell_{\Phi}}}{(x_{12}^2)^{2j_{\Phi}}} \nonumber\\
         &\expval{\phi^{(2)}_a(x_1)\bar{\phi}^{(2)}_b(x_1)}= \delta_{ab} \left(-4iC_{\Phi}j_{\Phi}\right)\frac{(x_{12}^-)^{\ell_{\Phi}+1}}{(x_{12}^2)^{2j_{\Phi}+1}} \nonumber\\
         &\expval{\psi^{(1)}_a(x_1)\bar{\psi}^{(1)}_b(x_1)}=\delta_{ab}C_{\Psi}\frac{(x_{12}^-)^{\ell_{\Psi}}}{(x_{12}^2)^{2j_{\Psi}}} \nonumber\\
         &\expval{\psi^{(2)}_a(x_1)\bar{\psi}^{(2)}_b(x_1)}=\delta_{ab}(+4iC_{\Psi}j_{\Psi})\frac{(x_{12}^-)^{\ell_{\Psi}+1}}{(x_{12}^2)^{2j_{\Psi}+1}}
\end{align}
We hope that the similarity of this symbol with the standard translation invariant interval $x_{12}^{\mu}=x_1^{\mu}-x_2^{\mu}$ will not be a source of confusion for the reader.

\subsection{Superconformal operators}
We now construct the superconformal operators. Since we are interested mostly in the applications to $\mathcal{N}=1$ SYM (see section \ref{sec:sym}) we will not construct superconformal operators that mix the bosonic and fermionic superfields. \\

We have three towers of superconformal operators made of bosonic superfields
\begin{align}
\label{eq:bostower}
    ^{\Phi}\mathbb{O}_n^{\alpha}(x,\theta,\bar{\theta})=& C_{\Phi}^{-1} i^{n+|t^{\alpha}|} (t^{\alpha})^{ab} \  \bar{\Phi}_a(x_{R},\bar{\theta})\ \mathbb{C}_{2n}^{j_{\Phi}j_{\Phi}}\left(\overleftarrow{D}_1+i\overleftarrow{\bar{D}}_{\dot{1}}, \overrightarrow{D}_1+i\overrightarrow{\bar{D}}_{\dot{1}}\right)\Phi_b(x_{L},\theta) \nonumber\\
    ^{\Phi}\mathbb{S}_n^{\alpha}(x,\theta,\bar{\theta})=& C_{\Phi}^{-1} i^{n+|t^{\alpha}|}(t^{\alpha})^{ab} \ \Phi_a(x_{L},\theta)\ \mathbb{C}_{2n+1}^{j_{\Phi}j_{\Phi}}\left(\overleftarrow{D}_1+i\overleftarrow{\bar{D}}_{\dot{1}}, \overrightarrow{D}_1+i\overrightarrow{\bar{D}}_{\dot{1}}\right)\Phi_b(x_{L},\theta) \nonumber\\
    ^{\Phi}\bar{\mathbb{S}}_n^{\alpha}(x,\theta,\bar{\theta})=&C_{\Phi}^{-1} i^{n+|t^{\alpha}|}(t^{\alpha})^{ab} \ \bar{\Phi}_a(x_{R},\bar{\theta})\ \mathbb{C}_{2n+1}^{j_{\Phi}j_{\Phi}}\left(\overleftarrow{D}_1+i\overleftarrow{\bar{D}}_{\dot{1}}, \overrightarrow{D}_1+i\overrightarrow{\bar{D}}_{\dot{1}}\right)\bar{\Phi}_b(x_{R},\bar{\theta})
\end{align}
and three towers of superconformal operators made of fermionic superfields
\begin{align}
\label{eq:fermtower}
    ^{\Psi}\mathbb{O}_n^{\alpha}(x,\theta,\bar{\theta})=&C_{\Psi}^{-1}  i^{n+|t^{\alpha}|}(t^{\alpha})^{ab} \  \bar{\Psi}_a(x_{R},\bar{\theta})\ \mathbb{C}_{2n}^{j_{\Psi}j_{\Psi}}\left(\overleftarrow{D}_1+i\overleftarrow{\bar{D}}_{\dot{1}}, \overrightarrow{D}_1+i\overrightarrow{\bar{D}}_{\dot{1}}\right)\Psi_b(x_{L},\theta) \nonumber\\
    ^{\Psi}\mathbb{S}_n^{\alpha}(x,\theta,\bar{\theta})=&C_{\Psi}^{-1} i^{n+|t^{\alpha}|}(t^{\alpha})^{ab} \ \Psi_a(x_{L},\theta)\ \mathbb{C}_{2n+1}^{j_{\Psi}j_{\Psi}}\left(\overleftarrow{D}_1+i\overleftarrow{\bar{D}}_{\dot{1}}, \overrightarrow{D}_1+i\overrightarrow{\bar{D}}_{\dot{1}}\right)\Psi_b(x_{L},\theta) \nonumber\\
    ^{\Psi}\bar{\mathbb{S}}_n^{\alpha}(x,\theta,\bar{\theta})=&C_{\Psi}^{-1} i^{n+|t^{\alpha}|}(t^{\alpha})^{ab} \ \bar{\Psi}_a(x_{R},\bar{\theta})\ \mathbb{C}_{2n+1}^{j_{\Psi}j_{\Psi}}\left(\overleftarrow{D}_1+i\overleftarrow{\bar{D}}_{\dot{1}}, \overrightarrow{D}_1+i\overrightarrow{\bar{D}}_{\dot{1}}\right)\bar{\Psi}_b(x_{R},\bar{\theta})
\end{align}
We refer to the operators made of the two fields of opposite chirality as \emph{balanced}, and to those made of two fields with the same chirality as \emph{unbalanced} \cite{Bochicchio:2021nup}. The matrices $(t^{\alpha})_{ab}$ act on the discrete indices of the elementary superfields, and they are chosen to have definite parity 
\begin{equation}
    (t^{\alpha})^{ba}=(-1)^{|t^{\alpha}|}(t^{\alpha})^{ab}
\end{equation}
and satisfy the conditions \eqref{eq:tssbarbos}, \eqref{eq:tssbarferm} so that none of the operators in Eqs. \eqref{eq:bostower}, \eqref{eq:fermtower} vanishes. Thanks to the factors $i^{n+|t^{\alpha}|}$ we have the hermiticity relations
\begin{align}
     &\left(^{\Phi}\mathbb{O}_n^{\alpha}\right)^{\dagger}= ^{\Phi}\mathbb{O}_n^{\alpha} \ , \qquad && \left(^{\Phi}\mathbb{S}_n^{\alpha}\right)^{\dagger} = ^{\Phi}\bar{\mathbb{S}}_n^{\alpha} \nonumber \\
     &\left(^{\Psi}\mathbb{O}_n^{\alpha}\right)^{\dagger}= ^{\Psi}\mathbb{O}_n^{\alpha}\ , \qquad && \left(^{\Psi}\mathbb{S}_n^{\alpha}\right)^{\dagger} = ^{\Psi}\bar{\mathbb{S}}_n^{\alpha}
\end{align}
The collinear superconformal charges of these operators are shown on the table \eqref{tab:collcharges}.

\begin{table}[h!]
\centering
    \begin{tabular}{C|CCCC}
        & \ell & \bar{\ell} & j & b  \\
        \hline
        ^{\Phi}\mathbb{O}_n & \ell_{\Phi}+n  & \ell_{\Phi}+n & 2j_{\Phi}+n & 0 \\
        ^{\Phi}\mathbb{S}_n & 2\ell_{\Phi}+n+1 & n & 2j_{\Phi}+n+\frac{1}{2} & -2j_{\Phi}+\frac{1}{2}  \\
        ^{\Phi}\bar{\mathbb{S}}_n  & n & 2\ell_{\Phi}+n+1 & 2j_{\Phi}+n+\frac{1}{2} & +2j_{\Phi}-\frac{1}{2}  \\
         \hline
         ^{\Psi}\mathbb{O}_n & \ell_{\Psi}+n  & \ell_{\Psi}+n & 2j_{\Psi}+n & 0   \\
        ^{\Psi}\mathbb{S}_n & 2\ell_{\Psi}+n+1 & n & 2j_{\Psi}+n+\frac{1}{2} & -2j_{\Psi}+\frac{1}{2}  \\
        ^{\Psi}\bar{\mathbb{S}}_n & n &  2\ell_{\Psi}+n+1 & 2j_{\Psi}+n+\frac{1}{2} & +2j_{\Psi}-\frac{1}{2}  \\
    \end{tabular}
    \caption{Collinear superconformal charges of the operators in Eqs. \eqref{eq:bostower}, \eqref{eq:fermtower}}
    \label{tab:collcharges}
\end{table}

\subsection{Components}
\label{sec:components}
The components of the operators in \eqref{eq:bostower} and \eqref{eq:fermtower} evaluated along the light-cone \eqref{eq:lc} form superconformal multiplets. According to the analysis of subsections \ref{sec:sl21intro} and \ref{sec:interp} it is possible to obtain the component fields of our operators evaluated on the light-cone by starting from the generating functions \eqref{eq:genfun++}, \eqref{eq:genfun--}, \eqref{eq:genfun+-}, \eqref{eq:genfun-+} by replacing
\begin{align}
        &  s \longrightarrow x^+ && \partial_s \longrightarrow \partial_+\qquad && \nonumber\\
        &\eta \longrightarrow \frac{2}{\varrho}\theta^1 && \bar{\eta} \longrightarrow \frac{2i}{\varrho}\bar{\theta}^{\dot{1}}\nonumber\\
        & \mathcal{D} \longrightarrow \frac{\rho}{2}D_1 && \bar{\mathcal{D}} \longrightarrow \frac{i\rho}{2}\bar{D}_{\dot{1}} \nonumber\\
        & \mathcal{F}^{(0)}(s) \longrightarrow \phi^{(1)},\psi^{(1)} && \bar{\mathcal{F}}^{(0)}(s) \longrightarrow \bar{\phi}^{(1)},\bar{\psi}^{(1)}  \nonumber\\
        & \mathcal{F}^{(1)}(s) \longrightarrow \phi^{(2)},\psi^{(2)}  && \bar{\mathcal{F}}^{(1)}(s) \longrightarrow \bar{\phi}^{(2)},\bar{\psi}^{(2)}
\end{align}
where $\varrho=2^{1/4}$. In each of the operators \eqref{eq:bostower}, \eqref{eq:fermtower} the two fields appearing on the left and on the right have the same collinear superconformal spin. Hence, it is convenient to express their components not in terms of the $\mathbb{P}^{j_1,j_2}_n$ introduced in Eq. \eqref{eq:symbol1}, but through the Jacobi and Gegenbauer polynomials, whose relation with the $\mathbb{P}^{j_1,j_2}_n$ is shown in Eqs. \eqref{eq:jacgeg0}. Hence, we define the quantities
\begin{equation}
    2j_{\Phi}=\alpha_{\Phi}+\frac{1}{2}\ , \qquad 2j_{\Psi}=\alpha_{\Psi}+\frac{1}{2}
\end{equation}
To obtain clearer expressions, we omit all the discrete indices of our fields and use the condensed notation
\begin{align}
\label{eq:condensed}
      &  \bar{\phi}^{(1)} C^{\alpha_{\Phi}}_n \phi^{(1)} \equiv \bar{\phi}^{(1)}(0) (i\overleftarrow{\partial}_++i\overrightarrow{\partial}_+)^nC^{\alpha_{\Phi}}_n\left(\frac{\overleftarrow{\partial}_+-\overrightarrow{\partial}_+}{\overleftarrow{\partial}_++\overrightarrow{\partial}_+}\right) \phi^{(1)}(0)\nonumber \\
        &\bar{\phi}^{(1)} P^{(2j_{\Phi}-1,2j_{\Phi})}_n \psi^{(2)} \equiv \bar{\phi}^{(1)}(0) (i\overleftarrow{\partial}_++i\overrightarrow{\partial}_+)^nP^{(2j_{\Phi}-1,2j_{\Phi})}_n\left(\frac{\overleftarrow{\partial}_+-\overrightarrow{\partial}_+}{\overleftarrow{\partial}_++\overrightarrow{\partial}_+}\right) \phi^{(2)}(0) 
\end{align}
and similarly for all the other possible combinations of component fields. The $i$'s inside the factors $(i\overleftarrow{\partial}+i\overrightarrow{\partial})^n$ have been inserted to make contact with the previous literature \cite{Bochicchio:2021nup,Bochicchio:2022uat,BPS41,BPS42}. One has to be careful in keeping track of the $i$'s that are absent from the definition of $\mathbb{P}^{j_1,j_2}_n$ but appear in Eq. \eqref{eq:condensed}. After these warnings, we are ready to show the results. For the bosonic sector, we have
\begin{subequations}
\label{eq:osphi}
\begin{equation}
\begin{split}
    ^{\Phi}\mathbb{O}_n=&C_{\Phi}^{-1}\frac{(-1)^ni^{|t^{\alpha}|}2^{\frac{3}{2}n}\Gamma(2\alpha_{\Phi})}{\Gamma\left(n+\alpha_{\Phi}+\frac{1}{2}\right)\Gamma\left(2\alpha_{\Phi}+n\right)\Gamma\left(\alpha_{\Phi}+\frac{1}{2}\right)} \\
    \Bigg\{ & \left(\bar{\phi}^{(1)}C_n^{\alpha_{\Phi}}\phi^{(1)}-\frac{4\alpha_{\Phi}}{2\alpha_{\Phi}+n}\bar{\phi}^{(2)}C_{n-1}^{\alpha_{\Phi}+1}\phi^{(2)}\right) \\
    +&\frac{2}{\varrho}\theta^1 \frac{\Gamma\left(2\alpha_{\Phi}+n\right)\Gamma\left(\alpha_{\Phi}+\frac{1}{2}\right)}{\Gamma\left(n+\alpha_{\Phi}+\frac{1}{2}\right)\Gamma(2\alpha_{\Phi})} \bar{\phi}^{(1)}P_n^{\left(\alpha_{\Phi}-\frac{1}{2},\alpha_{\Phi}+\frac{1}{2}\right)}\phi^{(2)} \\
    -&\frac{2}{\varrho} \bar{\theta}^{\dot{1}} \frac{\Gamma\left(2\alpha_{\Phi}+n\right)\Gamma\left(\alpha_{\Phi}+\frac{1}{2}\right)}{\Gamma\left(n+\alpha_{\Phi}+\frac{1}{2}\right)\Gamma(2\alpha_{\Phi})}\ \bar{\phi}^{(2)}P_n^{\left(\alpha_{\Phi}+\frac{1}{2},\alpha_{\Phi}-\frac{1}{2}\right)}\phi^{(1)} \\
    -& 2\sqrt{2}\theta\bar{\theta}^{\dot{1}} \frac{(n+1)}{2(n+2\alpha_{\Phi})} \left(\bar{\phi}^{(1)}C_{n+1}^{\alpha_{\Phi}}\phi^{(1)}+\frac{4\alpha_{\Phi}}{n+1}\bar{\phi}^{(2)}C_{n}^{\alpha_{\Phi}+1}\phi^{(2)}\right)\Bigg\}
\end{split}
\end{equation}
\begin{equation}
		\resizebox{1.00\textwidth}{!}{
		$
    \begin{split}
        &^{\Phi}\mathbb{S}_n= -C_{\Phi}^{-1}\frac{(-1)^ni^{|t^{\alpha}|}2^{\frac{3}{4}(2n+1)}}{\Gamma\left(\alpha_{\Phi}+n+\frac{1}{2}\right)\Gamma\left(\alpha_{\Phi}+n+\frac{3}{2}\right)} \\
        \Bigg\{&  \left(\phi^{(2)}P_{n}^{\left(\alpha_{\Phi}+\frac{1}{2},\alpha_{\Phi}-\frac{1}{2}\right)}\phi^{(1)}- \phi^{(1)}P_{n}^{\left(\alpha_{\Phi}-\frac{1}{2},\alpha_{\Phi}+\frac{1}{2}\right)}\phi^{(2)}\right) \\
        - & \frac{2}{\varrho}\theta^1 \frac{\Gamma(2\alpha_{\Phi}+2)\Gamma\left(\alpha_{\Phi}+n+\frac{1}{2}\right)}{\Gamma\left(2\alpha_{\Phi}+n+1\right)\Gamma\left(\alpha_{\Phi}+\frac{3}{2}\right)} \phi^{(2)}C_{n}^{\alpha_{\Phi}+1}\phi^{(2)} \\
        +& \frac{2}{\varrho}\bar{\theta} ^{\dot{1}} \frac{(n+1)\Gamma(2\alpha_{\Phi})\Gamma\left(\alpha_{\Phi}+n+\frac{1}{2}\right)}{\Gamma\left(2\alpha_{\Phi}+n+1\right)\Gamma\left(\alpha_{\Phi}+\frac{1}{2}\right)}\ \phi^{(1)} C_{n+1}^{\alpha_{\Phi}} \phi^{(1)} \\
        +& i2\sqrt{2}\theta^1\bar{\theta}^{\dot{1}}\Bigg[ i\frac{(2\alpha_{\Phi}+n+1)(n+1)}{2(\alpha_{\Phi}+n+1)(\alpha_{\Phi}+n+\frac{1}{2})}\left(\phi^{(2)}P_{n+1}^{\left(\alpha_{\Phi}+\frac{1}{2},\alpha_{\Phi}-\frac{1}{2}\right)}\phi^{(1)}+ \phi^{(1)}P_{n+1}^{\left(\alpha_{\Phi}-\frac{1}{2},\alpha_{\Phi}+\frac{1}{2}\right)}\phi^{(2)}\right) \\
        -&\frac{\alpha_{\Phi}}{2(\alpha_{\Phi}+n+1)}\partial_+\left(\phi^{(2)}P_{n}^{\left(\alpha_{\Phi}+\frac{1}{2},\alpha_{\Phi}-\frac{1}{2}\right)}\phi^{(1)}- \phi^{(1)}P_{n}^{\left(\alpha_{\Phi}-\frac{1}{2},\alpha_{\Phi}+\frac{1}{2}\right)}\phi^{(2)}\right) \Bigg]\Bigg\}    \end{split}
    $}
\end{equation}
\end{subequations}
while for the fermionic sector we have
\begin{subequations}
\label{eq:ospsi}
\begin{equation}
\begin{split}
    ^{\Psi}\mathbb{O}_n=&C_{\Psi}^{-1}\frac{(-1)^ni^{|t^{\alpha}|}2^{\frac{3}{2}n}\Gamma(2\alpha_{\Psi})}{\Gamma\left(n+\alpha_{\Psi}+\frac{1}{2}\right)\Gamma\left(2\alpha_{\Psi}+n\right)\Gamma\left(\alpha_{\Psi}+\frac{1}{2}\right)} \\
    \Bigg\{ & \left(\bar{\psi}^{(1)}C_n^{\alpha_{\Psi}}\psi^{(1)}+\frac{4\alpha_{\Psi}}{2\alpha_{\Psi}+n}\bar{\psi}^{(2)}C_{n-1}^{\alpha_{\Psi}+1}\psi^{(2)}\right) \\
    -&\frac{2}{\varrho}\theta^1 \frac{\Gamma\left(2\alpha_{\Psi}+n\right)\Gamma\left(\alpha_{\Psi}+\frac{1}{2}\right)}{\Gamma\left(n+\alpha_{\Psi}+\frac{1}{2}\right)\Gamma(2\alpha_{\Psi})} \bar{\psi}^{(1)}P_n^{\left(\alpha_{\Psi}-\frac{1}{2},\alpha_{\Psi}+\frac{1}{2}\right)}\psi^{(2)} \\
    -&\frac{2}{\varrho} \bar{\theta}^{\dot{1}} \frac{\Gamma\left(2\alpha_{\Psi}+n\right)\Gamma\left(\alpha_{\Psi}+\frac{1}{2}\right)}{\Gamma\left(n+\alpha_{\Psi}+\frac{1}{2}\right)\Gamma(2\alpha_{\Psi})}\ \bar{\psi}^{(2)}P_n^{\left(\alpha_{\Psi}+\frac{1}{2},\alpha_{\Psi}-\frac{1}{2}\right)}\psi^{(1)} \\
    -& 2\sqrt{2}\theta\bar{\theta}^{\dot{1}} \frac{(n+1)}{2(n+2\alpha_{\Psi})} \left(\bar{\psi}^{(1)}C_{n+1}^{\alpha_{\Psi}}\psi^{(1)}-\frac{4\alpha_{\Psi}}{n+1}\bar{\psi}^{(2)}C_{n}^{\alpha_{\Psi}+1}\psi^{(2)}\right)\Bigg\}
\end{split}
\end{equation}
\begin{equation}
		\resizebox{1.00\textwidth}{!}{
		$
    \begin{split}
        &^{\Psi}\mathbb{S}_n= C_{\Psi}^{-1}\frac{(-1)^ni^{|t^{\alpha}|}2^{\frac{3}{4}(2n+1)}}{\Gamma\left(\alpha_{\Psi}+n+\frac{1}{2}\right)\Gamma\left(\alpha_{\Psi}+n+\frac{3}{2}\right)} \\
        \Bigg\{&  \left(\psi^{(2)}P_{n}^{\left(\alpha_{\Psi}+\frac{1}{2},\alpha_{\Psi}-\frac{1}{2}\right)}\psi^{(1)}+ \psi^{(1)}P_{n}^{\left(\alpha_{\Psi}-\frac{1}{2},\alpha_{\Psi}+\frac{1}{2}\right)}\psi^{(2)}\right) \\
        + & \frac{2}{\varrho}\theta^1 \frac{\Gamma(2\alpha_{\Psi}+2)\Gamma\left(\alpha_{\Psi}+n+\frac{1}{2}\right)}{\Gamma\left(2\alpha_{\Psi}+n+1\right)\Gamma\left(\alpha_{\Psi}+\frac{3}{2}\right)} \psi^{(2)}C_{n}^{\alpha_{\Psi}+1}\psi^{(2)} \\
        +& \frac{2}{\varrho}\bar{\theta} ^{\dot{1}} \frac{(n+1)\Gamma(2\alpha_{\Psi})\Gamma\left(\alpha_{\Psi}+n+\frac{1}{2}\right)}{\Gamma\left(2\alpha_{\Psi}+n+1\right)\Gamma\left(\alpha_{\Psi}+\frac{1}{2}\right)}\ \psi^{(1)} C_{n+1}^{\alpha_{\Psi}} \psi^{(1)} \\
        +& i2\sqrt{2}\theta^1\bar{\theta}^{\dot{1}}\Bigg[ i\frac{(2\alpha_{\Psi}+n+1)(n+1)}{2(\alpha_{\Psi}+n+1)(\alpha_{\Psi}+n+\frac{1}{2})}\left(\psi^{(2)}P_{n+1}^{\left(\alpha_{\Psi}+\frac{1}{2},\alpha_{\Psi}-\frac{1}{2}\right)}\psi^{(1)}- \psi^{(1)}P_{n+1}^{\left(\alpha_{\Psi}-\frac{1}{2},\alpha_{\Psi}+\frac{1}{2}\right)}\psi^{(2)}\right) \\
        -&\frac{\alpha_{\Psi}}{2(\alpha_{\Psi}+n+1)}\partial_+\left(\psi^{(2)}P_{n}^{\left(\alpha_{\Psi}+\frac{1}{2},\alpha_{\Psi}-\frac{1}{2}\right)}\psi^{(1)}+ \psi^{(1)}P_{n}^{\left(\alpha_{\Psi}-\frac{1}{2},\alpha_{\Psi}+\frac{1}{2}\right)}\psi^{(2)}\right) \Bigg]\Bigg\}    \end{split}
    $}
\end{equation}
\end{subequations}
with $\varrho=2^{1/4} $. The operators $^{\Phi}\bar{\mathbb{S}}_n$ and $^{\Psi}\bar{\mathbb{S}}_n$ can be obtained from $^{\Phi}\mathbb{S}_n$ and $^{\Psi}\mathbb{S}_n$ with the substitutions
\begin{align}
    & \theta^1 \to i\bar{\theta}^{\dot{1}}\ , && \bar{\theta}^{\dot{1}} \to -i\theta^1 \nonumber \\
    & \phi^{(1)}\to \bar{\phi}^{(1)}\ ,  && \phi^{(2)} \to i\bar{\phi}^{(2)} \nonumber \\
    & \psi^{(1)} \to  \bar{\psi}^{(1)} \ ,  && \psi^{(2)} \to i\bar{\psi}^{(2)}
\end{align}

\subsection{Two-point correlators}
\label{sec:2pt}
From the values in table \eqref{tab:collcharges} and the results in appendix \ref{app:supconfcorr}, it is immediate to infer the form of the two-point correlators between superconformal operators. They are
\begin{align}
\label{eq:twopt}
        \expval{^{\Phi}\mathbb{O}_n^{\alpha}(Z_1)^{\Phi}\mathbb{O}^{\beta}_m(Z_2)}=&  ^{\Phi}\mathcal{C}_{\mathbb{O}_n} (-1)^ni^{|t^{\alpha}|+|t^{\beta}|}\mathrm{tr}\left(t^{\alpha}t^{\beta}\right)\delta_{nm}\frac{(x_{1\bar{2}}^-)^{\ell_{\Phi}+n}(x_{\bar{1}2}^-)^{\ell_{\Phi}+n}}{(x_{1\bar{2}}^2)^{2j_{\Phi}+n}(x_{\bar{1}2}^2)^{2j_{\Phi}+n}}
        \nonumber\\
        \expval{^{\Psi}\mathbb{O}^{\alpha}_n(Z_1)^{\Psi}\mathbb{O}^{\beta}_m(Z_2)}=& ^{\Psi}\mathcal{C}_{\mathbb{O}_n} (-1)^ni^{|t^{\alpha}|+|t^{\beta}|} \mathrm{tr}\left(t^{\alpha}t^{\beta}\right)\delta_{nm}\frac{(x_{1\bar{2}}^-)^{\ell_{\Psi}+n}(x_{\bar{1}2}^-)^{\ell_{\Psi}+n}}{(x_{1\bar{2}}^2)^{2j_{\Psi}+n}(x_{\bar{1}2}^2)^{2j_{\Psi}+n}}
        \nonumber\\
        \expval{^{\Phi}\mathbb{S}_n^{\alpha}(Z_1)^{\Phi}\bar{\mathbb{S}}^{\beta}_m(Z_2)}=&  ^{\Phi}\mathcal{C}_{\mathbb{S}_n} (-1)^ni^{|t^{\alpha}|+|t^{\beta}|} \mathrm{tr}\left(t^{\alpha}t^{\beta}\right)\delta_{nm}\frac{(x_{1\bar{2}}^-)^{2\ell_{\Phi}+n+1}(x_{\bar{1}2}^-)^{n}}{(x_{1\bar{2}}^2)^{4j_{\Phi}+n}(x_{\bar{1}2}^2)^{n+1}}
        \nonumber\\
        \expval{^{\Psi}\mathbb{S}^{\alpha}_n(Z_1)^{\Psi}\bar{\mathbb{S}}^{\beta}_m(Z_2)}=& ^{\Psi}\mathcal{C}_{\mathbb{S}_n} (-1)^ni^{|t^{\alpha}|+|t^{\beta}|} \mathrm{tr}\left(t^{\alpha}t^{\beta}\right)\delta_{nm}\frac{(x_{1\bar{2}}^-)^{2\ell_{\Phi}+n+1}(x_{\bar{1}2}^-)^{n}}{(x_{1\bar{2}}^2)^{4j_{\Psi}+n}(x_{\bar{1}2}^2)^{n+1}}
\end{align}
where the factor $(-1)^ni^{|t^{\alpha}|+|t^{\beta}|}$ comes from the factors $i^{n+|t^{\alpha}|}$ in Eqs. \eqref{eq:bostower} and \eqref{eq:fermtower}. The computation of the normalization constants in Eq. \eqref{eq:twopt} relies on the following identity \cite{Bochicchio:2021nup}: Let $\chi(x),\bar{\chi}(x)$ and $\xi(x),\bar{\xi}(x)$ be pairs of hermitian conjugate ordinary fields with two-point correlators
\begin{equation}
    \expval{\chi\bar{\chi}}=\frac{(x_{12}^-)^{\ell_{\chi}}}{(x_{12}^2)^{2j_{\chi}}}\ , \qquad \expval{\xi\bar{\xi}}=\frac{(x_{12}^-)^{\ell_{\xi}}}{(x_{12}^2)^{2j_{\xi}}}
\end{equation}
We omitted the argument of the fields inside the correlator for brevity. Then, by using the same condensed notation of Eq. \eqref{eq:condensed}, we have
\begin{align}
\label{eq:chixi}
        &\expval{\bar{\chi}P_n^{(2j_{\chi}-1,2j_{\xi}-1)}\xi\ \bar{\xi}P_n^{(2j_{\xi}-1,2j_{\chi}-1)}\chi} =\delta_{nm}(-4)^n\left(2j_{\chi}\right)_n\left(2j_{\xi}\right)_n\binom{2j_{\chi}+2j_{\xi}+n}{n}\frac{(x_{12}^-)^{\ell{\chi}+\ell_{\xi}}}{(x_{12}^2)^{2(j_{\chi}+j_{\xi}+n)}}\nonumber \\
        &\expval{\bar{\chi}C_n^{\alpha_{\chi}}\chi\ \bar{\chi}C_m^{\alpha_{\chi}}\chi}=\delta_{nm}\delta_{nm}(-4)^n\left[\left(2\alpha_{\chi}\right)_n\right]^2\binom{2\alpha_{\chi}+n+1}{n}\frac{(x_{12}^-)^{\ell{\chi}+\ell_{\xi}}}{(x_{12}^2)^{2(\alpha_{\chi}+n+\frac{1}{2})}}
\end{align}
where, as in section \ref{sec:sl21}, we used the symbol $(2a)_n\equiv \Gamma(2a+n)/\Gamma(2a)$ and the binomial coefficients in these and in the following expressions must be read as $\binom{\alpha}{\beta}=\frac{\Gamma(\alpha+1)}{\Gamma(\beta+1)\Gamma(\alpha-\beta+1)}$, since we cannot assume \emph{a priori} that the arguments are nonnegative integers. This identity can be proven by evaluating the correlators \eqref{eq:chixi} on the light-cone, and using the Schwinger parametrization
\begin{equation}
    \frac{\Gamma(2a)}{(x_{12}^+)^{2a}}=\int_0^{\infty} du\, u^{2a-1} \,e^{-ux_{12}^+}\ , \qquad \text{for}\ x_{12}^+>0
\end{equation}
Thanks to this parametrization, the Jacobi and Gegenbauers polynomials take, as arguments, the Schwinger parameters, and after a change of variables it is possible to apply on them the orthogonality relations
\begin{align}
        &\int_{-1}^{+1}dz\ (1-z)^{j_1}(1+z)^{j_2}P_n^{j_1,j_2}(z)P_m^{j_1,j_2}(z)=\nonumber \\
        &\delta_{nm}\frac{2^{j_1+j_2+1}}{2n+j_1+j_2+1}\frac{\Gamma\left(n+j_1+1\right)\Gamma\left(n+j_2+1\right)}{n!\Gamma\left(n+j_1+j_2+1\right)}
    \end{align}
        and
        \begin{align}
        \int_{-1}^{+1}(1-x^2)^{\alpha-\frac{1}{2}}C_n^{\alpha}(x)C_m^{\alpha}(x)=\delta_{nm}\frac{\pi 2^{1-2\alpha}\Gamma(n+2\alpha)}{n!(n+\alpha)\left[\Gamma(\alpha)\right]^2}
\end{align}
One then must substitute everything back into the original expression and use Lorentz invariance to lift the correlator from the light-cone to the whole spacetime.\par
In our case, the identities \eqref{eq:chixi} must be applied to the lowest component of the operators \eqref{eq:osphi}, \eqref{eq:osphi}, using the propagators for the components shown in Eq. \eqref{eq:propcomp}. The resulting expression must be compared with Eq. \eqref{eq:twopt} evaluated at $\theta^{\alpha}=\bar{\theta}^{\dot{\alpha}}=0$. We stress the fact that there is no loss of generality in this procedure, since the form of the correlators in Eq. \eqref{eq:twopt} is dictated by superconformal invariance alone. The results of the matching are
\begin{align}
\label{eq:csco}
        ^{\Phi}\mathcal{C}_{\mathbb{O}_n}=&2^{5n}\frac{1}{\left[\Gamma(2j_{\Phi})\Gamma(2j_{\Phi}+n)\right]^2}\binom{4j_{\Phi}+2n+1}{n+1} \nonumber\\
        ^{\Psi}\mathcal{C}_{\mathbb{O}_n}=& 2^{5n}\frac{1}{\left[\Gamma(2j_{\Psi})\Gamma(2j_{\Psi}+n)\right]^2}\binom{4j_{\Psi}+2n+1}{n+1} \nonumber\\
        ^{\Phi}\mathcal{C}_{\mathbb{S}_n}=&  -2^{5n+\frac{9}{2}}\frac{1}{(2j_{\Phi}+n)}\frac{1}{[\Gamma(2j_{\Phi})\Gamma(2j_{\Phi}+n)]^2}\binom{4j_{\Phi}+2n+1}{n} \nonumber\\
        ^{\Psi}\mathcal{C}_{\mathbb{S}_n}=&-2^{5n+\frac{9}{2}}\frac{1}{(2j_{\Psi}+n)}\frac{1}{[\Gamma(2j_{\Psi})\Gamma(2j_{\Psi}+n)]^2}\binom{4j_{\Psi}+2n+1}{n}
\end{align}

\subsection{Generating functionals}
\label{sec:genfunct1}
In this subsection, we summarize the results of section \ref{sec:genfreescft} applied to the our case. Since the $\Phi$- and the $\Psi$-sector are decoupled, the generating functional for the connected correlators is then a sum of a "bosonic" and a "fermionic" generating functional
\begin{equation}
    \mathcal{W}[J]=\mathcal{W}_{\Phi}[J]+\mathcal{W}_{\Psi}[J]
\end{equation}
where
\begin{subequations}
\label{eq:wphipsi}
{\footnotesize
\begin{align}
    &\mathcal{W}_{\Phi}[J]=\nonumber\\
    &-\frac{1}{2}\mathrm{str}\ \mathrm{log}\left[\delta^{(8)}(Z_1,Z_2)\delta_{ab}\mathds{1}_{2\times 2}\delta_{n_1k_1,n_2k_2}-2\ ^{\Phi}\mathbf{\Delta}^{-1}_{n_1k_1,n'k'}(Z_1,Z_2)\  ^{\Phi}\mathcal{M}_{n'k',n_2k_2}i^{\lfloor\frac{n_2}{2}\rfloor+|t^{\alpha}|}(\mathbf{t}^{\alpha})^{ab}\ {^{\Phi}J^{\alpha}_{n_2}}(Z_2)\right]
\end{align}}
and
{\footnotesize
\begin{align}
    &\mathcal{W}_{\Psi}[J]=\nonumber\\
    &+\frac{1}{2}\mathrm{str}\ \mathrm{log}\left[\delta^{(8)}(Z_1,Z_2)\delta^{ab}\mathds{1}_{2\times 2}\delta_{n_1k_1,n_2k_2}-2\  ^{\Psi} \mathbf{\Delta}^{-1}_{n_1k_1,n'k'}(Z_1,Z_2)\ ^{\Psi}\mathcal{M}_{n'k',n_2k_2}i^{\lfloor\frac{n_2}{2}\rfloor+|t^{\alpha}|}(\mathbf{t}^{\alpha})^{ab}\ {^{\Psi}J_{n_2}^{\alpha}}(Z_2)\right]
\end{align}}
\end{subequations}
The factor $i^{\lfloor\frac{n_2}{2}\rfloor+|t^{\alpha}|}$ comes from the normalization of the operators in Eqs. \eqref{eq:bostower} and \eqref{eq:fermtower}. For brevity we omitted the sum over $n',k'$. We remind the reader that $\mathrm{deg}(n,k)=(-1)^{n-k}$ and the partial supertrace over the $2\times 2$ indices is computed by means of the matrix identities in the appendix \ref{app:supermatrix}. The matrices $^{\Phi}\mathcal{M}$ and $^{\Psi}\mathcal{M}$ are
    \begin{align}
    \label{eq:phipsimatrix}
        &^{\Phi}\mathcal{M}_{nk,n'k'}=\begin{cases}
        (-1)^{k+1} \delta_{nn'}\delta_{kk'}\mathds{1}_{2\times 2}\ , & n,n'\ \text{odd} \\
        \begin{pmatrix}
            (-1)^{\lfloor\frac{n}{2}\rfloor} & 0 \\
            0 &  1
        \end{pmatrix}\delta_{nn'}\delta_{kk'} \ , & n,n'\ \text{even} \\
        0\ , & \text{otherwise}
    \end{cases}\nonumber\\
        &^{\Psi}\mathcal{M}_{nk,n'k'}=\begin{cases}
        \delta_{nn'}\delta_{kk'}\mathds{1}_{2\times 2}\ , & n,n'\ \text{odd} \\
        \begin{pmatrix}
            (-1)^{\lfloor\frac{n}{2}\rfloor+k+1} & 0\\
            0 &  (-1)^k
        \end{pmatrix}\delta_{nn'}\delta_{kk'} \ , & n,n'\ \text{even} \\
        0\ , & \text{otherwise}
        \end{cases}\nonumber\\
    \end{align}
The currents $^{\Phi}J_n$ and $^{\Psi}J_n$ are
\begin{subequations}
\label{eq:currents}
    \begin{equation}
        ^{\Phi}J_{2n+1}^{\alpha}(Z)=\begin{pmatrix}
        ^{\Phi}J_{\mathbb{S}_n^{\alpha}}(Z) & 0 \\
        0 & ^{\Phi}\bar{J}_{\mathbb{S}_n^{\alpha}}(Z)
    \end{pmatrix}\ , \qquad 
        ^{\Phi}J_{2n}^{\alpha}(Z)=\begin{pmatrix}
        0 & (-1)^{|t^{\alpha}|}\frac{^{\Phi}J_{\mathbb{O}_n^{\alpha}}(Z)}{2} \\
        \frac{^{\Phi}J_{\mathbb{O}_n^{\alpha}}(Z)}{2}  & 0 \\
    \end{pmatrix}    
    \end{equation}
and
        \begin{equation}
        ^{\Psi}J_{2n+1}^{\alpha}(Z)=\begin{pmatrix}
        ^{\Psi}J_{\mathbb{S}_n^{\alpha}}(Z) & 0 \\
        0 & ^{\Psi}\bar{J}_{\mathbb{S}_n^{\alpha}}(Z)
    \end{pmatrix}\ , \qquad 
        ^{\Psi}J_{2n}^{\alpha}(Z)=\begin{pmatrix}
        0 & (-1)^{|t^{\alpha}|}\frac{^{\Psi}J_{\mathbb{O}_n^{\alpha}}(Z)}{2} \\
        \frac{^{\Psi}J_{\mathbb{O}_n^{\alpha}}(Z)}{2}  & 0 \\
    \end{pmatrix}   
    \end{equation}
\end{subequations}
The matrix $\textbf{t}^{\alpha}$ is
\begin{equation}
\label{eq:tbold}
    \textbf{t}^{\alpha}=\begin{pmatrix}
        t^{\alpha} & \\
        & t^{\alpha}
    \end{pmatrix}
\end{equation}
We now turn to the kernels, which, with the same notation of section \ref{sec:functional}, are
\begin{align}
\label{eq:deltamatrix}
        &^{\Phi} \mathbf{\Delta}^{-1}_{nk,n'k'}(Z_1,Z_2)=\begin{pmatrix}
        0 & ^{\Phi}\Delta^{-1}_{nk,n'k'}(Z_1,Z_2) \\
        ^{\Phi}\Delta^{-1}_{nk,n'k'}(Z_1,Z_2) & 0
    \end{pmatrix} \nonumber \\
    &^{\Psi} \mathbf{\Delta}^{-1}_{nk,n'k'}(Z_1,Z_2)=\begin{pmatrix}
        0 & ^{\Psi}\Delta^{-1}_{nk,n'k'}(Z_1,Z_2) \\
        ^{\Psi}\Delta^{-1}_{nk,n'k'}(Z_1,Z_2) & 0
    \end{pmatrix}
\end{align}
where the entries $^{\Phi}\Delta^{-1}_{nk,n'k'}(Z_1,Z_2)$ and $^{\Psi}\Delta^{-1}_{nk,n'k'}(Z_1,Z_2)$ are computed from Eq. \eqref{eq:deltank} applied to Eq. \eqref{eq:prop}. The final result for the entries is quite involved
\begin{subequations}
    \begin{align}
    \label{eq:phidelta}
            ^{\Phi}\Delta^{-1}_{nk,n'k'}=&\frac{1}{\Gamma(1+\lfloor\frac{n-k}{2}\rfloor)\Gamma(2j_{\Phi}+\lfloor\frac{n-k+1}{2}\rfloor)}\frac{1}{\Gamma(1+\lfloor\frac{k'}{2}\rfloor)\Gamma(2j_{\Phi}+\lfloor\frac{k'+1}{2}\rfloor)}(-1)^{\lfloor\frac{k'+1}{2}\rfloor} \nonumber\\
             & (-1)^{\lfloor\frac{k'}{2}\rfloor}\frac{\Gamma\left(2j_{\Phi}+\lfloor\frac{n-k+1}{2}\rfloor+\lfloor\frac{k'+1}{2}\rfloor\right)}{\Gamma\left(2j_{\Phi}\right)} \frac{\left(2^{\frac{5}{2}}(x_{1\bar{2}}^-)\right)^{\lfloor\frac{n-k}{2}\rfloor+\lfloor\frac{k'}{2}\rfloor}}{(x_{1\bar{2}}^2)^{2j_{\Phi}+\lfloor\frac{n-k+1}{2}\rfloor+\lfloor\frac{k'+1}{2}\rfloor}}    \nonumber\\
             &\Big[ \left(4i(x_{1\bar{2}}\bar{\theta}_{12})_1\right)^{(n-k)\ \mathrm{mod}\ 2}\left(4i(\theta_{12}x_{1\bar{2}})_{\dot{1}}\right)^{k' \mathrm{mod}\ 2}+\nonumber\\
            +& \frac{2^{\frac{5}{2}}\left[(n-k) \mathrm{mod}\ 2\right]\ \left[k'\mathrm{mod}\ 2\right]}{2j_{\Phi}+\lfloor\frac{n-k}{2}\rfloor+\lfloor\frac{k'+1}{2}\rfloor-1}(x_{1\bar{2}}^2)\left((x_{1\bar{2}})_+-i2\sqrt{2}(\theta_{12})_1(\bar{\theta}_{12})_{\dot{1}}\right) \Big]
    \end{align}
    \begin{align}
    \label{eq:psidelta}
            ^{\Psi}\Delta^{-1}_{nk,n'k'}=&\frac{1}{\Gamma(1+\lfloor\frac{n-k}{2}\rfloor)\Gamma(2j_{\Psi}+\lfloor\frac{n-k+1}{2}\rfloor)}\frac{1}{\Gamma(1+\lfloor\frac{k'}{2}\rfloor)\Gamma(2j_{\Psi}+\lfloor\frac{k'+1}{2}\rfloor)}(-1)^{\lfloor\frac{k'}{2}\rfloor} \nonumber\\
             & (-1)^{\lfloor\frac{k'}{2}\rfloor}\frac{\Gamma\left(2j_{\Psi}+\lfloor\frac{n-k+1}{2}\rfloor+\lfloor\frac{k'+1}{2}\rfloor\right)}{\Gamma\left(2j_{\Psi}\right)} \frac{\left(2^{\frac{5}{2}}(x_{1\bar{2}}^-)\right)^{\lfloor\frac{n-k}{2}\rfloor+\lfloor\frac{k'}{2}\rfloor}}{(x_{1\bar{2}}^2)^{2j_{\Psi}+\lfloor\frac{n-k+1}{2}\rfloor+\lfloor\frac{k'+1}{2}\rfloor}} \nonumber\\
             &\Big[ \left(4i(x_{1\bar{2}}\bar{\theta}_{12})_1\right)^{(n-k)\mathrm{mod}\ 2}\left(4i(\theta_{12}x_{1\bar{2}})_{\dot{1}}\right)^{k' \mathrm{mod}\ 2}+\nonumber\\
            +& \frac{2^{\frac{5}{2}}\left[(n-k) \mathrm{mod}\ 2\right]\ \left[k'\mathrm{mod}\ 2\right]}{2j_{\Psi}+\lfloor\frac{n-k}{2}\rfloor+\lfloor\frac{k'+1}{2}\rfloor-1} (x_{1\bar{2}}^2)\left((x_{1\bar{2}})_+-i2\sqrt{2}(\theta_{12})_1(\bar{\theta}_{12})_{\dot{1}}\right) \Big]
    \end{align}
\end{subequations}
where $x_{1\bar{2}}^{\mu}$, $\theta_{12}^{\alpha}$ and $\bar{\theta}_{12}^{\dot{\alpha}}$ are defined in Eq. \eqref{eq:interval}. We now show the same generating functionals in momentum space, which are
\begin{subequations}
\label{eq:wphipsimom}
    \begin{align}
         ^{\Phi}\mathcal{W}[J]&=  - \frac{1}{2}\mathrm{str}\ \mathrm{log} \Bigg[ (2\pi)^4\delta^{(4)}(p_1-p_2)\delta^{(4)}(\theta_1-\theta_2)\delta^{ab}\mathds{1}_{2\times 2}\delta_{n_1k_1,n_2k_2}- \nonumber\\
         -&2\  ^{\Phi}\widetilde{ \mathbf{\Delta}}^{-1}_{n_1k_1,n'k'}(p_1; \theta_1,\bar{\theta}_1, \theta_2,\bar{\theta}_2)\ ^{\Phi}\mathcal{M}_{n'k',n_2k_2}i^{\lfloor\frac{n_2}{2}\rfloor+|t^{\alpha}|}(\textbf{t}^{\alpha})^{ab}\ {^{\Phi}}\widetilde{J}_{n_2}^{\alpha}(p_1-p_2; \theta_1,\bar{\theta}_1, \theta_2,\bar{\theta}_2)
         \Bigg]
\end{align}
\begin{align}
         ^{\Psi}\mathcal{W}[J]&=  + \frac{1}{2}\mathrm{str}\ \mathrm{log} \Bigg[ (2\pi)^4\delta^{(4)}(p_1-p_2)\delta^{(4)}(\theta_1-\theta_2)\delta^{ab}\mathds{1}_{2\times 2}\delta_{n_1k_1,n_2k_2}- \nonumber\\
         -&2\ ^{\Psi}\widetilde{ \mathbf{\Delta}}^{-1}_{n_1k_1,n'k'}(p_1; \theta_1,\bar{\theta}_1, \theta_2,\bar{\theta}_2)\ ^{\Psi}\mathcal{M}_{n'k',n_2k_2}i^{\lfloor\frac{n_2}{2}\rfloor+|t^{\alpha}|}(\textbf{t}^{\alpha})^{ab}\ {^{\Psi}}\widetilde{J}_{n_2}^{\alpha}(p_1-p_2; \theta_1,\bar{\theta}_1, \theta_2,\bar{\theta}_2)
         \Bigg]
\end{align}
\end{subequations}
For brevity we omitted the sum over $n',k'$. Everything is the same as before, except for some currents appearing through their Fourier transform with respect to the even coordinates. The Fourier transforms of the kernels are
\begin{subequations}
    \begin{align}
            ^{\Phi}\widetilde{\Delta}^{-1}_{nk,n'k'}=& \frac{1}{\Gamma(1+\lfloor\frac{n-k}{2}\rfloor)\Gamma(2j_{\Phi}+\lfloor\frac{n-k+1}{2}\rfloor)}\frac{1}{\Gamma(1+\lfloor\frac{k'}{2}\rfloor)\Gamma(2j_{\Phi}+\lfloor\frac{k'+1}{2}\rfloor)}(-1)^{\lfloor\frac{k'+1}{2}\rfloor} \nonumber\\
             &(-1)^{\lfloor\frac{k'}{2}\rfloor}(-i2^{\frac{3}{2}}p_+)^{\lfloor\frac{n-k}{2}\rfloor+\lfloor\frac{k'}{2}\rfloor}\Big[\left(2(p\bar{\theta}_{12})_1^{(n-k)\ \mathrm{mod}\ 2}\right)\left(2(\theta_{12}p)_{\dot{1}}^{k'\mathrm{mod}\ 2}\right)- \nonumber\\
            -& \left[(n-k)\ \mathrm{mod}\ 2\right]\ \left[k'\mathrm{mod}\ 2\right]\  (\sqrt{2}p_+) \Big] \mathrm{exp}(a\cdot p)\widetilde{G}_{\Phi}(p)
    \end{align}
    \begin{align}
            ^{\Psi}\widetilde{\Delta}^{-1}_{nk,n'k'}=& \frac{1}{\Gamma(1+\lfloor\frac{n-k}{2}\rfloor)\Gamma(2j_{\Psi}+\lfloor\frac{n-k+1}{2}\rfloor)}\frac{1}{\Gamma(1+\lfloor\frac{k'}{2}\rfloor)\Gamma(2j_{\Psi}+\lfloor\frac{k'+1}{2}\rfloor)}(-1)^{\lfloor\frac{k'}{2}\rfloor} \nonumber\\
             &(-1)^{\lfloor\frac{k'}{2}\rfloor}(-i2^{\frac{3}{2}}p_+)^{\lfloor\frac{n-k}{2}\rfloor+\lfloor\frac{k'}{2}\rfloor}\Big[\left(2(p\bar{\theta}_{12})_1^{(n-k)\ \mathrm{mod}\ 2}\right)\left(2(\theta_{12}p)_{\dot{1}}^{k'\mathrm{mod}\ 2}\right)- \nonumber\\
            -& \left[(n-k)\ \mathrm{mod}\ 2\right]\ \left[k'\mathrm{mod}\ 2\right]\  (\sqrt{2}p_+) \Big] \mathrm{exp}(a\cdot p)\widetilde{G}_{\Psi}(p)
    \end{align}
\end{subequations}
where
\begin{equation}
    a^{\mu}=\theta_1\sigma^{\mu}\bar\theta_1+\theta_2\sigma^{\mu}\bar\theta_2-2\theta_1\sigma^{\mu}\bar\theta_2
\end{equation}
and the Fourier transforms $\widetilde{G}_{\Phi}$, $\widetilde{G}_{\Psi} $ of the propagators in Eq. \eqref{eq:prop} can be found in appendix \ref{app:supconfcorr}. \\

The Euclidean version of these generating functionals can be obtained through a Wick rotation according to the rules of appendix \ref{app:euclidean}. The Wick-rotated generating functionals for the bosonic and fermionic correlators in position space are
\begin{subequations}
\label{eq:ewphipsi}
{\footnotesize
\begin{align}
    &\mathcal{W}_{\Phi}^E[J^E]= \nonumber \\
    -&\frac{1}{2}\mathrm{str}\ \mathrm{log}\left[\delta^{(8)}(Z_1^E,Z_2^E)\delta^{ab}\mathds{1}_{2\times 2}\delta_{n_1k_1,n_2k_2}-2\  ^{\Psi} (\mathbf{\Delta}^{-1})_{n_1k_1,n'k'}^E(Z_1^E,Z_2^E)\ ^{\Phi}\mathcal{M}_{n'k',n_2k_2}i^{\lfloor\frac{n_2}{2}\rfloor+|t^{\alpha}|}(\textbf{t}^{\alpha})^{ab}\ {^{\Phi}J}_{n_2}^{E\ \alpha}(Z_2^E)\right]
\end{align}}
and
{\footnotesize
\begin{align}
    &\mathcal{W}_{\Psi}^E[J^E]=\nonumber\\
    +&\frac{1}{2}\mathrm{str}\ \mathrm{log}\left[\delta^{(8)}(Z_1^E,Z_2^E)\delta^{ab}\mathds{1}_{2\times 2}\delta_{n_1k_1,n_2k_2}-2\  ^{\Psi} (\mathbf{\Delta}^{-1})_{n_1k_1,n'k'}^E(Z_1^E,Z_2^E)\ ^{\Psi}\mathcal{M}_{n'k',n_2k_2}i^{\lfloor\frac{n_2}{2}\rfloor+|t^{\alpha}|}(\textbf{t}^{\alpha})^{ab}\ {^{\Psi}J_{n_2}^{E\ \alpha}}(Z_2^E)\right] 
\end{align}}
\end{subequations}
For brevity we omitted the sum over $n',k'$. The Euclidean currents are the same of Eqs. \eqref{eq:currents} except for the Wick rotation of their argument, and the matrices $^{\Phi}\mathcal{M}$ and $^{\Psi}\mathcal{M}$ are those in Eq. \eqref{eq:phipsimatrix}. The Euclidean kernels are obtained by a Wick rotation of the kernels \eqref{eq:ephidelta} times a factor $-i$ that comes from the rotation of the delta functions according to Eq. \eqref{eq:edelta}. We obtain
\begin{subequations}
    \begin{equation}
    \label{eq:ephidelta}
        \begin{split}
            ^{\Phi}(\Delta^{-1})_{nk,n'k'}^E=&-i\frac{1}{\Gamma(1+\lfloor\frac{n-k}{2}\rfloor)\Gamma(2j_{\Phi}+\lfloor\frac{n-k+1}{2}\rfloor)}\frac{1}{\Gamma(1+\lfloor\frac{k'}{2}\rfloor)\Gamma(2j_{\Phi}+\lfloor\frac{k'+1}{2}\rfloor)}(-1)^{\lfloor\frac{k'+1}{2}\rfloor} \\
             & (-1)^{\lfloor\frac{k'}{2}\rfloor}\frac{\Gamma\left(2j_{\Phi}+\lfloor\frac{n-k+1}{2}\rfloor+\lfloor\frac{k'+1}{2}\rfloor\right)}{\Gamma\left(2j_{\Phi}\right)} \frac{\left(2^{\frac{5}{2}}(-ix_{1\bar{2}}^E)^{\bar{z}}\right)^{\lfloor\frac{n-k}{2}\rfloor+\lfloor\frac{k'}{2}\rfloor}}{(-(x_{1\bar{2}}^E)^2)^{2j_{\Phi}+\lfloor\frac{n-k+1}{2}\rfloor+\lfloor\frac{k'+1}{2}\rfloor}} \\
             &\Big[ \left(4(x_{1\bar{2}}^E\bar{\theta}_{12}^E)_1\right)^{(n-k)\ \mathrm{mod}\ 2}\left(4(\theta_{12}^Ex_{1\bar{2}}^E)_{\dot{1}}\right)^{k' \mathrm{mod}\ 2}+\\
            +i& \frac{2^{\frac{5}{2}}\left[(n-k) \mathrm{mod}\ 2\right]\ \left[k'\mathrm{mod}\ 2\right]}{2j_{\Phi}+\lfloor\frac{n-k}{2}\rfloor+\lfloor\frac{k'+1}{2}\rfloor-1}(x_{1\bar{2}}^E)^2\left((x_{1\bar{2}}^E)_{z}+2\sqrt{2}(\theta_{12}^E)_1(\bar{\theta}_{12}^E)_{\dot{1}}\right) \Big]
        \end{split}
    \end{equation}
    \begin{equation}
    \label{eq:epsidelta}
        \begin{split}
            ^{\Psi}(\Delta^{-1})_{nk,n'k'}^E=&-i\frac{1}{\Gamma(1+\lfloor\frac{n-k}{2}\rfloor)\Gamma(2j_{\Psi}+\lfloor\frac{n-k+1}{2}\rfloor)}\frac{1}{\Gamma(1+\lfloor\frac{k'}{2}\rfloor)\Gamma(2j_{\Psi}+\lfloor\frac{k'+1}{2}\rfloor)}(-1)^{\lfloor\frac{k'}{2}\rfloor} \\
             & (-1)^{\lfloor\frac{k'}{2}\rfloor}\frac{\Gamma\left(2j_{\Psi}+\lfloor\frac{n-k+1}{2}\rfloor+\lfloor\frac{k'+1}{2}\rfloor\right)}{\Gamma\left(2j_{\Psi}\right)} \frac{\left(2^{\frac{5}{2}}(-ix_{1\bar{2}}^E)^{\bar{z}}\right)^{\lfloor\frac{n-k}{2}\rfloor+\lfloor\frac{k'}{2}\rfloor}}{(-(x_{1\bar{2}}^E)^2)^{2j_{\Psi}+\lfloor\frac{n-k+1}{2}\rfloor+\lfloor\frac{k'+1}{2}\rfloor}} \\
             &\Big[ \left(4(x_{1\bar{2}}^E\bar{\theta}_{12}^E)_1\right)^{(n-k)\mathrm{mod}\ 2}\left(4(\theta_{12}^Ex_{1\bar{2}}^E)_{\dot{1}}\right)^{k' \mathrm{mod}\ 2}+\\
            +i& \frac{2^{\frac{5}{2}}\left[(n-k) \mathrm{mod}\ 2\right]\ \left[k'\mathrm{mod}\ 2\right]}{2j_{\Psi}+\lfloor\frac{n-k}{2}\rfloor+\lfloor\frac{k'+1}{2}\rfloor-1} (x_{1\bar{2}}^E)^2\left((x_{1\bar{2}}^E)_z+2\sqrt{2}(\theta_{12}^E)_1(\bar{\theta}_{12}^E)_{\dot{1}}\right) \Big]
        \end{split}
    \end{equation}
\end{subequations}
The Euclidean generating functionals in momentum space are
\begin{subequations}
\label{eq:ewphipsimom}
{\small
    \begin{equation}
    \begin{split}
         &^{\Phi}\mathcal{W}^E[J^E]=  - \frac{1}{2}\mathrm{str}\ \mathrm{log} \Bigg[ (2\pi)^4\delta^{(4)}(p_1^E-p_2^E)\delta^{(4)}(\theta_1^E-\theta_2^E)\delta^{ab}\mathds{1}_{2\times 2}\delta_{n_1k_1,n_2k_2}- \\
         -&2\ ^{\Phi}(\widetilde{ \mathbf{\Delta}}^{-1})_{n_1k_1,n'k'}(p_1^E; \theta_1^E,\bar{\theta}_1^E, \theta_2^E,\bar{\theta}_2^E)\ ^{\Phi}\mathcal{M}_{n'k',n_2k_2}i^{\lfloor\frac{n_2}{2}\rfloor+|t^{\alpha}|}(\textbf{t}^{\alpha})^{ab}\ {^{\Phi}}\widetilde{J}_{n_2}^{E\ \alpha}(p_1^E-p_2^E; \theta_1^E,\bar{\theta}_1^E, \theta_2^E,\bar{\theta}_2^E)
         \Bigg]
    \end{split}
\end{equation}}
{\small
\begin{equation}
    \begin{split}
         &^{\Psi}\mathcal{W}^E[J^E]=  + \frac{1}{2}\mathrm{str}\ \mathrm{log} \Bigg[ (2\pi)^4\delta^{(4)}(p_1^E-p_2^E)\delta^{(4)}(\theta_1^E-\theta_2^E)\delta^{ab}\mathds{1}_{2\times 2}\delta_{n_1k_1,n_2k_2}- \\
         -&2\ ^{\Psi}(\widetilde{ \mathbf{\Delta}}^{-1})_{n_1k_1,n'k'}^E(p_1^E; \theta_1^E,\bar{\theta}_1^E, \theta_2^E,\bar{\theta}_2^E)\ ^{\Psi}\mathcal{M}_{n'k',n_2k_2}i^{\lfloor\frac{n_2}{2}\rfloor+|t^{\alpha}|}(\textbf{t}^{\alpha})^{ab}\ {^{\Psi}}\widetilde{J}_{n_2}^{E\ \alpha}(p_1^E-p_2^E; \theta_1^E,\bar{\theta}_1^E, \theta_2^E,\bar{\theta}_2^E)
         \Bigg]
    \end{split}
\end{equation}}
\end{subequations}
For brevity we omitted the sum over $n',k'$. Again, the Euclidean currents are the same of Eqs. \eqref{eq:currents} except for the Wick rotation of their argument, and the matrices $^{\Phi}\mathcal{M}$ and $^{\Psi}\mathcal{M}$ are those in Eq. \eqref{eq:phipsimatrix}. We obtain
\begin{subequations}
    \begin{equation}
        \begin{split}
            ^{\Phi}(\widetilde{\Delta}^{-1})_{nk,n'k'}^E=& \frac{1}{\Gamma(1+\lfloor\frac{n-k}{2}\rfloor)\Gamma(2j_{\Phi}+\lfloor\frac{n-k+1}{2}\rfloor)}\frac{1}{\Gamma(1+\lfloor\frac{k'}{2}\rfloor)\Gamma(2j_{\Phi}+\lfloor\frac{k'+1}{2}\rfloor)}(-1)^{\lfloor\frac{k'+1}{2}\rfloor} \\
             &(-1)^{\lfloor\frac{k'}{2}\rfloor}(2^{\frac{3}{2}}p_z^E)^{\lfloor\frac{n-k}{2}\rfloor+\lfloor\frac{k'}{2}\rfloor}\Big[\left(2(ip^E\bar{\theta}_{12}^E)_1^{(n-k)\ \mathrm{mod}\ 2}\right)\left(2(i\theta_{12}^Ep^E)_{\dot{1}}^{k'\mathrm{mod}\ 2}\right)- \\
            -& \left[(n-k)\ \mathrm{mod}\ 2\right]\ \left[k'\mathrm{mod}\ 2\right]\  (i\sqrt{2}p_z^E) \Big] \mathrm{exp}(a^E\cdot p^E)\widetilde{G}_{\Phi}^E(p^E)
        \end{split}
    \end{equation}
    \begin{equation}
        \begin{split}
            ^{\Psi}(\widetilde{\Delta}^{-1})_{nk,n'k'}^E=& \frac{1}{\Gamma(1+\lfloor\frac{n-k}{2}\rfloor)\Gamma(2j_{\Psi}+\lfloor\frac{n-k+1}{2}\rfloor)}\frac{1}{\Gamma(1+\lfloor\frac{k'}{2}\rfloor)\Gamma(2j_{\Psi}+\lfloor\frac{k'+1}{2}\rfloor)}(-1)^{\lfloor\frac{k'}{2}\rfloor} \\
             &(-1)^{\lfloor\frac{k'}{2}\rfloor}(2^{\frac{3}{2}}p_z^E)^{\lfloor\frac{n-k}{2}\rfloor+\lfloor\frac{k'}{2}\rfloor}\Big[\left(2(ip^E\bar{\theta}_{12}^E)_1^{(n-k)\ \mathrm{mod}\ 2}\right)\left(2(i\theta_{12}^Ep^E)_{\dot{1}}^{k'\mathrm{mod}\ 2}\right)- \\
            -& \left[(n-k)\ \mathrm{mod}\ 2\right]\ \left[k'\mathrm{mod}\ 2\right]\  (i\sqrt{2}p_z^E) \Big] \mathrm{exp}(a^E\cdot p^E)\widetilde{G}_{\Psi}^E(p^E)
        \end{split}
    \end{equation}
\end{subequations}
where
\begin{equation}
    (a^E)^{\mu}=i\theta_1^E(\sigma^E)^{\mu}\bar\theta_1^E+i\theta_2^E(\sigma^E)^{\mu}\bar\theta_2^E-2i\theta_1^E(\sigma^E)^{\mu}\bar\theta_2^E
\end{equation}
The $\widetilde{G}_{\Phi}^E$, $\widetilde{G}_{\Psi}^E$ are the Wick-rotated Fourier transforms of the propagators in Eq. \eqref{eq:prop}, that are shown in appendix \ref{app:supconfcorr}.

\section{$\mathcal{N}=1$ SYM theory}\label{sec:sym}
We apply the results of the previous sections to $\mathcal{N}=1$ SYM theory in the limit of zero coupling, where the theory is superconformal.

\subsection{Introduction and conventions}
\label{sec:symintro}
We adopt the same conventions of Ref. \cite{Shifman:2012zz}. The only elementary field of the theory is a vector superfield $V=V^aT^a$ in the adjoint representation of the gauge group. The generators of the gauge algebra $\mathfrak{su}(N)$ are normalized as $\mathrm{tr}(T^aT^b)=\frac{1}{2}\delta^{ab}$. The gauge transformation laws of this field is
\begin{equation}
    e^{V}\longmapsto e^{i\bar{\Lambda}}e^{V}e^{-i\Lambda} 
\end{equation}
where $\Lambda$, $\bar\Lambda$ are a left-handed and a right-handed chiral Lie algebra-valued functions. The spinorial field strengths that we can construct out of $V$ are
\begin{align}
       & W_{\alpha}=\frac{1}{8}\bar{D}^2\left(e^{-V}D_{\alpha} e^{+V}\right)        , \qquad &&  W_{\alpha}\longmapsto e^{i\Lambda}W_{\alpha}e^{-i\Lambda}\nonumber \\
        &   \bar{W}_{\dot\alpha}=\frac{1}{8}D^2\left(e^V\bar{D}_{\dot{\alpha}}e^{-V}\right)     , \qquad && \bar{W}_{\dot\alpha}\longmapsto e^{i\bar{\Lambda}}\bar{W}_{\dot\alpha}e^{-i\bar{\Lambda}}
\end{align}
The spinorial covariant derivatives $\nabla_{\alpha}$ and $\bar{\nabla}_{\dot{\alpha}}$ and the vectorial covariant derivative $\nabla_{\alpha\dot{\alpha}}=-\frac{i}{2}\{\nabla_{\alpha},\bar{\nabla}_{\dot{\alpha}}\}$ are constructed accordingly. The lagrangian of the theory is
\begin{align}
\label{eq:lagrangian}
    \mathcal{L}_{SYM}=&\left(\frac{N}{4g^2}\int d^2\theta\ W^{a\ \alpha}W^a_{\alpha}+\text{h.c.}\right)
\end{align}
where $g$ is the (real) 't Hooft coupling, with $g^2=g_{YM}^2N$. We take $g$ to be real, so that theta terms are absent. To express the lagrangian \eqref{eq:lagrangian} in terms of the component fields, we write the component expansion of the vector superfield $V$ in the Wess-Zumino gauge
\begin{equation}
\label{eq:vectorsuper}
    V^a(x,\theta,\bar{\theta})=-2\theta^{\alpha}(\sigma^{\mu})_{\alpha\dot{\alpha}}\bar{\theta}^{\dot{\alpha}}A_{\mu}^a(x)-2i\bar{\theta}^2\theta^{\alpha}\lambda^a_{\alpha}(x)+2i\theta^2\bar{\theta}_{\dot{\alpha}}\bar{\lambda}^{\dot{\alpha}a}(x)+\theta^2\bar{\theta}^2D^a(x)
\end{equation}
Inserting these fields in Eq. \eqref{eq:lagrangian}, integrating over the odd variables $\theta^{\alpha}$, $\bar{\theta}^{\dot{\alpha}}$, and eliminating the auxiliary field $D$ one finally obtains
\begin{equation}
    \mathcal{L}_{SYM}^{\text{(Wess-Zumino)}}= \frac{N}{g^2}\mathrm{tr}\left[-\frac{1}{2}F_{\mu\nu}F^{\mu\nu}+2i\lambda^{\alpha}\mathcal{D}_{\alpha\dot\alpha}\bar{\lambda}^{\dot{\alpha}}\right]
\end{equation}
Where the symbol $\mathcal{D}_{\mu}$ denotes a covariant derivative.
\par In the limit $g \to 0$, field strength is a free fields, and its only nonzero two-point functions is
\begin{equation}
\label{eq:ww2}
    \expval{g^{-1}W_{\alpha}^a(x_{L,1},\theta_1)\ g^{-1}\bar{W}_{\dot\beta}^b(x_{R,2},\bar\theta_2)}_0=\  2^{-\frac{1}{2}}C_W \delta^{ab}\frac{(x_{1\bar{2}})_{\alpha\dot\beta}}{(x_{1\bar{2}}^2)^2}
\end{equation}
where $C_W=-\frac{1}{\sqrt{2}\pi^2 N}$. The propagator between the two components $W_1$, $\bar{W}_{\dot{1}}$ is
\begin{align}
\label{eq:ww2second}
    \expval{g^{-1}W_{1}^a(x_{L,1},\theta_1)\ g^{-1}\bar{W}_{\dot{1}}^b(x_{R,2},\bar\theta_2)}_0= & -\frac{1}{\sqrt{2}N\pi^2}\frac{(x_{1\bar{2}})_+}{(x_{1\bar{2}}^2)^2} \nonumber \\
    =&\sqrt{2}\frac{1}{N}\partial_+\frac{1}{-\Box}(x_{1\bar{2}})
\end{align}
\newline

\subsection{Twist-$2$ operators}
\label{sec:twistsym}
We construct the twist-$2$ operators out of the components of maximal spin along the light-cone of $W_{\alpha}$ and $\bar{W}_{\dot{\alpha}}$. The collinear superconformal charges of these operators are shown in table \eqref{tab:chargessym1}.

\begin{table}[h!]
    \centering
    \begin{tabular}{C|CCCCC}
         & \ell & \bar{\ell} & j & b & \tau \\
        \hline
         W_1 & 1 & 0 & 1 & -\frac{3}{2} & 1 \\
         \bar{W}_{\dot{1}} & 0 & 1 & 1 & +\frac{3}{2} & 1 \\
    \end{tabular}
    \caption{Collinear superconformal charges of the building blocks of the twist-$2$ operators}
    \label{tab:chargessym1}
\end{table}
The gauge-invariant twist-$2$ operators constructed out of the gluon superfields are
\begin{align}
\label{eq:ww}
        & \mathbb{W}_n= 2(g^2N)^{-1}C_W^{-1} i^n \mathrm{tr}\left[ (e^{-V}\ \bar{W}_{\dot{1}} e^{V})\ \mathbb{C}^{1,1}_{2n}\left(\overleftarrow{\nabla}_1+i\overleftarrow{\bar{\nabla}}_{\dot{1}}, \overrightarrow{\nabla}_1+i\overrightarrow{\bar{\nabla}}_{\dot{1}}\right)\  W_1\right] \nonumber\\
        & \mathbb{W}_n^+= 2(g^2N)^{-1}C_W^{-1}N^{-1} i^{2n} \mathrm{tr}\left[ W_1\ \mathbb{C}^{1,1}_{4n+1}\left(\overleftarrow{\nabla}_1+i\overleftarrow{\bar{\nabla}}_{\dot{1}}, \overrightarrow{\nabla}_1+i\overrightarrow{\bar{\nabla}}_{\dot{1}}\right) \ W_1 \right] \nonumber\\
        & \mathbb{W}_n^-= 2(g^2N)^{-1}C_W^{-1}N^{-1} i^{2n}\mathrm{tr}\left[ \bar{W}_{\dot{1}}\ \mathbb{C}^{1,1}_{4n+1}\left(\overleftarrow{\nabla}_1+i\overleftarrow{\bar{\nabla}}_{\dot{1}}, \overrightarrow{\nabla}_1+i\overrightarrow{\bar{\nabla}}_{\dot{1}}\right) \ \bar{W}_{\dot{1}}\right]
\end{align}
The factors $2$ in front of the them have been added to compensate for the $\frac{1}{2}$ appearing in the normalization of the $\mathfrak{su}(N)$ generators. The operators satisfy the hermiticity relations
\begin{equation}
    \left(\mathbb{W}_n\right)^{\dagger}=\mathbb{W}_n\ , \qquad  \left(\mathbb{W}_n^+\right)^{\dagger}=\mathbb{W}_n^-
\end{equation}
Note that although the lagrangian \eqref{eq:lagrangian} is written in Wilsonian normalization and the two-point function of the field strength in Eq. \eqref{eq:ww2} vanishes in the limit of zero coupling, the correlators of the operators in \eqref{eq:ww} are well-defined for $g\to 0$ thanks to our choice of the normalization. The collinear superconformal charges of these operators are shown in table \eqref{tab:chargessym2}.
\begin{table}[h!]
\centering
    \begin{tabular}{C|CCCCC}
        & \ell & \bar{\ell} & j & b & \tau  \\
        \hline
        \mathbb{W}_n & n+1  & n+1 & n+2 & 0 & 2  \\
        \mathbb{W}_n^+ & 2n+3 & 2n & 2n+\frac{5}{2} & -\frac{3}{2} & 2 \\
        \mathbb{W}_n^- & 2n &  2n+3 & 2n+\frac{5}{2} & +\frac{3}{2} & 2 \\
    \end{tabular}
    \caption{Collinear superconformal charges of the twist-$2$ operators of $\mathcal{N}=1$ SYM theory.}
    \label{tab:chargessym2}
\end{table}
In the Wess-Zumino and light-cone gauge, the components of the gluon fields in the light-cone directions of superspace are
\begin{align}
        &W_1\rvert_{\text{l.c.}}= +i\varrho\left(\lambda + \frac{2}{\varrho}\theta^1\partial_+\bar{A}\right) \nonumber\\
        &\bar{W}_{\dot{1}}\rvert_{\text{l.c.}}=-i\varrho\left(\bar{\lambda}-\frac{2}{\varrho}\bar{\theta}^{\dot{1}}\partial_+A\right)
\end{align}
We used the notation
\begin{align}
    &\lambda= \varrho^{-1}\lambda_1\ , \qquad \bar{\lambda}= \varrho^{-1}\bar{\lambda}_{\dot{1}}\nonumber \\
    &A=\frac{A_1+iA_2}{\sqrt{2}}\ , \qquad \bar{A}=\frac{A_1-iA_2}{\sqrt{2}} 
\end{align}
with $\varrho=2^{1/4}$.

\subsection{Components}
\label{eq:symcomp}
We write the explicit component expansion for the twist-$2$ operators  \eqref{eq:ww} in terms of the Jacobi and Gegenbauer polynomials. We use the same notation and conventions of subsection \ref{sec:components}. The operators are expressed in the Wess-Zumino and light-cone gauge, which is the reason why no covariant derivatives appear. We have
\begin{subequations}
\label{eq:wwcomponents}
    \begin{align}
            \mathbb{W}_n=(g^2N)^{-1}C_W^{-1}&\frac{(-1)^n2^{\frac{3}{2}n+\frac{3}{2}}}{(n+1)!(n+2)!}\Bigg\{\left(\bar{\lambda}^aC_n^{3/2}\lambda^a+\frac{6}{n+3}\partial_+ A^aC_{n-1}^{5/2}\partial_+\bar{A}^a\right) \nonumber\\
            -&\frac{2}{\varrho}\theta^1 \left(\frac{n+2}{2}\right) \bar{\lambda}^aP_n^{(1,2)}\partial_+\bar{A}^a-\frac{2}{\varrho}\bar{\theta}^{\dot{1}}\left(\frac{n+2}{2}\right)\partial_+ A^a P_n^{(2,1)}\lambda^a \nonumber\\
            -&\sqrt{2}\theta^1\bar{\theta}^{\dot{1}}\frac{n+1}{n+3}\left(\bar{\lambda}^aC_{n+1}^{3/2}\lambda^a-\frac{6}{n+1}\partial_+ A^aC_{n}^{5/2}\partial_+\bar{A}^a\right)\Bigg\}
    \end{align}
    \begin{align}
            \mathbb{W}_n^+=-(g^2N)^{-1}C_W^{-1}&\frac{2^{3n+\frac{5}{4}}}{(2n+1)!(2n+2)!}\Bigg\{\left(\partial_+\bar{A}^aP_{2n}^{(2,1)}\lambda^a+\lambda^a P_{2n}^{(1,2)}\partial_+\bar{A}^a\right) \nonumber\\
            +& \frac{2}{\varrho}\theta^1\ \frac{12}{(2n+2)(2n+3)}\partial_+\bar{A}^a C_{2n}^{5/2}\partial_+\bar{A}^a+\frac{2}{\varrho}\bar{\theta}^{\dot{1}}\frac{2(2n+1)}{(2n+2)(2n+3)}\ \lambda^a C_{2n+1}^{3/2}\lambda^a \nonumber\\
            +&i2\sqrt{2}\theta^1\bar{\theta}^{\dot{1}}\Bigg[i\frac{(2n+1)(2n+4)}{2(2n+\frac{5}{2})(2n+2)}\left(\partial_+\bar{A}^aP_{2n+1}^{(2,1)}\lambda^a-\lambda^a P_{2n+1}^{(1,2)}\partial_+\bar{A}^a\right) \nonumber\\
            -&\frac{3}{2(4n+5)}\partial_+\left(\partial_+\bar{A}^aP_{2n}^{(2,1)}\lambda^a+\lambda^a P_{2n}^{(1,2)}\partial_+\bar{A}^a\right)\Bigg]\Bigg\}
    \end{align}
\end{subequations}
where $\varrho=2^{1/4}$. The components of $\mathbb{W}^-$ can be obtained from those of $\mathbb{W}^+$ with the substitutions
\begin{equation}
    \begin{gathered}
        \theta^1 \to i\bar{\theta}^{\dot{1}} \qquad \bar{\theta}^{\dot{1}}\to -i\theta^1 \qquad  \lambda \to \bar{\lambda}^a  \qquad \partial_+\bar{A}^a\to i \partial_+ A^a
    \end{gathered}
\end{equation}
These result coincide with those in the literature. In particular, the operators $\mathcal{S}_{j\ell}^{1,2}$, $\mathcal{P}_{j\ell}^{1,2}$, $\mathcal{U}_{j\ell}$, $\mathcal{V}_{j\ell}$ of Ref. \cite{Belitsky:1998gu} are components of the supermultiplets
\begin{align}
     \partial_+^{\ell-j}\mathbb{W}_{j}\ \sim\ & \mathcal{P}^2_{j\ell}+\theta^1\left(\mathcal{U}_{j\ell}+\mathcal{V}_{j\ell}\right) + \bar{\theta}^{\dot{1}}\left(\mathcal{U}_{j\ell}-\mathcal{V}_{j\ell}\right)+\theta^1\bar{\theta}^{\dot{1}}\mathcal{S}^1_{j+1\ \ell} \qquad && (j \text{ even}) \nonumber\\
     \partial_+^{\ell-j}\mathbb{W}_{j}\ \sim\ & \mathcal{S}^2_{j\ell}+\theta^1\left(\mathcal{U}_{j\ell}+\mathcal{V}_{j\ell}\right) + \bar{\theta}^{\dot{1}}\left(\mathcal{U}_{j\ell}-\mathcal{V}_{j\ell}\right)+\theta^1\bar{\theta}^{\dot{1}}\mathcal{P}^1_{j+1\ \ell} \qquad && (j \text{ odd})
\end{align}
It can also be shown, using the identities of section \ref{sec:compsl2}, that in the language of Ref. \cite{Belitsky:2004sc} the supermultiplets $\mathbb{W}_n$, $\mathbb{W}_n^+$, $\mathbb{W}_n^-$ correspond to the $\Phi\Psi$, $\Psi\Psi$ and $\Phi\Phi$ sectors respectively.

\subsection{Two-point functions}
\label{sec:sym2pt}
In the limit of zero coupling, the two-point functions and their normalizations can be obtained by simply substituting the values of the table \eqref{tab:chargessym2} into the results of subsection \ref{sec:2pt}. One obtains
\begin{align}
        \expval{\mathbb{W}_n(Z_1)\bar{\mathbb{W}}_m(Z_2)}=&\left(1-\frac{1}{N^2}\right)  \mathcal{C}_{\mathbb{W}_n}(-1)^n \delta_{nm}\frac{(x_{1\bar{2}}^-)^{n+1}(x_{\bar{1}2}^-)^{n+1}}{(x_{1\bar{2}}^2)^{n+2}(x_{\bar{1}2}^2)^{n+1}} \nonumber\\
        \expval{\mathbb{W}_n^+(Z_1)\mathbb{W}^-_m(Z_2)}=&\left(1-\frac{1}{N^2}\right)    \mathcal{C}_{\mathbb{W}_n^{\pm}}\delta_{nm}\frac{(x_{1\bar{2}}^-)^{2n}(x_{\bar{1}2}^-)^{2n+3}}{(x_{1\bar{2}}^2)^{2n+4}(x_{\bar{1}2}^2)^{2n+1}}
\end{align}
with the normalizations
\begin{align}
    \mathcal{C}_{\mathbb{W}_n}= & +2^{5n}\frac{1}{(n+1)!^2}\binom{2n+5}{n+1} \nonumber\\
    \mathcal{C}_{\mathbb{W}_n^{\pm}}= & -2^{10n+\frac{9}{2}}\frac{1}{(2n+1)!(2n+2)!}\binom{4n+5}{2n}
\end{align}

\subsection{Minkowskian generating functionals}
\label{sec:genfunctsym}
Again, the results of this subsection are just a special case of those of subsection \ref{sec:genfunct1}. In the present case, the generating functional of the connected correlators of the twist-$2$ operators of Eq. \eqref{eq:ww} in the zero-coupling is given by
\begin{equation}
\label{eq:symgenfun}
\begin{split}
    &\mathcal{W}[J]= \\
    &\quad\frac{N^2-1}{2}\mathrm{str}\ \mathrm{log}\left[\delta^{(8)}(Z_1,Z_2)\mathds{1}_{2\times 2}\delta_{n_1k_1,n_2k_2}-2\  \mathbf{\Delta}^{-1}_{n_1k_1,n'k'}(Z_1,Z_2)\mathcal{M}_{n'k',n_2k_2}i^{\lfloor\frac{n_2}{2}\rfloor} \frac{J_{n_2}(Z_2)}{N}\right]
\end{split}
\end{equation}
that has obviously the structure of the logarithm of a functional superdeerminant. For brevity we omitted the sum over $n',k'$. The matrix $\mathbf{\Delta}^{-1}_{n_1k_1,n'k'}(Z_1,Z_2)$ is
\begin{equation}
\label{eq:matrixsym}
      \mathbf{\Delta}^{-1}_{nk,n'k'}(Z_1,Z_2)=\begin{pmatrix}
        0 & \Delta^{-1}_{nk,n'k'}(Z_1,Z_2) \\
        \Delta^{-1}_{nk,n'k'}(Z_1,Z_2) & 0
    \end{pmatrix}
\end{equation}
whose entries are the kernels in \eqref{eq:psidelta} with $j_{\Psi}=1$
\begin{align}
\label{eq:wdelta}
        \Delta^{-1}_{nk,n'k'}=&\frac{1}{\Gamma(1+\lfloor\frac{n-k}{2}\rfloor)\Gamma(2+\lfloor\frac{n-k+1}{2}\rfloor)}\frac{1}{\Gamma(1+\lfloor\frac{k'}{2}\rfloor)\Gamma(2+\lfloor\frac{k'+1}{2}\rfloor)}(-1)^{\lfloor\frac{k'}{2}\rfloor} \nonumber\\
         & (-1)^{\lfloor\frac{k'}{2}\rfloor} \Gamma\left(2+\lfloor\frac{n-k+1}{2}\rfloor+\lfloor\frac{k'+1}{2}\rfloor\right) \frac{\left(2^{\frac{5}{2}}(x_{1\bar{2}}^-)\right)^{\lfloor\frac{n-k}{2}\rfloor+\lfloor\frac{k'}{2}\rfloor}}{(x_{1\bar{2}}^2)^{2+\lfloor\frac{n-k+1}{2}\rfloor+\lfloor\frac{k'+1}{2}\rfloor}} \nonumber\\
         &\Big[ \left(4i(x_{1\bar{2}}\bar{\theta}_{12})_1\right)^{(n-k)\mathrm{mod}\ 2}\left(4i(\theta_{12}x_{1\bar{2}})_{\dot{1}}\right)^{k' \mathrm{mod}\ 2}+\nonumber\\
        +& \frac{2^{\frac{5}{2}}\left[(n-k) \mathrm{mod}\ 2\right]\ \left[k'\mathrm{mod}\ 2\right]}{\lfloor\frac{n-k}{2}\rfloor+\lfloor\frac{k'+1}{2}\rfloor+1} (x_{1\bar{2}}^2)\left((x_{1\bar{2}})_+-i2\sqrt{2}(\theta_{12})_1(\bar{\theta}_{12})_{\dot{1}}\right) \Big]
\end{align}
An alternative way to write the kernel using \eqref{eq:deltank} and \eqref{eq:ww2second} is
\begin{align}
    \Delta^{-1}_{nk,n'k'}=&-(2\pi^2)\frac{1}{\Gamma(1+\lfloor\frac{n-k}{2}\rfloor)\Gamma(2+\lfloor\frac{n-k+1}{2}\rfloor)}\frac{1}{\Gamma(1+\lfloor\frac{k'}{2}\rfloor)\Gamma(2+\lfloor\frac{k'+1}{2}\rfloor)}(-1)^{\lfloor\frac{k'}{2}\rfloor} && \nonumber \\
     & (D_1^{(1)}+i\bar{D}_{\dot{1}}^{(1)})^{n-k}(D_1^{(2)}+i\bar{D}_{\dot{1}}^{(2)})^{k'}\partial_+\frac{1}{-\Box}(x_{1\bar{2}}) &&
\end{align}
The matrix $\mathcal{M}_{n'k',n_2k_2}$ is just the matrix $^{\Psi}\mathcal{M}_{n'k',n_2k_2}$ defined in Eq. \eqref{eq:phipsimatrix}
\begin{equation}
\label{eq:mnk}
    \mathcal{M}_{nk,n'k'}=\begin{cases}
        \delta_{nn'}\delta_{kk'}\mathds{1}_{2\times 2}\ , & n,n'\ \text{odd} \\
        \begin{pmatrix}
            (-1)^{\lfloor\frac{n}{2}\rfloor+k+1} & 0\\
            0 &  (-1)^k
        \end{pmatrix}\delta_{nn'}\delta_{kk'} \ , & n,n'\ \text{even} \\
        0\ , & \text{otherwise}
        \end{cases}
\end{equation}
The entries of the currents are $J_{n}(Z)$ are
    \begin{equation}
    \label{eq:symcurrents}
        J_{4n+1}(Z)=\begin{pmatrix}
        J_{\mathbb{W}_n^+}(Z) & 0 \\
        0 & J_{\mathbb{W}_n^-}(Z)
    \end{pmatrix}\ , \quad 
        J_{2n}(Z)=\begin{pmatrix}
        0 & \frac{J_{\mathbb{W}_n}(Z)}{2} \\
        \frac{J_{\mathbb{W}_n}(Z)}{2}  & 0 \\
    \end{pmatrix}   \ , \quad J_{4n+3}(Z)=0
    \end{equation}
The factor $N^2-1$ in $\mathcal{W}[J]$ appears because of the trace over the $\mathfrak{su}(N)$ indices in the adjoint representations. In subsection \ref{sec:check} we provide a check of the generating functional of the correlators of the $\mathbb{W}_n$ superfield in the super Yang-Mills sector, thus showing that our methods are compatible with the ordinary space techniques used in \cite{BPS41,BPS42}.

\subsection{Euclidean generating functionals}
\label{sec:eucgenfunctsym}

The notion of Wick rotation in superspace is explained in details in appendix \ref{app:euclidean} and its effect on our generating functionals is discussed int subsection \ref{sec:genfunct1}. The notation is explained in appendix \ref{app:euclidean} and in subsection \ref{sec:genfunctsym}. In this subsection, we will show only the final results of this procedure.
\par The Euclidean generating functional of the connected correlators of the twist-$2$ operators in the zero-coupling limit is
{\small
\begin{equation}
\label{eq:eucsymgenfun}
\begin{split}
    &\mathcal{W}^E[J^E]= \\
    &\quad\frac{N^2-1}{2}\mathrm{str}\ \mathrm{log}\left[\delta^{(8)}(Z_1^E,Z_2^E)\mathds{1}_{2\times 2}\delta_{n_1k_1,n_2k_2}-2\ (\mathbf{\Delta}^{-1})_{n_1k_1,n'k'}^E(Z_1^E,Z_2^E)\mathcal{M}_{n'k',n_2k_2}i^{\lfloor\frac{n_2}{2}\rfloor} \frac{J_{n_2}^E(Z_2^E)}{N}\right]
\end{split}
\end{equation}}
The external currents and the matrices $\mathcal{M}$ are defined as in Eq. \eqref{eq:mnk}. The matrix $(\mathbf{\Delta}^{-1})^E_{n_1k_1,n'k'}(Z_1,Z_2)$ is
\begin{equation}
      (\mathbf{\Delta}^{-1}_{nk,n'k'})^E(Z_1^E,Z_2^E)=\begin{pmatrix}
        0 & (\Delta^{-1}_{nk,n'k'})^E(Z_1^E,Z_2^E) \\
        (\Delta^{-1}_{nk,n'k'})^E(Z_1^E,Z_2^E) & 0
    \end{pmatrix}
\end{equation}
whose entries are the kernels in \eqref{eq:epsidelta} with $j_{\Psi}=1$
\begin{align}
\label{eq:ewdelta}
        (\Delta^{-1})_{nk,n'k'}^E(Z_1^E,Z_2^E)=&-i\frac{1}{\Gamma(1+\lfloor\frac{n-k}{2}\rfloor)\Gamma(2+\lfloor\frac{n-k+1}{2}\rfloor)}\frac{1}{\Gamma(1+\lfloor\frac{k'}{2}\rfloor)\Gamma(2+\lfloor\frac{k'+1}{2}\rfloor)}(-1)^{\lfloor\frac{k'}{2}\rfloor} \nonumber\\
         & (-1)^{\lfloor\frac{k'}{2}\rfloor}\left(2^{\frac{5}{2}}(-ix_{1\bar{2}}^E)^{\bar{z}}\right)^{\lfloor\frac{n-k}{2}\rfloor+\lfloor\frac{k'}{2}\rfloor}\frac{\Gamma\left(2+\lfloor\frac{n-k+1}{2}\rfloor+\lfloor\frac{k'+1}{2}\rfloor\right)}{(-(x_{1\bar{2}}^E)^2)^{2+\lfloor\frac{n-k+1}{2}\rfloor+\lfloor\frac{k'+1}{2}\rfloor}} \nonumber\\
         &\Big[ \left(4(x_{1\bar{2}}^E\bar{\theta}_{12}^E)_1\right)^{(n-k)\mathrm{mod}\ 2}\left(4(\theta_{12}^Ex_{1\bar{2}}^E)_{\dot{1}}\right)^{k' \mathrm{mod}\ 2}+\nonumber\\
        +i& \frac{2^{\frac{5}{2}}\left[(n-k) \mathrm{mod}\ 2\right]\ \left[k'\mathrm{mod}\ 2\right]}{\lfloor\frac{n-k}{2}\rfloor+\lfloor\frac{k'+1}{2}\rfloor+1} (x_{1\bar{2}}^E)^2\left((x_{1\bar{2}}^E)_z+2\sqrt{2}(\theta_{12}^E)_1(\bar{\theta}_{12}^E)_{\dot{1}}\right) \Big]
\end{align}
The entries of the currents are $J_{n}^E(Z^E)$ are
    \begin{equation}
    \label{eq:esymcurrents}
        J_{4n+1}^E(Z^E)=\begin{pmatrix}
        J_{\mathbb{W}_n^+}^E(Z^E) & 0 \\
        0 & J_{\mathbb{W}_n^-}^E(Z^E)
    \end{pmatrix}\ , \quad 
        J_{2n}^E(Z^E)=\begin{pmatrix}
        0 & \frac{J_{\mathbb{W}_n}^E(Z^E)}{2} \\
        \frac{J_{\mathbb{W}_n}^E(Z^E)}{2}  & 0 \\
    \end{pmatrix}   \ , \quad J_{4n+3}^E(Z^E)=0
 \end{equation}

\subsection{Matching with the conformal generating functional in the component formalism}
\label{sec:check}
In this subsection we apply the rules of subsection \ref{sec:ordinary} to find the generating functional of the conformal connected correlators of the lowest component of the balanced operators $\mathbb{W}_n$ defined in Eq. \eqref{eq:ww}. This computation shows that in this special case our results exactly coincide with those in Refs. \cite{BPS41, BPS42} by the component formalism.
\par To find contact with Refs. \cite{BPS41, BPS42}, we define the rescaled superfield
\begin{align}
\label{eq:wsmall}
    w_s=&\frac{N^2C_W}{2}\frac{s!(s+1)!}{(-1)^{s-1}2^{\frac{3}{2}s}}\mathbb{W}_{s-1} \nonumber \\
    =&\frac{N}{g^2}\frac{s!(s+1)!}{(-1)^{s-1}2^{\frac{3}{2}s}}\ \mathrm{tr}\ (e^{-V}\ \bar{W}_{\dot{1}} e^{V})\ \mathbb{C}^{\frac{3}{2},\frac{3}{2}}_{2(s-1)}\left(\overleftarrow{\nabla}_1+i\overleftarrow{\bar{\nabla}}_{\dot{1}}, \overrightarrow{\nabla}_1+i\overrightarrow{\bar{\nabla}}_{\dot{1}}\right)\  W_1
\end{align}
where we have used the definition of $\mathbb{W}_{s-1}$ in Eq. \eqref{eq:ww}. In the language of Refs. \cite{BPS41, BPS42}, its lowest component in the Wess-Zumino and light-cone gauge is
\begin{align}
\label{eq:stilde}
        w_s\rvert_{\theta=\bar{\theta}=0}=&\frac{N}{2g^2}\left(\bar{\lambda}^aC_{s-1}^{3/2}\lambda^a+\frac{6}{s+2}\partial_+A^aC_{s-2}^{5/2}\partial_+\bar{A}^a\right)\nonumber \\
        =&\begin{cases}
            S_{s}^{(2)}=O^{\lambda}_{s}+\frac{6}{s+2}O^A_{s} \qquad & s\ \text{even} \\
            \tilde{S}_{s}^{(2)}=\tilde{O}^{\lambda}_{s}-\frac{6}{s+2}\tilde{O}^A_{s} \qquad & s\ \text{odd}
        \end{cases}
\end{align}
where $s$ is the collinear spin of the lowest component. We omitted the spacetime indices and, contrarily to Refs. \cite{BPS41, BPS42}, the operators are expressed in Wilsonian normalization. For simplicity, we will consider only the operators $w_s$ with even $s$.
\par We rewrite the $\mathcal{N}=1$ SYM theory generating functional of Eq. \eqref{eq:symgenfun} in the limit $g\to 0$ in ordinary space by using the results of subsection \ref{sec:ordinary}
\begin{equation}
\label{eq:simplified}
\begin{split}
    &\mathcal{W}[K]=\frac{N^2-1}{2}\mathrm{str}\ \mathrm{log}\Big[\delta^{(4)}(x_1-x_2){\delta^A}_B\mathds{1}_{2\times 2}\delta_{n_1k_1,n_2k_2} \\
    -&2\sum_{A',C,n',k'}(-1)^{|e_{B}|(|K^{n_2}_C|+|e_{C}|)}  (\mathbf{\Delta}^{-1}_{n_1k_1,n'k'})^{AA'}(x_1,x_2)\mathcal{M}_{n'k',n_2k_2}{T_{A'B}}^Ci^{\lfloor\frac{n_2}{2}\rfloor}\frac{K_{n_2,C}(x_2)}{N}\Big]
\end{split}
\end{equation}
We are interested only in the lowest component of $w_s$ with even $s$. Hence, we must put to zero all sources except $K_{n_2, 1}$, where, according to the second line of Eq. \eqref{eq:wsmall}
\begin{equation}
    n_2=2(s_2-1)\ , \qquad (s_2\ \text{even})
\end{equation}
Since the matrix $\mathcal{M}$ is diagonal in $n$, the indices that survive in the sums are
\begin{equation}
\begin{split}
    n_1=2(s_1-1)\ , \quad n'=2(s'-1)\  , \qquad (s_1, s'\ \text{even})
\end{split}
\end{equation}
In the rest of this subsection, we will always take $s_1, s_2$ and $s'$ to be even. Given that we only consider $K_{2(s_2-1), 1}$, and hence the index $^C$ in Eq. \eqref{eq:simplified} is equal to $1$, the only structure constant we need (see Eq. \eqref{eq:Talgebra}) is
\begin{equation}
    {T_{AB}}^1=\begin{pmatrix}
        +1 & 0 & 0 & 0 \\
        0 & 0 & 0 & 0 \\
        0 & 0 & 0 & 0 \\
        0 & 0 & 0 & 0 \\
    \end{pmatrix}
\end{equation}
Therefore, by inspection of ${T_{AB}}^1$ the indices $A,A',B$ in the expression \eqref{eq:simplified} can be only equal to $1$. This means that
\begin{equation}
    (-1)^{|e_B|(|K_C^{n_2}|+|e_C|)}=1
\end{equation}
because $(-1)^{|e_1|}=1$ as shown in Eq. \eqref{eq:esmall}. The generating functional now takes the form
 {\small
\begin{equation}
\label{eq:simplified1}
\begin{split}
    \mathcal{W}[K]=\frac{N^2-1}{2}&\mathrm{str}\ \mathrm{log}\Big[\delta^{(4)}(x_1-x_2)\mathds{1}_{2\times 2}\delta_{s_1k_1,s_2k_2} \\
    -&2\sum_{s',k'}(\mathbf{\Delta}^{-1}_{2(s_1-1)\ k_1,2(s'-1)\ k'})^{11}(x_1,x_2)\mathcal{M}_{2(s'-1)\ k',2(s_2-1)\ k_2}i^{s_2-1}\frac{K_{2(s_2-1),1}(x_2)}{N}\Big]
\end{split}
\end{equation}}
We recall that by Eq. \eqref{eq:phipsimatrix}
\begin{equation}
    \mathcal{M}_{2(s'-1)\ k',2(s_2-1)\ k_2}=(-1)^{k'}\delta_{s's_2}\delta_{k'k_2}\mathds{1}_{2\times 2}
\end{equation}
Looking at the definition Eq. \eqref{eq:kdef} and Eq. \eqref{eq:symcurrents}, we see that the matrix $K_{2(s-1),1}(x)$ can be decomposed as
\begin{equation}
    K_{2(s-1),1}(x)=\frac{NC_W}{2}\frac{s!(s+1)!}{(-1)^{s-1}2^{\frac{3}{2}s}}\begin{pmatrix}
        0 & \frac{K_s(x)}{2} \\
        \frac{K_s(x)}{2} & 0
    \end{pmatrix}
\end{equation}
The c-number current $K_s$ appearing in the rhs is \emph{defined} by the above expression. The normalizaton has been chosen so that $K_s$ is coupled to the $w_s$ defined in Eq. \eqref{eq:wsmall}.
\par Since the operators $w_s$ are balanced, we can take the determinant over the $2\times 2$ indices by using Eq. \eqref{eq:ogenfun} and by recalling that the indices $s_1,s_2$ are even
{\small
\begin{align}
    \mathcal{W}[K]=(N^2-1)\ &\mathrm{str}\ \mathrm{log}\Bigg[\delta^{(4)}(x_1-x_2)\delta_{\tilde{s}_1k_1,\tilde{s}_2k_2} \nonumber\\
    -&(-1)^{k_1}(\Delta^{-1})^{11}_{2\tilde{s}_1\ k_1,2\tilde{s}_2\ k_2}(x_1,x_2)\frac{NC_W}{2}\frac{(\tilde{s}_2+1)!(\tilde{s}_2+2)!}{i^{\tilde{s}_2}2^{\frac{3}{2}\tilde{s}_2+\frac{3}{2}}}K_{\tilde{s}_2+1}(x_2)\Bigg]
\end{align}}
For clarity, we introduced the new indices
\begin{equation}
    \tilde{s}_1=s_1-1\ , \qquad \tilde{s}_2=s_2-1
\end{equation}
that are always taken to be odd. We will restore the original indices $s_1, s_2$ near the end of this subsection. Since the
{\footnotesize
\begin{equation}
\label{eq:wk}
\begin{split}
    &\mathcal{W}[K]=(N^2-1)\\
    \Bigg( &\mathrm{tr}\ \mathrm{log}\left[\delta^{(4)}(x_1-x_2)\delta_{\tilde{s}_1\ell_1,\tilde{s}_2\ell_2}-(\Delta^{-1})^{11}_{2\tilde{s}_1\ 2\ell_1,2\tilde{s}_2\ 2\ell_2}(x_1,x_2)\frac{NC_W}{2}\frac{(\tilde{s}_2+1)!(\tilde{s}_2+2)!}{i^{\tilde{s}_2}2^{\frac{3}{2}\tilde{s}_2+\frac{3}{2}}}K_{\tilde{s}_2+1}(x_2)\right] \\
    &-\mathrm{tr}\ \mathrm{log}\left[\delta^{(4)}(x_1-x_2)\delta_{\tilde{s}_1 \ell_1,\tilde{s}_2 \ell_2}+(\Delta^{-1})^{11}_{2\tilde{s}_1\ 2\ell_1+1,2\tilde{s}_2\ 2\ell_2+1}(x_1,x_2)\frac{NC_W}{2}\frac{(\tilde{s}_2+1)!(\tilde{s}_2+2)!}{i^{\tilde{s}_2}2^{\frac{3}{2}\tilde{s}_2+\frac{3}{2}}}K_{\tilde{s}_2+1}(x_2)\right]\Bigg)
\end{split}
\end{equation}}
Now we need to determine the kernels $(\Delta^{-1})^{11}_{2\tilde{s}_1\ 2\ell_1,2\tilde{s}_2\ 2\ell_2}(x_1,x_2)$ and $(\Delta^{-1})^{11}_{2\tilde{s}_1\ 2\ell_1+1,2\tilde{s}_2\ 2\ell_2+1}(x_1,x_2)$. They can be found using Eqs. \eqref{eq:deltank}, \eqref{eq:ww} and \eqref{eq:wsmall}. We obtain
\begin{subequations}
    \begin{align}
        &(\Delta^{-1})^{11}_{2\tilde{s}_1\ 2\ell_1,2\tilde{s}_2\ 2\ell_2}(x_1,x_2)= \nonumber \\
        &-(2\pi^2)\frac{1}{\Gamma(1+\lfloor\frac{2\tilde{s}_1-2\ell_1}{2}\rfloor)\Gamma(2+\lfloor\frac{2\tilde{s}_1-2\ell_1+1}{2}\rfloor)}\frac{1}{\Gamma(1+\lfloor\frac{2\ell_2}{2}\rfloor)\Gamma(2+\lfloor\frac{2\ell_2+1}{2}\rfloor)}(-1)^{\lfloor\frac{2\ell_2}{2}\rfloor}\nonumber\\
        & (D_1^{(1)}+i\bar{D}_{\dot{1}}^{(1)})^{2\tilde{s}_1-2\ell_1}(D_1^{(2)}+i\bar{D}_{\dot{1}}^{(2)})^{2\ell_2}\partial_+\frac{1}{-\Box}(x_{1\bar{2}})\Bigg\rvert_{\theta=\bar{\theta}=0}
    \end{align}
    \begin{align}
        &(\Delta^{-1})^{11}_{2\tilde{s}_1\ 2\ell_1+1,2\tilde{s}_2\ 2\ell_2+1}(x_1,x_2)=   \nonumber\\
        &-(2\pi^2)\frac{1}{\Gamma(1+\lfloor\frac{2\tilde{s}_1-2\ell_1-1}{2}\rfloor)\Gamma(2+\lfloor\frac{2\tilde{s}_1-2\ell_1}{2}\rfloor)}\frac{1}{\Gamma(1+\lfloor\frac{2\ell_2+1}{2}\rfloor)\Gamma(2+\lfloor\frac{2\ell_2+2}{2}\rfloor)}(-1)^{\lfloor\frac{2\ell_2+1}{2}\rfloor}\nonumber\\
        & (D_1^{(1)}+i\bar{D}_{\dot{1}}^{(1)})^{2\tilde{s}_1-2\ell_1-1}(D_1^{(2)}+i\bar{D}_{\dot{1}}^{(2)})^{2\ell_2+1}\partial_+\frac{1}{-\Box}(x_{1\bar{2}})\Bigg\rvert_{\theta=\bar{\theta}=0}
    \end{align}
\end{subequations}
where $x_{1\bar{2}}$ is the supertranslation-invariant interval defined in Eq. \eqref{eq:interval}. The properties of the floor function allow us to write
\begin{align}
        & \Big\lfloor\frac{2\tilde{s}_1-2\ell_1}{2}\Big\rfloor=\Big\lfloor\frac{2\tilde{s}_1-2\ell_1+1}{2}\Big\rfloor=\tilde{s}_1-\ell_1 \nonumber\\ 
        & \Big\lfloor\frac{2\tilde{s}_1-2\ell_1-1}{2}\Big\rfloor= \tilde{s}_1-\ell_1-1 \nonumber\\
        & \Big\lfloor\frac{2\ell_2}{2}\Big\rfloor=\Big\lfloor\frac{2\ell_2+1}{2}\Big\rfloor=\ell_2 \nonumber\\
        & \Big\lfloor\frac{2\ell_2+2}{2}\Big\rfloor=\ell_2+1
\end{align}
We also have the identities
\begin{align}
\label{eq:id}
    &\displaystyle\left(D_1^{(1)}+i\bar{D}_{\dot{1}}^{(1)}\right)^{m}\left(D_1^{(2)}+i\bar{D}_{\dot{1}}^{(2)}\right)^{n}\frac{1}{-\Box}(x_{1\bar{2}})\Bigg\rvert_{\theta=\bar{\theta}=0}\nonumber\\
    =&\begin{cases}
        \displaystyle(-1)^{\tilde{s}_1-\ell_1}2^{\frac{3}{2}(\tilde{s}_1-\ell_1+\ell_2)}\partial_+^{\tilde{s}_1-\ell_1+\ell_2}\frac{1}{-\Box}(x_{12})\ , \quad & \text{if } m=2(\tilde{s}_1-\ell_1),\ n=2\ell_2 \\
        \displaystyle(-1)^{\tilde{s}_1-\ell_1}2^{\frac{3}{2}(\tilde{s}_1-\ell_1+\ell_2)}\partial_+^{\tilde{s}_1-\ell_1+\ell_2}\frac{1}{-\Box}(x_{12}) \ , \quad & \text{if } m=2(\tilde{s}_1-\ell_1-1)+1,\ n=2\ell_2+1
    \end{cases}
\end{align}
that follow trivially from the very definition of the supertranslation invariant interval $x_{1\bar{2}}$. In this way the kernels become
\begin{subequations}
    \begin{align}
        &(\Delta^{-1})^{11}_{2\tilde{s}_1\ 2\ell_1,2\tilde{s}_2\ 2\ell_2}(x_1,x_2)= \nonumber \\
        &\quad-(2\pi^2)\frac{1}{(\tilde{s}_1-\ell_1)!(\tilde{s}_1-\ell_1+1)!}\frac{1}{\ell_2!(\ell_2+1)!}(-1)^{\tilde{s}_1+1-\ell_1+\ell_2}\partial_+^{\tilde{s}_1+1-\ell_1+\ell_2}\frac{1}{-\Box}(x_{12})
    \end{align}
    \begin{align}
        &(\Delta^{-1})^{11}_{2\tilde{s}_1\ 2\ell_1+1,2\tilde{s}_2\ 2\ell_2+1}(x_1,x_2)= \nonumber \\
        &\quad-(2\pi^2)\frac{1}{(\tilde{s}_1-\ell_1-1)!(\tilde{s}_1-\ell_1+1)!}\frac{1}{\ell_2!(\ell_2+2)!}(-1)^{\tilde{s}_1+1-\ell_1+\ell_2}\partial_+^{\tilde{s}_1+1-\ell_1+\ell_2}\frac{1}{-\Box}(x_{12})
    \end{align}
\end{subequations}
or, with a slight change
\begin{subequations}
    \begin{equation}
        \begin{split}
       \frac{NC_W}{2}\frac{(\tilde{s}_2+1)!(\tilde{s}_2+2)!}{i^{\tilde{s}_2}2^{\frac{3}{2}\tilde{s}_2+\frac{3}{2}}}&(\Delta^{-1})^{11}_{2\tilde{s}_1\ 2\ell_1,2\tilde{s}_2\ 2\ell_2}(x_1,x_2)= \nonumber \\
       =&\frac{r_{\tilde{s}_1+1,\ell_1}}{r_{\tilde{s}_2+1,\ell_2}}\frac{i^{\tilde{s}_2}}{2^2}(\tilde{s}_2+2)\binom{\tilde{s}_1+1}{\ell_1+1}\binom{\tilde{s}_2+1}{\ell_2}\partial_+^{\tilde{s}_1+1-\ell_1+\ell_2}\frac{1}{-\Box}(x_{12})
         \end{split}
    \end{equation}
    \begin{equation}
    \begin{split}
     \frac{NC_W}{2}\frac{(\tilde{s}_2+1)!(\tilde{s}_2+2)!}{i^{\tilde{s}_2}2^{\frac{3}{2}\tilde{s}_2+\frac{3}{2}}}&(\Delta^{-1})^{11}_{2\tilde{s}_1\ 2\ell_1,2\tilde{s}_2\ 2\ell_2}(x_1,x_2)= \nonumber \\
     =&\frac{r_{\tilde{s}_1+1,\ell_1}'}{r_{\tilde{s}_2+1,\ell_2}'}\frac{i^{\tilde{s}_2}}{2^2}(\tilde{s}_2+2)\binom{\tilde{s}_1+1}{\ell_1+2}\binom{\tilde{s}_2+1}{\ell_2}\partial_+^{\tilde{s}_1+1-\ell_1+\ell_2}\frac{1}{-\Box}(x_{12})
         \end{split}
    \end{equation}
\end{subequations}
where we introduced the sequences of nonzero numbers
\begin{equation}
    r_{s,\ell}=\frac{(-1)^{\ell}2^{\frac{3}{2}(s-\ell-1)}(\ell+1)!}{(s-\ell)!s!}\ , \qquad r_{s,\ell}'=\frac{(-1)^{\ell}2^{\frac{3}{2}(s-\ell-1)}(\ell+2)!}{(s-\ell)!s!}
\end{equation}
and we repeatedly used the fact that $\tilde{s}_1$, $\tilde{s}_2$ are odd. We can finally plug everything into the expression \eqref{eq:wk} and restore the original indices $s_1, s_2$. We find
\begin{align}
        \mathcal{W}[K]=&(N^2-1)\ \mathrm{tr}\ \mathrm{log}\Bigg[\delta_{s_1\ell_1,s_2\ell_2}I_{\mathbb{R}^4}-\frac{r_{s_1,\ell_1}}{r_{s_2,\ell_2}}\frac{i^{s_2}}{2^2}(s_2+1)\binom{s_1}{\ell_1+1}\binom{s_2}{\ell_2}\partial_+^{s_1+\ell_1+\ell_2}\frac{-i}{-\Box}\  K_{s_2} \Bigg] \nonumber\\
        -&(N^2-1)\ \mathrm{tr}\ \mathrm{log}\Bigg[\delta_{s_1\ell_1,s_2\ell_2}I_{\mathbb{R}^4}+\frac{r_{s_1,\ell_1}'}{r_{s_2,\ell_2}'}\frac{i^{s_2}}{2^2}(s_2+1)\binom{s_1}{\ell_1+2}\binom{s_2}{\ell_2}\partial_+^{s_1-\ell_1+\ell_2}\frac{-i}{-\Box}\  K_{s_2} \Bigg]
\end{align}
where we omitted the spacetime indices for brevity and introduce the symbol $I_{\mathbb{R}^4}=\delta^{(4)}(x_1-x_2)$. We can now use the possibility to redefine the kernel according to Eq. \eqref{eq:rescaling} to eliminate the factors containing $r_{s,\ell}$ and $r_{s',\ell'}$. We finally obtain
\begin{align}
        \mathcal{W}[K]=&(N^2-1)\ \mathrm{tr}\ \mathrm{log}\Bigg[\delta_{s_1\ell_1,s_2\ell_2}I_{\mathbb{R}^4}-\frac{i^{s_2}}{2^2}(s_2+1)\binom{s_1}{\ell_1+1}\binom{s_2}{\ell_2}\partial_+^{s_1+\ell_1+\ell_2}\frac{-i}{-\Box}\  K_{s_2} \Bigg] \nonumber\\
        -&(N^2-1)\ \mathrm{tr}\ \mathrm{log}\Bigg[\delta_{s_1\ell_1,s_2\ell_2}I_{\mathbb{R}^4}+\frac{i^{s_2}}{2^2}(s_2+1)\binom{s_1}{\ell_1+2}\binom{s_2}{\ell_2}\partial_+^{s_1-\ell_1+\ell_2}\frac{-i}{-\Box}\  K_{s_2} \Bigg]
\end{align}
This formula can be compared with the corresponding result of Ref. \cite{BPS41}. We start from the conformal generating functional of Eq. (62) of Ref. \cite{BPS41}. We put al the external currents to zero except for $j_{O^{A}_{sk}}$ and $j_{O^{\lambda}_{sk}}$. The resulting generating functional is
\begin{align}
    &\Gamma_{\text{conf}}\left[j_{O^{A}},0,j_{O^{\lambda}},0,0,0\right]= \nonumber\\
    &+(N^2-1)\log\det \left[\delta_{s_1k_1, s_2k_2}\delta^{(4)}(x_1-x_2)-\mathcal{D}^{-1}_{\lambda\, s_1k_1,s_2k_2}(x_1-x_2) j_{O^\lambda_{s_2k_2}}(x_2)\right] \nonumber \\
	&-(N^2-1)\log\det \left[\delta_{s_1k_1, s_2k_2}\delta^{(4)}(x_1-x_2)+\mathcal{D}^{-1}_{A\, s_1k_1,s_2k_2}(x_1-x_2) j_{O^A_{s_2k_2}}(x_2)\right] 
\end{align}
The kernels are defined as
\begin{align}
    &\mathcal{D}^{-1}_{\lambda\, s_1k_1, s_2k_2}(x)=\frac{1}{2}\frac{s_1+1}{2}{s_1\choose k_1}{s_2\choose k_2+1}(-i\partial_{+})^{s_1-k_1+k_2-1}(-i\partial_{+})\frac{-i}{-\square}(x) \nonumber \\
    &\mathcal{D}^{-1}_{A\, s_1k_1, s_2k_2}(x) =\frac{1}{2}\frac{\Gamma(3)\Gamma(s_1+3)}{\Gamma(5)\Gamma(s_1+1)}{s_1\choose k_1}{s_2\choose k_2+2}(-i\partial_{+})^{s_1-k_1+k_2}\frac{-i}{-\square}(x)
\end{align}
We recall that in these expressions the indices $s_1, s_2$ are always even. Substituting the kernels into $\Gamma_{\text{conf}}$ and omitting the spacetime indices, one finds
{\small
\begin{align}
    &\Gamma_{\text{conf}}\left[j_{O^{A}},0,j_{O^{\lambda}},0,0,0\right]= \nonumber\\
    &+(N^2-1)\log\det \left[\delta_{s_1k_1, s_2k_2}I_{\mathbb{R}^4}-\frac{(-i)^{s_1-k_1+k_2}}{2^2}(s_1+1){s_1\choose k_1}{s_2\choose k_2+1}\partial_{+}^{s_1-k_1+k_2}\frac{-i}{-\square}j_{O^\lambda_{s_2k_2}}\right] \nonumber \\
	&-(N^2-1)\log\det \left[\delta_{s_1k_1, s_2k_2}I_{\mathbb{R}^4}+\frac{(-i)^{s_1-k_1+k_2}}{24} (s_1+1)(s_1+2){s_1\choose k_1}{s_2\choose k_2+2}\partial_{+}^{s_1-k_1+k_2}\frac{-i}{-\square}j_{O^A_{s_2k_2}}\right]
\end{align}}
Since the currents $j_{O^\lambda_{sk}}$ and $j_{O^A_{sk}}$ generate the insertion of the $k$-th monomial in $O^\lambda_{s}$ and $O^A_{s}$ respectively, we choose
\begin{align}
\label{eq:j=K}
    &j_{O^\lambda_{si}}(x)=K_s(x) \ , \qquad && i=0,1,...,s-1 \nonumber \\
    &j_{O^A_{si}}(x)=\frac{6}{s+2}K_s(x)\ ,\qquad && i=0,1,...,s-2
\end{align}
where $K_s$ is the current appearing in Eq. \eqref{eq:wk} that for even $s$ generates the insertions of the operator $S_{s}^{(2)}$ defined in Eq. \eqref{eq:stilde}. With this choice, the generating funtional becomes
\begin{align}
\label{eq:gammaconf}
    &\Gamma_{\text{conf}}\left[j_{O^{A}},0,j_{O^{\lambda}},0,0,0\right]\rvert_{j=K}= \nonumber\\
    &+(N^2-1)\tr\log \left[\delta_{s_1k_1, s_2k_2}I_{\mathbb{R}^4}-\frac{(-i)^{s_1-k_1+k_2}}{2^2}(s_1+1){s_1\choose k_1}{s_2\choose k_2+1}\partial_{+}^{s_1-k_1+k_2}\frac{-i}{-\square}K_{s_2}\right] \nonumber \\
	&-(N^2-1)\tr\log \left[\delta_{s_1k_1, s_2k_2}I_{\mathbb{R}^4}+\frac{(-i)^{s_1-k_1+k_2}}{2^2} (s_1+1){s_1\choose k_1}{s_2\choose k_2+2}\partial_{+}^{s_1-k_1+k_2}\frac{-i}{-\square}K_{s_2}\right]
\end{align}
where the symbol $\rvert_{j=K}$ denotes the condition \eqref{eq:j=K} and used the property log det = tr log. The equality between Eq. \eqref{eq:gammaconf} and Eq. \eqref{eq:wk} follows by applying the identity \eqref{eq:rescaling}.

\section{Renormalization-group improvement}\label{sec:rgimprove}
In this section, we follow Ref. \cite{Bochicchio:2022uat} to improve the results of sections \ref{sec:appliedcorrelators} and \ref{sec:sym} with the aid of the renormalization group. This method allows us to show that for asymptotically free theories that are free in the zero-coupling limit the generating functional of those asymptotic correlators that do not vanish in the conformal limit retains the form of a logarithm of a functional determinant. In this section, all operators and correlators are assumed to be Wick-rotated. Euclidean objects will be denoted with a superscript $^{E}$. Except when the subscript $_{\text{bare}}$ is present, operators are intended to be renormalized. We denote their infinite renormalization constant with the letter $\mathscr{Z}$ and the finite renormalization constant arising from the Callan-Symanzik equation as $\mathcal{Z}$.
\par The formulation of supersymmetric field theories in Euclidean superspace is extensively described in Ref. \cite{Lukierski:1982hr} and summarized in Ref. \cite{Morris:1985hi} and in appendix \ref{app:euclidean}. We assume the existence of a gauge-invariant regularization procedure that preserves four-dimensional supersymmetry\footnote{Since the operators we are interested in have nice transformation properties under the collinear superconformal algebra, one may also employ a regularization procedure that preserves supersymmetry along the light-cone directions only, in analogy to Ref. \cite{Belitsky:2005qn}.}\cite{Stockinger:2005gx}.

\subsection{Operator mixing: generalities}
\label{sec:mixing1}
In order to keep the notation light, we  work in ordinary space following Ref. \cite{Bochicchio:2022uat}. Everything that we say also applies  to theories defined on superspace \cite{Gates:1983nr}. Consider the Euclidean \emph{connected} correlator 
\begin{equation}
\label{eq:oi}
    G^{E}_{I_1...I_n}(x_1^E,...,x_n^E; \mu, g(\mu))\equiv \expval{O_{I_1}^{E}(x_1^E),...,O_{I_n}^{E}(x_n^E)}_{\mathrm{conn}}
\end{equation}
where the local operators $O_{I}^{E}(x)$ form a basis of operators that mix under renormalization, and have canonical dimension $D_{I}$ and an anomalous dimension matrix ${\gamma_{I}}^{J}(g)$. These connected correlators satisfy the Callan-Symanzik equation
\begin{align}
       & \Big(\sum_{i=1}^nx_i^E\cdot \frac{\partial}{\partial x_i^E}+\beta(g)\frac{\partial}{\partial g}+\sum_{i=1}^n D_{I_i}\Big)G^{E}_{I_1I_2...I_n}(x_i^E;\mu, g(\mu))+\nonumber \\
        &+\sum_{J}\left({\gamma_{I_1}}^{J}(g)G^{E}_{JI_2...I_n}(x_i^E;\mu, g(\mu)) +...+{\gamma_{I_n}}^{J}(g)G^{E}_{I_1I_2...J}(x_i^E;\mu, g(\mu))\right)=0
\end{align}
whose solution is
\begin{equation}
\label{eq:cssolution}
\begin{split}
    G^{E}_{I_1I_2...I_n}(\lambda x_i^E&;\mu, g(\mu))= \\
    =&\lambda^{-\sum_i D_{I_i}}\sum_{J_1,J_2,...,J_n}{\mathcal{Z}_{I_1}}^{J_1}(\lambda){\mathcal{Z}_{I_2}}^{J_2}(\lambda)...{\mathcal{Z}_{I_n}}^{J_n}(\lambda)G^{E}_{J_1J_2...J_n}(x_i^E;\mu, g(\mu/\lambda))
\end{split}
\end{equation}
The matrices ${\mathcal{Z}_{I}}^{J}(\lambda)$ satisfy the matrix differential equation
\begin{equation}
    \left(\frac{\partial}{\partial g}+\frac{\gamma(g)}{\beta(g)}\right)\mathcal{Z}(\lambda)=0
\end{equation}
whose solution is
\begin{equation}
    \mathcal{Z}(\lambda)=\mathrm{P}\ \mathrm{exp}\left(\int_{g(\mu)}^{g(\mu/\lambda)}dg\frac{\gamma(g)}{\beta(g)}\right)
\end{equation}
where $\mathrm{P}$ denotes path-ordering.
\par Suppose now that there is a renormalization scheme in which $\gamma(g)/\beta(g)$ is diagonal and one-loop exact (we will refer to such a scheme as \emph{non-resonant diagonal}, and we will justify its existence at the end of this subsection). In this scheme
\begin{equation}
    {\mathcal{Z}_I}^J(\lambda)=\mathcal{Z}_I(\lambda){\delta_I}^J
\end{equation}
with
\begin{equation}
\label{eq:renfact}
    \mathcal{Z}_{I}(\lambda)=\left(\frac{g(\mu)}{g\left(\frac{\mu}{\lambda}\right)}\right)^{\frac{\gamma_0^{(I)}}{\beta_0}}
\end{equation}
In an asymptotically free theory $\beta_0 > 0$ that implies
\begin{equation}
    \quad g^2(\mu/\lambda)\underset{\lambda\to 0}{\sim}\frac{1}{\beta_0\ \mathrm{log}\left(\frac{1}{\lambda^2}\right)}\left(1-\frac{\beta_1}{\beta_0^2}\frac{\mathrm{log}\ \mathrm{log}\left(\frac{1}{\lambda^2}\right)}{\mathrm{log}^2\left(\frac{1}{\lambda^2}\right)}\right)
\end{equation}
Then, in the non-resonant diagonal scheme, Eq. \eqref{eq:cssolution} reduces to
\begin{equation}
\label{eq:csdiagonal}
\begin{split}
    G^{E}_{I_1I_2...I_n}(\lambda x_i^E&;\mu, g(\mu))=\lambda^{-\sum_i D_{I_i}}\mathcal{Z}_{I_1}(\lambda)\mathcal{Z}_{I_2}(\lambda)...\mathcal{Z}_{I_n}(\lambda)G^{E}_{I_1I_2...I_n}(x_i^E;\mu, g(\mu/\lambda))
\end{split}
\end{equation}
In perturbation theory, the correlator $G^{E}_{I_1I_2...I_n}(x_i^E;\mu, g(\mu/\lambda))$ in the rhs of Eq. \eqref{eq:csdiagonal} admits the asymptotic expansion in powers of the running coupling $g^2(\mu/\lambda)$
\begin{align}
\label{eq:expansion}
    G^{E}_{I_1I_2...I_n}(x_i^E;\mu, g(\mu/\lambda))=G^{E}_{\mathrm{conf}\ I_1I_2...I_n}(x_i^E)+\sum_{k=1}^{\infty}g^{2k}(\mu/\lambda)\ G^{E}_{2k; I_1I_2...I_n}(x_i^E;\mu)
\end{align}
The first coefficient of this expansion $G^{E}_{\mathrm{conf}\ I_1I_2...I_n}(x_i)$, being independent of the coupling, coincides with the conformal correlator at zero coupling computed as in sections \ref{sec:genfreescft}, \ref{sec:appliedcorrelators}. If $G_{\text{conf}}^E$ does not vanish, since in asymptotically free theories the coupling goes to zero in the limit $\lambda\to 0$, for fixed coordinates all the remaining terms in Eq. \eqref{eq:csdiagonal} are subleading with respect to the conformal one. As a consequence, asymptotically
\begin{equation}
    G^{E}_{I_1I_2...I_n}(x_i^E;\mu, g(\mu/\lambda)) \underset{\lambda \to 0}{\sim} G_{\mathrm{conf}\ I_1I_2...I_n}^{E}(x_i^E)
\end{equation}
Hence it follows that, asymptotically
\begin{equation}
\label{eq:asymp0}
    \begin{split}
        G^{E}_{I_1I_2...I_n}(\lambda x_i^E;\mu, g(\mu))\underset{\lambda \to 0}{\sim} \lambda^{-\sum_{i}D_{I_i}}\mathcal{Z}_{I_1}(\lambda)\mathcal{Z}_{I_2}(\lambda)...\mathcal{Z}_{I_n}(\lambda)G^{E}_{\mathrm{conf}\ I_1I_2...I_n}(x_i^E)
    \end{split}
\end{equation}
Consequently we define the asymptotic correlator
\begin{align}
\label{eq:asymp}
    G^{E}_{\mathrm{asymp}\ I_1I_2...I_n}(\lambda x_i^E;\mu, g(\mu))\equiv \lambda^{-\sum_{i}D_{I_i}}\mathcal{Z}_{I_1}(\lambda)\mathcal{Z}_{I_2}(\lambda)...\mathcal{Z}_{I_n}(\lambda)G^{E}_{\mathrm{conf}\ I_1I_2...I_n}(x_i^E)
\end{align}
that is the object we are interested in.

\par To conclude this section, we state three theorems that allow us to establish whether or not there is a scheme where $\gamma(g)/\beta(g)$ is diagonal and one-loop exact.

The anomalous dimension matrix of twist-$2$ operators in QCD was first calculated in Refs. \cite{Gross:1973ju, Gross:1974cs, PhysRevD.9.416}. Subsequently, the problem of its diagonalizability was tackled in Refs. \cite{Efremov:1978rn, Efremov:1979qk, Lepage:1979zb, Lepage:1980fj, Makeenko:1980bh, Shifman:1980dk}. In these works, it was noticed by direct computations, that conformal symmetry played a significant role in the construction of a diagonal basis of twist-$2$ operators for the one-loop anomalous dimension matrix $\gamma_0$. The role of conformal symmetry and unitarity was systematically understood by the formulation of the following theorems. The first of which has been proven in Refs. \cite{Ohrndorf:1981qv, Craigie:1983fb}, and establishes the diagonalizability of $\gamma_0$ under certain conditions.
\begin{theorem}
\label{thm:ctd}
    Consider a massless asymptotically free QCD-like theory that is conformal up to order $g^2$. Let $O_s^{E}$ be a gauge-invariant operator of given collinear twist $\tau$ that at $g=0$ for each $s=0,1,2,...$ reduces to a collinear primary conformal field of scaling dimension $D_s=\tau+s$. Let us assume that the one-loop mixing matrix in an minimal subtraction (MS) scheme\footnote{We recall that in the minimal subtraction scheme the conformal symmetry is lost at one loop. The conformal renormalization scheme can be reached through a finite scheme change at order $g^2$ \cite{Belitsky:2007jp} that does not affect the diagonal form of $\gamma_0$.} reads
    \begin{equation}
    \label{eq:mixing}
        O_s^{E}=\sum_{k=0}^{s}\mathscr{Z}_{s,k}\ (i\partial_z^E)^{s-k} O_{k\ \mathrm{bare}}^{E}
    \end{equation}
    where $\mathscr{Z}_{s,k}$ are the divergent multiplicative renormalization factors.
    \par Then, $(\gamma_0)_{s,k}$ and $\mathscr{Z}_{s,k}$ are diagonal at one loop and $O_s^E$ are multiplicatively renormalizable at order $g^2$.
\end{theorem}
Theorem \ref{thm:ctd} applies to twist-$2$ operators \cite{Braun:2008ia} in pure YM theory \cite{Bochicchio:2022uat} and in $\mathcal{N}=1$ SYM theory \cite{BPS41, BPS42}. \par More recently, a stronger version of theorem \ref{thm:ctd} was proven by noticing that, beyond conformal symmetry, also unitarity in the gauge-invariant sector poses further constraints on the one-loop renormalization properties of the twist-$2$ operators \cite{Becchetti:2021for}.
\begin{theorem}
\label{thm:bb}
    Consider a massless quantum field theory that is conformal up to order $g^2$ in perturbation theory (specifically, a massless, asymptotically free, QCD-like theory). Let $O_I$ a set gauge-invariant hermitian operators. Up to order $g^2$, conformal symmetry allows us to construct a set of states $\ket{O_{in}}$ and $\bra{O_{out}}$ by means of the operator-state correspondence (see Appendix \ref{app:supconfinnprod}). Let
    \begin{align}
        \mathcal{G}=\bra{O_{in}}\ket{O_{out}}=G_0+g^2G_1+...
    \end{align}
    Conformal symmetry up to order $g^2$ implies
    \begin{equation}
        \gamma_0G_0-G_0\gamma_0^T=0
    \end{equation}
    \par Then, if $\gamma_0$ is diagonalizable, $\gamma_0$ commutes with $G_0$ in the diagonal basis, and thus $\gamma_0$ and $G_0$ are simultaneously diagonalizable.
    \par Moreover, if $\gamma_0$ is nondiagonalizable, $G_0$ has necessarily both negative and positive eigenvalues, and the theory cannot be unitary in its free conformal limit.
\end{theorem}

Finally, the criteria for the existence of the non-resonant diagonal scheme are established by the following theorem.

\begin{theorem}
\label{thm:non-resonant}
    Let $\gamma(g)$ be the anomalous dimension matrix of a set of gauge-invariant operators that mix under renormalization in a massless, asymptotically free, QCD-like theory with beta function $\beta(g)$. Suppose that the matrix $\gamma_0/\beta_0$ is diagonal and non-resonant, i.e. the sequence of its eigenvalues in nonincreasing order $\lambda_1,\lambda_2,...$ satisfies
\begin{equation}
\label{eq:non-resonant}
    \lambda_i-\lambda_j\neq 2k\ , \qquad i>j
\end{equation}
for any nonvanishing integer $k$. Then, there exists a scheme in which the matrix $\gamma(g)/\beta(g)$ is diagonal and one-loop exact to all orders of perturbation theory.
\end{theorem}

It was verified numerically up to $s=10^4$ in Ref. \cite{Bochicchio:2022uat} and Refs. \cite{BPS41, BPS42} that theorem \ref{thm:non-resonant} respectively applies to twist-$2$ operators in pure YM theory and $\mathcal{N}=1$ SYM theory, so that the non-resonant diagonal scheme exists and the asymptotic estimates in Eq. \eqref{eq:asymp0} hold for the above operators.
\par In Ref. \cite{Scardino:2024bgr} it was theoretically demonstrated that $\gamma_0/\beta_0$ is non-resonant in pure YM theory for twist-$2$ operators. In an upcoming publication \cite{Scardino2025} it will be proven that $\gamma_0/\beta_0$ is non-resonant in $\mathcal{N}=1$ SYM theory for twist-$2$ operators.
\par In the next section we will describe how the above theorems intertwine with supersymmetry.

\subsection{Operator mixing: supersymmetric field theories}
\label{sec:susyopmix}
In supersymmetric theories, renormalization mixes superfields only with other superfields. Consider a set of superfields $\mathcal{O}^E_I(x^E,\theta^E,\bar{\theta}^E)$ that mix under renormalization and that have canonical dimension $D_I$. The solution of the Callan-Symanzik equation for their Euclidean $n$-point correlators is
\begin{align}
    &G^{E}_{I_1...I_n}( \lambda x_i^E, \lambda^{1/2}\theta_i^E, \lambda^{1/2}\bar{\theta}_i^E;\mu, g(\mu))= \nonumber \\
    &\quad\lambda^{-\sum_i D_{I_i}}\sum_{J_1,J_2,...,J_n}{\mathcal{Z}_{I_1}}^{J_1}(\lambda){\mathcal{Z}_{I_2}}^{J_2}(\lambda)...{\mathcal{Z}_{I_n}}^{J_n}(\lambda)G^{E}_{J_1...J_n}(x_i^E, \theta_i^E, \bar{\theta}_i^E;\mu, g(\mu/\lambda))
\end{align}
Theorem \ref{thm:bb} also applies to superfields with no additional assumption. If $\gamma_0/\beta_0$ satisfies the non-resonance condition \eqref{eq:non-resonant}, the asymptotic correlators for the superfields in the non-resonant diagonal scheme take the form
\begin{align}
\label{eq:diagcorrsusy}
    G^{E}_{\mathrm{asymp\ }I_1I_2...I_n}(\lambda x_i^E,& \lambda^{1/2}\theta_i^E, \lambda^{1/2}\bar{\theta}_i^E;\mu, g(\mu))\equiv \nonumber \\
    \equiv&\lambda^{-\sum_{i}D_{I_i}}\mathcal{Z}_{I_1}(\lambda)\mathcal{Z}_{I_2}(\lambda)...\mathcal{Z}_{I_n}(\lambda)\ G^{E}_{\mathrm{conf}\ I_1I_2...I_n}(x_i^E,\theta_i^E, \bar{\theta}_i^E)
\end{align}
As in the nonsupersymmetric case, this relation is valid as long as $G_{\mathrm{conf}}$ does not vanish. 
\par Our goal is now to extend Theorem \ref{thm:ctd} to supersymmetric field theories as well. Consider a set of Euclidean superfields $\mathcal{O}_s^{E}(Z^E)$ with the component expansion
\begin{align}
\label{eq:mathcalo}
    \mathcal{O}_s^{E}(Z^E)=O^{(1)E}_s(x^E)+&(\theta^E)^1 O^{(2)E}_s(x^E)+(\bar{\theta}^E)^{\dot{1}}O^{(3)E}_s(x^E) \nonumber \\
    +&(\theta^E)^1(\bar{\theta}^E)^{\dot{1}} O^{(4)E}_s(x^E)+(\theta^E)^1(\bar{\theta}^E)^{\dot{1}}\frac{b_s}{2j_s}(i\partial_z^E)O^{(1)E}_s(x^E)
\end{align}
To avoid irrelevant technical complications, we neglect the terms of the superfield component expansion that include $(\theta^E)^2$ and $(\bar{\theta}^E)^{\dot{2}}$. We assume that the superfields $\mathcal{O}_s^{E}(Z^E)$ for each $s$ at $g=0$ reduce to a superconformal primary that in Minkowski signature transforms under the irreducible representations $[j_s,b_s]$ of the collinear superconformal group. Notice that under this assumption each component of the superfield has a definite $R$ charge compatible with the irreducible representation $[j_s,b_s]$ (section \ref{sec:sl21}). We also assume that each component of the superfields $\mathcal{O}_s^{E}(Z^E)$ mixes with the other components with the same spin and $R$ charge, and with the derivatives of the components of the superfields with lower spin. \par 
Since super-Poincaré-covariant objects are allowed to mix only with super-Poincaré-covariant objects, the only possible mixings consistent with the symmetries are
\begin{align}
        &\mathcal{O}_s^{E}= \sum_{k=0}^{s}\mathscr{Z}^{(1)}_{s,k}(i\partial_z^E)^{s-k}\mathcal{O}_{k\ \mathrm{bare}}^E+\sum_{k=0}^{s-1}\mathscr{Z}_{s,k}^{(2)}(i\partial_z^E)^{s-1-k}\left(\frac{b_k-j_k}{2j_k}D^E_1\bar{D}^E_{\dot{1}}+\frac{b_k+j_k}{2j_k}\bar{D}^E_{\dot{1}}D^E_{1}\right)\mathcal{O}_{k\ \mathrm{bare}}^{E} \\
         \label{eq:do}
        &D_1^E\mathcal{O}_s^{E}=\sum_{k=0}^{s}\mathscr{Z}^{(3)}_{s,k}(i\partial_z^E)^{s-k}D_1^E\mathcal{O}_{k\ \mathrm{bare}}^{E} \\
        \label{eq:dobar}
        &\bar{D}^E_{\dot{1}}\mathcal{O}_s^{E}= \sum_{k=0}^{s}\mathscr{Z}^{(4)}_{s, k}(i\partial_z^E)^{s-k}\bar{D}^E_{\dot{1}}\mathcal{O}_{k\ \mathrm{bare}}^{E} \\ 
       & \bigg(\frac{b_s-j_s}{2j_s}D_1^E\bar{D}^E_{\dot{1}}+\frac{b_s+j_s}{2j_s}\bar{D}^E_{\dot{1}}D^E_1\bigg)\mathcal{O}_s^{E}=\sum_{k=0}^{s+1} \mathscr{Z}^{(5)}_{s,k}(i\partial_z^E)^{s+1-k}\mathcal{O}_{k\ \mathrm{bare}}^{E} \nonumber \\
        &\hspace{4.5cm}+\sum_{k=0}^{s}\mathscr{Z}_{s,k}^{(6)}(i\partial_z^E)^{s-k}\left(\frac{b_k-j_k}{2j_k}D^E_1\bar{D}^E_{\dot{1}}+\frac{b_k+j_k}{2j_k}\bar{D}^E_{\dot{1}}D^E_1\right)\mathcal{O}_{k\ \mathrm{bare}}^{E}
\end{align}
where all the superfields and their spinor derivatives are evaluated at $\theta=0$. Moreover, applying the differential operators $D_1^E$, $\bar{D}_{\dot{1}}^E$ on both sides of each of these relations, we find the constraints that follow from supersymmetry
\begin{equation}
\label{eq:relations}
    \begin{split}
        \mathscr{Z}_{s,k}^{(3)}=&\begin{cases}
            \mathscr{Z}_{s,s}^{(1)}\ , \qquad & (k=s) \\
            \mathscr{Z}_{s,k}^{(1)}+i2\sqrt{2}\left(\frac{b_k+j_k}{2j_k}\right)\mathscr{Z}_{s,k}^{(2)}\ , & (k<s)
        \end{cases} \\
        \mathscr{Z}_{s,k}^{(4)}=&\begin{cases}
            \mathscr{Z}_{s,s}^{(1)}\ , \qquad & (k=s) \\
            \mathscr{Z}_{s,k}^{(1)}+i2\sqrt{2}\left(\frac{b_k-j_k}{2j_k}\right)\mathscr{Z}_{s,k}^{(2)}\ , & (k<s)
        \end{cases} \\
        \mathscr{Z}_{s,k}^{(5)}=&\begin{cases}
            0\ , \qquad & (k=s+1) \\
            \frac{1}{i2\sqrt{2}}\left[\left(\frac{b_k+j_k}{2j_k}\right)\left(\frac{b_s-j_s}{2j_s}\right)\mathscr{Z}_{s,k}^{(3)}-\left(\frac{b_k-j_k}{2j_k}\right)\left(\frac{b_s+j_s}{2j_n}\right)\mathscr{Z}_{s,k}^{(4)}\right]\ , & (k<s+1)
        \end{cases} \\
        \mathscr{Z}_{s,k}^{(6)}=&\left(\frac{b_s+j_s}{2j_s}\right)\mathscr{Z}_{s,k}^{(3)}-\left(\frac{b_s-j_s}{2j_s}\right)\mathscr{Z}_{s,k}^{(4)}
    \end{split}
\end{equation}
Thanks to the $R$ symmetry, the conformal primaries in Eqs. \eqref{eq:do} and \eqref{eq:dobar} automatically satisfy the assumptions of theorem \ref{thm:ctd}. 
As a consequence, $\mathscr{Z}^{(3)}$ and $\mathscr{Z}^{(4)}$ are in fact diagonal that by \eqref{eq:relations} also implies that $\mathscr{Z}^{(1)}$, $\mathscr{Z}^{(2)}$, $\mathscr{Z}^{(5)}$ and $\mathscr{Z}^{(6)}$ are diagonal as well. Therefore, the combination of supersymmetry and theorem \ref{thm:ctd} implies that each superfield $\mathcal{O}_s$ is multiplicatively renormalizable at one loop in the MS scheme
\begin{equation}
\label{eq:diagonal}
    \mathcal{O}_s^{E}=\mathscr{Z}_s\mathcal{O}_{s\ \mathrm{bare}}^{E}
\end{equation}
As a consequence, $\gamma_0$ is diagonal
\begin{equation}
    (\gamma_0)_{ss'}=(\gamma_0)_s\delta_{ss'}
\end{equation}

\subsection{Renormalization-group improved generating functional}
In the non-resonant diagonal scheme the generating functional of the UV-asymptotic connected functions of the operators under consideration retains the same functional form of the generating functional of the generating function of conformal connected correlators, as can be seen by \cite{Bochicchio:2022uat}
\begin{align}
\label{eq:wasymp}
    &\mathcal{W}_{\mathrm{asymp}}^E[J^I;\lambda]\nonumber\\
    &\equiv \sum_{n=0}^{\infty}\sum_{I_i}\int[dx_i^E]\ G^{E}_{\mathrm{asymp}\ I_1...I_n}(\lambda x_i^E;\mu, g(\mu/\lambda))J^{I_1}(x_1^E)...J^{I_n}(x_n^E) \nonumber\\
    &=\sum_{n=0}^{\infty}\sum_{I_i}\int[dx_i^E]\ \lambda^{-\sum_I D_I}\mathcal{Z}_{I_1}(\lambda)...\mathcal{Z}_{I_n}(\lambda)G^{E}_{\mathrm{conf}\ I_1...I_n}(x_i^E)J^{I_1}(x_1^E)...J^{I_n}(x_n^E) \nonumber\\
    &=W_{\mathrm{conf}}^E\left[J^I\mathcal{Z}_I(\lambda)\lambda^{-D_I}\right]
\end{align}
\par The results and observations of this subsection including Eq. \eqref{eq:wasymp} apply to supersymmetric field theories defined in superspace with no modification. Suppose that in some supersymmetric field theory we find a set of superfields $\mathcal{O}^E_I(Z^E)$ to which the results of this section apply. Suppose also that the generating functional of conformal connected correlators in superspace has the form of the logarithm of a functional superdeterminant as those in sections \ref{sec:genfreescft}, \ref{sec:appliedcorrelators}, \ref{sec:sym}
\begin{equation}
    \mathcal{W}^E_{\mathrm{conf.}}[J]=\mathrm{const.}\times \mathrm{log}\ \mathrm{sdet}\left[\delta_{IJ}\delta^{(8)}(Z_1^E,Z_2^E)-(\Delta^{-1})^E_{IJ}(Z_1^E,Z_2^E)J^J(Z_2^E)\right]
\end{equation}
Then Eq. \eqref{eq:wasymp} implies that in the non-resonant diagonal scheme there the generating functional of asymptotic connected correlators takes the form
\begin{equation}
    \mathcal{W}^E_{\mathrm{asymp.}}[J]=\mathrm{const.}\times \mathrm{log}\ \mathrm{sdet}\left[\delta_{IJ}\delta^{(8)}(Z_1^E,Z_2^E)-(\Delta^{-1})^E_{IJ}(Z_1^E,Z_2^E)\lambda^{-D_J}\mathcal{Z}_J(\lambda)J^J(Z_2^E)\right]
\end{equation}
where $D_I$ is the canonical dimension of $\mathcal{O}_I^E$.

\subsection{Application to $\mathcal{N}=1$ SYM theory}
\label{sec:rgsym}
This theoretical machinery allows us to write the superspace form of the generating functional of the Euclidean UV asymptotic, connected correlators of twist-$2$ operators in $\mathcal{N}=1$ SYM theory. The ordinary spacetime version of this object was first worked out in Refs. \cite{BPS41, BPS42}.
\par The superfield twist-$2$ operators in $\mathcal{N}=1$ SYM theory are those of Eq. \eqref{eq:ww} with the light-cone components in Eqs. \eqref{eq:wwcomponents}. The correspondence between our superfields in Eq. \eqref{eq:ww} and their components in Refs. \cite{BPS41, BPS42} is shown in table \eqref{tab:bps}.

\begin{table}[h!]
    \centering
    \begin{tabular}{l|cccc}
         Superfield & \multicolumn{4}{c}{Content}   \\
         \hline
         $\mathbb{W}_n$ ($n$ even) & $\tilde{S}^{(2)}_{n+1} $ & $M_{n+1}$ & $\bar{M}_{n+1}$ & $S^{(1)}_{n+2} $   \\
         $\mathbb{W}_n$ ($n$ odd) & $S^{(2)}_{n+1} $ & $M_{n+1}$ &$\bar{M}_{n+1}$ & $\tilde{S}^{(1)}_{n+2}$   \\
         $\mathbb{W}_n^+$ & $T_{2n}$ & $S_{2n+2}^A $ & $S_{2n+2}^{\lambda}$ & $T_{2n+1}$  \\
         $\mathbb{W}_n^-$ & $\bar{T}_{2n}$ & $\bar{S}_{2n+2}^A $ & $\bar{S}_{2n+2}^{\lambda}$ & $\bar{T}_{2n+1}$   \\
    \end{tabular}
    \caption{Twist-$2$ operators of Refs. \cite{BPS41, BPS42} contained in the superfields of Eq. \eqref{eq:ww}.}
    \label{tab:bps}
\end{table}

In agreement with the results of section \ref{sec:susyopmix}, the one-loop anomalous dimensions of $\mathbb{W}_n$ and $\mathbb{W}_n^{\pm}$ given by\footnote{$\psi(z)=\Gamma'(z)/\Gamma(z)$ is the Digamma function}
\begin{align}
        \gamma_0^{(\mathbb{W}_n)}=& \frac{1}{4\pi^2}\left(\psi(n+4)+\psi(n+1)-2\psi(1)-\frac{2(-1)^n}{(n+1)(n+2)(n+3)}-\frac{3}{2}\right) \nonumber\\
         \gamma_0^{(\mathbb{W}_n^{\pm})}=& \frac{1}{4\pi^2}\left(2\psi(2n+3)-2\psi(1)-\frac{3}{2}\right) \nonumber\\
\end{align}
are constant along each supermultiplet \cite{Belitsky:2004sc, BPS41, BPS42}. The canonical dimensions are
\begin{equation}
    D^{(\mathbb{W}_n)}=n+3\ , \qquad D^{(\mathbb{W}_n^{\pm})}=2n+\frac{7}{2}
\end{equation}
by table \eqref{tab:chargessym2}. This result shows that Eq. \eqref{eq:diagonal} is satisfied as expected from superconformal symmetry.
\par To lighten the notation, we denote the renormalized superfields in the non-resonant diagonal scheme with the same symbol of the bare operators i.e. $\mathbb{W}_n$ and $\mathbb{W}_n^{\pm}$ with the renormalization factors
\begin{equation}
    \mathcal{Z}^{(\mathbb{W}_n)}(\lambda)=\left(\frac{g(\mu)}{g\left(\frac{\mu}{\lambda}\right)}\right)^{\frac{\gamma_0^{(\mathbb{W}_n)}}{\beta_0}}\ , \qquad \mathcal{Z}^{(\mathbb{W}_n^{\pm})}(\lambda)=\left(\frac{g(\mu)}{g\left(\frac{\mu}{\lambda}\right)}\right)^{\frac{\gamma_0^{(\mathbb{W}_n^{\pm})}}{\beta_0}}
\end{equation}
the asymptotic behaviour of $g\left(\frac{\mu}{\lambda}\right)$ being given in Eq. \eqref{eq:renfact} with $\beta_0=\frac{3}{(4\pi)^2}$ and $\beta_1=\frac{6}{(4\pi)^4}$ \cite{Shifman:2012zz}.
\par We now construct the generating functional of Euclidean UV asymptotic, connected correlators in this scheme. This can be done applying the formula \eqref{eq:wasymp} to the Euclidean generating functionals of subsection \ref{sec:eucgenfunctsym}. To write the generating functional we define the $2\times 2$ matrices
\begin{equation}
    \mathcal{Z}_{2n}(\lambda)=\mathcal{Z}^{(\mathbb{W}_n)}(\lambda)\mathds{1}_{2\times 2} \ , \qquad \mathcal{Z}_{4n+1}(\lambda)=\mathcal{Z}^{(\mathbb{W}_n^{\pm})}(\lambda)\mathds{1}_{2\times 2}\ , \qquad \mathcal{Z}_{4n+1}(\lambda)=0
\end{equation}
and the quantities
\begin{equation}
    D_{2n}=n+3\ , \qquad D_{4n+1}=2n+\frac{7}{2}\ , \qquad D_{4n+3}=0
\end{equation}
In this notation, the asymptotic generating functional of the connected correlators of the twist-$2$ operators in $\mathcal{N}=1$ SYM theory is simply
\begin{align}
\label{eq:rgimproved}
    &\mathcal{W}_{\text{asymp}}^E[J_{\mathbb{W}}, J_{\mathbb{W}^+},J_{\mathbb{W}^-};\lambda]= \nonumber\\
    &\quad\frac{N^2-1}{2}\mathrm{str}\ \mathrm{log}\left[\mathds{1}_{2\times 2}\delta_{n_1k_1,n_2k_2}I^E_{\mathbb{R}^{4|4}}-2\  (\mathbf{\Delta}^{-1})_{n_1k_1,n'k'}^E\ \mathcal{M}_{n'k',n_2k_2}\ \lambda^{-D_{n_2}}\ \mathcal{Z}_{n_2}(\lambda)i^{\lfloor\frac{n_2}{2}\rfloor}\frac{J_{n_2}^E}{N}\right]
\end{align}
\sloppy where we omitted the dependence from the Euclidean superspace coordinates $Z^E=(x_{\mu}^E,\theta_{\alpha}^E,\bar{\theta}_{\dot{\alpha}}^E)$ to lighten the notation and introduced the symbol
\begin{equation}
\label{eq:superid}
    I^E_{\mathbb{R}^{4|4}}=\delta^{(4)}(x_1^E-x_2^E)\delta^{(2)}(\theta_1^E-\theta_2^E)\delta^{(2)}(\bar{\theta}_1^E-\bar{\theta}_2^E)
\end{equation}
The rhs of Eq. \eqref{eq:rgimproved} inherits from the corresponding superconformal object in Eq. \eqref{eq:eucsymgenfun} the structure of the logarithm of a functional superdeterminant, and it is the main result of this work.
\par We explain once again the meaning of the symbols appearing in Eq. \eqref{eq:rgimproved}, that were first defined in subsection \ref{sec:genfunct1} and specialized to $\mathcal{N}=1$ SYM theory in the free limit in subsections \ref{sec:genfunctsym} and \ref{sec:eucgenfunctsym}. The matrix $(\mathbf{\Delta}^{-1})^E_{n_1k_1,n'k'}(Z_1,Z_2)$ is
\begin{equation}
      (\mathbf{\Delta}^{-1}_{nk,n'k'})^E(Z_1^E,Z_2^E)=\begin{pmatrix}
        0 & (\Delta^{-1}_{nk,n'k'})^E(Z_1^E,Z_2^E) \\
        (\Delta^{-1}_{nk,n'k'})^E(Z_1^E,Z_2^E) & 0
    \end{pmatrix}
\end{equation}
whose entries are the kernels in \eqref{eq:epsidelta} with $j_{\Psi}=1$
\begin{align}
        (\Delta^{-1})_{nk,n'k'}^E(Z_1^E,Z_2^E)=&-i\frac{1}{\Gamma(1+\lfloor\frac{n-k}{2}\rfloor)\Gamma(2+\lfloor\frac{n-k+1}{2}\rfloor)}\frac{1}{\Gamma(1+\lfloor\frac{k'}{2}\rfloor)\Gamma(2+\lfloor\frac{k'+1}{2}\rfloor)}(-1)^{\lfloor\frac{k'}{2}\rfloor} \nonumber\\
         & (-1)^{\lfloor\frac{k'}{2}\rfloor}\left(2^{\frac{5}{2}}(-ix_{1\bar{2}}^E)^{\bar{z}}\right)^{\lfloor\frac{n-k}{2}\rfloor+\lfloor\frac{k'}{2}\rfloor}\frac{\Gamma\left(2+\lfloor\frac{n-k+1}{2}\rfloor+\lfloor\frac{k'+1}{2}\rfloor\right)}{(-(x_{1\bar{2}}^E)^2)^{2+\lfloor\frac{n-k+1}{2}\rfloor+\lfloor\frac{k'+1}{2}\rfloor}} \nonumber\\
         &\Big[ \left(4(x_{1\bar{2}}^E\bar{\theta}_{12}^E)_1\right)^{(n-k)\mathrm{mod}\ 2}\left(4(\theta_{12}^Ex_{1\bar{2}}^E)_{\dot{1}}\right)^{k' \mathrm{mod}\ 2}+\nonumber\\
        +i& \frac{2^{\frac{5}{2}}\left[(n-k) \mathrm{mod}\ 2\right]\ \left[k'\mathrm{mod}\ 2\right]}{\lfloor\frac{n-k}{2}\rfloor+\lfloor\frac{k'+1}{2}\rfloor+1} (x_{1\bar{2}}^E)^2\left((x_{1\bar{2}}^E)_z+2\sqrt{2}(\theta_{12}^E)_1(\bar{\theta}_{12}^E)_{\dot{1}}\right) \Big]
\end{align}
The matrix $\mathcal{M}_{n'k',n_2k_2}$ is
\begin{equation}
    \mathcal{M}_{nk,n'k'}=\begin{cases}
        \delta_{nn'}\delta_{kk'}\mathds{1}_{2\times 2}\ , & n,n'\ \text{odd} \\
        \begin{pmatrix}
            (-1)^{\lfloor\frac{n}{2}\rfloor+k+1} & 0\\
            0 &  (-1)^k
        \end{pmatrix}\delta_{nn'}\delta_{kk'} \ , & n,n'\ \text{even} \\
        0\ , & \text{otherwise}
        \end{cases}
\end{equation}
The entries of the currents are $J_{n}^E(Z^E)$ are
    \begin{equation}
        J_{4n+1}^E(Z)=\begin{pmatrix}
        J_{\mathbb{W}_n^+}^E(Z^E) & 0 \\
        0 & J_{\mathbb{W}_n^-}^E(Z^E)
    \end{pmatrix}\ , \quad 
        J_{2n}^E(Z^E)=\begin{pmatrix}
        0 & \frac{J_{\mathbb{W}_n}^E(Z^E)}{2} \\
        \frac{J_{\mathbb{W}_n}^E(Z^E)}{2}  & 0 \\
    \end{pmatrix}   \ , \quad J_{4n+3}^E(Z^E)=0
 \end{equation}
Using the properties of subsection \ref{sec:correlators}, we can also write the RG-improved generators of connected correlators for the bosonic superfield twist-$2$ operators $\mathbb{W}_n$ and the fermionic superfield twist-$2$ operators $\mathbb{W}_n^{\pm}$. We find, for the bosonic superfields
\begin{align}
\label{eq:asympbos0}
    &\mathcal{W}^E_{\text{asymp}}[J_{\mathbb{W}}, J_{\mathbb{W}^+}=0,J_{\mathbb{W}^-}=0;\lambda]= \nonumber \\
    &+\frac{N^2-1}{2}\mathrm{str}\ \mathrm{log}\left[\delta_{n_1k_1,n_2k_2}I^E_{\mathbb{R}^{4|4}}-(-1)^{k_2}i^{n_2}(\Delta^{-1}_{2n_1\ k_1,2n_2\ k_2})^E\lambda^{-n-3}\mathcal{Z}^{(\mathbb{W}_{n_2})}(\lambda)\frac{J_{n_2}^E}{N}\right] \nonumber \\
    &+\frac{N^2-1}{2}\mathrm{str}\ \mathrm{log}\left[\delta_{n_1k_1,n_2k_2}I^E_{\mathbb{R}^{4|4}}-(-1)^{k_2+n_2+1}i^{n_2}(\Delta^{-1}_{2n_1\ k_1,2n_2\ k_2})^E\lambda^{-n_2-3}\mathcal{Z}^{(\mathbb{W}_{n_2})}(\lambda)\frac{J_{n_2}^E}{N}\right]
\end{align}
and for the fermionic superfields
\begin{equation}
\label{eq:asympferm0}
				\resizebox{0.98\textwidth}{!}{%
	$\begin{aligned}
    &\mathcal{W}^E_{\text{asymp}}[J_{\mathbb{W}}=0, J_{\mathbb{W}^+},J_{\mathbb{W}^-};\lambda]=  \\
    &\quad\frac{N^2-1}{2}\mathrm{str}\ \mathrm{log}\left[ \delta_{n_1k_1,n_2k_2}I^E_{\mathbb{R}^{4|4}}- 4(\Delta^{-1}_{4n_1+1\ k_1,4n'+1\ k'})^E\lambda^{-2n'-\frac{7}{2}}\mathcal{Z}^{(\mathbb{W}^-_{n'})}(\lambda)\frac{J_{\mathbb{W}_{n'}^{-}}^E}{N}(\Delta^{-1}_{4n'+1\ k',4n_2+1\ k_2})^E\lambda^{-2n_2-\frac{7}{2}}\mathcal{Z}^{(\mathbb{W}^+_{n_2})}(\lambda) \frac{J_{\mathbb{W}_{n'}^{+}}^E}{N}\right]
    				\end{aligned}$
} 
\end{equation}
We can write the generating functionals Eqs. \eqref{eq:rgimproved}, \eqref{eq:asympbos0} and \eqref{eq:asympferm0} in a slightly different way that will turn out to be useful for their nonperturbative interpretation in section \ref{sec:interpretation}. In the space of the indices $(n,k)$ on which the supertraces are taken, we define the subspaces of positive and negative $\mathbb{Z}_2$-grading as defined below Eq. \eqref{eq:supertrace}
\begin{align}
    &\mathcal{S}_+=\left\{(n,k)\in \mathbb{Z}^2 | n\in \mathbb{Z},\ k\in\{0,1,...,n\},\ \mathrm{deg}(n,k)=(-1)^{n-k}=+1\right\} \nonumber \\
    &\mathcal{S}_-=\left\{(n,k)\in{\mathbb{Z}^2} | n\in \mathbb{Z},\ k\in\{0,1,...,n\},\ \mathrm{deg}(n,k)=(-1)^{n-k}=-1\right\}
\end{align}
We define the traces $\mathrm{tr}_{\pm}$ as the traces over the subspaces $\mathcal{S}_{\pm}$. By the definition of Eq. \eqref{eq:supertrace}, the supertraces in Eqs. \eqref{eq:rgimproved}, \eqref{eq:asympbos0} and \eqref{eq:asympferm0} uniquely split into
\begin{align}
    \mathrm{str}(X)=\mathrm{tr}_{+}(X)-\mathrm{tr}_{-}(X)
\end{align}
where
\begin{align}
   & \mathrm{tr}_+(X)=\sum_{(n,k)\in\mathcal{S}_+}\int d^8Z^E \mathrm{tr}_{2\times 2}X_{nk,nk}(Z^E,Z^E) \nonumber \\
    &\mathrm{tr}_-(X)=\sum_{(n,k)\in\mathcal{S}_-}\int d^8Z^E \mathrm{tr}_{2\times 2}X_{nk,nk}(Z^E,Z^E)
\end{align}
for some graded matrix $X$ in the space of the $(n,k)$. The superspace integration measure is defined as $\int d^8Z^E=\int d^4x^E d^2\theta^E d^2\bar{\theta}^E$. With these definitions, we find
\begin{align}
    \label{eq:rgimprovetrace}
    &\mathcal{W}_{\text{asymp}}^E[J_{\mathbb{W}}, J_{\mathbb{W}^+},J_{\mathbb{W}^-};\lambda]= \nonumber\\
    &+\frac{N^2-1}{2}\mathrm{tr}_+\ \mathrm{log}\left[\mathds{1}_{2\times 2}\delta_{n_1k_1,n_2k_2}I^E_{\mathbb{R}^{4|4}}-2\  (\mathbf{\Delta}^{-1})_{n_1k_1,n'k'}^E\ \mathcal{M}_{n'k',n_2k_2}\ \lambda^{-D_{n_2}}\ \mathcal{Z}_{n_2}(\lambda)i^{\lfloor\frac{n_2}{2}\rfloor}\frac{J_{n_2}^E}{N}\right]\nonumber \\
    &-\frac{N^2-1}{2}\mathrm{tr}_-\ \mathrm{log}\left[\mathds{1}_{2\times 2}\delta_{n_1k_1,n_2k_2}I^E_{\mathbb{R}^{4|4}}-2\  (\mathbf{\Delta}^{-1})_{n_1k_1,n'k'}^E\ \mathcal{M}_{n'k',n_2k_2}\ \lambda^{-D_{n_2}}\ \mathcal{Z}_{n_2}(\lambda)i^{\lfloor\frac{n_2}{2}\rfloor}\frac{J_{n_2}^E}{N}\right]
\end{align}
and
\begin{align}
    \label{eq:asympbostr}
    &\mathcal{W}^E_{\text{asymp}}[J_{\mathbb{W}}, J_{\mathbb{W}^+}=0,J_{\mathbb{W}^-}=0;\lambda]= \nonumber \\
    &+\frac{N^2-1}{2}\mathrm{tr}_+\ \mathrm{log}\left[\delta_{n_1k_1,n_2k_2}I^E_{\mathbb{R}^{4|4}}-(-1)^{k_2}i^{n_2}(\Delta^{-1}_{2n_1\ k_1,2n_2\ k_2})^E\lambda^{-n-3}\mathcal{Z}^{(\mathbb{W}_{n_2})}(\lambda)\frac{J_{n_2}^E}{N}\right] \nonumber \\
    &-\frac{N^2-1}{2}\mathrm{tr}_-\ \mathrm{log}\left[\delta_{n_1k_1,n_2k_2}I^E_{\mathbb{R}^{4|4}}-(-1)^{k_2}i^{n_2}(\Delta^{-1}_{2n_1\ k_1,2n_2\ k_2})^E\lambda^{-n-3}\mathcal{Z}^{(\mathbb{W}_{n_2})}(\lambda)\frac{J_{n_2}^E}{N}\right] \nonumber \\
    &+\frac{N^2-1}{2}\mathrm{tr}_+\ \mathrm{log}\left[\delta_{n_1k_1,n_2k_2}I^E_{\mathbb{R}^{4|4}}-(-1)^{k_2+n_2+1}i^{n_2}(\Delta^{-1}_{2n_1\ k_1,2n_2\ k_2})^E\lambda^{-n_2-3}\mathcal{Z}^{(\mathbb{W}_{n_2})}(\lambda)\frac{J_{n_2}^E}{N}\right] \nonumber \\
    &-\frac{N^2-1}{2}\mathrm{tr}_-\ \mathrm{log}\left[\delta_{n_1k_1,n_2k_2}I^E_{\mathbb{R}^{4|4}}-(-1)^{k_2+n_2+1}i^{n_2}(\Delta^{-1}_{2n_1\ k_1,2n_2\ k_2})^E\lambda^{-n_2-3}\mathcal{Z}^{(\mathbb{W}_{n_2})}(\lambda)\frac{J_{n_2}^E}{N}\right]
\end{align}
and
\begin{equation}
    \label{eq:asympfermtr}
    				\resizebox{0.98\textwidth}{!}{%
	$\begin{aligned}
    &\mathcal{W}^E_{\text{asymp}}[J_{\mathbb{W}}=0, J_{\mathbb{W}^+},J_{\mathbb{W}^-};\lambda]=  \\
    &+\frac{N^2-1}{2}\mathrm{tr}_+\ \mathrm{log}\left[ \delta_{n_1k_1,n_2k_2}I^E_{\mathbb{R}^{4|4}}- 4(\Delta^{-1}_{4n_1+1\ k_1,4n'+1\ k'})^E\lambda^{-2n'-\frac{7}{2}}\mathcal{Z}^{(\mathbb{W}^-_{n'})}(\lambda)\frac{J_{\mathbb{W}_{n'}^{-}}^E}{N}(\Delta^{-1}_{4n'+1\ k',4n_2+1\ k_2})^E\lambda^{-2n_2-\frac{7}{2}}\mathcal{Z}^{(\mathbb{W}^+_{n_2})}(\lambda) \frac{J_{\mathbb{W}_{n'}^{+}}^E}{N}\right] \\
    &-\frac{N^2-1}{2}\mathrm{tr}_-\ \mathrm{log}\left[ \delta_{n_1k_1,n_2k_2}I^E_{\mathbb{R}^{4|4}}- 4(\Delta^{-1}_{4n_1+1\ k_1,4n'+1\ k'})^E\lambda^{-2n'-\frac{7}{2}}\mathcal{Z}^{(\mathbb{W}^-_{n'})}(\lambda)\frac{J_{\mathbb{W}_{n'}^{-}}^E}{N}(\Delta^{-1}_{4n'+1\ k',4n_2+1\ k_2})^E\lambda^{-2n_2-\frac{7}{2}}\mathcal{Z}^{(\mathbb{W}^+_{n_2})}(\lambda) \frac{J_{\mathbb{W}_{n'}^{+}}^E}{N}\right]
    				\end{aligned}$
} 
\end{equation}

\section{Nonperturbative matching}
\label{sec:interpretation}
As explained in section \ref{sec:nonperturbative}, the generating functional of the connected correlators of glueball and gluinoball composite superfields $\tilde{O}$, $\tilde{M}$ in Euclidean $\mathcal{N}=1$ SU($N$) SYM theory admits the large-$N$ expansion
\begin{equation}
\label{eq:spheretorus}
    \mathcal{W}^E[J_{\tilde{O}},J_{\tilde{M}}]=\mathcal{W}^E_{\text{sphere}}[J_{\tilde{O}},J_{\tilde{M}}]+\mathcal{W}^E_{\text{torus}}[J_{\tilde{O}},J_{\tilde{M}}]+ \cdots
\end{equation}
In section \ref{sec:sym} we specialized to the non-chiral superfields $\mathbb{W}_n$, $\mathbb{W}_n^+$ and $\mathbb{W}_n^-$ defined in \eqref{eq:ww} and computed the corresponding asymptotic generating functional in Eq. \eqref{eq:rgimprovetrace}. The corresponding objects in the decomposition furnished by Eq. \eqref{eq:spheretorus} read as follows
\begin{align}
\label{eq:torus}
    &\mathcal{W}_{\text{torus, asymp}}^E[J_{\mathbb{W}}, J_{\mathbb{W}^+},J_{\mathbb{W}^-};\lambda]= \nonumber \\
    &-\frac{1}{2}\mathrm{tr}_+\ \mathrm{log}\left[\mathds{1}_{2\times 2}\delta_{n_1k_1,n_2k_2}I^E_{\mathbb{R}^{4|4}}-2\  (\mathbf{\Delta}^{-1})_{n_1k_1,n'k'}^E\ \mathcal{M}_{n'k',n_2k_2}\ \lambda^{-D_{n_2}}\ \mathcal{Z}_{n_2}(\lambda)i^{\lfloor\frac{n_2}{2}\rfloor}\frac{J_{n_2}^E}{N}\right]\nonumber \\
    &+\frac{1}{2}\mathrm{tr}_-\ \mathrm{log}\left[\mathds{1}_{2\times 2}\delta_{n_1k_1,n_2k_2}I^E_{\mathbb{R}^{4|4}}-2\  (\mathbf{\Delta}^{-1})_{n_1k_1,n'k'}^E\ \mathcal{M}_{n'k',n_2k_2}\ \lambda^{-D_{n_2}}\ \mathcal{Z}_{n_2}(\lambda)i^{\lfloor\frac{n_2}{2}\rfloor}\frac{J_{n_2}^E}{N}\right]\nonumber \\
\end{align}
with
\begin{align}
    \mathcal{W}_{\text{sphere, asymp}}^E[J_{\mathbb{W}}, J_{\mathbb{W}^+},J_{\mathbb{W}^-};\lambda]=-N^2\mathcal{W}_{\text{torus, asymp}}^E[J_{\mathbb{W}}, J_{\mathbb{W}^+},J_{\mathbb{W}^-};\lambda]
\end{align}
We recall that the identity in Euclidean superspace $I^E_{\mathbb{R}^{4|4}}$ is defined in Eq. \eqref{eq:superid}.
\par In the nonperturbative solution of $\mathcal{N}=1$ SU($N$) SYM theory, $\mathcal{W}^E_{\text{sphere}}$ is the sum of glueball and gluinoball tree diagram, while $\mathcal{W}^E_{\text{torus}}$ is the sum of glueball and gluinoball one-loop diagrams. Because of this correspondence, $\mathcal{W}^E_{\text{torus}}$ can be read nonperturbatively from Eq. \eqref{eq:glueballW}
\begin{align}
    &\mathcal{W}^E_{\text{glueball/gluinoball, one-loop}}[J_{\widetilde{\Phi}},J_{\widetilde{\Psi}}] = \nonumber \\
    &\quad \frac{1}{2}\text{str}\log
    \begin{pmatrix}\ast'_2(-\Delta+M^2)+\frac{1}{N}\ast'_3\widetilde{\Phi}_J\ast'_3& \frac{1}{N}\ast'_3\ast'_3\widetilde{\Psi}_J\\ 
        \frac{1}{N}\ast'_3\ast'_3\widetilde{\Psi}_J&\ast_2(-\Delta+M^2)+\frac{1}{N}\ast_3\widetilde{\Phi}_J\ast_3 \end{pmatrix}
\end{align}
Therefore, Eq. \eqref{eq:torus} proves that, in the short-distance limit, the generating functional $\mathcal{W}_{\text{torus}}$ does have the structure predicted by the large-$N$ arguments of section \ref{sec:nonperturbative}. Indeed, because of asymptotic freedom, it should hold for some choice of the interpolating superfields $\widetilde{\Phi},\ \widetilde{\Psi}$
\begin{align}
    \mathcal{W}_{\text{torus, asymp}}^E[J_{\mathbb{W}}, J_{\mathbb{W}^+},J_{\mathbb{W}^-};\lambda] \underset{\lambda \to 0}{\sim} \mathcal{W}^E_{\text{glueball/gluinoball, one-loop}}[J_{\widetilde{\Phi}},J_{\widetilde{\Psi}};\lambda] 
\end{align}
where $\mathcal{W}^E_{\text{glueball/gluinoball, one-loop}}[J_{\widetilde{\Phi}},J_{\widetilde{\Psi}};\lambda]$ is $\mathcal{W}^E_{\text{glueball/gluinoball, one-loop}}[J_{\widetilde{\Phi}},J_{\widetilde{\Psi}}]$ with the coordinates rescaled by a factor $\lambda$. More specifically, we verify the following structures
\begin{align}
\label{eq:asympbos}
    \mathcal{W}^E_{\text{torus, asymp}}[J_{\mathbb{W}}, J_{\mathbb{W}^+}=0,J_{\mathbb{W}^-}=0;\lambda]\underset{\lambda \to 0}{\sim}  \mathcal{W}^E_{\text{glueball/gluinoball, one-loop}}[J_{\widetilde{\Phi}},J_{\widetilde{\Psi}}=0;\lambda] 
\end{align}
and
\begin{align}
\label{eq:asympferm}
    \mathcal{W}^E_{\text{torus, asymp}}[J_{\mathbb{W}}=0, J_{\mathbb{W}^+},J_{\mathbb{W}^-};\lambda] \underset{\lambda \to 0}{\sim}   \mathcal{W}^E_{\text{glueball/gluinoball, one-loop}}[J_{\widetilde{\Phi}}=0,J_{\widetilde{\Psi}};\lambda] 
\end{align}
where
\begin{align}
\label{eq:philogdet}
	\mathcal{W}^E_{\text{glueball/gluinoball 1-loop }}[J_{\widetilde{\Phi}},0] 
	=&+\frac{1}{2}\text{tr}\log
	\left(\mathcal{I}+(\ast'_2(-\Delta+M^2))^{-1}\frac{1}{N}\ast'_3\widetilde{\Phi}_J\ast'_3\right)\nonumber\\
	&-\frac{1}{2}\text{tr}\log
	\left(\mathcal{I}+(\ast_2(-\Delta+M^2))^{-1}\frac{1}{N}\ast_3\widetilde{\Phi}_J\ast_3\right)
\end{align}
and
\begin{align}
	&\mathcal{W}^E_{\text{torus, asymp}}[J_{\mathbb{W}}, J_{\mathbb{W}^+}=0,J_{\mathbb{W}^-}=0;\lambda]= \nonumber \\
	&\quad -\frac{1}{2}\mathrm{tr}_+\ \mathrm{log}\left[\delta_{n_1k_1,n_2k_2}I^E_{\mathbb{R}^{4|4}}-(-1)^{k_2}i^{n_2}(\Delta^{-1}_{2n_1\ k_1,2n_2\ k_2})^E\lambda^{-n-3}\mathcal{Z}^{(\mathbb{W}_{n_2})}(\lambda)\frac{J_{n_2}^E}{N}\right] \nonumber \\
    &\quad +\frac{1}{2}\mathrm{tr}_-\ \mathrm{log}\left[\delta_{n_1k_1,n_2k_2}I^E_{\mathbb{R}^{4|4}}-(-1)^{k_2}i^{n_2}(\Delta^{-1}_{2n_1\ k_1,2n_2\ k_2})^E\lambda^{-n-3}\mathcal{Z}^{(\mathbb{W}_{n_2})}(\lambda)\frac{J_{n_2}^E}{N}\right] \nonumber \\
	&\quad-\frac{1}{2}\mathrm{tr}_+\ \mathrm{log}\left[\delta_{n_1k_1,n_2k_2}I^E_{\mathbb{R}^{4|4}}-(-1)^{k_2+n_2+1}i^{n_2}(\Delta^{-1}_{2n_1\ k_1,2n_2\ k_2})^E\lambda^{-n_2-3}\mathcal{Z}^{(\mathbb{W}_{n_2})}(\lambda)\frac{J_{n_2}^E}{N}\right] \nonumber \\
    &\quad+\frac{1}{2}\mathrm{tr}_-\ \mathrm{log}\left[\delta_{n_1k_1,n_2k_2}I^E_{\mathbb{R}^{4|4}}-(-1)^{k_2+n_2+1}i^{n_2}(\Delta^{-1}_{2n_1\ k_1,2n_2\ k_2})^E\lambda^{-n_2-3}\mathcal{Z}^{(\mathbb{W}_{n_2})}(\lambda)\frac{J_{n_2}^E}{N}\right] 
\end{align}
while for the fermionic superfields
\begin{align}
\label{eq:psilogdet}
    &\mathcal{W}^E_{\text{glueball/gluinoball 1-loop }}[0,J_{\widetilde{\Psi}}]=  \nonumber \\
     &\quad-\frac{1}{2}\text{tr}\log
	\left[\mathcal{I}-\left(\ast_2(-\Delta+M^2)\right)^{-1}\frac{1}{N}\ast'_3\ast'_3\widetilde{\Psi}_J\left(\ast'_2(-\Delta+M^2)\right)^{-1}\frac{1}{N}\ast'_3\ast'_3\widetilde{\Psi}_J\right] 
\end{align}
and
\begin{equation}
	\resizebox{0.98\textwidth}{!}{%
		$\begin{aligned}
			&\mathcal{W}^E_{\text{torus, asymp}}[J_{\mathbb{W}}=0, J_{\mathbb{W}^+},J_{\mathbb{W}^-};\lambda]=  \\
			&\quad -\frac{1}{2}\mathrm{tr}_+\ \mathrm{log}\left[ \delta_{n_1k_1,n_2k_2}I^E_{\mathbb{R}^{4|4}}- 4(\Delta^{-1}_{4n_1+1\ k_1,4n'+1\ k'})^E\lambda^{-2n'-\frac{7}{2}}\mathcal{Z}^{(\mathbb{W}^-_{n'})}(\lambda)\frac{J_{\mathbb{W}_{n'}^{-}}^E}{N}(\Delta^{-1}_{4n'+1\ k',4n_2+1\ k_2})^E\lambda^{-2n_2-\frac{7}{2}}\mathcal{Z}^{(\mathbb{W}^+_{n_2})}(\lambda) \frac{J_{\mathbb{W}_{n'}^{+}}^E}{N}\right] \\
            &\quad +\frac{1}{2}\mathrm{tr}_-\ \mathrm{log}\left[ \delta_{n_1k_1,n_2k_2}I^E_{\mathbb{R}^{4|4}}- 4(\Delta^{-1}_{4n_1+1\ k_1,4n'+1\ k'})^E\lambda^{-2n'-\frac{7}{2}}\mathcal{Z}^{(\mathbb{W}^-_{n'})}(\lambda)\frac{J_{\mathbb{W}_{n'}^{-}}^E}{N}(\Delta^{-1}_{4n'+1\ k',4n_2+1\ k_2})^E\lambda^{-2n_2-\frac{7}{2}}\mathcal{Z}^{(\mathbb{W}^+_{n_2})}(\lambda) \frac{J_{\mathbb{W}_{n'}^{+}}^E}{N}\right]
		\end{aligned}$
	} 
\end{equation}
The generating functionals on the rhs of Eqs. \eqref{eq:asympbos} and \eqref{eq:asympferm} were defined in Eqs. \eqref{eq:asympbostr} and \eqref{eq:asympfermtr} respectively. Eqs. \eqref{eq:philogdet} and \eqref{eq:psilogdet} were computed from Eq. \eqref{eq:glueballW} using the definition of superdeterminant given in Eq. \eqref{eq:sdet}. Hence, the matching of the $\log \sdet$ structure of the above nonperturbative and UV-asymptotic RG-improved generating functionals of correlators of twist-$2$ operators in the large-$N$ expansion to the leading-nonplanar order sets strong qualitative and quantitative UV constraints on the yet-to-come nonperturbative solution of large-$N$ $\mathcal{N} = 1$ SYM theory and it may be an essential guide for the search of such a solution.

\section{Conclusions}
\label{sec:conclusions}
We now summarize the results of this work.
\par In section \ref{sec:nonperturbative}, the general structure of the generating functional of connected correlators of glueball and gluinoball operators in large-$N$ $\mathcal{N}=1$ SU($N$) SYM theory is discussed in the superfield formalism, following the analogous treatment in the component formalism \cite{BPS41, BPS42}. A general fundamental nonperturbative argument \cite{Bochicchio:2016toi, Bochicchio:2024gtn, BPSL} did show that its large-$N$ leading nonplanar part should have the form of the logarithm of a functional superdeterminant \cite{BPS41, BPS42}. In section \ref{sec:interpretation}, employing the superfield formalism introduced in the present paper, and following the techniques of Refs. \cite{Bochicchio:2021nup, Bochicchio:2022uat,BPS41, BPS42} developed in the component formalism, this statement is successufully verified in the short-distance limit confirming the previous result in the component formalism \cite{BPS41, BPS42}. In order to perform this test in the superfield formalism the following technical innovations (sections \ref{sec:sl21}, \ref{sec:genfreescft}, \ref{sec:appliedcorrelators}, \ref{sec:sym}, \ref{sec:rgimprove}) are necessary.
\par In section \ref{sec:sl21}, we construct gauge-invariant superfield twist-$2$ operators that in the free limit of the theory transform under irreducible representations of the collinear superconformal group. A group-theoretical recipe to construct the above superfield twist-$2$ operators from superconformal primaries in general superconformal field theories is provided. A complete classification of the infinite-dimensional irreducible representations of the superalgebra $\mathfrak{sl}(2|1)$ is given, and the direct-sum decomposition rule for tensor products of irreducible representations is worked out.
\par In section \ref{sec:genfreescft}, it is shown how to compute the generating functionals of correlators of bilinear operators made of free elementary (super)fields.
\par In section \ref{sec:appliedcorrelators}, the above formalism is applied to superfield twist-$2$ operators in $\mathcal{N}=1$ superconformal free field theories.
\par In section \ref{sec:sym}, we specialize to $\mathcal{N}=1$ SYM theory. We construct the gauge-invariant twist-$2$ operators, for the first time in a manifestly gauge-invariant and supersymmetric-covariant fashion. Using the rules of sections \ref{sec:genfreescft} and \ref{sec:appliedcorrelators}, we compute the generating functional of their connected correlators in the zero-coupling limit and demonstrate that it has the structure of the logarithm of a functional superdeterminant.
\par In section \ref{sec:rgimprove}, we extend the conformal results of Refs. \cite{Ohrndorf:1981qv, Craigie:1983fb} to $\mathcal{N}=1$ supersymmetric theories and apply it to superfield twist-$2$ operators $\mathcal{N}=1$ SU($N$) SYM theory. We prove that the superfield twist-$2$ operators are multiplicatively renormalizable at one-loop by a new argument based on our manifestly supersymmetric construction. We also compute the short-distance limit of the renormalization group-improved generating functional of connected correlators of superfield twist-$2$ operators in Euclidean $\mathcal{N}=1$ SU($N$) SYM theory, finding that it also inherits the structure of the logarithm of a functional superdeterminant from the corresponding superconformal object.
\par Finally, as alluded to at the beginning of this section, we verify in section \ref{sec:interpretation} that the UV-asymptotic generating functional computed in section \ref{sec:rgimprove} in terms of gluons and gluinos matches the structure of the logarithm of a functional superdeterminant of the corresponding nonperturbative object -- which should be asymptotic to according to asymptotic freedom -- computed in terms of glueballs and gluinoballs in section \ref{sec:nonperturbative}. This matching sets strong qualitative and quantitative UV constraints on the yet-to-come nonperturbative solution of large-$N$ $\mathcal{N} = 1$ SYM theory and it may be an essential guide for the search of such a solution.

\section*{Acknowledgements} The authors would like to thank M. Bochicchio for reading and improving the manuscript. G.S. would like to thank G. Korchemsky for useful conversations at the XIII Workshop on Geometric Correspondences of Gauge Theories (SISSA, Trieste).

\appendix

\section{Conventions}\label{app:conventions}

\subsection{Spinors}
\label{app:spinors}
The Pauli four-vectors are defined as
\begin{equation}
    \sigma^{\mu}=(\mathds{1},\sigma_i)\ , \qquad \bar\sigma^{\mu}=(\mathds{1},-\sigma_i)
\end{equation}
and the Dirac matrices as
\begin{equation}
\label{eq:dirac}
    \gamma^{\mu}=\begin{pmatrix}
        0 & \sigma^{\mu} \\
        \bar\sigma^{\mu} & 0
    \end{pmatrix}
\end{equation}
Note that, in this representation
\begin{equation}
   ( \gamma^{\mu})^T=(\gamma^0,-\gamma^1,\gamma^2,-\gamma^3)\ , \qquad (\gamma^{\mu})^*=(\gamma^0,\gamma^1,-\gamma^2,\gamma^3)
\end{equation}
From which it follows that
\begin{equation}
    \gamma^0\gamma^{\mu}\gamma^0=(\gamma^{\mu})^{\dagger}\ , \qquad C^{-1}\gamma^{\mu}C=-(\gamma^{\mu})^T
\end{equation}
where
\begin{equation}
    C=\begin{pmatrix}
        -i\sigma_2 & 0 \\
        0 & +i\sigma_2
    \end{pmatrix}\ , \qquad 
    C^{-1}=\begin{pmatrix}
        +i\sigma_2 & 0 \\
        0 & -i\sigma_2
    \end{pmatrix}
\end{equation}
is the charge conjugation matrix. One can also define a fifth Dirac matrix
\begin{equation}
    \gamma_\chi=\gamma_0\gamma_1\gamma_2\gamma_3=\begin{pmatrix}
        -i\mathds{1} & 0 \\
        0 & +i\mathds{1}
    \end{pmatrix}
\end{equation}
that anticommutes with all the other $\gamma^{\mu}$. The Dirac matrices can be used to construct the generators of the $(\frac{1}{2},0)\oplus(0,\frac{1}{2})$ representation of the Lorentz group
\begin{equation}
    \Sigma_{\mu\nu}=\frac{i}{2}[\gamma_{\mu},\gamma_{\nu}]=\begin{pmatrix}
        \frac{i}{2}(\sigma_{\mu}\bar\sigma_{\nu}-\sigma_{\nu}\bar\sigma_{\mu}) & 0 \\
        0 & \frac{i}{2}(\bar\sigma_{\mu}\sigma_{\nu}-\bar\sigma_{\nu}\sigma_{\mu})
    \end{pmatrix}
\end{equation}
This representation is, of course, reducible, and pseudounitary. We introduce also the matrices
\begin{equation}
    \sigma_{\mu\nu}= \frac{i}{4}(\sigma_{\mu}\bar\sigma_{\nu}-\sigma_{\nu}\bar\sigma_{\mu})\ , \qquad \bar\sigma_{\mu\nu}=\frac{i}{4}(\bar\sigma_{\mu}\sigma_{\nu}-\bar\sigma_{\nu}\sigma_{\mu})
\end{equation}
It is easy to see that they are self-dual and anti-self-dual respectively. If $\Lambda=\mathrm{exp}\left(-\frac{i}{2}\theta_{\mu\nu}J^{\mu\nu}\right)$ is an element of the Lorentz group and $D(\Lambda)=\mathrm{exp}\left(-\frac{i}{2}\theta_{\mu\nu}\Sigma^{\mu\nu}\right)$, then
\begin{equation}
    D^{-1}(\Lambda)=\gamma^0D^{\dagger}(\Lambda)\gamma^0
\end{equation}
A Dirac spinor $\Psi$ is an objects transforming under this representation. Given a spinor $\Psi$, one can define the adjoint spinor $\bar\Psi$ and the charge conjugated spinor $\Psi^C$ as
\begin{equation}
    \bar\Psi\equiv \Psi^{\dagger}\gamma^0\ , \qquad \Psi^C\equiv C\bar\Psi^T
\end{equation}
With these definitions, $\bar\Psi\Psi$ and $\Psi^C\Psi$ are Lorentz scalars. The spinors satisfying the condition $\Psi^C=\Psi$ are called \emph{Majorana spinors}. We now decompose a generic spinor $\Psi$ as
\begin{equation}
    \Psi=\begin{pmatrix}
        \lambda_{\alpha} \\
        \bar\chi^{\dot\alpha}
    \end{pmatrix}
\end{equation}
where $\lambda_{\alpha}$ and $\bar\chi^{\dot\alpha}$ are Weyl spinors that transform as $(\frac{1}{2},0)$ and $(0,\frac{1}{2})$ respectively. One can pass from the $(\frac{1}{2},0)$ to the $(0,\frac{1}{2})$ representation and vice versa through hermitian conjugation
\begin{equation}
    \bar\lambda=\lambda^{\dagger}\ , \qquad \chi=\bar\chi^{\dagger}
\end{equation}
or, component by component
\begin{equation}
    \bar\lambda_{\dot\alpha}=(\lambda_{\alpha})^*\ , \qquad \chi^{\alpha}=(\bar\chi^{\dot\alpha})^*
\end{equation}
With these definitions, the decomposition of the adjoint spinor in Weyl spinors looks like
\begin{equation}
    \bar\Psi=\begin{pmatrix}
    \chi^{\alpha} & \bar\lambda_{\dot\alpha}
    \end{pmatrix}
\end{equation}
We can raise and lower the indices of Weyl spinors through the operation of charge conjugation
\begin{equation}
    \Psi^C\equiv
        \begin{pmatrix}
        \chi_{\alpha} \\
        \bar\lambda^{\dot\alpha}
    \end{pmatrix}
    =
    \begin{pmatrix}
        (-i\sigma_2)_{\alpha\beta}\chi^{\beta} \\
        (+i\sigma_2)^{\dot\alpha\dot\beta}\bar\lambda_{\dot\beta}
        \end{pmatrix}
\end{equation}
The quantities $\lambda_{\alpha}\lambda^{\alpha}$ and $\bar\chi^{\dot\alpha}\bar\chi_{\dot\alpha}$ are clearly Lorentz-invariant, so the matrices that raise and lower spinor indices can be seen as a matrix in the space of Weyl spinors. These matrices charge conjugation matrices for Weyl spinors can be rewritten in covariant form as two-dimensional Levi-Civita symbols that raise and lower spinor indices
\begin{align}
\label{eq:raislow}
    &\lambda^{\alpha}=\varepsilon^{\alpha\beta}\lambda_{\beta}\ , && \lambda_{\alpha}=\varepsilon_{\alpha\beta}\lambda^{\beta} \nonumber \\
    &\bar\chi_{\dot\alpha}=\varepsilon_{\dot\alpha\dot\beta}\bar\chi^{\dot\beta}\ , && \bar\chi^{\dot\alpha}=\varepsilon^{\dot\alpha\dot\beta}\bar\chi_{\dot\beta}
\end{align}
with
\begin{equation}
    \varepsilon^{12}=-\varepsilon_{12}=\varepsilon^{\dot{1}\dot{2}}=-\varepsilon_{\dot{1}\dot{2}}=1
\end{equation}
The matrices $(\sigma^{\mu})_{\alpha\dot\alpha}$ and $(\bar{\sigma}^{\mu})^{\dot\alpha \alpha}$ allow us to express any tensor $V_{\mu_1...\mu_n}$ in spinor notation with the rules \cite{Shifman:2012zz}
\begin{align}
	\label{eq:dictonary}
        &V_{\alpha_1...\alpha_n;\dot{\alpha}_1...\dot{\alpha}_n}=V_{\mu_1...\mu_n}(\sigma^{\mu_1})_{\alpha_1\dot{\alpha}_1}...(\sigma^{\mu_n})_{\alpha_n\dot{\alpha}_n} \nonumber\\
        &V_{\mu_1...\mu_n}=\frac{1}{2^{n}}V_{\alpha_1...\alpha_n;\dot{\alpha}_1...\dot{\alpha}_n}(\bar{\sigma}_{\mu_1})^{\dot\alpha_1 \alpha_1}...(\bar{\sigma}_{\mu_n})^{\dot\alpha_n \alpha_n}
\end{align}
Hence, the most general tensor structure for a quantity $V$ transforming in the Lorentz representation $\left(\frac{\ell}{2}, \frac{\bar{\ell}}{2}\right)$ is
\begin{equation}
\label{eq:llbar}
    V_{\alpha_1...\alpha_{\ell};\dot{\alpha}_1...\dot{\alpha}_{\bar{\ell}}}
\end{equation}
In the expressions for the kernels in the subsections \ref{sec:genfunct1}, \ref{sec:genfunctsym}, \ref{sec:eucgenfunctsym}, we will often use the notation $(v\bar\theta)_{\alpha}=v_{\mu}(\sigma^{\mu})_{\alpha\dot{\alpha}}\bar{\theta}^{\dot{\alpha}}$ and $(\theta v)_{\dot\alpha}=v_{\mu}\theta^{\alpha}(\sigma^{\mu})_{\alpha\dot{\alpha}}$, where $v$ is some vector.

\subsection{Light-cone notation\label{app:notation}}

We mostly follow the notation in \cite{Braun:2003rp}. We define the Minkowskian metric as:
\begin{equation}
	(g_{\mu\nu}) = \text{diag}(+1,-1,-1,-1)
\end{equation}
The light-cone coordinates are:
\begin{equation}
\label{eq:xpm}
	x^{\pm} = \frac{x^0\pm x^3}{\sqrt{2}}=x_{\mp}
\end{equation}
The corresponding Minkowskian (squared) distance is:
\begin{equation}
	\label{mod2}
	\rvert x \rvert^2 = 2 x^+ x^- -x_{\perp}^2
\end{equation}
where
\begin{equation}
	x^2_\perp=(x^1)^2+(x^2)^2
\end{equation}
We denote the derivative with respect to $x^+$ by:
\begin{equation}
	\partial_+ = \frac{\partial}{\partial x^+} = \partial_{x^+} =\frac{\partial}{\partial x_-} = \partial_{x_-}
\end{equation}
We define the light-like vectors $n^\mu$ and $\bar{n}^\mu$
\begin{equation}
	n_\mu n^\mu = \bar{n}_\mu \bar{n}^\mu = 0 \qquad n_\mu \bar{n}^\mu = 1
\end{equation}
that can be parametrized as $(n^{\mu}) = \frac{1}{\sqrt{2}}(1,0,0,1)$ and $(\bar{n}^{\mu}) = \frac{1}{\sqrt{2}}(1,0,0,-1)$.
More broadly, we define the light-cone components of a vector $V^{\mu}$ as
\begin{equation}
	V_+=V^{\mu}n_{\mu}=\frac{V_0+V_3}{\sqrt{2}}\ , \qquad V_-=V^{\mu}\bar{n}_{\mu}=\frac{V_0-V_3}{\sqrt{2}}
\end{equation}
and the two transverse components
\begin{equation}
    V=\frac{V_1+iV_2}{\sqrt{2}}\ , \qquad \bar{V}=\frac{V_1-iV_2}{\sqrt{2}}
\end{equation}
With this notation, we can write concisely
\begin{equation}
    V_{\mu}\sigma^{\mu}=\sqrt{2}\begin{pmatrix}
        V_+ & \bar{V} \\
        V & V_-
    \end{pmatrix}
\end{equation}
By using Eq. \eqref{eq:dictonary} we see that the $+$ indices are related to the $1\dot{1}$ indices in the spinor representation.
For four-spinors, the projectors onto the light-cone are \cite{Braun:2003rp}
\begin{equation}
\begin{gathered}
    \Pi_+=\frac{1}{2}\gamma_-\gamma_+=\begin{pmatrix}
        1 & & & \\
        & 0 & & \\
        & & 0 & & \\
        & & & 1
    \end{pmatrix} \qquad
    \Pi_-=\frac{1}{2}\gamma_+\gamma_-=\begin{pmatrix}
        0 & & & \\
        & 1 & & \\
        & & 1 & & \\
        & & & 0
        \end{pmatrix}
\end{gathered}
\end{equation}
where the Dirac matrices are defined below in Eq. \eqref{eq:dirac}.

\subsection{The superconformal algebra}
\label{app:repchir}
The full superconformal algebra in the four-spinor notation is
\begin{equation}
\label{eq:supconf}
    \begin{gathered}
        [\mathbf{M}_{\mu\nu},\mathbf{M}_{\rho\sigma}]=-i(g_{\mu\rho}\mathbf{M}_{\nu\sigma}-g_{\nu\rho}\mathbf{M}_{\mu\sigma} -g_{\mu\sigma}\mathbf{M}_{\nu\rho}+g_{\nu\sigma}\mathbf{M}_{\mu\rho}) \\
        [\mathbf{M}_{\mu\nu},\mathbf{P}_{\rho}]=-i(g_{\mu\rho}\mathbf{P}_{\nu}-g_{\nu\rho}\mathbf{P}_{\mu})\ , \qquad [\mathbf{M}_{\mu\nu},\mathbf{K}_{\rho}]=-i(g_{\mu\rho}\mathbf{K}_{\nu}-g_{\nu\rho}\mathbf{K}_{\mu}) \\
        [\mathbf{P}_{\mu},\mathbf{P}_{\nu}]=[\mathbf{D},\mathbf{M}_{\mu\nu}]=[\mathbf{K}_{\mu},\mathbf{K}_{\nu}]=0 \\
        [\mathbf{D},\mathbf{P}_{\mu}]=-i\mathbf{P}_{\mu}\ , \qquad [\mathbf{D},\mathbf{K}_{\mu}]=+i\mathbf{K}_{\mu}\ , \qquad         [\mathbf{P}_{\mu}, \mathbf{K}_{\nu}]=2i(g_{\mu\nu}\mathbf{D}-\mathbf{M}_{\mu\nu}) \\
        [\mathbf{M}_{\mu\nu},\mathbf{Q}]=-\frac{1}{2}\Sigma_{\mu\nu}\mathbf{Q}\ , \qquad [\mathbf{M}_{\mu\nu},\mathbf{S}]=-\frac{1}{2}\Sigma_{\mu\nu}\mathbf{S}\\
        [\mathbf{P}_{\mu}, \mathbf{Q}]=[\mathbf{K}_{\mu},\mathbf{S}]=0\ , \qquad
        [\mathbf{P}_{\mu},\mathbf{S}]=-\gamma_{\mu}\mathbf{Q}\ , \qquad [\mathbf{K}_{\mu}, \mathbf{Q}]=-\gamma_{\mu}\mathbf{S} \\
        [\mathbf{R},\mathbf{M}_{\mu\nu}]=[\mathbf{R},\mathbf{K}_{\mu}]=[\mathbf{R},\mathbf{P}_{\mu}]=[\mathbf{R},\mathbf{D}]=0 \\
        [\mathbf{D},\mathbf{Q}]=-\frac{i}{2}\mathbf{Q}\ , \qquad  [\mathbf{D},\mathbf{S}]=+\frac{i}{2}\mathbf{S}\ , \qquad[\mathbf{R},\mathbf{Q}]=-i\gamma_\chi\mathbf{Q}\ , \qquad [\mathbf{R},\mathbf{S}]=+i\gamma_\chi\mathbf{S} \\
        \{\mathbf{Q},\bar{\mathbf{Q}} \}=2\gamma^{\mu}\mathbf{P}_{\mu}\ , \qquad \{\mathbf{S},\bar{\mathbf{S}} \}=2\gamma^{\mu}\mathbf{K}_{\mu} \\
        \{\mathbf{S},\bar{\mathbf{Q}}\}=\Sigma_{\mu\nu}\mathbf{M}^{\mu\nu}+2i\mathbf{D}+3i\gamma_\chi\mathbf{R}
    \end{gathered}
\end{equation}
the Dirac matrices and the chirality matrix $\gamma_{\chi}$ are defined in appendix \ref{app:spinors}. The four-spinors $\mathbf{Q}$ and $\mathbf{S}$ satisfy the Majorana condition. The representations of the conformal algebra on superspace coordinates are found through the induced representations technique \cite{Sohnius:1985qm}. In a field theory, we define a local operators $\Phi(0)$ with support in the origin, satisfying
\begin{equation}
\begin{gathered}
    [\Phi(0),\mathbf{M}_{\mu\nu}]= \mathcal{S}_{\mu\nu}\Phi(0)\ , \qquad 
        [\Phi(0),\mathbf{D}]= i D \Phi(0)\ , \qquad
        [\Phi(0),\mathbf{R}]= r \Phi(0)
\end{gathered}
\end{equation}
under the stability subgroup of the origin. We then define
\begin{equation}
    \Phi(x,\theta,\bar\theta)\equiv e^{-i(x^{\mu}\mathbf{P}_{\mu}+\theta^{\alpha}\mathbf{Q}_{\alpha}+\bar{\mathbf{Q}}_{\dot{\alpha}}\bar{\theta}^{\dot{\alpha}})}\Phi(0) e^{+i(x^{\mu}\mathbf{P}_{\mu}+\theta^{\alpha}\mathbf{Q}_{\alpha}+\bar{\mathbf{Q}}_{\dot{\alpha}}\bar{\theta}^{\dot{\alpha}})}
\end{equation}
where $x^{\mu}$ is a vectorial even coordinate, and $\theta^{\alpha}$, $\bar{\theta}^{\dot{\alpha}}$ are spinorial odd coordinates. The action $G$ of some superconformal generator $\mathbf{G}$ on these coordinates is defined as
\begin{equation}
    G\ \Phi(x,\theta,\bar\theta)=\left[\Phi(x,\theta,\bar\theta), \mathbf{G}\right\}
\end{equation}
where $[\cdot, \cdot\}$ denotes the anticommutator when both arguments are fermionic and the commutator otherwise. We will not need the representations of the full superconformal algebra, that is found in Ref. \cite{AIHPA_1977__27_4_425_0} with slightly different conventions. In our conventions, the generators of supersymmetry are represented by
\begin{align}
\label{eq:susy}
    &P_{\mu}= i\partial_{\mu}  \nonumber \\
    &Q_{\alpha}= i\frac{\partial}{\partial\theta^{\alpha}}-(\sigma^{\mu}\bar\theta)_{\alpha}\partial_{\mu} \nonumber \\
    &\bar{Q}_{\dot\alpha}= -i\frac{\partial}{\partial\bar\theta^{\dot\alpha}}+(\theta\sigma^{\mu})_{\dot\alpha}\partial_{\mu} \nonumber \\
    &M_{\mu\nu}= i(x_{\mu}\partial_{\nu}-x_{\nu}\partial_{\mu})+\frac{1}{2}(\theta\sigma_{\mu\nu})^{\alpha}\frac{\partial}{\partial\theta^{\alpha}}-\frac{1}{2}(\bar\theta\bar\sigma_{\mu\nu})^{\dot\alpha}\frac{\partial}{\partial\bar\theta^{\dot\alpha}}+\mathcal{S}_{\mu\nu}
\end{align}
This group action on the coordinates arises from a left group action of the generators on the group element $g(x,\theta,\bar{\theta})=e^{+i(x\cdot\mathbf{P}+\theta\cdot\mathbf{Q}+\bar{\mathbf{Q}}\cdot\bar{\theta})}$. The right group action of the supertranslations allows to define the chiral covariant derivatives $D_{\alpha}$, $\bar{D}_{\dot\alpha}$
\begin{align}
\label{eq:dalpha}
    &D_{\alpha}\Phi(x,\theta,\bar\theta)=g(x,\theta,\bar\theta)\left[\Phi(0), -i\mathbf{Q}_{\alpha}\right\}g^{-1}(x,\theta,\bar\theta)\ , && D_{\alpha}= +\frac{\partial}{\partial\theta^{\alpha}}-i(\sigma^{\mu}\bar\theta)_{\alpha}\partial_{\mu} \nonumber \\
    &\bar{D}_{\dot\alpha}\Phi(x,\theta,\bar\theta)=g(x,\theta,\bar\theta)\left[\Phi(0), -i\bar{\mathbf{Q}}_{\dot\alpha}\right\}g^{-1}(x,\theta,\bar\theta)\ , && \bar{D}_{\dot\alpha} =-\frac{\partial}{\partial\bar\theta^{\dot\alpha}}+ i(\theta\sigma^{\mu})_{\dot\alpha}\partial_{\mu}
\end{align}
The chirality conditions $D_{\alpha}\bar\Phi=0$ and $\bar{D}_{\dot\alpha}\Phi=0$ are compatible with the action of the superconformal algebra only if
\begin{align}
    &D_{\alpha}\bar{\Phi}=0\  \implies\ r=+\frac{2}{3}D\ , && \mathcal{S}^{\mu\nu}\ \text{anti-selfdual} \nonumber \\
    &\bar{D}_{\dot\alpha}\Phi=0\ \implies \  r=-\frac{2}{3}D\ , && \mathcal{S}^{\mu\nu}\ \text{selfdual}
\end{align}
When this conditions are satisfied, the dependence of chiral fields $\Phi$ and $\bar{\Phi}$ on the coordinates is constrained to be
\begin{align}
\label{eq:supchir}
    &\bar{D}_{\dot\alpha}\Phi=0  \implies \Phi=\Phi(x_L,\theta)\ , &&x_{L}^{\mu}=x^{\mu}-i\theta\sigma^{\mu}\bar{\theta} \nonumber \\
    &D_{\alpha}\bar\Phi=0  \implies \bar\Phi=\bar\Phi(x_R,\bar\theta)\ , &&x_{R}^{\mu}=x^{\mu}+i\theta\sigma^{\mu}\bar{\theta}
\end{align}
This kind of chirality is stronger than that of super-Poincaré-invariant theories, in which the only conditions that chiral fields $\Phi$ and $\bar{\Phi}$ are required to satisfy are $D_{\alpha}\bar\Phi=0$ and $\bar{D}_{\dot\alpha}\Phi=0$. Furthermore, the definition of chirality given in Eq. \eqref{eq:chirality} translated in the field theory language is
\begin{equation}
    \bar{D}_{\dot{1}}\Phi=0\ , \qquad D_1\bar{\Phi}=0
\end{equation}
which is weaker than \eqref{eq:supchir} because the vanishing of the chiral fields under the action of $\bar{D}_{\dot{2}}$ and $D_2$ is not required. Also, the condition $j\pm b=0$ is implied by \eqref{eq:supchir}, but the converse is not true. We conclude that the chirality condition in the superconformal sense is stronger than the chirality condition in the collinear superconformal sense, which is stronger than the chirality condition in the super-Poincaré sense. The existence of this hierarchy is not a surprise, since the algebras with which the chirality conditions are required to be compatible are different. However, it is common in literature to use the word "chirality" with no further specification to denote any of these three notions (see e.g. Ref. \cite{Belitsky:2006cp}). In sections \ref{sec:genfreescft}, \ref{sec:appliedcorrelators} and \ref{sec:sym}, the elementary fields considered are assumed to be chiral in the strongest sense.

\section{Euclidean superspace}
\label{app:euclidean}

\subsection{Wick rotation in ordinary spacetime}
\label{sec:wickordinary}
We define the Euclidean metric
\begin{equation}
    (\delta_{\mu\nu})=\mathrm{diag}(+1,+1,+1,+1)
\end{equation}
The Wick rotation for the positions ad momenta is defined as
\begin{align}
\label{eq:wick}
   &  x^0=x_0\to -ix_{4}^E\ , \qquad  && x^i=-x_i\to x_{i}^E \nonumber \\
   &  p_0=p^0\to +ip_{4}^E\ , \qquad && p_i=-p^i \to p_{i}^E
\end{align}
With these choices we have
\begin{equation}
\label{eq:pxwick}
\begin{gathered}
    p\cdot x = p_{\mu}x^{\mu}\to p^E\cdot x^E \\
    x^2\to -(x^E)^2\ , \qquad p^2\to -(p^E)^2
\end{gathered}
\end{equation}
The transformation \eqref{eq:pxwick} acts on the light-cone coordinates \eqref{eq:xpm} as
\begin{align}
\label{eq:wicklc}
    & x^+=\frac{x^0+x^3}{\sqrt{2}}\to -i\frac{x_4^E+ix_3^E}{\sqrt{2}}=-i(x^E)^z=-ix^E_{\bar{z}} \nonumber \\
    & x^-=\frac{x^0-x^3}{\sqrt{2}}\to -i\frac{x_4^E-ix_3^E}{\sqrt{2}}=-i(x^E)^{\bar{z}}=-ix^E_{z}
\end{align}
The Euclidean Dirac matrices are defined by the relations
\begin{equation}
\label{eq:egamma}
    \gamma^{0}=\gamma_{4}^E\ , \qquad \gamma^i=i\gamma_{i}^E
\end{equation}
We also define the Wick rotation for the gluon field
\begin{equation}
\label{eq:gluonwick}
    A_0=A^0\to +iA_{4}^E\ , \qquad A_i=-A^i\to A_{i}^E
\end{equation}
while spinor fields are Wick-rotated trivially
\begin{equation}
\label{eq:quarkwick}
    \Psi \to \Psi^E\ , \qquad \bar{\Psi}\to \bar{\Psi}^E
\end{equation}
We recall that after the Wick rotation the two Dirac spinors $\Psi^E$ and $\bar{\Psi}^E\gamma_4$ are not related by hermitian conjugation. Let us consider the QCD action in Minkowski space
\begin{equation}
    S_M=\int d^4x\left[-\frac{N}{2g^2}\mathrm{Tr}\ F_{\mu\nu}F^{\mu\nu}+\bar{\Psi} i \gamma^{\mu}(\partial_{\mu}-iA_{\mu})\Psi\right]
\end{equation}
Our Wick rotation acts on this object as
\begin{equation}
        iS_M\to -S_E 
    \end{equation}
with
\begin{equation}
        S_E=\int d^4x^E\left[+\frac{N}{2g^2}\mathrm{Tr}\ F_{\mu\nu}^EF^{E\mu\nu}+\bar{\Psi}^E(\gamma^E)^{\mu}(\partial_{\mu}^E-iA_{\mu}^E)\Psi^E \right]
\end{equation}
Contrarily to Ref. \cite{Shifman:2012zz}, our conventions allow to pass from the Minkowski to the Euclidean action without any redefinition of the spinor fields.

\subsection{Euclidean spinors}
From the definition \eqref{eq:egamma} it follows that the Euclidean Pauli four-vectors are
\begin{align}
\label{eq:epauli}
    &\sigma_{\mu}^E=(-i\sigma_i, 1)\ , &&   \bar{\sigma}_{\mu}^E=(+i\sigma_i, 1)
\end{align}
The new generators of the $(\frac{1}{2},0)$ and $(0,\frac{1}{2})$ representations of the Euclidean Lorentz group i.e. $SO(4)$ are
\begin{align}
    &\sigma_{\mu\nu}^E=\frac{i}{4}\left(\sigma_{\mu}^E\bar{\sigma}_{\nu}^E-\sigma_{\nu}^E\bar{\sigma}_{\mu}^E\right) \nonumber \\
    &\bar{\sigma}_{\mu\nu}^E=\frac{i}{4}\left(\bar{\sigma}_{\mu}^E\sigma_{\nu}^E-\bar{\sigma}_{\nu}^E\sigma_{\mu}^E\right)
\end{align}
We define two finite left-handed and right-handed $SO(4)$ spinor rotations as
\begin{equation}
    L=\mathrm{exp}\left(-\frac{i}{2}\theta^{\mu\nu}\sigma_{\mu\nu}^E\right)\ , \qquad R=\mathrm{exp}\left(-\frac{i}{2}\theta^{\mu\nu}\bar{\sigma}_{\mu\nu}^E\right)
\end{equation}
Inspection of these formulae reveals the following properties
\begin{align}
    & (\sigma_{\mu\nu}^E)^*=-\sigma_2\sigma_{\mu\nu}^E\sigma_2=(\sigma_{\mu\nu}^E)^T \nonumber \\
    & (\bar{\sigma}_{\mu\nu}^E)^*=-\sigma_2\bar{\sigma}_{\mu\nu}^E\sigma_2=(\bar{\sigma}_{\mu\nu}^E)^T
\end{align}
which for finite $SO(4)$ rotations translate into
\begin{equation}
\label{eq:conjugation}
    L^*=\sigma_2 L\sigma_2=(L^{-1})^T\ , \qquad R^*=\sigma_2 R\sigma_2=(R^{-1})^T
\end{equation}
Let us consider a left-handed and a right-handed Euclidean spinors $\lambda_{\alpha}$ and $\chi^{\dot{\alpha}}$ (the right-handed spinors has not been denoted with a bar for reasons that will be clear soon) transforming as
\begin{equation}
    \lambda_{\alpha}\longmapsto {L_{\alpha}}^{\beta}\lambda_{\beta}\ , \qquad \chi^{\dot{\alpha}}\longmapsto {R^{\dot{\alpha}}}_{\dot{\beta}}\chi^{\dot{\beta}}
\end{equation}
From the relations \eqref{eq:conjugation} it follows that the complex conjugated spinors transform under the dual representations of $(\frac{1}{2},0)$ and $(0,\frac{1}{2})$
\begin{align}
    & (\lambda_{\alpha})^*=\bar{\lambda}^{\alpha} \longmapsto \bar{\lambda}^{\beta}{(L^{-1})_{\beta}}^{\alpha} \nonumber \\
    & (\chi^{\dot{\alpha}})^*=\bar{\chi}_{\dot{\alpha}}\longmapsto \bar{\chi}^{\dot{\beta}}{(L^{-1})_{\dot{\beta}}}^{\dot{\alpha}}
\end{align}
The relations \eqref{eq:conjugation} also tell us that the spinor indices can be raised and lowered with the exactly the same rules of appendix \ref{app:conventions}. It may be tempting to identify
\begin{equation}
    \lambda^{\alpha}=\bar{\lambda}^{\alpha}\ , \qquad \chi^{\dot{\alpha}}=\bar{\chi}^{\dot{\alpha}}
\end{equation}
but these Majorana conditions would prevent the Euclidean spinors to carry any $U(1)$ charge. Because of this, we will consider them to be independent complex variables.
\par The correct way to relate spinors belonging to representations of opposite chirality is by the Osterwalder-Schrader (OS) conjugation \cite{Osterwalder:1972vwp, Osterwalder:1973zr, vanNieuwenhuizen:1996tv, Mountain:1999tt} which is defined as the product of a Euclidean time reversal and a hermitian conjugation
\begin{align}
\label{eq:osconjugation}
    & \lambda^{\alpha} \overset{\text{OS}}{\longleftrightarrow} \bar{\chi}_{\dot{\alpha}} \nonumber \\
    & \chi^{\dot{\alpha}} \overset{\text{OS}}{\longleftrightarrow} \bar{\lambda}_{\alpha} 
\end{align}

\subsection{Euclidean superspace}
From the previous discussion it is evident that a Euclidean superspace must have at least four independent complex Grassmann coordinates \cite{Lukierski:1982hr, Morris:1985hi}, namely
\begin{equation}
    \theta_{\alpha}^E\ , \quad (\bar{\theta}^E)^{\alpha}\ , \quad (\theta^E)^{\dot{\alpha}}\ , \quad \bar{\theta}_{\dot{\alpha}}^E
\end{equation}
Spinor indices can be raised and lowered as in Eq. \eqref{eq:raislow}. The pairs related by OS conjugation are
\begin{align}
    & (\theta^E)^{\alpha} \overset{\text{OS}}{\longleftrightarrow} \bar{\theta}_{ \dot{\alpha}}^E  \nonumber \\
    & (\theta^E)^{\dot{\alpha}} \overset{\text{OS}}{\longleftrightarrow} \bar{\theta}_{ \alpha}^E 
\end{align}
To construct a $\mathcal{N}=1$ Euclidean superalgebra, we can define the OS-self-conjugate subspaces
\begin{align}
    &\mathcal{S}^+=(x_{\mu}^E, \theta_{\alpha}^E, \bar{\theta}_{\dot{\alpha}}^E) \nonumber \\
    &\mathcal{S}^-=(x_{\mu}^E, \bar{\theta}_{\alpha}^E, \theta_{\dot{\alpha}}^E) 
\end{align}
From now on, we will focus on the subspace $\mathcal{S}^+$, but analogous results follow for $\mathcal{S}^-$ too.
\par We can define a $\mathcal{N}=1$ Euclidean supersymmetry algebra acting on $\mathcal{S}^+$. In two-spinor notation, the commutation rules of the algebra are
\begin{align}
\label{eq:esusyalg}
        & \left[\mathbf{M}_{\mu\nu}^E,\mathbf{M}_{\rho\sigma}^E\right]=-i(\delta_{\mu\rho}\mathbf{M}_{\nu\sigma}^E-\delta_{\nu\rho}\mathbf{M}_{\mu\sigma}^E -\delta_{\mu\sigma}\mathbf{M}_{\nu\rho}^E+\delta_{\nu\sigma}\mathbf{M}_{\mu\rho}^E) \nonumber\\
        & \left[\mathbf{M}_{\mu\nu}^E,\mathbf{P}_{\rho}^E\right]=-i(\delta_{\mu\rho}^E\mathbf{P}_{\nu}^E-\delta_{\nu\rho}\mathbf{P}_{\mu}^E) \nonumber\\
        & \left[\mathbf{Q}_{\alpha}^E,\mathbf{M}_{\mu\nu}^E\right]= {(\sigma_{\mu\nu}^E)_{\alpha}}^{\beta}\mathbf{Q}_{\beta}^E  \nonumber\\
        & \left[(\bar{\mathbf{Q}}^E)^{\dot\alpha},\mathbf{M}_{\mu\nu}^E\right]={(\bar{\sigma}_{\mu\nu}^E)^{\dot\alpha}}_{\dot\beta}\bar{\mathbf{Q}}_{E}^{\dot\beta}  \nonumber\\
        & \left[\mathbf{P}_{\mu}^E,\mathbf{P}_{\nu}^E\right]=\left\{\mathbf{Q}_{\alpha}^E,\mathbf{Q}_{\beta}^E\right\}=\left\{\bar{\mathbf{Q}}_{\dot{\alpha}}^E,\bar{\mathbf{Q}}_{\dot{\beta}}^E\right\}=0 \nonumber\\
        & \left\{\mathbf{Q}_{\alpha}^E, \bar{\mathbf{Q}}_{\dot{\alpha}}^E \right\}=2i(\sigma^{E \mu})_{\alpha\dot{\alpha}}\mathbf{P}_{\mu}^E
\end{align}
One can also define Euclidean superfields depending only on the coordinates in $\mathcal{S}^+$. This latter condition is usually referred in literature as \emph{Grassmann analiticity}. An example of Grassmann-analytic superfield is the Euclidean vector multiplet
\begin{equation}
    V^E(x^E,\theta^E,\bar{\theta}^E)
\end{equation}
Contrarily to its Minkowskian counterpart, this superfield is not real, but OS-self-conjugate, and becomes real only after analytic continuation to Minkowski space. The action of the algebra \eqref{eq:esusyalg} on $\mathcal{S}^+$ and the Euclidean superfields defined thereof can be found by means of the method of induced representations as in appendix \ref{app:conventions}. We quote only the results for the generators
\begin{align}
\label{eq:esusy}
        P_{\mu}^E=& i\partial_{\mu}^E \nonumber\\
        Q_{\alpha}^E=& i\frac{\partial}{\partial\theta^{E\alpha}}-i(\sigma^{E\mu}\bar\theta^E)_{\alpha}\partial_{\mu}^E \nonumber\\
        \bar{Q}_{\dot\alpha}^E=& -i\frac{\partial}{\partial\bar\theta^{E\dot\alpha}}+i(\theta^E\sigma^{E\mu})_{\dot\alpha}\partial_{\mu}^E \nonumber\\
        M_{\mu\nu}^E=& i(x_{\mu}^E\partial_{\nu}^E-x_{\nu}^E\partial_{\mu}^E)+\frac{1}{2}(\theta^E\sigma_{\mu\nu}^E)^{\alpha}\frac{\partial}{\partial\theta^{E\alpha}}-\frac{1}{2}(\bar\theta^E\bar\sigma_{\mu\nu}^E)^{\dot\alpha}\frac{\partial}{\partial\bar\theta^{E\dot\alpha}}+\mathcal{S}_{\mu\nu}^E
\end{align}
and the Euclidean chiral covariant derivatives
\begin{align}
    & D_{\alpha}^E= +\frac{\partial}{\partial\theta^{E\alpha}}+(\sigma^{E\mu}\bar\theta^E)_{\alpha}\partial_{\mu}^E \nonumber \\
    & \bar{D}_{\dot\alpha}^E =-\frac{\partial}{\partial\bar\theta^{E\dot\alpha}}-(\theta^E\sigma^{E\mu})_{\dot\alpha}\partial_{\mu}^E
\end{align}
In analogy to their Minkowskian counterparts, Euclidean chiral covariant derivatives can be employed to define Euclidean chiral superfields through the conditions
\begin{align}
\label{eq:esupchir}
    &\bar{D}_{\dot\alpha}^E\Phi^E=0  \implies \Phi^E=\Phi^E(x_L^E,\theta^E)\ , \qquad (x_{L}^E)^{\mu}=(x^E)^{\mu}+\theta^E(\sigma^E)^{\mu}\bar{\theta}^E \nonumber \\
    &D_{\alpha}^E\bar\Phi^E=0  \implies \bar\Phi^E=\bar\Phi^E(x_R^E,\bar\theta^E)\ , \qquad (x_{R}^{E})^{\mu}=(x^E)^{\mu}-\theta^E(\sigma^E)^{\mu}\bar{\theta}^E
\end{align}
Some examples of Euclidean chiral superfields are the quark superfields $Q^E$, $\bar{Q}^E$ and the spinorial field strength $W_{\alpha}^E$, $\bar{W}_{\dot{\alpha}}^E$. In Euclidean superspace, these pairs of chiral superfields are no more hermitian conjugates, but rather OS-conjugates.

\subsection{Wick rotation in superspace}
The Wick rotation in superspace maps Minkowski superspace into $\mathcal{S}^+$. The even coordinates and the momenta are mapped as in Eq. \eqref{eq:wick}. The odd coordinates are rotated as
\begin{equation}
    \theta_{\alpha}\longrightarrow \theta_{\alpha}^E\ , \qquad \bar\theta_{\dot\alpha}\longrightarrow \bar\theta_{\dot\alpha}^E
\end{equation}
Whenever they occur, the Pauli four-vectors must be expressed in terms of their Euclidean counterparts defined in Eq. \eqref{eq:epauli}
\begin{align}
   & \sigma^0=\sigma_{4}^E\ , \qquad \sigma^i=i\sigma_{i}^E \nonumber \\
   & \bar{\sigma}^0=\bar{\sigma}_{4}^E\ , \qquad \bar{\sigma}^i=i\bar{\sigma}_{i}^E
\end{align}
The chiral left-handed and right-handed coordinates in Eq. \eqref{eq:supchir} transform into those of Eq. \eqref{eq:esupchir} as
\begin{align}
    & x_L^{0}\longrightarrow -i(x_{L}^E)_4\ , \qquad x_L^i\longrightarrow (x_{L}^E)_i  \nonumber \\
    & x_R^{0}\longrightarrow -i(x_{R}^E)_4\ , \qquad x_R^i\longrightarrow (x_{R}^E)_i
\end{align}
The supertranslation-invariant interval in Eq. \eqref{eq:interval} transforms as
\begin{align}
    & x_{1\bar{2}}^0=-x_{2\bar{1}}^0\longrightarrow -i(x_{1\bar{2}}^E)_4=+i(x_{\bar{1}2}^E)_4 \nonumber  \\
    & x_{1\bar{2}}^i=-x_{2\bar{1}}^i\longrightarrow (x_{1\bar{2}}^E)_i=-(x_{\bar{1}2}^E)_i
\end{align}
where we introduced the Euclidean supertranslation invariant interval
\begin{equation}
    x_{1\bar{2}}^E=-x_{\bar{1}2}^E=x_{12}^E+\theta_{1}^E\sigma^E\bar{\theta}_{1}^E+\theta_{2}^E\sigma^E\bar{\theta}_{2}^E-2\theta_{1}^E\sigma^E\bar{\theta}_{2}^E
\end{equation}
The light-cone components of these coordinates and their scalar products are rotated according to the same rules in ordinary space in Eqs. \eqref{eq:wicklc} and \eqref{eq:wicklc}. The integration measure over superspace is rotated as
\begin{equation}
    \int d^8Z=\int d^4xd^2\theta d^2\bar{\theta}\longrightarrow -i\int d^4x^Ed^2\theta^E d^2\bar{\theta}^E =-i\int d^8Z^E
\end{equation}
Consequently, delta functions are rotated as
\begin{equation}
\label{eq:edelta}
    \delta^{(8)}(Z_1,Z_2) \longrightarrow +i\delta^{(8)}(Z_1^E,Z_2^E)
\end{equation}
Although in the present work we do not need it in the present work, we show the Euclidean lagrangian of $\mathcal{N}=1$ SYM theory
\begin{align}
\label{eq:elagrangian}
    \mathcal{L}_E=-&\left(\frac{N}{2g^2}\mathrm{Tr}\int d^2\theta^E (W^E)^2+\text{OS.c.}\right)
\end{align}
where "OS.c." denotes the Osterwalder-Schrader conjugation.
This lagrangian can be obtained by performing the Wick rotations described in this subsection on the superfields and on the superspace integration measure, and transforming their physical components according to Eqs. \eqref{eq:gluonwick} and \eqref{eq:quarkwick}. The details on this procedure are described in Refs. \cite{Lukierski:1982hr, Morris:1985hi}.

\section{Two-point correlators}
\label{app:supconfcorr}

\subsection{Solution of the superconformal Ward identities}
The superconformal Ward identities severely constrain the structure of Minkowskian correlators. Let us consider two conjugate operators $\mathcal{O}_{\alpha_1...\alpha_{\ell};\dot{\alpha}_1...\dot{\alpha}_{\bar{\ell}}}$, $\bar{\mathcal{O}}_{\beta_1...\beta_{\bar{\ell}};\dot{\beta}_1...\dot{\beta_{\ell}}}$ of dimension $D$, transforming under the representations $\left(\frac{\ell}{2},\frac{\bar{\ell}}{2}\right)$ and $\left(\frac{\bar{\ell}}{2},\frac{\ell}{2}\right)$ of the Lorentz group, and with $R$-charges $\pm r$ respectively. Their components with maximal spin projections along the light-cone will be denoted as $\mathcal{O}_+$, $\bar{\mathcal{O}}_+$. Their spin projection and heliticy are
\begin{equation}
    s=\frac{\ell+\bar{\ell}}{2}\ , \qquad h=\frac{\bar{\ell}-\ell}{2}
\end{equation}
which can be substituted into the definitions in Eq. \eqref{eq:twist}. We use the following notation for the supertranslation-invariant intervals
\begin{equation}
\label{eq:interval}
\begin{gathered}
    x_{1\bar{2}}=-x_{\bar{1}2}=x_{12}-i\theta_1\sigma\bar{\theta}_1-i\theta_2\sigma\bar{\theta}_2+2i\theta_1\sigma\bar{\theta}_2 \\
    \theta_{12}=\theta_1-\theta_2\ , \qquad \bar{\theta}_{12}=\bar{\theta}_1- \bar{\theta}_2
\end{gathered}
\end{equation}
The Euclidean two-point correlators can be found by applying the rules of appendix \ref{app:euclidean}.

\paragraph{Position space.} According to Ref. \cite{Li:2014gpa}, the two-point correlator of $\mathcal{O}$ and $\bar{\mathcal{O}}$ is
\begin{equation}
\label{eq:oolc}
    \expval{\mathcal{O}_{\alpha_1...\alpha_{\ell};\dot{\alpha}_1...\dot{\alpha}_{\bar{\ell}}}\bar{\mathcal{O}}_{\beta_1...\beta_{\bar{\ell}};\dot{\beta}_1...\dot{\beta}_{\ell}}}=2^{-s}C_{\mathcal{O}}\frac{{x_{1\bar{2}}}_{(\alpha_1\dot{\beta}_1}...{x_{1\bar{2}}}_{\alpha_\ell)\dot{\beta}_{\ell}}{x_{\bar{1}2}}_{(\beta_1\dot{\alpha}_1}...{x_{\bar{1}2}}_{\beta_{\bar{\ell}})\dot{\alpha}_{\bar{\ell}}}}{(x_{1\bar{2}}^2)^{j-b}(x_{\bar{1}2}^2)^{j+b}}
\end{equation}
where the brackets denote the symmetrization with respect to the undotted indices only. The factor $2^{-s}$ in front of the normalization constant has been inserted for convenience. Projecting all the Lorentz indices on the light-cone, we obtain
\begin{equation}
    \expval{\mathcal{O}_+\bar{\mathcal{O}}_+}=C_{\mathcal{O}}\frac{(x_{1\bar{2}}^-)^{\ell}(x_{\bar{1}2}^-)^{\bar{\ell}}}{(x_{1\bar{2}}^2)^{j-b}(x_{\bar{1}2}^2)^{j+b}}
\end{equation}
Putting the coordinates on the light-cone \eqref{eq:lc} one finds
\begin{equation}
    \expval{\mathcal{O}_+\bar{\mathcal{O}}_+}\rvert_{\text{l.c.}}=2^{-2j}C_{\mathcal{O}}\frac{1}{(x_{12}^-)^{\tau}}\frac{1}{(x_{1\bar{2}}^+)^{j-b}(x_{\bar{1}2}^+)^{j+b}}
\end{equation}
At $\theta_{1,2}^{\alpha}=\bar{\theta}^{\dot{\alpha}}=0$, this two-point function takes the values
\begin{equation}
\label{eq:twoptconf}
    \expval{\mathcal{O}_+\bar{\mathcal{O}}_+}\rvert_{\substack{\text{l.c.} \\ \theta=0} }=2^{-s}C_{\mathcal{O}}\frac{1}{(2x_{12}^-x_{12}^+)^D}\left(\frac{x_{12}^-}{x_{12}^+}\right)^s
\end{equation}
that coincides with the conformal results of Ref. \cite{Braun:2003rp}.

\paragraph{Momentum space.} We now specialize to the two-point correlator between a left-handed and a right-handed chiral operators transforming under the representations $(\ell/2,0)$ and $(0,\ell/2)$ of the Lorentz group respectively. Neglecting the normalization, their two-point correlator in position space is
\begin{equation}
    G(x_{1\bar{2}})=\frac{(x_{1\bar{2}}^-)^{\ell}}{(x_{1\bar{2}}^2)^{2j}}=(-2)^{-\ell}\frac{\Gamma(\tau)}{\Gamma(2j)}e^{-ia\cdot\partial}\partial_+^{\ell}\frac{1}{(x^2_{12})^{\tau}}
\end{equation}
with $\ell=2s$, while
\begin{equation}
    \int d^4x_1\ d^4x_2\ G(x_{1\bar{2}})\ e^{-ip_1\cdot x_1-ip_2\cdot x_2}=(2\pi)^4\delta^{(4)}(p_1+p_2)\ \widetilde{G}(p_1)e^{a\cdot p_1}
\end{equation}
where we have defined
\begin{equation}
\label{eq:amu}
\begin{gathered}
     \widetilde{G}(p)=\int d^4x\ G(x) e^{-ip\cdot x}   \\
     a^{\mu}=\theta_1\sigma^{\mu}\bar\theta_1+\theta_2\sigma^{\mu}\bar\theta_2-2\theta_1\sigma^{\mu}\bar\theta_2
\end{gathered}
\end{equation}
and $\tau$ is the collinear superconformal twist defined in Eq. \eqref{eq:twist}. All the dependence on the odd coordinates in contained in the factor $e^{a\cdot p}$. $\widetilde{G}(p)$ can be obtained by a standard Fourier transform in Minkowski space, that, we recall, is defined as the analytic continuation of the Fourier transform of the corresponding Euclidean two-point function. Since the Fourier integral may be UV-divergent at the origin, we need to analytically continue the spacetime dimension to $d=4-2\varepsilon$. After subtracting the divergence as $\varepsilon \to 0$ we obtain
\begin{equation}
    \widetilde{G}(p)=-i\frac{(4\pi)^2}{(-4)^D\Gamma(2j)\Gamma(\tau-1)}e^{p\cdot a}(ip_+)^{\ell}(-p^2)^{\tau-2}\mathrm{log}\left(\frac{-p^2}{4\pi\mu^2}\right)
\end{equation}
After Wick-rotation, this expression turns into
\begin{equation}
    \widetilde{G}^E(p^E)=\frac{(4\pi)^2}{(-4)^D\Gamma(2j)\Gamma(\tau-1)}e^{p^E\cdot a^E}(-p_z^E)^{\ell}(({p^E})^2)^{\tau-2}\mathrm{log}\left(\frac{{(p^E})^2}{4\pi\mu^2}\right)
\end{equation}
where
\begin{equation}
    (a^E)^{\mu}=i\theta_1^E(\sigma^E)^{\mu}\bar\theta_1^E+i\theta_2^E(\sigma^E)^{\mu}\bar\theta_2^E-2i\theta_1^E(\sigma^E)^{\mu}\bar\theta_2^E
\end{equation}
The meaning of the breaking of the conformal symmetry due to the appearance of the new mass scale $\mu$ is extensively discussed in Refs. \cite{Bochicchio:2013tfa, LimadeSouza:2016hcs}.

\subsection{Superconformal inner product}
\label{app:supconfinnprod}
The operator-state correspondence holds in superconformal field theories. Given a local Euclidean superfield $\mathcal{O}^E(x^E,\theta^E)$, we can create in and out states by acting on the vacuum $\ket{0}$ as follows \cite{Buchbinder:1998qv, Gates:1983nr, Park:1997bq}
\begin{align}
\label{eq:inout}
   &  \ket{\mathcal{O}}=\mathcal{O}^E(0, 0)\ket{0} \nonumber \\
    & \bra{\mathcal{O}}=\lim\limits_{(x^E,\theta^E)\to 0}\bra{0}\textbf{I}\mathcal{O}^E(x^E,\theta^E)\textbf{I}
\end{align}
where the operator $\textbf{I}$ is the inversion operator, that is idempotent and acts on the Euclidean superconformal generators as
\begin{align}
    \textbf{I}\textbf{P}_{\mu}^E\textbf{I}=\textbf{K}_{\mu}^E\ , \qquad \textbf{I}\textbf{D}^E\textbf{I}=-\textbf{D}^E\ , \qquad \textbf{I}\textbf{M}^E_{\mu\nu}\textbf{I}=\textbf{M}^E_{\mu\nu} \nonumber \\
    \textbf{I}\textbf{R}^E\textbf{I}=-\textbf{R}^E\ , \qquad \textbf{I}\textbf{Q}^E_{\alpha}\textbf{I}=\bar{\textbf{S}}^E_{\dot{\alpha}}\ , \qquad \textbf{I}\bar{\textbf{Q}}^E_{\dot{\alpha}}\textbf{I}=\textbf{S}^E_{\alpha}
\end{align}
The action of this operator on coordinates and superfields can be found through the method of induced representation as in section \ref{sec:sl21}. From the definition \eqref{eq:inout} it immediately follows that
\begin{align}
    \bra{\mathcal{O}_{I}}\ket{\mathcal{O}_{J}}=\lim\limits_{(x^E,\theta^E)\to 0}\expval{\textbf{I}\mathcal{O}^E_I(x^E,\theta^E)\textbf{I}\mathcal{O}^E_J(0, 0)}
\end{align}
is an inner product on the Hilbert space.
\par We now see how to compute the matrix $ \bra{\mathcal{O}_{I}}\ket{\mathcal{O}_{J}}$ from the Euclidean version of the two-point functions \eqref{eq:oolc}. Thanks to superconformal symmetry, we can restrict to the anaylitic continuation to Euclidean spacetime of the light-cone coordinates
\begin{equation}
    Z^E_{\text{l.c.}}=\left(x_{\bar{z}}^E, x_z^E\right)
\end{equation}
all the remaining coordinates being $0$, including the odd ones. Let $\mathcal{O}_I^E$ be a spin $s$ Euclidean superfield, and let $\mathcal{O}_{I+}$ be its component of maximal spin along the Euclidean light-cone. From Refs. \cite{Buchbinder:1998qv, Gates:1983nr, Park:1997bq} we know that the action of the inversion operator $\textbf{I}$ on Euclidean coordinates and superfields on the light-cone is
\begin{align}
    &\textbf{I}\mathcal{O}_{I, +}(Z_{\text{l.c.}})\textbf{I}=\left(\frac{x_z^E}{x_{\bar{z}}^E}\right)^s{\left(e^{-2\Delta \mathrm{log}\sqrt{2\mu^2x_z^Ex_{\bar{z}}^E}}\right)_I}^J\bar{\mathcal{O}}_{J, +}(IZ_{\text{l.c.}}^E) \nonumber \\
    &IZ_{\text{l.c.}}=\left(Ix_{\bar{z}}^E,Ix_z^E\right)=\left(\frac{1}{x_{\bar{z}}^E},\frac{1}{x_z^E}\right)
\end{align}
where $\Delta$ is the matrix of scaling dimensions, possibly nondiagonal in logCFTs \cite{Becchetti:2021for}. We then write the generalization of Eq. \eqref{eq:twoptconf} on the Euclidean light-cone with nondiagonal $\Delta$
\begin{equation}
    \expval{\bar{\mathcal{O}}_{I+}^E(Z_{\text{l.c.}}^E)\mathcal{O}_{J+}^E(0)} =\left(\frac{x_z^E}{x_{\bar{z}}^E}\right)^s{\left(e^{-\Delta \mathrm{log}\sqrt{2\mu^2x_z^Ex_{\bar{z}}^E}}\right)_I}^{I'}\mathcal{G}_{I'J'}{\left(e^{-\Delta^T \mathrm{log}\sqrt{2\mu^2x_z^Ex_{\bar{z}}^E}}\right)^{J'}}_J
\end{equation}
where $\mathcal{G}$ is a constant matrix. This expression follows from the Callan-Symanzik equation, as shown in Ref. \cite{Becchetti:2021for}. It then follows that
\begin{align}
    &\bra{\mathcal{O}_{I+}}\ket{\mathcal{O}_{J+}} \nonumber \\
    =&\lim\limits_{2x_z^Ex_{\bar{z}}^E\to0}\expval{\textbf{I}\mathcal{O}_{I+}^E(Z_{\text{l.c.}}^E)\textbf{I}\mathcal{O}_{J+}^E(0)} \nonumber \\
    =&\lim\limits_{2x_z^Ex_{\bar{z}}^E\to0} {\left(e^{-\Delta \mathrm{log}\sqrt{2\mu^2x_z^Ex_{\bar{z}}^E}}\right)_I}^{I'}\mathcal{G}_{I'J'}{\left(e^{+\Delta^T \mathrm{log}\sqrt{2\mu^2x_z^Ex_{\bar{z}}^E}}\right)^{J'}}_J
\end{align}
From the independence of coordinates in the lhs of the above equation it follows that \cite{Becchetti:2021for}
\begin{equation}
    \Delta\mathcal{G}-\mathcal{G}\Delta^T=0
\end{equation}
that implies
\begin{equation}
    \gamma_0G_0-G_0\gamma_0^T=0
\end{equation}
according to theorem \ref{thm:bb}. Then, the derivation of the unitarity constraint in theorem \ref{thm:bb} follows step by step as in Ref. \cite{Becchetti:2021for}.

\section{Super-matrix identities}
\label{app:supermatrix}
Let $(A_a)_{ij}$ and $(B^a)^{ij}$ be two sets of supermatrices such that ${(A_aB^a)_{i}}^j$ is always even. Consider the object
\begin{equation}
    I=-\mathrm{log}\ \mathrm{det}\left({\delta_i}^j-\sum_{a=1}^N{(A_aB^a)_i}^j\right)
\end{equation}
where the determinant is taken over the $ij$ indices. Using the identity $\mathrm{log}\ \mathrm{det}(M)=\mathrm{tr} \ \mathrm{log}(M)$ and Taylor-expanding the logarithm, we find
\begin{equation}
    I=\sum_{M=1}^{\infty}\frac{1}{M}\sum_{\{a_i\}}\mathrm{tr}\left(A_{a_1}B^{a_1}...A_{a_M}B^{a_M}\right)
\end{equation}
We can use the cyclicity of the trace to displace $B^{a_M}$ on the left. However, since $B^{a_M}$ has $\mathbb{Z}_2$-grading $(-1)^{|B^{a_M}|}$ and $I$ is even, the monomials acquire a factor $(-1)^{|B^{a_M}|}$
\begin{equation}
    I=\sum_{M=1}^{\infty}\frac{1}{M}\sum_{\{a_i\}}(-1)^{|B^{a_M}|}\mathrm{tr}\left(B^{a_{M}}A_{a_1}B^{a_1}...B^{a_{M-1}}A_{a_M}\right)
\end{equation}
We can see ${(B^{a}A_{b})^i}_j$ as a matrix $B\otimes A$ with two distinct pairs of indices $ab$ and $ij$. In this way, we can write
\begin{equation}
\label{eq:tr}
    I=\sum_{M=1}^{\infty}\frac{1}{M}\mathrm{tr}\left((-1)^F(B\otimes A)^M\right)
\end{equation}
where this time the trace is taken over both the $ab$ and $ij$ indices, and $(-1)^F$ is an operator with eigenvalues $\pm 1$ defined as $[(-1)^F]_{ai,a'i'}=(-1)^{|B_a|}\delta_{ii'}\delta_{aa'}$. Because of this factor, we must resum the series as
\begin{equation}
    I=-\mathrm{str}\ \mathrm{log}\left({\delta^i}_j{\delta^{a}}_{b}-{(B^aA_b)^i}_j\right)
\end{equation}
The $\mathbb{Z}_2$-grading of the indices is assigned by hand as $\mathrm{det}(a,i)=(-1)^{|B_a|}$. One can again use the identity $\mathrm{log}\ \mathrm{sdet}(X)=\mathrm{str} \ \mathrm{log}(X)$. The 'kernel' ${(B^aA_b)^i}_j$ is, to some degree, arbitrary. Given a sequence of nonzero numbers $r_a$ it is always possible to perform the rescaling
\begin{equation}
\label{eq:rescaling}
    {(B^aA_b)^i}_j \longrightarrow \frac{r_a}{r_b}{(B^aA_b)^i}_j
\end{equation}
leaving $I$ invariant. This property follows from the expansion \eqref{eq:tr}.
\par We also report here two useful formulae. The first, is the definition of the determinant of an ordinary block matrix $M$. If the matrix has the form
\begin{equation}
    M=\begin{pmatrix} 
        A & B \\
        C & D
    \end{pmatrix}
\end{equation}
where $A$, $B$, $C$, $D$ are bosonic submatrices, its determinant is \cite{powell2011calculating}
\begin{align}
    \label{eq:block}
        & \mathrm{det}(M)=\mathrm{det}(A)\ \mathrm{det}\left(D-CA^{-1}B\right)\ , && \text{if $A$ is invertible}  \nonumber \\
        & \mathrm{det}(M)=\mathrm{det}(D)\ \mathrm{det}\left(A-BD^{-1}C\right)\ , && \text{if $D$ is invertible} \\
\end{align}
The second formula, is the definition of the superdeterminant of a graded matrix $X$. If the matrix has the form
\begin{equation}
    X=\begin{pmatrix} 
        A & B \\
        C & D
    \end{pmatrix}
\end{equation}
where $A$, $D$ are bosonic submatrices and $C$, $D$ are fermionic submatrices, its superdeterminant is \cite{ZinnJustin2021QuantumFT}
\begin{align}
\label{eq:sdet}
	&\text{sdet}(X)= \frac{\det(A)}{\det(D-CA^{-1}B)}\ ,  && \text{if $A$ is invertible} \nonumber \\
	&\text{sdet}(X)=\frac{\det(A-BD^{-1}C)}{\det(D)}\ ,  && \text{if $D$ is invertible}
\end{align}
The supertrace is defined as
\begin{align}
\label{eq:strapp}
    \mathrm{str}(X)=\mathrm{tr}(A)-\mathrm{tr}(D)
\end{align}
The superdeterminant and the supertrace satisfy the property
\begin{align}
    \mathrm{log}\ \mathrm{sdet}(X)=\mathrm{str}\ \mathrm{log}(X)
\end{align}
for any supermatrix $X$ that admits a logarithm.

\section{The $\mathfrak{sl}(2)$ algebra}
\label{app:sl2}
In this appendix, we will repeat the analysis of section \ref{sec:sl21} for the algebra $\mathfrak{sl}(2)$, which is isomorphic the collinear conformal algebra. This appendix is only pedagogical, and has the aim to show what our method of section \ref{sec:sl21} looks like when it is applied to an already well-known situation (see \cite{Braun:2003rp} and references therein).

\subsection{Generators and commutators}
The Lie algebra $\mathfrak{sl}(2)$ consists of three generators $\mathbf{L}_\pm,\ \mathbf{L}$ satisfying the commutation rules
\begin{equation}
\label{eq:commutators}
    [\mathbf{L}_+,\mathbf{L}_-]=2\mathbf{L} \ , \qquad [\mathbf{L},\mathbf{L}_\pm]=\pm \mathbf{L}_\pm
\end{equation}
This algebra has a quadratic Casimir element
\begin{equation}
\label{eq:casimir}
    \mathbb{L}^2=\mathbf{L}_+\mathbf{L}_-+\mathbf{L}^2-\mathbf{L}=\mathbf{L}_-\mathbf{L}_++\mathbf{L}^2+\mathbf{L}
\end{equation}
The Lie algebras $\mathfrak{sl}(2)$ is isomorphic $\mathfrak{su}(2)$ although the groups $SL(2)$ and $SU(2)$ are not.

\subsection{Representations}

\paragraph{Abstract construction}
We are looking for representations of $\mathfrak{sl}(2)$ with a highest weight vector $\Psi$ satisfying
\begin{equation}
    \mathbf{L}_-\Psi=0\ , \qquad \mathbf{L}\Psi=j\Psi
\end{equation}
As a consequence of this definition, in each representation the quadratic Casimir takes the value
\begin{equation}
    \mathbb{L}^2\Psi=j(j-1)\Psi
\end{equation}
which means that each representation is univocally identified with its highest weight $j$. We denote each representation as $[j]$. Descendants can be obtained by repeatedly acting with $\mathbf{L}_+$ on the highest weight vector
\begin{equation}
\label{eq:sl2desc}
    \Psi_{j,j+n}\propto \mathbf{L}_+^n\Psi_{j,j}
\end{equation}
We label each state as $\Psi_{j,m}$, where $j$ is the highest weight and $m$ is the eigenvalue of $\mathbf{L}$. From now on, we will choose the proportionality constant in Eq. \eqref{eq:sl2desc} to be $1$. This completely fixes the action of the other generators on the vectors. Note that the representations here defined cannot be unitary, as unitary representations of $\mathfrak{su}(2)\cong \mathfrak{sl}(2)$ must be finite-dimensional.

\paragraph{Representation by differential operators}
The abstract representations $[j]$ can be used to construct a representation in the space of holomorphic functions on the complex plane. The action of the algebra is defined as follows. Let $s\in\mathbb{C}$ and
\begin{equation}
    \mathcal{F}_{j}(s)=e^{-s\mathbf{L}_+}\Psi_{j,j}
\end{equation}
This object is a vector in the representation $[j]$ and can be seen as a generating function of its elements according to the rule
\begin{equation}
    \mathbf{L}_+^n\Psi_{j,j}=(-\partial_s)^n\mathcal{F}_j(s)\rvert_{s=0}
\end{equation}
The action of the generators $\mathbf{L}_{\pm},\mathbf{L}$ on $\phi_j(z)$ can be written as a differential action on $s$
\begin{align}
\label{eq:differentialsl2}
    & \mathbf{L}\mathcal{F}_j(s)=L\mathcal{F}_j(s) \qquad && L=s\partial_s+j \nonumber \\
    & \mathbf{L}_+\mathcal{F}_j(s)=L_-\mathcal{F}_j(s) \qquad && L_-=(-\partial_s) \nonumber \\
    & \mathbf{L}_-\mathcal{F}_j(s)=L_+\mathcal{F}_j(s) \qquad && L_+=s^2\partial_s+2js
\end{align}
We use the boldface letters to denote the generators acting on vectors to distinguish them from the generators acting on coordinates. The correspondence $\mathbf{L}_{\pm}\leftrightarrow L_\mp$ is needed to leave unchanged the commutation rules between the generators of the differential representation. This redefinition is also employed in Ref. \cite{Braun:2003rp}. These infinitesimal transformations integrate to
\begin{align}
    & e^{\lambda L_+}\mathcal{F}_j(s)=(1-\lambda s)^{-2j}\mathcal{F}_j\left(\frac{s}{1-\lambda s}\right) \nonumber \\
    & e^{\lambda L}\mathcal{F}_j(s)=e^{\lambda j}\mathcal{F}_j(e^{\lambda}s) \nonumber \\
    & e^{\lambda L_-}\mathcal{F}_j(s)=\mathcal{F}_j(s-\lambda)
\end{align}

\subsection{Direct sum decomposition}
\label{sec:compsl2}

\paragraph{Abstract construction}
Let us consider two representations $[j_1]$ and $[j_2]$. We want to find the expression of a vector $\Psi_{j+n,j+n}^{j_1;j_2}\in [j_1]\otimes [j_2]$ satisfying
\begin{equation}
\label{eq:sl2lowestw}
     \mathbf{L}_-\Psi_{j+n,j+n}^{j_1;j_2}=0\ , \qquad \mathbf{L}\Psi_{j+n,j+n}^{j_1;j_2}=(j_1+j_2+n)\Psi_{j+n,j+n}^{j_1;j_2}
\end{equation}
where, for brevity, we labelled $j=j_1+j_2$. In other words, $\Psi_{j+n,j+n}$ must be the highest weight vector of the $[j_1+j_2+n]\subset [j_1]\otimes [j_2]$ representation, if it exists. The second condition means that this state must be of the form
\begin{equation}
    \Psi_{j+n,j+n}^{j_1;j_2}=\sum_{k=0}^na_k \mathbf{L}_+^{k}\Psi_{j_1,j_1}\otimes \mathbf{L}_+^{n-k}\Psi_{j_2,j_2}
\end{equation}
For the moment, let us assume $j_1,j_2\geq 1$. Applying $\mathbf{L}_-$ on the left, one obtains
\begin{equation}
    \sum_{k=0}^{n-1}[-a_{k+1}(k+1)(2j_1+k)+a_k(k-n)(2j_2+n-k-1)]\ \mathbf{L}_+^{k}\Psi_{j_1,j_1}\otimes \mathbf{L}_+^{n-k-1}\Psi_{j_2,j_2}
\end{equation}
This condition is satisfied only if
\begin{equation}
\label{eq:recursion}
    \frac{a_{k+1}}{a_k}=-\frac{n-k}{k+1}\frac{2j_2+n-k-1}{2j_1+k} \implies a_k=a_0\binom{n}{k}\frac{(-1)^n}{\Gamma(2j_1+k)\Gamma(2j_2+n-k)}
\end{equation}
Which, up to an arbitrary multiplicative constant, yields the expression
\begin{align}
\label{eq:highestweight}
    \Psi_{j+n,j+n}^{j_1;j_2}=&\sum_{n_1+n_2=n}\binom{n}{n_1}\frac{(-1)^{n_1}}{\Gamma(2j_1+n_1)\Gamma(2j_2+n_2)}\mathbf{L}_+^{n_1}\Psi_{j_1,j_1}\otimes \mathbf{L}_+^{n_2}\Psi_{j_2,j_2} \nonumber \\
    = &\Psi_{j_1,j_1} \mathbb{P}^{(j_1,j_2)}_n(\overleftarrow{\mathbf{L}}_+,\overrightarrow{\mathbf{L}}_+) \Psi_{j_2,j_2}
\end{align}
where, for brevity, we introduced the symbol
\begin{equation}
    \label{eq:symbol}
    \mathbb{P}^{a,b}_n(x_1,x_2)=\sum_{n_1+n_2=n}\binom{n}{n_1}\frac{(-1)^{n_1}}{\Gamma(2a+n_1)\Gamma(2b+n_2)}x_1^{n_1}x_2^{n_2}
\end{equation}
Descendants can be obtained simply by multiplying both sides by powers of $\mathbf{L}_+$
\begin{equation}
\label{eq:descendants}
    \Psi_{j+n,j+n+k}^{j_1;j_2}=\Psi_{j_1,j_1}(\overleftarrow{\mathbf{L}}_++\overrightarrow{\mathbf{L}}_+)^{k} \ \mathbb{P}^{(j_1,j_2)}_n\left(\overleftarrow{\mathbf{L}}_+,\overrightarrow{\mathbf{L}}_+\right)\Psi_{j_2,j_2}
\end{equation} 
When at least one between $j_1$ and $j_2$ is negative, the solution of the recursion in \eqref{eq:recursion} must be constructed by taking the initial condition for $\mathrm{max}(0,-2j_1+1) \leq k \leq \mathrm{min}(n,n+2j_2-1)$. The resulting solution for the highest weight vectors have the form
\begin{equation}
    \Psi_{j+n,j+n}^{j_1;j_2}=\sum_{k=\mathrm{max}(0,-2j_1+1)}^{\mathrm{min}(n,n+2j_2-1)}\binom{n}{k}\frac{(-1)^{k}}{\Gamma(2j_1+k)\Gamma(2j_2+n-k)}\mathbf{L}_+^{k}\Psi_{j_1,j_1}\otimes \mathbf{L}_+^{n-k}\Psi_{j_2,j_2}
\end{equation}
If $j_1\leq 0$ and $j_2>0$, we have
\begin{equation}
    \Psi_{j+n,j+n}^{j_1;j_2}=\frac{\Gamma(n+1)}{\Gamma(n+2j_1)}\Psi_{j_1,j_1}(-\overleftarrow{\mathbf{L}}_+)^{1-2j_1}\mathbb{P}^{1-j_1,j_2}_{n+2j_1-1}(\overleftarrow{\mathbf{L}}_+,\overrightarrow{\mathbf{L}}_+)\Psi_{j_2,j_2}
\end{equation}
If both $j_1,j_2\leq 0$, we have
\begin{equation}
    \Psi_{j+n,j+n}^{j_1;j_2}=\frac{\Gamma(n+1)}{\Gamma(n+2j_1+2j_2-2)}\Psi_{j_1,j_1}(-\overleftarrow{\mathbf{L}}_+)^{1-2j_1}\mathbb{P}^{1-j_1,1-j_2}_{n+2j_1+2j_2-2}(\overleftarrow{\mathbf{L}}_+,\overrightarrow{\mathbf{L}}_+)(+\overrightarrow{\mathbf{L}}_+)^{1-2j_2}\Psi_{j_2,j_2}
\end{equation}
We tacitly defined $\mathbb{P}^{a,b}_n(x_1,x_2)=0$ whenever $n< 0$. We can thus state the direct sum decomposition
\begin{equation}
\label{eq:sl2comp}
    [j_1]\otimes [j_2]=\bigoplus_{n=0}^{\infty}\ [j_1+j_2+n]
\end{equation}
where the two bases are connected by the Clebsch-Gordan coefficients shown above. \\
The polynomials $\mathbb{P}_n$ are related to the Jacobi polynomials $P_n$ and the Gegenabuer polynomials $C_n$ by
\begin{subequations}
\label{eq:jacgeg}
\begin{equation}
    \mathbb{P}^{a,b}_n(x_1,x_2)=\frac{n!}{\Gamma(2a+n)\Gamma(2b+n)}(x_1+x_2)^nP_n^{(2a-1,2b-1)}\left(\frac{x_2-x_1}{x_2+x_1}\right)
\end{equation}
\begin{equation}
    \mathbb{P}^{a,a}_n(x_1,x_2)=\frac{n!\Gamma(2\alpha)}{\Gamma\left(n+\alpha+\frac{1}{2}\right)\Gamma(2\alpha+n)\Gamma\left(\alpha+\frac{1}{2}\right)}(x_1+x_2)^nC_n^{\alpha}\left(\frac{x_2-x_1}{x_2+x_1}\right)
\end{equation}
\end{subequations}
where in the second equation $2a=\alpha+\frac{1}{2}$. The properties of the Jacobi and Gegenbauer polynomials are extensively discussed in Refs. \cite{Braun:2003rp, Bochicchio:2021nup, Bochicchio:2022uat}.

\paragraph{Polynomial realization}
A shortcut for the previous results makes use of the \emph{polynomial realization} of $\mathfrak{sl}(2)$, in which the generators act as the differential operators \eqref{eq:differentialsl2} in the space of polynomials of the $s$. In a representation, say $[j]$, we denote the polynomial corresponding to the vector $\Psi_{j,j+n}$ as $\mathcal{P}_{j,j+n}(s)$. A vector belonging to a tensor product of two representations, say $\Psi_{j_1,j_1+n_1}\otimes \Psi_{j_2,j_2+n_2}\in [j_1]\otimes [j_2]$ is simply the product of the two states i.e. $\mathcal{P}_{j_1,j_1+n_1}(s_1)\mathcal{P}_{j_2,j_2+n_2}(s_2)$.
\par Let us construct the representation $[j]$ in the space of polynomials. The highest weight vector corresponds to a polynomial $\mathcal{P}_{j,j}(s)$ satisfying
\begin{equation}
   L_-\mathcal{P}_{j,j}(s)=-\partial_s\mathcal{P}_{j,j}(s)=0
\end{equation}
The only solution to this equation is the constant polynomial, which we may normalize to unity. Hence
\begin{equation}
    \mathcal{P}_{j,j}(s)=1
\end{equation}
The descendants can be obtained applying repeatedly $L_+$. We find
\begin{equation}
\label{eq:conversion}
    \mathcal{P}_{j,j+n}(s)=(2j)_ns^n
\end{equation}
where $(2j)_n\equiv \Gamma(2j+n)/\Gamma(2j)$.
\par We now use the polynomial realization to find the direct sum decomposition of the tensor product $[j_1]\otimes [j_2]$ with $j_1,j_2>0$. A polynomial $\mathcal{P}(s)$ corresponding to a primary satisfies the condition
\begin{equation}
    L_-\mathcal{P}(s_1,s_2)\equiv (L_-^{(1)}+L_-^{(2)})\mathcal{P}(s_1,s_2)=-(\partial_{s_1}+\partial_{s_2})\mathcal{P}(s_1,s_2)=0
\end{equation}
Hence, the condition $L_-\mathcal{P}(s_1,s_2)=0$ tells us that $\mathcal{P}$ must be invariant under simultaneous translations of $s_1,s_2$ i.e. can depend only on $s_1-s_2$. The only possibilities are then
\begin{equation}
    \mathcal{P}_{n}(s_1,s_2)=(s_1-s_2)^n\ , \qquad n\in\mathbb{N}
\end{equation}
The meaning of the index $n$ can be understood by applying to each $\mathcal{P}_n$ the generator
\begin{equation}
    L=L^{(1)}+L^{(2)}=s_1\partial_{s_1}+j_1+s_2\partial_{s_1}+j_2
\end{equation}
What is found is
\begin{equation}
    L\mathcal{P}_n(s_1,s_2)=(j_1+j_2+n)\mathcal{P}_n(s_1,s_2)
\end{equation}
We conclude that the polynomial $\mathcal{P}_n(s_1,s_2)$ corresponds to the primary $\Psi_{j_1+j_2+n,j_1+j_2+n}$ of the representation $[j_1+j_2+n]\subset [j_1]\otimes [j_2]$. Hence, we choose the label
\begin{equation}
    \mathcal{P}^{j_1;j_2}_{j_1+j_2+n,j_1+j_2+n}(s_1,s_2)=(s_1-s_2)^n
\end{equation}
Thanks to this technique, we proved \eqref{eq:sl2comp} with almost no effort. We now use this same technique to write the vector $\Psi_{j_1+j_2+n,j_1+j_2+n}$ in the basis $\Psi_{j_1,j_1+k}\otimes \Psi_{j_2,j_2+\ell}$. Let us expand the corresponding polynomial as
\begin{equation}
    \mathcal{P}^{j_1;j_2}_{j_1+j_2+n,j_1+j_2+n}(s_1,s_2)=\sum_{k_1+k_2=n}\binom{n}{n_1}(-1)^{n_1}s_1^{n_1}s_2^{n_2}
\end{equation}
Substituting Eq. \eqref{eq:conversion} into this expression we find
\begin{align}
\label{eq:ppp}
    &\mathcal{P}^{j_1;j_2}_{j_1+j_2+n,j_1+j_2+n}(s_1,s_2)=\nonumber\\
    &\sum_{k_1+k_2=n}\binom{n}{n_1}\frac{\Gamma(2j_1)\Gamma(2j_2)}{\Gamma(2j_1+n_1)\Gamma(2j_2+n_2)}(-1)^{n_1}\mathcal{P}_{j_1,j_1+n_1}(s_1)\mathcal{P}_{j_2,j_2+n_2}(s_2)
\end{align}
From which it follows
\begin{equation}
    \Psi_{j_1+j_2+n,j_1+j_2+n}^{j_1;j_2}=\sum_{n_1+n_2=n}\binom{n}{n_1}\frac{\Gamma(2j_1)\Gamma(2j_2)}{\Gamma(2j_1+n_1)\Gamma(2j_2+n_2)}(-1)^{n_1}\Psi_{j_1,j_1+n_1}\otimes \Psi_{j_2,j_2+n_2}
\end{equation}
which coincide to the result of Eq. \eqref{eq:highestweight} up to an irrelevant normalization factor.

\section{The polynomials $\mathbb{C}^{j_1,j_2}_n$}
\label{app:cjj}

\subsection{Example of proof}
In this appendix we show how to obtain Eqs. \eqref{eq:cjj} from Eqs. \eqref{eq:samechir++}, \eqref{eq:samechir--}, \eqref{eq:oppchir+-}, \eqref{eq:oppchir-+} using the relations \eqref{eq:uvw} and the notations \eqref{eq:condensed0}, \eqref{eq:jjbar}. Since the proofs for each case are very similar, we will only show the equivalence of Eq. \eqref{eq:oppchir+-} and Eq. \eqref{eq:cjjb}. Let us write
\begin{align}
        &\Psi^{j_1,- j_1;j_2, +j_2}_{j+n,-\bar{j};j+n,j_1+j_2+n,-\bar{j}}= \nonumber\\
        =&n!\ \Psi_1^+\mathbb{C}^{j_1,j_2}_{2n}(\overleftarrow{\mathbf{U}}_+,\overrightarrow{\mathbf{U}}_+)\Psi_2^- \nonumber\\
        =& n!\sum_{k_1+k_2=2n}\frac{(-1)^{\lfloor\frac{k_1+1-|\Psi_1|}{2}\rfloor}}{\Gamma\left(1+\lfloor\frac{k_1}{2}\rfloor\right)\Gamma\left(1+\lfloor\frac{k_2}{2}\rfloor\right)\Gamma\left(2j_1+\lfloor\frac{k_1+1}{2}\rfloor\right)\Gamma\left(2j_2+\lfloor\frac{k_2+1}{2}\rfloor\right)}\nonumber\\
        &\textbf{U}_+^{k_1} \Psi_{1}^+\otimes \textbf{U}_+^{k_2} \Psi_2^-
\end{align}
Since $k_1+k_2=2n$, $k_1$ and $k_2$ must be simultaneously even or odd. Hence, we write
\begin{align}
        &\Psi^{j_1,- j_1;j_2, +j_2}_{j+n,j+n,-\bar{j}}= \nonumber\\
        =&n!\sum_{\ell_1+\ell_2=n}\frac{(-1)^{\lfloor\frac{2\ell_1+1-|\Psi_1|}{2}\rfloor}}{\Gamma\left(1+\lfloor\frac{2\ell_1}{2}\rfloor\right)\Gamma\left(1+\lfloor\frac{2\ell_2}{2}\rfloor\right)\Gamma\left(2j_1+\lfloor\frac{2\ell_1+1}{2}\rfloor\right)\Gamma\left(2j_2+\lfloor\frac{2\ell_2+1}{2}\rfloor\right)}\nonumber\\
        &\textbf{U}_+^{2\ell_1} \Psi_1^+\otimes \textbf{U}_+^{2\ell_2} \Psi_2^-+ \nonumber\\
        +&n!\sum_{\ell_1+\ell_2=n-1}\frac{(-1)^{\lfloor\frac{2\ell_1+2-|\Psi_1|}{2}\rfloor}}{\Gamma\left(1+\lfloor\frac{2\ell_1+1}{2}\rfloor\right)\Gamma\left(1+\lfloor\frac{2\ell_2+1}{2}\rfloor\right)\Gamma\left(2j_1+\lfloor\frac{2\ell_1+2}{2}\rfloor\right)\Gamma\left(2j_2+\lfloor\frac{2\ell_2+2}{2}\rfloor\right)}\nonumber\\
        &\textbf{U}_+^{2\ell_1+1} \Psi_1^+\otimes \textbf{U}_+^{2\ell_2+1} \Psi_2^-
\end{align}
where in the first line we chose $k_1=2\ell_1$, $k_2=2\ell_2$ and in the second line we chose $k_1=2\ell_1+1$, $k_2=2\ell_2+1$. Using the properties of the floor function $\lfloor\cdot\rfloor$ we find
\begin{align}
    & (-1)^{\lfloor\frac{2\ell_1+1-|\Psi_1|}{2}\rfloor}= (-1)^{\ell_1} \qquad && (-1)^{\lfloor\frac{2\ell_1+2-|\Psi_1|}{2}\rfloor}=(-1)^{\ell_1+|\Psi_1|+1} \nonumber \\
    & \Big\lfloor\frac{2\ell_1}{2}\Big\rfloor=\Big\lfloor\frac{2\ell_1+1}{2}\Big\rfloor=\ell_1 \qquad && \Big\lfloor\frac{2\ell_2}{2}\Big\rfloor=\Big\lfloor\frac{2\ell_2+1}{2}\Big\rfloor=\ell_2 \nonumber \\
    &  \Big\lfloor\frac{2\ell_1+2}{2}\Big\rfloor=\ell_1+1 && \Big\lfloor\frac{2\ell_2+2}{2}\Big\rfloor=\ell_2+1
\end{align}
Hence, we obtain
\begin{equation}
    \begin{split}
        &\Psi^{j_1,- j_1;j_2, +j_2}_{j+n,j+n,-\bar{j}}= \\
        =&\sum_{\ell_1+\ell_2=n}\binom{n}{\ell_1}\frac{(-1)^{\ell_1}}{\Gamma\left(2j_1+\ell_1\right)\Gamma(2j_2+\ell_2)}\textbf{U}_+^{2\ell_1} \Psi_1^+\otimes \textbf{U}_+^{2\ell_2} \Psi_2^--\\
        -&(-1)^{|\Psi_1|}n\sum_{\ell_1+\ell_2=n-1}\binom{n-1}{\ell_1}\frac{(-1)^{\ell_1}}{\Gamma\left(2j_1+1+\ell_1\right)\Gamma(2j_2+1+\ell_2)}\textbf{U}_+^{2\ell_1+1} \Psi_1^+\otimes \textbf{U}_+^{2\ell_2+1} \Psi_2^-
    \end{split}
\end{equation}
From Eq. \eqref{eq:uvw} we have
    \begin{align}
    & \textbf{U}_+^{2\ell_1} \Psi_1^+=\textbf{L}_+^{\ell_1}\Psi_1^+ \nonumber \\
    & \textbf{U}_+^{2\ell_1+1} \Psi_1^+=\textbf{L}_+^{\ell_1}\textbf{V}_+\Psi_1^+ \nonumber \\
    & \textbf{U}_+^{2\ell_2} \Psi_2^-=\textbf{L}_+^{\ell_2}\Psi_2^- \nonumber \\
    &\textbf{U}_+^{2\ell_2+1} \Psi_2^-=\textbf{L}_+^{\ell_2}\textbf{W}_+\Psi_2^- 
    \end{align}
Comparing with Eq. \eqref{eq:symbol} we finally arrive to
\begin{align}
         &\Psi^{j_1,-j_1;j_2,j_2}_{j+n,j+n,-\bar{j}}=\nonumber \\
         =&\Psi_{1}^+\left[\mathbb{P}^{j_1,j_2}_n(\overleftarrow{\mathbf{L}}_+,\overrightarrow{\mathbf{L}}_+)-(-1)^{|\Psi_1|} n \overleftarrow{\mathbf{V}}_+\mathbb{P}^{j_1+\frac{1}{2},j_2+\frac{1}{2}}_{n-1}(\overleftarrow{\mathbf{L}}_+,\overrightarrow{\mathbf{L}}_+)\ \overrightarrow{\mathbf{W}}_+\right]\Psi_2^-
\end{align}
as we wanted to show.

\subsection{Symmetry properties}
\label{sec:csymm}
Let $V_1\otimes V_2$ be the tensor product of two graded vector spaces and let $\chi$ be the map
\begin{align}
        \chi: & V_1\otimes V_2 \longrightarrow V_2\otimes V_1 \nonumber \\
        & v_1\otimes v_2 \longmapsto (-1)^{|v_1||v_1|}v_2\otimes v_1
    \end{align}
that in the field theory language corresponds to the exchange of two fields inside a product. We want to determine the behavior under the action of $\chi$ on a
    \begin{equation}
        \Psi_1\mathbb{C}^{j_1,j_2}_{n}(\overleftarrow{\mathbf{U}}_+,\overrightarrow{\mathbf{U}}_+)\Psi_2
    \end{equation}
where $\Psi_1$ and $\Psi_2$ are highest weight vectors in some chiral representation. After applying $\chi$ and exchanging the indices $k_1$ and $k_2$ (defined in Eq. \eqref{eq:polynomialC}) in the resulting expression we find
\begin{equation}
    \sum_{k_1+k_2=n}\frac{(-1)^{\lfloor\frac{k_2+1-|\Psi_1|}{2}\rfloor+(k_1+|\Psi_1|)(k_2+|\Psi_2|)}}{\Gamma\left(1+\lfloor\frac{k_2}{2}\rfloor\right)\Gamma\left(1+\lfloor\frac{k_2}{2}\rfloor\right)\Gamma\left(2j_2+\lfloor\frac{k_1+1}{2}\rfloor\right)\Gamma\left(2j_1+\lfloor\frac{k_2+1}{2}\rfloor\right)}\mathbf{U}_+^{k_1}\Psi_2\otimes\mathbf{U}_+^{k_2}\Psi_1
\end{equation}
We now consider the phase factor $(-1)^{\lfloor\frac{k_2+1-|\Psi_1|}{2}\rfloor+(k_1+|\Psi_1|)(k_2+|\Psi_2|)}$. The reader can check case by case that
\begin{equation}
    (-1)^{\lfloor\frac{k_2+1-|\Psi_1|}{2}\rfloor+(k_1+|\Psi_1|)(k_2+|\Psi_2|)}=(-1)^{\lfloor\frac{n}{2}\rfloor+(n+|\Psi_1|)(n+|\Psi_2|)}(-1)^{\lfloor\frac{k_1+1-|\Psi_2|}{2}\rfloor}
\end{equation}
It follows that
\begin{equation}
\label{eq:csymmetry}
    \Psi_1\mathbb{C}^{j_1,j_2}_{n}(\overleftarrow{\mathbf{U}}_+,\overrightarrow{\mathbf{U}}_+)\Psi_2 \overset{\chi}{\longmapsto} (-1)^{\lfloor\frac{n}{2}\rfloor+(n+|\Psi_1|)(n+|\Psi_2|)}\ \Psi_2\mathbb{C}^{j_2,j_1}_{n}(\overleftarrow{\mathbf{U}}_+,\overrightarrow{\mathbf{U}}_+)\Psi_1
\end{equation}

\section{Twist-$2$ quark operators in $\mathcal{N}=1$ SQCD}
\label{app:sqcd}

In this appendix, we show how to construct twist-$2$ quark operators in $\mathcal{N}=1$ supersymmetric QCD (SQCD). We closely follow subsections \ref{sec:symintro}, \ref{sec:twistsym}, \ref{eq:symcomp}.
\par In addition to the vector superfield $V$, SQCD possesses also $N_f$ chiral scalar superfields $Q^{iI}$ in the fundamental representation, and $N_f$ chiral scalar superfields $\tilde{Q}_{iI}$. We denoted the (anti)fundamental color indices of the fields as a lowercase $^i$, and the flavor indices as an uppercase $^I$. The gauge transformation laws of the superfields are
\begin{align}
  &  Q\longmapsto e^{i\Lambda} Q\ , \qquad \bar{Q}\longmapsto \bar{Q}e^{-i\bar\Lambda}\nonumber \\
   & \tilde{Q}\longmapsto \tilde{Q}e^{-i\Lambda}\ , \qquad \bar{\tilde{Q}}\longmapsto e^{i\bar{\Lambda}}\bar{\tilde{Q}}
\end{align}
where $\Lambda$, $\bar\Lambda$ are a left-handed and a right-handed chiral Lie algebra-valued functions. The lagrangian of the theory is\footnote{For simplicity we omit any possible mass term and superpotential.}
\begin{align}
\label{eq:LagrangianSQCD}
    \mathcal{L}_{SQCD}=&\left(\frac{N}{4g^2}\int d^2\theta\ W^{a\ \alpha}W^a_{\alpha}+\text{h.c.}\right)+\sum_{I=1}^{N_f}\int d^4\theta\  \bar{Q}_Ie^{V}Q^I+\sum_{I=1}^{N_f}\int d^4\theta\  \tilde{Q}_I e^{-V}\bar{\tilde{Q}}^I 
\end{align}
where $g$ is the (real) 't Hooft coupling, with $g^2=g_{YM}^2N$. Again, we take $g$ to be real, so that theta terms are absent.
\par To express this lagrangian in ordinary spacetime, we write the component expansion of the quark superfields
\begin{align}
    & Q^{iI}(x_L,\theta)=q^{iI}(x_L)+\sqrt{2}\theta^{\alpha}\psi_{\alpha}^{iI}(x_L)+\theta^2F^{iI}(x_L) \nonumber \\
    & \tilde{Q}_{iI}(x_L,\theta)=\tilde{q}_{iI}(x_L)+\sqrt{2}\theta^{\alpha}\tilde{\psi}_{\alpha iI}(x_L)+\theta^2\tilde{F}_{iI}(x_L) \nonumber \\
    & \bar{Q}_{iI}(x_R,\bar\theta)=\bar{q}_{iI}(x_R)+\sqrt{2}\bar{\theta}_{\dot{\alpha}}\bar{\psi}^{\dot{\alpha}}_{iI}(x_R)+\bar\theta^2\bar{F}_{iI}(x_R)  \nonumber \\
    & \bar{\tilde{Q}}^{iI}(x_R,\bar\theta)=\bar{\tilde{q}}^{iI}(x_R)+\sqrt{2}\bar{\theta}_{\dot{\alpha}}\bar{\tilde{\psi}}^{\dot{\alpha}iI}(x_R)+\bar\theta^2\bar{\tilde{F}}^{iI}(x_R)
\end{align}
and insert it in the lagrangian \eqref{eq:LagrangianSQCD} together with the vector superfield in the Wess-Zumino gauge in Eq. \eqref{eq:vectorsuper}. Integrating over the odd variables $\theta^{\alpha}$, $\bar{\theta}^{\dot{\alpha}}$, and eliminating the auxiliary fields $F, \tilde{F}, \bar{F}, \bar{\tilde{F}}$ one finally obtains
\begin{align}
\label{eq:offshell}
    \mathcal{L}_{SQCD}^{\text{(Wess-Zumino)}}= & \frac{N}{g^2}\mathrm{tr}\left[-\frac{1}{2}F_{\mu\nu}F^{\mu\nu}+2i\lambda^{\alpha}\mathcal{D}_{\alpha\dot\alpha}\bar{\lambda}^{\dot{\alpha}}+D^2\right] \nonumber \\
    +&\sum_{I=1}^{N_f}\left[\mathcal{D}_{\mu}\bar{q}_I\mathcal{D}^{\mu}q^I+i\bar{\psi}_{\dot\alpha I}\mathcal{D}^{\dot\alpha \alpha}\psi_{\alpha}^I+i\sqrt{2}\bar{q}_I\lambda^{\alpha}\psi_{\alpha}^I-i\sqrt{2}\bar{\psi}_{\dot{\alpha} I}\bar{\lambda}^{\dot{\alpha}}q^I \right]\nonumber \\
    +&\sum_{I=1}^{N_f}\left[\mathcal{D}_{\mu}\tilde{q}_I\mathcal{D}^{\mu}\bar{\tilde{q}}^I+i\tilde{\psi}^{\alpha}_I\mathcal{D}_{\alpha \dot\alpha}\bar{\tilde{\psi}}^{\dot\alpha I}-i\sqrt{2}\tilde{\psi}^{\alpha}_I\lambda_{\alpha}\bar{\tilde{q}}^I+i\sqrt{2}\tilde{q}^I\bar{\lambda}_{\dot{\alpha}}\bar{\tilde{\psi}}^{\dot{\alpha} I}\right] \nonumber \\
    +&D^a\sum_{I=1}^{N_f}\left(\bar{q}_IT^aq^I-\tilde{q}_IT^a\bar{\tilde{q}}^I\right)
\end{align}
where the symbol $\mathcal{D}_{\mu}$ denotes a covariant derivative. The quark propagators are
\begin{align}
    \expval{Q^{iI}(x_{L,1},\theta_1)\bar{Q}_{jJ}(x_{R,2},\bar\theta_2)}= {\delta^i}_j{\delta^I}_JC_Q\frac{1}{(x_{1\bar{2}})^2}\nonumber \\
    \expval{\bar{\tilde{Q}}^{iI}(x_{R,1},\bar\theta_1)\tilde{Q}_{jJ}(x_{L,2},\theta_2)}= {\delta_i}^j{\delta_I}^JC_Q\frac{1}{(x_{\bar{1}2})^2}
\end{align}
where $C_Q=i/4\pi^2$.
\par The twist-$2$ gluon operators of $\mathcal{N}=1$ SQCD are the same in Eq. \eqref{eq:ww}. The quark operators are constructed from the building blocks in table \eqref{tab:chargessqcd1}.
\begin{table}[h!]
    \centering
    \begin{tabular}{C|CCCCC}
          & \ell & \bar{\ell} & j & b & \tau \\
         \hline
         Q & 0 & 0 & \frac{1}{2} & -\frac{1}{2} & 1 \\
         \bar{Q} & 0 & 0 & \frac{1}{2} & +\frac{1}{2} & 1 \\
         \tilde{Q} & 0 & 0 & \frac{1}{2} & -\frac{1}{2} & 1 \\
         \bar{\tilde{Q}} & 0 & 0 & \frac{1}{2} & +\frac{1}{2} & 1 \\
    \end{tabular}
    \caption{Collinear superconformal charges of the building blocks of the twist-$2$ quark operators of $\mathcal{N}=1$ SQCD. Their gluon counterparts are shown in table \eqref{tab:chargessym1}.}
    \label{tab:chargessqcd1}
\end{table}
We have four distinct towers of operators
\begin{align}
\label{eq:qq}
    &\mathbb{Q}_n^{\alpha}= N^{-\frac{1}{2}}C_Q^{-1}i^{n+|t^{\alpha}|} {(t^{\alpha})^{I}}_{J} \bar{Q}_I\ e^{V} \mathbb{C}^{\frac{1}{2},\frac{1}{2}}_{2n}\left(\overleftarrow{\nabla}_1+i\overleftarrow{\bar{\nabla}}_{\dot{1}}, \overrightarrow{\nabla}_1+i\overrightarrow{\bar{\nabla}}_{\dot{1}}\right)\ Q^J \nonumber\\
    &\mathbb{\tilde{Q}}_n^{\alpha}= N^{-\frac{1}{2}}C_Q^{-1}i^{n+|t^{\alpha}|}  {(t^{\alpha})^{I}}_{J} \tilde{Q}_I\ e^{-V} \mathbb{C}^{\frac{1}{2},\frac{1}{2}}_{2n}\left(\overleftarrow{\nabla}_1+i\overleftarrow{\bar{\nabla}}_{\dot{1}}, \overrightarrow{\nabla}_1+i\overrightarrow{\bar{\nabla}}_{\dot{1}}\right)\ \bar{\tilde{Q}}^J \nonumber \\
    & \mathbb{Q}_n^{\alpha+}=  N^{-\frac{1}{2}}C_Q^{-1} i^{n+|t^{\alpha}|}{(t^{\alpha})^{I}}_{J} \tilde{Q}_I \mathbb{C}^{\frac{1}{2},\frac{1}{2}}_{2n+1}\left(\overleftarrow{\nabla}_1+i\overleftarrow{\bar{\nabla}}_{\dot{1}}, \overrightarrow{\nabla}_1+i\overrightarrow{\bar{\nabla}}_{\dot{1}}\right)\ Q^J \nonumber \\
    & \mathbb{Q}_n^{\alpha-}= N^{-\frac{1}{2}}C_Q^{-1} i^{n+|t^{\alpha}|}{(t^{\alpha})^{I}}_{J} \bar{Q}_I \mathbb{C}^{\frac{1}{2},\frac{1}{2}}_{2n+1}\left(\overleftarrow{\nabla}_1+i\overleftarrow{\bar{\nabla}}_{\dot{1}}, \overrightarrow{\nabla}_1+i\overrightarrow{\bar{\nabla}}_{\dot{1}}\right)\ \bar{\tilde{Q}}^J 
\end{align}
where the $t^{\alpha}$ are a complete set of $N_f\times N_f$ matrices and the spinor covariant derivatives $\nabla_{\alpha}, \bar{\nabla}_{\dot{\alpha}}$ have been introduced in section \ref{sec:sym}. The operators satisfy the hermiticity relations
\begin{align}
    &\left(\mathbb{Q}_n^{\alpha}\right)^{\dagger} =\mathbb{Q}_n^{\alpha}\ , \qquad && \left(\mathbb{Q}_n^{\alpha+}\right)^{\dagger}=\mathbb{Q}_n^{\alpha-} \nonumber \\
    &\left(\tilde{\mathbb{Q}}_n^{\alpha}\right)^{\dagger} =\tilde{\mathbb{Q}}_n^{\alpha}\ , \qquad && \left(\tilde{\mathbb{Q}}_n^{\alpha+}\right)^{\dagger}=\tilde{\mathbb{Q}}_n^{\alpha-} 
\end{align}
The charges of the elementary quark operators are shown in table \eqref{tab:chargessqcd2}.
\begin{table}[h!]
\centering
    \begin{tabular}{C|CCCCC}
        & \ell & \bar{\ell} & j & b & \tau  \\
        \hline
        \mathbb{Q}_n & n  & n & n+1 & 0 & 2 \\
        \tilde{\mathbb{Q}}_n & n  & n & n+1 & 0 & 2 \\
        \mathbb{Q}_n^+ & n+1  & n & n+\frac{3}{2} & -\frac{1}{2} & 2 \\
        \mathbb{Q}_n^- & n  & n+1 & n+\frac{3}{2} & +\frac{1}{2} & 2 \\
    \end{tabular}
    \caption{Collinear superconformal charges of the twist-$2$ quark operators of $\mathcal{N}=1$ SQCD. Their gluon counterparts are shown in table \eqref{tab:chargessym2}.}
    \label{tab:chargessqcd2}
\end{table}
The components of the quark fields in the light-cone directions of superspace are
\begin{align}
        &Q\rvert_{\text{l.c.}}=q+\frac{2}{\varrho}\theta^1\psi \qquad && \tilde{Q}\rvert_{\text{l.c.}}=\tilde{q}+\frac{2}{\varrho}\theta^1\tilde{\psi} \nonumber\\
        &\bar{Q}\rvert_{\text{l.c.}}=\bar{q}-\frac{2}{\varrho}\bar{\theta}^{\dot{1}}\bar{\psi} \qquad  && \bar{\tilde{Q}}\rvert_{\text{l.c.}}=\bar{\tilde{q}}-\frac{2}{\varrho}\bar{\theta}^{\dot{1}}\bar{\tilde{\psi}}
\end{align}
where we used the notation
\begin{align}
    &\psi=\varrho^{-1}\psi_1 \ , \qquad \bar{\psi}=\varrho^{-1}\bar{\psi}_{\dot{1}} \nonumber \\
     &\tilde{\psi}=\varrho^{-1}\psi_1 \ , \qquad \bar{\tilde{\psi}}=\varrho^{-1}\bar{\psi}_{\dot{1}}
\end{align}
with $\varrho=2^{1/4}$.
\par We now show the component expansion of the operators \eqref{eq:qq} in the light-cone directions. We work the Wess-Zumino and light-cone gauge, and use the same notation and conventions of subsection \ref{sec:components}, omitting discrete indices. The components of $\mathbb{Q}_n$ are
\begin{align}
    \mathbb{Q}_n=N^{-\frac{1}{2}}C_Q^{-1}\frac{(-1)^ni^{|t^{\alpha}|}2^{\frac{3}{2}n}}{n!^2}\Bigg\{&\left(\bar{q}C_n^{1/2}q-\frac{2}{n+1}\bar{\psi}C_{n-1}^{3/2}\psi\right) +\frac{2}{\varrho}\theta^1  \bar{q}P_n^{(0,1)}\psi -\frac{2}{\varrho}\bar{\theta}^{\dot{1}}\ \bar{\psi}P_n^{(1,0)}q \nonumber\\
    -&\frac{\sqrt{2}}{2}\theta^1\bar{\theta}^{\dot{1}}\left(\bar{q}C_{n+1}^{1/2}q+\frac{2}{n+1}\bar{\psi}C_n^{3/2}\psi\right) \Bigg\}
\end{align}
The components of $\tilde{\mathbb{Q}}$ can be obtained from those of $\mathbb{Q}$ with the substitutions
\begin{align}
    & \theta^1\to i\bar{\theta}^{\dot{1}}\ ,  && \bar{\theta}^{\dot{1}}\to-i\theta^1 \nonumber \\
    & q\to \bar{\tilde{q}}\ , && \bar{q}\to \tilde{q} \nonumber \\
    & \psi\to i\bar{\tilde{\psi}}\ ,  &&  i\bar{\psi}\to \tilde{\psi}
\end{align}
The components of $\mathbb{Q}_n^+$ are
\begin{align}
    \mathbb{Q}_n^+=-&N^{-\frac{1}{2}}\frac{(-1)^ni^{|t^{\alpha}|}2^{\frac{3}{4}(2n+1)}}{n!(n+1)!}\Bigg\{ \left(\tilde{\psi}P_n^{(1,0)}q-\tilde{q}P_n^{(0,1)}\psi\right) -\frac{2\theta^1}{\varrho}\frac{2}{n+1}\tilde{\psi}C_n^{3/2}\psi+\frac{2\bar{\theta}^{\dot{1}}}{\varrho}\tilde{q}C_{n+1}^{1/2}q \nonumber \\
    +&2i\sqrt{2}\theta^1\bar{\theta}^{\dot{1}}\Bigg[i\frac{n+2}{2n+3}\left(\tilde{\psi}P_{n+1}^{(1,0)}q+\tilde{q}P_{n+1}^{(0,1)}\psi\right)-\frac{1}{2(2n+3)}\partial_+\left(\tilde{\psi}P_n^{(1,0)}q-\tilde{q}P_n^{(0,1)}\psi\right)\Bigg]\Bigg\}
\end{align}
with $\varrho=2^{1/4}$. The components of $\mathbb{Q}^-$ can be obtained from those of $\mathbb{Q}^+$ with the substitutions
\begin{align}
    & \theta^1\to i\bar{\theta}^{\dot{1}}\ , && \bar{\theta}^{\dot{1}}\to-i\theta^1 \nonumber \\
    & \tilde{q}\to \bar{q}  && q \to \bar{\tilde{q}} \nonumber \\
    & \tilde{\psi} \to i \bar{\psi} &&  \psi \to i\bar{\tilde{\psi}}
\end{align}

\bibliographystyle{JHEP}
\bibliography{References}

\end{document}